\documentclass[11pt]{article}
\usepackage[report]{ssl-report}
\usepackage{booktabs}
\usepackage{array}
\usepackage{longtable}
\usepackage{adjustbox}
\usepackage{environ}
\usepackage{enumitem}
\usepackage[numbers,sort&compress]{natbib}
\usepackage[colorlinks=true,linkcolor=sslblue,citecolor=sslblue,urlcolor=sslblue]{hyperref}
\usepackage{placeins}   % floats stay within their section via the barrier below
\usepackage{needspace}
\AddToHook{begindocument/end}{%
  \pretocmd{\section}{\FloatBarrier\vspace*{0pt}\Needspace{8\baselineskip}}{}{\errmessage{Cannot protect section headings}}}

\makeatletter\let\bb@table\table
\renewcommand{\table}{\setlength{\belowcaptionskip}{\medskipamount}\bb@table}\makeatother
\makeatletter\renewcommand{\@pnumwidth}{2em}\makeatother

\patchcmd{\sslmaketitle}
  {\begin{minipage}{.88\textwidth}}
  {\begin{minipage}{\textwidth}\linespread{1.0}}
  {}{\errmessage{Cannot adjust report abstract width}}

\patchcmd{\sslmaketitle}{\sslcoversheet}
  {\sslcoversheet\newgeometry{hmargin=1.15in,top=.72in,bottom=.78in,footskip=18pt}}
  {}{\errmessage{Cannot adjust title page margins}}

\patchcmd{\sslcoversheet}
  {\restoregeometry}
  {\par\vskip .6em{\scriptsize\color{black!70}%
    AI tools were used in the preparation of this manuscript.\par}%
    \restoregeometry}
  {}{\errmessage{Cannot add cover preparation disclosure}}

\newcommand{\E}{\mathbb{E}}
\newcommand{\Gam}{\operatorname{Gamma}}
\newcommand{\Normal}{\mathcal{N}}

\newcommand{\code}[1]{\texttt{#1}}
\DeclareRobustCommand{\frag}[1]{#1\unskip}
\title{Drive Failure and Lifetime Simulation in the Backblaze Drive Statistics Corpus, 2013--2026}
\author{Ahmed Amer\textsuperscript{1} \and Darrell D. E. Long\textsuperscript{1,2}}
\date{Draft of 2026-10-17}
\sslreportnumber{SSL-TR-2026-001}
\sslstatus{Preprint v0.1}
\sslaffiliation{\textsuperscript{1}Santa Clara University\\\textsuperscript{2}University of California, Santa Cruz}
\sslabstract{Storage simulators estimate the chance of data loss from assumptions about drive lifetimes,
differences between drive models, and possible dependencies between failures. 
Common lifetime models include a simple exponential distribution, 
a single Weibull curve, and a fixed three-component bathtub distribution. 
We test these assumptions against Backblaze's daily records from
April 2013 to June 2026. From this 13-year trove of data, we identify 
504{,}624 data drives with a usable power-on age, drawn from 99 drive
models (after reconciling equivalent names), which amounts to 
2.0 million drive-years. The records include 31{,}881 failures: removals because a drive
physically failed or was judged likely to fail.
Accounting for age and drives already in service when observation begins, we fit models
to earlier records and test their predictions against nine successive later years. We also
forecast the drives in service at each forecast date.

We find that accounting for differences between drive models helps more than any 
fine distinctions among the leading lifetime distributions. 
We assess forecasts with a log score, which rewards
assigning greater probability to the failure counts that actually occur. Adding drive-model
differences and attributes to a two-Weibull baseline
improves this score about 4.3~times as much as replacing a single Weibull with two competing
Weibulls while retaining the same drive-model treatment. 

Fleet failure risk falls in early life and rises at older ages, giving the fitted age curve
the familiar bathtub shape.\footnote{The usual intuition is that drives with latent defects
would tend to fail early. Those that remain in service without failing early could therefore
be assumed to face a period of lower failure risk before wear-out becomes more important
as they reach older ages.} The operational failure records do not identify the underlying
physical mechanisms. Two competing Weibull hazard components describe this pattern well,
but several alternatives predict similarly. The fixed bathtub benchmark
predicts 2.4 times the observed failures. Allowing unusually good or bad years matters too:
on 169 well-observed combinations of drive model and year, selected using earlier records,
forecast ranges intended to contain 95\% of annual failure counts contained only 44\% when
annual variation was omitted, but 92\% when we allowed both fleet-wide changes and unusually
good or bad years for individual drive models.
Useful average failure forecasts can nonetheless substantially understate uncertainty.

For an existing fleet, current condition adds information beyond model and age. A SMART
(Self-Monitoring, Analysis and Reporting Technology) warning improves failure-risk ranking:
the one-year area under the receiver operating characteristic curve (AUC) rises from 0.64
to 0.73. Gradient-boosted trees using monthly SMART records perform best on our measure of
individual failure probabilities, with AUC 0.74 at one year and 0.83 at 90 days, yet
predict about 19\% and 22\% too many failures, respectively. In other words, 
identifying individual high-risk drives, and forecasting replacement demand, 
require separate checks.

Long-term survival is harder to predict.
Drives age as calendar time advances, and newer drive models enter a changing fleet, 
making these effects difficult to separate. Different fitted
accounts of aging and calendar time imply fifteen-year survival probabilities up to
18 percentage points apart; assumptions about future calendar trends produce differences
of up to 27 points. It is therefore critical that any simulations that depend on long lifetimes 
should test their sensitivity to these assumptions. 
We supply implementations that generate simulated lifetimes,
represent shared variation, and model changes between SMART warning states, failure, and
retirement. Allowing failure risk to change from year to year is computationally inexpensive:
in our benchmark, generating a drive's lifetime takes about \frag{154}~nanoseconds
of computation for each year of simulated service.
}
\makeatletter
\sslcitation{Ahmed Amer and Darrell D. E. Long. \emph{\@title}. \ssl@series, \ssl@number.
  \sslname{}, Santa Clara University, \@date. \ssl@status. \ssl@urllink}
\makeatother

\begin{document}
% The unnumbered cover precedes the style's restart at page 1.
\hypersetup{pageanchor=false}
\sslmaketitle
\hypersetup{pageanchor=true}
\restoregeometry

% Fit manuscript tables to the report's narrower measure without editing their data.
% Install this after the title, whose author block also uses tabular.
\let\sslOriginalTabular\tabular
\let\sslOriginalEndTabular\endtabular
\RenewEnviron{tabular}[2][c]{%
  \begin{adjustbox}{max width=\linewidth}%
    \sslOriginalTabular[#1]{#2}\BODY\sslOriginalEndTabular
  \end{adjustbox}}

\newpage
\begingroup
\linespread{1.0}\selectfont
\setlength{\parskip}{.3em}
\section*{Summary of Findings}
\addcontentsline{toc}{section}{Summary of Findings}

Storage planning needs to anticipate replacement demand, identify vulnerable drives, and
assess whether several failures could overwhelm redundancy. We use Backblaze's April
2013--June 2026 records: 504{,}624 drives from 99 models and 31{,}881 failures (including
removals because failure was judged likely). 
Forecast models were tested against nine successive
later years, which also informed model development.
Our main findings can be summarized as follows:

\begin{itemize}[leftmargin=1.2em,topsep=.25em,itemsep=.25em,parsep=0pt]
\item Knowing which drive models are installed matters more than fine distinctions among the
leading lifetime models. We used a log score that rewards forecasts which assign greater
probability to the failure counts that actually occur. Adding drive-model differences and
attributes to a two-Weibull baseline
improves this score about 4.3 times as much as replacing a single Weibull with a more 
capable model of two competing Weibulls that retained the same drive-model treatment.
Meanwhile, the fixed three-component ``bathtub'' distribution commonly 
used in storage simulations predicts 2.4 times the observed failures. 
These results favor estimating rates for the actual drives being deployed
over assigning every drive a generic lifetime curve.

\item Planning for an average year is not enough. 
In well-observed drive-model groups, forecast
ranges intended to contain 95\% of annual failure counts contain only
44\unskip\% when shared annual variation is omitted.
Allowing unusually good or bad years for each drive model, together with changes in the
fleet's overall failure rate, raises coverage to
92\unskip\%. This gives a more credible account of
fluctuations in replacement demand. In other words, it's important to be able to account 
for potentially good vs bad drive-model years.

\item Greater flexibility does not guarantee more useful forecasts. Allowing each drive model
its own wear-out pattern improves the statistical score, yet the median forecasts summed
across test periods exceed observed failures by 50--73\%. The corresponding forecasts
from a common wear-out pattern with shared annual variation are 4\% below the total.
Statistical scores and replacement-demand forecasts need separate checks.

\item For drives already in service, knowing their current condition 
adds useful information. SMART warnings improve the ranking of drives by 
failure risk: one-year AUC rises from 0.64 to 0.73 (where
0.5 represents pure chance and 1 represents a perfect ranking). 
Machine learning using monthly SMART records
performs best on our measure of individual failure probabilities, but predicts about 
19\% too many failures at one year out. 
This means that identifying individual high-risk drives, and estimating how many
replacements to stock, are distinctly different tasks.

\item Long-term survival depends strongly on assumptions. Only 2\% of recorded operating time
is at ages above eight years. Models with similar three-year forecasts imply fifteen-year
survival probabilities up to 18 percentage points apart; assumptions about future changes
in fleet failure rates produce differences of up to 27 points.

\end{itemize}

We also provide implementations for generating simulated lifetimes. In our illustrative
disk-array experiment, changing the drive model changes estimated data-loss risk by four
orders of magnitude. Sharing elevated failure risks across drives can also greatly increase
array risk. These are sensitivity results: the fleet records do not establish dependence
within an array or Backblaze's actual data-loss probability. Drive selection, shared risks,
and replacement and repair policies need joint attention.
\par\endgroup

\newpage
\tableofcontents
\clearpage

% Part 1: questions, data
\section{Introduction}\label{sec:intro}

A storage operator needs enough replacement drives for routine failures and enough redundancy
to survive several failures before repairs finish. Both decisions depend on how long drives last, how risks differ between drive models,
and whether failures occur independently. We ask which assumptions about these risks are
supported by observed drive histories, and which remain uncertain.

Backblaze's Drive Stats corpus provides daily records of a fleet of several hundred thousand
drives, published since 2013. We use these records to compare commonly used lifetime models,
including a constant failure rate, a Weibull curve that allows risk to change with age, and
a fixed three-component ``bathtub'' distribution intended to describe early failures, a
relatively stable middle period, and wear-out. We test the value of distinguishing drive models, allowing rates to change over calendar
time, and using a drive's reported condition.

Throughout, a \emph{failure} means that Backblaze recorded a drive as failed or judged it
likely to fail and removed it (Section~\ref{sec:reconcile}). Drives also leave the fleet
through retirement without a failure flag. The records describe this operational outcome, not the physical mechanisms of failure;
applying the rates elsewhere requires attention to the operator's replacement policy.

Three related questions need different evaluations:
\begin{enumerate}[leftmargin=1.6em]
\item \emph{How does failure risk vary?} We estimate how the failure rate changes with age,
  drive model, and calendar time. For these comparisons, we use the amount of time drives
  actually operated in each test period and evaluate predictions of their failure counts.
  This tests the estimated rates; it does not forecast future fleet size or retirement.
\item \emph{Which drives will fail, and how many replacements will be needed?} Starting
  with the drives in service on a given date, we forecast failures over a specified period.
  We check both individual failure probabilities and the total number of predicted failures.
\item \emph{What do these assumptions imply for data-loss risk?} We generate simulated
  lifetimes, check that the implementations reproduce the intended distributions, measure
  their computational cost, and compare their consequences in an illustrative disk array.
\end{enumerate}

For the principal comparisons, we fit observations through the end of each year and test
predictions against the following year, over nine successive years. We examine whether
forecasts assign high probability to the counts that occurred, whether they over- or
under-predict totals, and whether their uncertainty ranges contain outcomes as often as
intended. We also used these years during model development, so they are not an untouched
final test. Section~\ref{sec:methods} gives the evaluation details.

Differences between drive models and shared annual variation deserve attention before
fine distinctions among the leading lifetime distributions. Current condition improves
forecasts for an existing fleet. Yet a better statistical score need not mean a better
estimate of replacement demand, and similar short-term forecasts can imply very different
long-term survival. Section~\ref{sec:practical} discusses these practical distinctions.

The contribution combines these comparisons with implementations that make the fitted
models usable in simulation. We supply R and C++ lifetime samplers, measure their costs,
and compare their implications for data loss in a redundant disk array (RAID). Several
components have precedents in studies of this corpus. In the literature we reviewed, we
did not find an analysis separating aging, calendar time, and the introduction of newer
drive models, or one combining Backblaze-fitted lifetime models with a redundancy simulation
and measured sampling costs. Section~\ref{sec:related} discusses those precedents and gaps.

After the data and evaluation methods, we compare failure-rate models and forecasts for
drives already in service, then examine long-term survival and simulation. The appendices
provide additional models, diagnostics, timing results, the disk-array example, and
reproducibility details.

\section{Related Work}\label{sec:related}

We review field studies of drive reliability, Elerath's lifetime models and the
three-component bathtub used in storage simulations, and the lifetime models that others have
fitted to this corpus.
Table~\ref{tab:priorwork} sets the closest of these studies beside this one.

\subsection{Field Studies of Drive Reliability}

Schroeder and Gibson~\cite{schroeder2007fast} studied about 100{,}000 drives at high-performance
computing and internet-service sites. Replacement rates did not settle after the first year.
They rose steadily with age, and early wear-out mattered more than infant mortality. Their
often-quoted Weibull fit has exponent $\beta$ between 0.7 and 0.8, where $\beta$ is the
power of time in the cumulative hazard. This fit describes intervals between replacements
in a system, not drive lifetimes. Pinheiro, Weber, and Barroso~\cite{pinheiro2007fast} reported
annualized failure rates in a Google fleet from 1.7\% in the first year to more than 8.6\% in
the third. Citing earlier work, they noted that failure rates are known to be highly
correlated with drive model, manufacturer and vintage, and they showed no breakdown of their own. Jiang et al.~\cite{jiang2008fast} found that failures of storage subsystems in NetApp
systems are correlated and bursty. Disk failures among them were correlated within a shelf but
much less bursty than the other failure types. Ford et al.~\cite{ford2010availability} found that
a large share of the times a storage node in Google clusters became unavailable occur in
correlated bursts, so that models assuming independence overstate availability. Disk failures
were a small part of those events. Greenan, Plank, and
Wylie~\cite{greenan2010mttm} argued against mean time to data loss from Markov models and for
simulation with non-exponential lifetimes.

\subsection{Elerath's Models and the Three-Component Bathtub}\label{sec:elerathlit}

Elerath and Pecht~\cite{elerath2007dsn,elerath2009tc} simulate RAID groups with four
distributions per drive: time to operational failure, time to restore, time to latent defect,
and time to scrub. Elerath's dissertation~\cite{elerath2007diss} uses the same model. For time
to operational failure it gives a single Weibull with characteristic life 461{,}386~h and
$\beta=1.12$, from a field population of more than 120{,}000 drives observed for up to 6{,}000 hours. A
comparison there with field data on ATA drives uses 75{,}000~h and $\beta=1.49$ (its
summary gives 1.4). A secondary
source~\cite{karmakar2015scalable} quotes the conference paper with the same 461{,}386~h and
$\beta=1.12$. For
Elerath and Schindler~\cite{elerath2014tos}, a secondary source~\cite{kishani2018human}
tabulates three drive models, each a single Weibull. In a section titled ``The Bathtub
Myth'', the dissertation also argues that drive field data do not follow the textbook bathtub.
It shows vintages of one drive model whose Weibull scales differ by a factor of about eight.
Elerath's 2007 magazine article~\cite{elerath2007queue} gives no lifetime parameters. It says
that some drive populations ``are composed of multiple failure distributions from causes that
dominate at different points in time'', that failure rates differ across suppliers and between
lots of one drive model, and that subpopulations ``should not be grouped together for
analyses''.

Simulation studies of erasure-coded storage, including our own earlier ones, have used a
three-component bathtub: competing Weibulls with exponents 0.5, 1 and 3 and scales
$2.2\times10^{7}$, $8.77\times10^{5}$ and $1.2\times10^{5}$~h. We have not found these constants
in the Elerath sources we could read, nor in secondary accounts of the others. The functional
form is the additive Weibull of Xie and Lai~\cite{xie1996additive}, which has two terms,
extended to three, and the poly-Weibull of
Berger and Sun~\cite{berger1993polyweibull}. We score it below with its constants fixed, as a
benchmark, and fit its form freely in Appendix~\ref{sec:laws}. We call it the fixed
three-component benchmark; the software keeps the legacy name \code{elerath} for it. A result
against these constants is not a test of Elerath's published models, which are single Weibulls
fitted to his own data.

\subsection{Lifetime Models Fitted to Backblaze Data}

The closest prior work is by Mittman, Lewis-Beck, and
Meeker~\cite{mittman2019hierarchical,mittman2017diss}. They used Drive Stats through the first
quarter of 2016 (63 drive models, 75{,}297 drives; 44 models with at least three failures). They
fit a left-truncated, right-censored likelihood (drives enter observation already aged and may
leave without failing; Section~\ref{sec:lik}) with the general limited failure population (GLFP)
model of Chan and Meeker~\cite{chan1999glfp}: two Weibulls, an early mode that affects a fraction
$\pi$ of drives and a wear-out mode that affects all. Their model is hierarchical across drive
models and fitted in Stan. In the model they select, the early-failure Weibull is shared, while
the defect fraction $\pi$ and the wear-out parameters vary by drive model. They compared nested
models by approximate leave-one-out cross-validation. Lewis-Beck, Tian, and
Meeker~\cite{lewisbeck2022prediction} froze the data at the first quarter of 2016, set aside the data from the second
quarter of 2016 to the first quarter of 2019, and checked one-week forecasts of failures in each
of the 26 weeks after the freeze. Before each week they updated the risk set, removing the
drives that had failed or left service. They stopped at 26 weeks because most drive models then
had few drives left. Each of their forecasts is thus one week ahead for the drives still in
service; ours are fixed-origin forecasts over 90 and 365 days for the drives in service at the
origin, a different evaluation. As an example
of a miss they give one drive model whose failures their predictions significantly
underestimated, and they attribute it to possible batch effects. A two-mode Weibull with partial pooling across drive models,
validated on later data, is therefore not new for this corpus. Ruan et al.~\cite{ruan2025subsampling} use the
ST4000DM000 records of 2016 to 2018, under the GLFP model, as a case study for subsampling heavily
censored data. Our work differs in five ways.
Both modes act on every drive (competing risks without a susceptible fraction). The drive-model
effect is a gamma-distributed hazard multiplier with covariates, whose posterior (a gamma again, by
conjugacy) and negative-multinomial predictive distribution are textbook results that we use rather than
contribute. We model calendar-time shocks (Section~\ref{sec:time}). Our data run ten years
longer. And we score a wider set of candidates, including non-parametric ones, on nine
successive held-out years. Our contribution is the analysis of the corpus, the comparative
evidence across those candidates and evaluation designs, and the combination of the components
into samplers, not any one component. We read the dissertation chapter and the preprint of the
second paper, not the journal versions, and did not extract their per-model estimates.
We fit the GLFP model with a gamma drive-model effect in Appendix~\ref{sec:laws}. We do not fit their
hierarchical form, with parameters per drive model.

Siemroth and Park~\cite{siemroth2026manufacturer} fit Cox and Weibull proportional-hazards
regressions to 443{,}156 drives through mid-2025, at the level of the manufacturer, with
installation-year indicators, capacity, form factor and mean temperature among their
covariates. Their Cox hazard ratios relative to Seagate are 0.40 for HGST, 0.51 for Western
Digital, and 1.07 for Toshiba (0.41, 0.52 and 1.07 in their Weibull model). They estimate that
each degree Celsius of mean drive temperature raises the hazard by 2.1\%, and they report that
controlling for a drive's position in the data center does not change the manufacturer
differences. Ours, from
$\exp(\gamma)$ in Table~\ref{tab:finalpar}, are 0.20, 0.46, and 0.35. The order agrees for HGST
and Western Digital, and the answer differs for Toshiba. Appendix~\ref{sec:mfrcontrast}
helps explain the difference. Their specification fitted to our data gives Toshiba
0.85\unskip, close to theirs. The two numbers answer different questions.
Theirs is an adjusted hazard ratio between drives of the two manufacturers, conditional on
their covariates. Our $\exp(\gamma)$ compares the prior means of drive models of equal
capacity, vintage and helium status, and Seagate has many small drive models with high failure rates.
Weighting our posterior multipliers by exposure instead gives Toshiba
0.59\unskip. Toshiba's drives were in service mainly in the years of
low fleet-wide hazard, and in the earlier B3 fit the exposure-weighted ratio including year effects is
0.86\unskip, consistent with calendar time accounting for much of the
remaining difference. HeART~\cite{kadekodi2019heart}
drives redundancy decisions from empirical failure-rate curves per drive group in the Backblaze
logs, and Tiger~\cite{kadekodi2022tiger} from such curves in Google and Backblaze logs. HeART
reports that the highest useful-life failure rate among six drive groups is more than 3.5 times
the lowest, and Tiger states, citing HeART, that the difference between groups ``can be over an
order of magnitude''. D-Rex~\cite{gonthier2025drex}
sets the failure rates of the nodes in a distributed-storage reliability framework from
Backblaze's published annualized rates. We say that drive models differ by a
factor of about 95 (Section~\ref{sec:hier}). That is the ratio of the largest to the smallest
posterior-mean multiplier among drive models with at least 500 drive-years, 0.072 to 6.9. Among
the 33 drive models with at least 100 failures it is also 95. Across all 99, where the largest
multiplier belongs to a small drive model, it is 327. It is a range of extremes, not a typical
spread. Modeling calendar-year effects reduces the range for models
with at least 100 failures to 32 (Section~\ref{sec:bayesfit}). These ranges should not be
set beside the published spreads without those qualifications.

Backblaze's own reports trace the same history. In 2013 they described three piecewise rates
forming a bathtub~\cite{bb2013howlong}. In 2021 they wrote that ``there is almost no
left side of the bathtub''~\cite{bb2021bathtub}. In 2025 they list the assumptions of the
bathtub (identical devices, independent failures, constant environment), say that they ``have
really only seen model-by-model variation'', and note that healthy drives are sometimes taken out of service~\cite{bb2025bathtub}. The ST3000DM001 episode~\cite{bb2015csi} is a
documented cluster by drive model and year; Pecht and Elburn~\cite{pecht2020circuitworld}
analyse the reported failures of that model in the corpus and the limits of its SMART record
(we could read only the abstract).

Two further studies fit this corpus with the kinds of model we use. Monaco, Gorfine and
Hsu~\cite{monaco2018frailtysurv} ship with their frailtySurv package a data set drawn from about two years of the records, one
row per drive (52{,}422 drives, 2{,}885 failures). They fit a semiparametric Cox model with a
gamma frailty shared within drive model, with temperature and three SMART-derived indicators as
covariates, to the Western Digital subset (3{,}530 drives, 40 models, 178 failures). A gamma
frailty by drive model on this corpus is therefore not new either. What differs here is a parametric
baseline with a prior mean predicted from drive-model attributes, the use of the closed-form
posterior and predictive distribution for forecasting, and evaluation on later years. Oakley,
Forshaw, Philipson and Wilson~\cite{oakley2023multistate} extend Mittman et al.\ with the SMART
record. They start from the five attributes that Backblaze uses (5, 187, 188, 197, 198), drop 188
because a jump in it did not appear to change the failure rate of their drive model, and treat a
drive's first non-zero value of the other four as entry to a critical state. They fit Bayesian
illness-death and four-state models for left-truncated, right-censored semi-competing-risks
data to one drive model (4{,}707 drives deployed 2013--2016). Their survival predictions update
with the drive's state, and they assess them by time-dependent AUC and prediction error under
Monte Carlo cross-validation on held-out drives. This is
the transition model that a simulator of an existing fleet would need (Section~\ref{sec:sim}).
We fit such a model across the fleet in Section~\ref{sec:mstate}.

Other analyses of the corpus fit lifetime distributions to failed drives only, or predict
failure from the SMART record. Arslan and Zeydan~\cite{arslan2021heavytailed} take two years of
records (2016--2017, 2{,}987 failures) and fit exponential, Weibull, Gaussian hypergeometric
and Argus distributions to the lifetimes of the drives that failed, each measured from the start
of operation in the two-year data to failure. We did not find a treatment of drives still
running, or of drives that entered the record in service, so their Weibull estimate
$\beta=1.30$ describes the failed drives alone. Pinciroli, Yang,
Alter and Smirni~\cite{pinciroli2023lifespan} study five Seagate models from 2014 to 2019
beside SSD data from Google. They relate failures to workload, such as the time a drive spends
positioning its heads, report that young and old drives fail in similar ways, and train
classifiers on the SMART record. Amram, Dunn, Toledano and Zhuo~\cite{amram2021interpretable}
study one drive model, the ST12000NM0007. They fit an optimal survival tree to the drives that
failed between the first quarters of 2019 and 2020, a second one with censored drives included,
and interpretable classifiers to the SMART record. Mohapatra, Coursey and
Sengupta~\cite{mohapatra2023eol} estimate the remaining life of a drive from windows of its SMART
history with a recurrent network. Ahmed and Green~\cite{ahmed2024survival} train a one-dimensional
convolutional classifier of failure and, separately, a Cox regression on the SMART record of
37{,}037 ST4000DM000 drives from 2013 to 2022. Da Silva and Alvares~\cite{dasilva2026cindex}
reproduce that Cox regression's concordance (0.9595 against the reported 0.958) once they
restrict the healthy drives to those older than seven years, as they read the original
paper's filter to imply, and report that the model fails an in-sample calibration test under
the assumption of non-informative censoring, which they note they cannot test; a calibrated
forecast is what we score. Hu, So and Zhuang~\cite{hu2024sequential} estimate a mean
lifetime per drive model under an assumed exponential distribution, for the drive models with at
least 100 failures (26; one is reported in the main text and the rest in a supplement we did
not read), to illustrate a sequential censoring design. Yang, Wang and
Hu~\cite{yang2025censored} describe reliability estimation per drive model and failure-count
prediction on the full corpus; we could read only the abstract. Nasyrov, Nasyrov and
Nasyrov~\cite{nasyrov2023distribution} tabulate the failed drives by time in service and brand
and find, for almost every brand, a large share of failures early in service beside the maxima
that correspond to the drives' declared characteristics; an earlier note of
theirs~\cite{nasyrov2022bigdata} counts 339 publications citing the corpus to May 2022, finds
that the largest group only mention the company, and observes that the studies which use the
data mostly take a small slice of it, often one drive model.
Christoffersen~\cite{christoffersen2021dynamichazard} uses the corpus (79{,}668 drives) to
illustrate survival models whose coefficients follow a state-space process in the drive's
power-on age, estimated by Kalman-type filters. Vasilev, Filimonova, Petrovskiy and
Mashechkin~\cite{vasilev2024stratified} fit survival trees whose leaves are Cox,
Kaplan--Meier or accelerated-failure-time models to the 2016--2018 and 2018--2021 records, and
Vasilev~\cite{vasilev2025censoring} compares the distributions of removal and failure times
with a quantile plot, finds the removals informative on this corpus and not on an Alibaba SSD
set, and reports that survival-tree ensembles with an allowance for informative censoring
improve on this corpus only (integrated Brier score 0.084 to 0.071), concluding that the
operator's replacement policy has to be modelled. We examine the same removals in
Section~\ref{sec:retire}. Table~\ref{tab:priorwork} sets the lifetime studies beside ours. We
did not find an age-period-cohort treatment of this corpus, nor a paper that feeds
Backblaze-fitted lifetime models to an erasure-code durability simulator with measured sampling
costs.

\begin{table}[h]
\centering\small
\caption{Prior Analyses of This Corpus Compared with This Paper. Truncation and censoring give how a study treats drives already in service when first seen and drives still running at the end of the data. Validation names the data on which a study checked its predictions.}\label{tab:priorwork}
\scriptsize\setlength{\tabcolsep}{3pt}
\begin{tabular}{@{}>{\raggedright\arraybackslash}p{1.7cm}>{\raggedright\arraybackslash}p{2.2cm}>{\raggedright\arraybackslash}p{1.6cm}>{\raggedright\arraybackslash}p{1.4cm}>{\raggedright\arraybackslash}p{2.1cm}>{\raggedright\arraybackslash}p{2.0cm}>{\raggedright\arraybackslash}p{2.0cm}>{\raggedright\arraybackslash}p{2.0cm}@{}}
\toprule
Study & Data & Truncation and censoring & Age scale & Drive-model effect & SMART & Validation & For simulation \\
\midrule
Mittman et al.~\cite{mittman2019hierarchical} & 2013 to early 2016, 63 drive models, 75{,}297 drives, 44 drive models fitted & Left-truncated, right-censored & Age in hours (source not stated) & GLFP with a shared early mode, and the defect fraction and wear-out parameters per drive model & None & Approximate leave-one-out & A lifetime distribution per drive model \\
Lewis-Beck et al.~\cite{lewisbeck2022prediction} & As Mittman et al., with data to early 2019 set aside & As Mittman et al. & As Mittman et al. & As Mittman et al. & None & One-week forecasts in each of 26 weeks, risk set updated weekly & As Mittman et al. \\
Monaco et al.~\cite{monaco2018frailtysurv} & About two years, 52{,}422 drives, a WDC subset fitted (3{,}530 drives, 40 drive models, 178 failures) & Right-censored, no entry time & Follow-up time as recorded & Semiparametric Cox model with a gamma frailty per drive model & Temperature and three error indicators as covariates & None & None (the package simulates frailty data, not drives) \\
Oakley et al.~\cite{oakley2023multistate} & 2013 to 2016, one drive model (4{,}707 drives) & Left-truncated, right-censored & Power-on age & None (one drive model) & Critical states from attributes 5, 187, 197 and 198, with transitions modelled & Time-dependent AUC and prediction error under Monte Carlo cross-validation on held-out drives & State-dependent survival for one drive model \\
Arslan and Zeydan~\cite{arslan2021heavytailed} & 2016 to 2017, 2{,}987 failures & Failed drives only & Time from the start of operation in the data & None & None & None & Fitted lifetime distributions \\
Siemroth and Park~\cite{siemroth2026manufacturer} & 2013 to mid-2025, 443{,}156 drives & Right-censored & Different from ours & Manufacturer effects in Cox and Weibull regressions, with installation year and capacity & Temperature as a covariate & None reported & None \\
This paper & 2013 to mid-2026, 504{,}656 data drives, 99 harmonized drive models & Left-truncated, right-censored, daily & Power-on age & Gamma multiplier per drive model with covariates, year effects and model-year shocks & Observed state on 1~January in the regression and the Bayesian model, and boosted classifiers on the monthly record at 90 and 365 days & Nine successive held-out years with calibration, and prospective cohorts of the drives in service on each 1~January and each quarter, 2018 to 2026, per drive and in total & Lifetime models and samplers with a shared system context, with timings. The classifiers are timed, not integrated. \\
\bottomrule
\end{tabular}
\end{table}
PACEMAKER~\cite{kadekodi2020pacemaker} is the closest prior use of field reliability data in a
storage simulation. Its evaluation (Section~7) replays the Backblaze log and three Google
cluster logs date by date: the simulated cluster changes with the logged additions, failures
and decommissionings, and the system learns an empirical AFR curve per drive group as the data
arrive, to decide when to change the redundancy of each group. It reports the IO of these
transitions and the space saved while holding MTTDL at or above a target. The failures in it are
the logged ones. It fits no lifetime distribution and draws no lifetimes, so it cannot simulate a drive
beyond the ages and dates in the log, which our samplers do, with measured costs. Frank et
al.~\cite{frank2025uncertain} argue, citing the Drive Stats, that consensus protocols should
replace the fixed count of tolerated faults with a probability of failure for each machine that
depends on its type and age; a fitted lifetime distribution would supply those probabilities. We
have not
obtained Hall's paper on tools for predicting the reliability of large storage
systems~\cite{hall2016tools} or Chen and Xiao's grey forecasting model of drive
failures~\cite{chen2024grey}.

\section{Data}\label{sec:data}

The record is one row per drive per day from Backblaze's storage fleet, April 2013 to June
2026: 744{,}525{,}683 drive-days of 512{,}600 drives, of which 504{,}656 are hard disk data
drives. Each row gives the serial number, the drive model, the capacity, a failure flag and
the SMART attributes. Ages, exposure and failures are all derived from these rows.

\subsection{The Corpus: Which Drives and Years Are Included}

Backblaze publishes one file per day, from 10~April 2013, with one row per drive in its
storage fleet~\cite{bb2026testdata}. We use the 4{,}830 daily files through 30~June 2026, as
published in 45 quarterly and yearly archives. Three files (2014-11-02, 2015-11-01,
2017-01-30) contain only a header. Through 2025 the files hold 681{,}960{,}203 rows, one per
drive and day, for 489{,}633 drives, of which 31{,}021 failed. One of our programs reduces the
files to one record per drive per month and to one record per drive. It keeps one row per
serial number per day and drops rows with an empty serial, which leaves 681{,}960{,}196
drive-days. A second program, written separately, counts directly from the daily files. The
two agree on the number of drives and of failed drives. We built yearly risk sets on age since
first sighting from each. They have the same numbers of intervals and failures in every year
from 2013 to 2025, and the same total exposure except for 15 days in 2022. These are checks of
totals, not of individual records.

\subsection{What the Daily Files Contain}\label{sec:smart}

Each daily file has one row per drive with its serial number, drive model, capacity, a failure
flag, and SMART attributes in raw and normalized form. SMART (Self-Monitoring, Analysis and
Reporting Technology) attributes are counters a drive reports about itself: power-on hours,
temperature, reallocated and pending sectors, errors, power cycles and others. Which attributes
a drive reports, and how the raw value is encoded, vary by manufacturer and drive model. An
unreported attribute is absent from the row, not zero. Our lifetime models use the failure flag
and power-on hours (SMART attribute~9). Section~\ref{sec:smartstate} and
Appendix~\ref{sec:cov} use the warning attributes, temperature and load counters. The set of
attributes reported has grown (Table~\ref{tab:smartcols}) from 40 raw attributes in 2013 to 93
by the end of 2024. The 40 of 2013 are present in every later year. They include the five that
Backblaze uses to judge a drive (5, 187, 188, 197, 198) and attribute 9. From July 2023 the
files also give each drive's datacenter, cluster, vault, pod, and slot. Appendix~\ref{sec:vaults}
uses the first four.

\begin{table}[h]
\centering\footnotesize
\caption{Columns of the Daily Files by Year. The counts are for the last daily file of each year. Every file has the date, serial number, drive model, capacity and failure flag. Each SMART attribute contributes a raw and a normalized column.}\label{tab:smartcols}
\begin{tabular}{@{}rrrl@{}}
\toprule
Year & Columns & SMART attributes & Other columns \\
\midrule
2013 & 85 & 40 & --- \\
2014 & 85 & 40 & --- \\
2015 & 95 & 45 & --- \\
2016 & 95 & 45 & --- \\
2017 & 95 & 45 & --- \\
2018 & 129 & 62 & --- \\
2019 & 131 & 63 & --- \\
2020 & 149 & 72 & --- \\
2021 & 179 & 87 & --- \\
2022 & 179 & 87 & --- \\
2023 & 193 & 91 & datacenter, cluster\_id, vault\_id, pod\_id, pod\_slot\_num, is\_legacy\_format \\
2024 & 197 & 93 & datacenter, cluster\_id, vault\_id, pod\_id, pod\_slot\_num, is\_legacy\_format \\
2025 & 197 & 93 & datacenter, cluster\_id, vault\_id, pod\_id, pod\_slot\_num, is\_legacy\_format \\
2026 & 197 & 93 & datacenter, cluster\_id, vault\_id, pod\_id, pod\_slot\_num, is\_legacy\_format \\
\bottomrule
\end{tabular}
\end{table}

Short records are rare (Table~\ref{tab:shortlived}): 185 of 512{,}600 drives are seen on a
single day, and 571 for a week or less.

\begin{table}[h]
\centering\small
\caption{Drives by the Span of Their Record. The span is the number of days over which a drive appears in the corpus (all drives, through June 2026). The last three columns give how the record ends.}\label{tab:shortlived}
\begin{tabular}{@{}lrrrrr@{}}
\toprule
Span & Drives & \% & Failure flag & Left without failure flag & Present at end \\
\midrule
Seen on 1 day only & 185 & 0.04 & 73 & 99 & 13 \\
Seen on 2 to 7 days & 386 & 0.08 & 221 & 117 & 48 \\
Seen on 8 to 30 days & 3{,}737 & 0.73 & 587 & 349 & 2{,}801 \\
Seen for more than 30 days & 508{,}292 & 99.16 & 32{,}685 & 123{,}435 & 352{,}172 \\
\bottomrule
\end{tabular}
\end{table}

\subsection{Drives and Drive Models}\label{sec:population}

We class drives by drive model and reported capacity. The \emph{data drives} are the hard
drives of 1.4~TB and more: 504{,}656 drives, 2.02 million calendar-span drive-years and
31{,}899 failures through June 2026. The other classes are 4{,}008 hard drives below 1.4~TB or
of unreported capacity (51 models) and 3{,}936 solid-state drives (25 models), mostly boot
drives (Appendix~\ref{sec:classes}). Of the data drives, 504{,}624 have a usable age
(Section~\ref{sec:age}). They carry 31{,}881 of the failures, and they are the population of
the main HDD lifetime fits. The descriptive comparison of lifetime distributions in
Section~\ref{sec:families} uses all drives through 2025. We recognize solid-state drives by
drive-model name, and class two models whose names do not say so as solid-state by their
capacity and family. The drive-model field is not constant: 3{,}322 of 512{,}600 serial numbers
carry more than one name, on 118{,}235 drive-days (0.016\%), mostly the same name with a doubled
space. We squeeze white space, merge three pairs of labels of one part (Hitachi and HGST
HDS5C4040ALE630 and HDS724040ALE640, and WUH721816ALE6L4 with and without its maker), and give
each drive one name throughout, its last. That leaves 99 drive models of data drives. Apart from five drives, the data drives come from four manufacturers
(Seagate, HGST, Toshiba and Western Digital). The Ultrastar WUH models are sold under Western
Digital's name but are helium drives of the HGST line, and we class them with HGST: the
Western Digital level holds only its eight air-filled models. We mark helium status using
SMART-attribute presence on eleven sampled days from 2015 to 2026. Ten drive models absent
from those samples are assigned air by the model-name rule, without measured confirmation
from the sampled days. HGST-line helium models report SMART attribute 22,
Toshiba's report 23 and 24, and every other model of those makers reports none of the three.
Seagate documents that its helium drives support attribute 200 and its air drives do not~\cite{seagate2024openseachest}.
On that basis 43 of the 99 drive models, with 66\% of the exposure, are helium. Nearly all
helium models are of 10~TB or more, so helium and capacity are confounded in these data.
A drive model's manufacturer, helium status, capacity and the year it first appears in the
record are the covariates available for every drive model. We call
the last the drive model's \emph{vintage} or year of introduction: the model's first observed
date in this corpus, as a fractional year. The files record neither a manufacturing date nor a
launch date.

Matching part numbers can conceal different observed populations. The raw labels
\code{WDC WUH721816ALE6L4} and \code{WUH721816ALE6L4} have 26{,}685 and 624 drives,
respectively. The latter group has 22 failures over 472 drive-years of calendar exposure,
compared with 452 over 84{,}504 drive-years for the WDC-prefixed group: 4.66\% and
0.53\% per year. These descriptive rates do not adjust for age or calendar time. We retain
the raw labels for cohort checks, including the SMART comparison in
Section~\ref{sec:smartstate}.

Table~\ref{tab:ledger} follows the records from the daily files to the populations and
exposure the analyses use. Every analysis uses one event history (\code{bb\_events.R}). A drive
enters at its first sighting. A failed drive leaves at its final failure flag, and the failure
is absorbing: 1{,}026\unskip\ failed data drives appear in the daily
files after that flag, over 398{,}009\unskip\ calendar days, and none
of those rows is exposure or puts the drive back in a risk set. Any other drive leaves at its
last sighting. The exposure of the Lexis table, the table of failures and exposure days by drive
model, calendar month and age bin that the main fits use (Section~\ref{sec:lexis}), is the days on
which a drive appears, up to and
including the day of its failure: 731{,}605{,}328\unskip\ drive-days for
the data drives with a usable age, against 739{,}067{,}489\unskip\ days of
calendar span from first sighting to exit. Analyses that use calendar spans instead of observed
days say so.

The files do not say why a failed drive reappears. A flag may have been transient, or a
technician may have returned a pulled drive to service; a third of the
1{,}025\unskip\ reappearing data drives with a usable age are ST3000DM001s
(348\unskip), the model Backblaze describes pulling,
bench-testing and reusing in 2014 (Section~\ref{sec:apc}). Whatever the cause, the choice of
reading does not move the fleet-level estimates. Table~\ref{tab:postfail} refits the two
competing Weibulls of Section~\ref{sec:families}, with a free multiplier per drive model, to the data drives with a usable age under
three readings of the rows after a failure flag: absorbing (the paper's rule), episodes (every
flag a failure, the rows between flags exposure; 44\unskip\
drives carry two flags) and dropping the reappearing drives altogether. The absorbing and
episode readings agree to 0.01\unskip\ points of
ten-year survival, and among the 33\unskip\ drive models
with at least 100 failures the per-model multiplier changes by at
most 4.6\unskip\%
(WDC WUH721816ALE6L0\unskip). Dropping the drives is not a neutral
choice: it removes failed drives selectively and lowers one drive model's multiplier
(WDC WUH721816ALE6L0\unskip) by
27\unskip\%.

\begin{table}[h]
\centering\footnotesize\setlength{\tabcolsep}{4pt}
\caption{Three Readings of the Rows After a Failure Flag. Two competing Weibulls on power-on age with a free multiplier per drive model, HDD data drives with a usable age, calendar-span intervals (\code{bb\_postfail\_sensitivity.R}). The paper's rule is absorbing. Under episodes every flag is a failure and the rows between flags are exposure. Dropping removes the drives seen after a flag. The parameters $\beta_1$ and $\beta_2$ are the age exponents in the two cumulative hazard components; hazard and survival are evaluated at ten years of power-on age.}\label{tab:postfail}
\begin{tabular}{@{}lrrrrrr@{}}
\toprule
Reading & Drives & Failures & $\beta_1$ & $\beta_2$ & $h(10)$ (\%/yr) & $S(10)$ (\%) \\
\midrule
Absorbing (the paper) & 504{,}624 & 31{,}881 & 0.339 & 1.597 & 3.11 & 81.19 \\
Episodes & 504{,}624 & 31{,}925 & 0.339 & 1.597 & 3.11 & 81.19 \\
Drives dropped & 503{,}599 & 30{,}856 & 0.401 & 1.600 & 3.00 & 81.79 \\
\bottomrule
\end{tabular}
\end{table}

\begin{table}[h]
\centering\footnotesize\setlength{\tabcolsep}{4pt}
\caption{Population Ledger. Counts from the daily files to the populations and exposure the analyses use, under the event history in which a failed drive leaves at its final failure flag. Data run from April 2013 to 30 June 2026, so 2026 has half a year of follow-up.}\label{tab:ledger}
\begin{tabular}{@{}llr@{}}
\toprule
Population & Quantity & Value \\
\midrule
All records & Rows in the daily files (drive-days) & 744{,}525{,}683 \\
All records & Unique serial numbers & 512{,}600 \\
\addlinespace
HDD data drives & Drives (class by drive-model name and capacity) & 504{,}656 \\
Small or unknown HDDs & Drives (class by drive-model name and capacity) & 4{,}404 \\
SSDs & Drives (class by drive-model name and capacity) & 3{,}540 \\
\addlinespace
Class appendix & Drives moved from the small or unknown HDDs to the SSDs (two named models) & 396 \\
Small or unknown HDDs & Drives after the move & 4{,}008 \\
SSDs & Drives after the move & 3{,}936 \\
\addlinespace
HDD data drives & Drives & 504{,}656 \\
HDD data drives & Failed drives (a failure flag) & 31{,}899 \\
HDD data drives & Drives with a usable power-on age & 504{,}624 \\
HDD data drives, usable age & Failed drives & 31{,}881 \\
\addlinespace
HDD data drives & Failed drives seen after their final failure flag & 1{,}026 \\
HDD data drives & Calendar days after the final failure flag (excluded) & 398{,}009 \\
HDD data drives, usable age & Calendar days after the final failure flag (excluded) & 396{,}255 \\
HDD data drives, usable age & Observed rows after the failure day (excluded) & 299{,}533 \\
\addlinespace
HDD data drives, usable age & Calendar span, first sighting to exit (days) & 739{,}067{,}489 \\
HDD data drives, usable age & Observed exposure in the Lexis table (days) & 731{,}605{,}328 \\
HDD data drives, usable age & Events in the Lexis table & 31{,}881 \\
HDD data drives, usable age & Observed exposure from 1 January 2026 (days; follow-up ends 30 June 2026) & 61{,}826{,}576 \\
\addlinespace
All records & Serial numbers with a failure flag & 33{,}566 \\
All records & Rows with a failure flag & 33{,}623 \\
All records & Serial numbers with more than one failure flag & 57 \\
\bottomrule
\end{tabular}
\end{table}

\subsection{What a Failure Is, and How Our Counts Compare with Backblaze's}\label{sec:reconcile}

Backblaze counts a drive as failed when it is removed and replaced because it has stopped
working or ``has shown evidence of failing soon''~\cite{bb2014smart}. Proactive removals are
failures. Their 2025 description is operational. A drive that vanishes from the reporting
population is classified from its repair ticket, and they acknowledge ``cosmetic'' failures
when work on a drive model spans a quarter~\cite{bb2025q3}. Retirement is by policy and by
drive model, so censoring carries information about the drive model's standing with its
operator.

Backblaze's lifetime table for the second quarter of 2026 lists 558{,}148{,}607 drive-days,
21{,}609 failures, and an annualized rate of 1.41\%~\cite{bb2026testdata}. Our corpus has more of
both. Table~\ref{tab:reconcile} uses our age-eligible Lexis population, not every raw record.
We select drive models by their June 2026 exposure divided by 30 days, a monthly average
count rather than the number present on 30~June. Selecting HDD data models with an average
of at least 100 gives totals within 1\% of Backblaze's. This numerical agreement does not
establish that the two tables use the same inclusion rules.

\begin{table}[h]
\centering\footnotesize\setlength{\tabcolsep}{4pt}
\caption{Age-Eligible Lexis Counts Against Backblaze's Published Lifetime Totals. Our rows use observed drive-days up to and including the day of the final failure flag. The June count is that month's exposure divided by 30; it is not an end-of-month roster. The last row is Backblaze's published table. The annualized rate is $100\times365\times$ failures per drive-day.}\label{tab:reconcile}
\begin{tabular}{@{}llrrrr@{}}
\toprule
Population & Drive models & Models & Drive-days & Failures & Rate (\%) \\
\midrule
All classes, age eligible & all & 175 & 743{,}337{,}459 & 33{,}545 & 1.65 \\
All classes, age eligible & exposure in June 2026 & 78 & 573{,}954{,}645 & 22{,}994 & 1.46 \\
All classes, age eligible & June mean at least 100 & 39 & 567{,}880{,}875 & 21{,}890 & 1.41 \\
All classes, age eligible & June mean at least 1000 & 24 & 517{,}579{,}173 & 18{,}860 & 1.33 \\
HDD data, age eligible & all & 99 & 731{,}605{,}328 & 31{,}881 & 1.59 \\
HDD data, age eligible & exposure in June 2026 & 49 & 563{,}275{,}377 & 21{,}774 & 1.41 \\
HDD data, age eligible & June mean at least 100 & 32 & 562{,}510{,}439 & 21{,}693 & 1.41 \\
HDD data, age eligible & June mean at least 1000 & 23 & 515{,}464{,}789 & 18{,}836 & 1.33 \\
\midrule
Backblaze lifetime table, Q2 2026 & as published & & 558{,}148{,}607 & 21{,}609 & 1.41 \\
\bottomrule
\end{tabular}
\end{table}

\subsection{How We Estimate Drive Age}\label{sec:age}

Age is power-on age, not time since a drive first appears in the record, because many drives
enter the record already in service. For each drive and each calendar month, we take the
SMART~9 reading on the drive's first observed day of that month: $s$ hours, $\Delta$ days
after first sighting. It gives a candidate for the age at first sighting
\begin{equation}
a^{(0)} = \frac{s}{24} - \Delta \quad\text{days},
\end{equation}
and we use the candidate from the earliest month that has a reading. For 49 drives with no
such reading (27 age-eligible HDD data drives) we use the month's minimum reading, placed at the middle of the drive's observed
days in that month. We set values between $-45$ and $0$ days to 0 and reject values outside
$[-45\text{ d},\,20\text{ yr}]$. Age thereafter advances with the calendar:
$a(t) = a^{(0)} + (t - f)$, where $f$ is the date of first sighting and $t-f$ is in days.

\textbf{Why only the earliest reading.} The counter is unreliable later in life. On 21{,}404 of
504{,}656 HDD data drives (through 2026-06) a later reading disagrees with the earliest by more
than 45 days and 10\% of the elapsed span. The largest group are Seagate 8~TB and 6~TB models
whose counter drops by about 3{,}258 days (78{,}200 hours) late in life, evidently a wrap.
Others advance more slowly than the calendar. Requiring agreement would have discarded the
longest-lived drives, so we use later readings only to mark drives, not to exclude them.

\textbf{Coverage.} Through 2025-12-31, 489{,}184 of 489{,}633 drives (99.9\%) and 31{,}000 of
31{,}021 failed drives have a usable $a^{(0)}$. Its median is 8 days. When first seen, 15.6\%
of drives were more than 30 days old, and among HDD data drives 2.8\% were more than a year
old.
Age after first sighting is calendar time, so for drives that were powered off for part of
their service it overstates power-on age.

\subsection{Drives That Leave Without a Failure Flag}\label{sec:retire}

A drive's record ends in one of three ways: with a failure flag, at the end of the data, or
with neither. Our models treat the third as censoring that carries no information about the
drive. That is wrong if drives are pulled because they are about to fail. The data give no
reason for a removal, so we examine the pattern of removals. Vasilev~\cite{vasilev2025censoring}
reached the same question from the SMART record: on this corpus the removal times are
distributed unlike the failure times, and survival models that allow for informative removal
score better (Section~\ref{sec:related}).

We call a drive removed if it has no failure flag and was last seen more than thirty days
before 30~June 2026. We call a removal bulk if at least twenty drives of the same drive model
were last seen on the same day, and single otherwise. The SMART warning is that of
Section~\ref{sec:smartstate}: any of SMART attributes 5, 187, 188, 197, 198 non-zero.

\begin{table}[h]
\centering
\caption{How the Records of the 504{,}656 Data Drives End. Warned is the share with a SMART warning on or before the last day. Age is power-on age.}\label{tab:retiretotal}
\small\begin{tabular}{@{}lrrrr@{}}
\toprule
End of record & Drives & Warned (\%) & Median years in service & Median age at exit (yr) \\
\midrule
Failure flag & 31{,}899 & 76.2 & 2.8 & 3.1 \\
In service at the end & 352{,}380 & 9.5 & 3.7 & 3.8 \\
Removed, with 19 or more others & 116{,}957 & 15.1 & 4.1 & 4.6 \\
Removed, with fewer & 3{,}420 & 31.8 & 2.2 & 3.1 \\
\bottomrule
\end{tabular}
\end{table}

\begin{table}[h]
\centering
\caption{Exits per Drive-Year by Warning State.}\label{tab:retirerates}
\small\begin{tabular}{@{}lrrrr@{}}
\toprule
State & Drive-years & Failure (\%/yr) & Single removal (\%/yr) & Bulk removal (\%/yr) \\
\midrule
Clear & 1{,}874{,}002 & 0.40 & 0.12 & 5.3 \\
Warned & 149{,}457 & 16.27 & 0.73 & 11.8 \\
\bottomrule
\end{tabular}
\end{table}

Bulk removals are 116{,}957 of the 120{,}377 (Table~\ref{tab:retiretotal}). There are 789 such
events. The largest took 2{,}107 drives of one model in a day, at a median age of 8.6 years.
The threshold matters little between 5 and 20 (118{,}280 and 116{,}957 bulk) and more at 100
(91{,}945). The year 2022 has no bulk removal at all and 116 single ones.

\emph{Single removals.} There are 3{,}420, against 31{,}899 failures. Of them 32\% were
warned, against 9.5\% of drives in service. A warned drive is removed singly at 5.9 times the
rate of a clear one (Table~\ref{tab:retirerates}). This association is consistent with some
removals prompted by a warning, but it does not identify their reasons. A warned drive is
22 times as likely to end with a failure flag as with a single removal.

\emph{Bulk removals.} A warned drive leaves in a bulk removal at 2.2 times the rate of a
clear one. That comparison mixes drive models and ages: the migrated models are old Seagate
models, which carry warnings more often. A within-event comparison is cleaner. For each of the 777
events that left drives of the same model behind, we set the share warned among the drives
removed against the share warned, on that day, among those that stayed. Pooled by
Mantel--Haenszel, the odds ratio is 1.24 (95\% interval 1.22 to 1.26). The events removed
17{,}412 warned drives where 15{,}224 would be expected if they took drives without regard to
the warning. The interval treats the events as independent strata. The same drives serve as those
that stayed in many events, so the interval is too narrow. The excess of 2{,}188 is 6.9\% of
the failure count, and it is concentrated. ST4000DM000 supplies 1{,}698 of it and ST12000NM0007
547, while ST8000DM002 has 204 fewer warned drives removed than expected.

Age differences may contribute to the excess.
Over all events the drives removed are older than those that stay by a median of 0.17 years
over events (a mean of 0.30 weighted by drives removed). Within the 161 events of ST4000DM000
they are older by a median of 1.42 years (weighted mean 0.75). Older drives of a model are
warned more often, but we did not compute an age-adjusted
ratio. Within the 210 events of ST12000NM0007 the age difference is 0.04 years.

In 42 events at least 30\% of the drives removed are warned, and that share is at least three
times the share among the drives that stay. They remove 3{,}023 drives, 1{,}430 of them
warned. Of the 42, 30 events, with 2{,}302 of the drives, are of ST4000DM000, so the fourth row of
Table~\ref{tab:retiresens} mostly concerns one drive model. These events motivate a sensitivity
analysis; the warning shares alone do not establish why either warned or clear drives left.

\begin{table}[h]
\centering
\caption{Sensitivity of the Fit to Removals Counted as Failures. Each row refits two competing Weibulls with a free multiplier per drive model, with more removals recoded as failures. Hazards are for the exposure-weighted average drive model. The last row is a deliberately severe recoding scenario, not a plausible endpoint definition. It is not a bound either, because removals outside the selected events are not recoded.}\label{tab:retiresens}
\footnotesize\setlength{\tabcolsep}{4pt}\begin{tabular}{@{}lrrrrrrrr@{}}
\toprule
& & & & \multicolumn{3}{c}{Hazard (\%/yr) at age} & \multicolumn{2}{c}{Survival to} \\
Counted as failures & Events & $\beta_1$ & $\beta_2$ & 1 yr & 5 yr & 10 yr & 5 yr & 10 yr \\
\midrule
Failure flags only & 31{,}881 & 0.339 & 1.597 & 1.04 & 2.11 & 3.11 & 0.926 & 0.812 \\
and warned drives removed singly & 32{,}970 & 0.347 & 1.628 & 1.05 & 2.21 & 3.32 & 0.924 & 0.804 \\
and all drives removed singly & 35{,}287 & 0.353 & 1.648 & 1.13 & 2.34 & 3.56 & 0.919 & 0.792 \\
and the 42 bulk events with a high warned share & 38{,}310 & 0.309 & 1.610 & 1.21 & 2.58 & 3.85 & 0.912 & 0.776 \\
and every warned drive removed & 50{,}606 & 0.354 & 2.231 & 0.84 & 4.33 & 10.02 & 0.898 & 0.630 \\
\bottomrule
\end{tabular}
\end{table}

\emph{Effect on the fit.} In Table~\ref{tab:retiresens} we refit the lifetime model with
removals recoded as failures. We use two competing Weibull failure processes, with cumulative hazard
$\Lambda(a)=a^{\beta_1}+r\,(a/4)^{\beta_2}$, with a free multiplier for each drive model. For
given $(\beta_1,\beta_2,r)$ the multiplier that maximizes the likelihood is the model's failures
over its expected count, so the profile log-likelihood is
\begin{equation}
\ell(\beta_1,\beta_2,r)=\sum_g\Bigl[d_g\log\frac{d_g}{E_g}-d_g\Bigr]+\sum_{i:\,\text{failed}}\log \lambda(a_i),
\qquad E_g=\sum_{i\in g}\bigl[\Lambda(a_i)-\Lambda(a_{0i})\bigr],
\end{equation}
with $\lambda(a)=\Lambda'(a)$, and $a_{0i}$ and $a_i$ the ages at which drive $i$ is first and last seen. This is a
fixed-effects version of the model of Section~\ref{sec:hier}. We use it here because it refits
in seconds. Its first row gives $\beta_2=1.60$ where the gamma-frailty fit gives 1.57. Counting single removals as
failures, warned or all, raises the five-year hazard by 5\% to 11\%. Adding every drive
in the 42 events above raises the hazard by 22\% at five years and 24\% at ten. Counting every
warned drive that was ever removed, in bulk or not, doubles the hazard at five years and
triples it at ten. That row assumes all 17{,}636 warned drives in bulk removals were about to
fail, where the within-event comparison finds 2{,}188 more warned drives than an indifferent
removal would take. That number is not a bound either, since it counts only selection that
acts through the warning.

Warnings and removals are associated. Recoding the removals in the fourth row raises the
fitted hazard by about a fifth, but that scenario does not estimate how many removed drives
would have failed under continued service. A policy that retires a whole drive model early because it is failing
cannot be excluded. It would remove the old-age exposure of the worst models, which is one
more reason the wear-out end of the curve is not identified (Section~\ref{sec:apc}).

% Part 2: models and evaluation design, before any score
\section{Methods}\label{sec:methods}

We fit lifetime models while accounting for drives already in service when observation
begins and drives whose failure is not observed before they leave the record. We then test
their predictions on later years. The calculations can use individual drive histories or
equivalent tables of operating time and failures; the notation is tabulated at the end.

\subsection{Likelihood}\label{sec:lik}

Each drive contributes an interval of power-on age $(a_0,a_1]$ over which we observed it. Here
$a_0$ is its age at the start of its first observed day (or of a training window), $a_1$ is its
age at the end of its last observed day, failure day, or window, and $\delta=1$ if the interval ends in
failure. A drive that enters the record already in service is left-truncated. It is in the data
only because it survived to $a_0$, so we condition its contribution on that survival. For a distribution with survival function
$S_\theta$ and cumulative hazard $H_\theta=-\log S_\theta$,
\begin{equation}\label{eq:lik}
\ell(\theta)=\sum_i\bigl[\delta_i\log h_\theta(a_{1i}) - H_\theta(a_{1i}) + H_\theta(a_{0i})\bigr]
\end{equation}
in continuous time. We group drives with the same $(a_0,a_1,\delta)$ and drive model, with a
weight $w_i$.

\textbf{Failures within a day.} Ages are recorded in whole days, so a failure recorded at age
$a_1$ occurred somewhere in $(a_1-\Delta t,\,a_1]$, with $\Delta t$ one day. When ages are in
years, as they are in every formula below unless days are stated, $\Delta t$ is $1/365.25$.
A density evaluated at $a_1$ would be inaccurate where the hazard changes quickly, which is
the early-life regime. We therefore use the likelihood of an interval $(a_0,a_1,\delta)$
\begin{equation}
L_i(\theta) = \frac{\bigl[S_\theta(a_{1i}-\Delta t) - S_\theta(a_{1i})\bigr]^{\delta_i}\,S_\theta(a_{1i})^{1-\delta_i}}{S_\theta(a_{0i})},
\qquad
\ell(\theta) = \sum_i w_i \log L_i(\theta).
\label{eq:lldisc}
\end{equation}
Because ages are integers, $\ell$ depends on the data only through three weight vectors on the
distinct ages (failures at $u$, censorings at $u$, entries at $u$). One evaluation therefore
costs a few thousand survival-function calls, regardless of the number of drives. We tested
the fit on 370{,}000 simulated left-truncated, right-censored lifetimes from two competing
Weibulls with $(\beta_1,\eta_1,\beta_2,\eta_2) = (0.6, 60, 2.5, 9)$, recorded in whole days. It
recovers $(0.595, 62.1, 2.485, 8.98)$ with standard errors $(0.005, 2.4, 0.021, 0.026)$.

\textbf{Standard errors.} We finish each fit by BFGS from the Nelder--Mead optimum and compute
the Hessian $\mathcal{H}$ of $-\ell$ in the working (log) parameters numerically. With $J$ the
Jacobian of the natural parameters with respect to the working parameters, the reported
standard errors are the square roots of the diagonal of $J\mathcal{H}^{-1}J^{\top}$. They assume
independent intervals. Failures are not independent (Section~\ref{sec:time}), so these are
independence-based estimates, not bounds. We compare models on held-out years instead.

\subsection{Lexis Table: Organizing Failures by Drive Age and Calendar Time}\label{sec:lexis}

A Lexis table cross-classifies exposure time and events by age and calendar time. Each cell
holds the drive-days lived at a given age in a given period and the events that occurred there.
It is the natural structure for a hazard that may depend on both~\cite{keiding1990,carstensen2007}.
We tabulate the exposure and failures of HDD data drives in this form by drive model, calendar
month and age bin, on power-on age and on age since first sighting. We count age in whole
days. There are 81 bins: each of days 1 to 30 separately, days 31--45, 46--60, 61--91, then
quarter-years to 12 years, and beyond 12 years. A bin holds the ages above its lower edge up to
and including its upper edge, in whole days. We use the same convention wherever bins appear.

We record the days on which a drive appears in each month as a bit mask and divide each run of
consecutive observed days exactly among the bins it falls in. Nothing is then assumed about
where in a bin exposure lies. This matters in the first weeks, where the hazard changes fastest
and where most drives are not yet under observation. It also matters because 4.3\% of
drive-months have at least one day on which the drive is absent from the daily file (median
gap one day, 99\% under ten). Placing a month's observed days consecutively from its first
would move exposure after a gap to earlier ages, while failures keep their dates.
The size of that difference is in Appendix~\ref{sec:repropacked}. We count a failure once,
in the month of the drive's final failure flag and the bin of its age that day.

For a distribution with cumulative hazard $H$, the expected count in a cell with exposure $E_c$
drive-years in age bin $(\ell_b,u_b]$ (endpoints converted to years) is
$\mu_c=E_c\,[H(u_b)-H(\ell_b)]/(u_b-\ell_b)$, times any multipliers of the model. With that
piecewise-constant hazard the survival likelihood and a Poisson likelihood on the cells are the
same up to a constant~\cite{holford1980,laird1981}.

\subsection{Testing Predictions on Later Years}\label{sec:validdesign}

For $Y = 2017,\dots,2025$ we fit every candidate to the data through December of year $Y$.
It then predicts every cell of year $Y+1$ (for 2026, January--June) given that cell's actual
exposure: nine folds, 24{,}975
 held-out failures. Drive-model attributes
(capacity, manufacturer, helium, date of first appearance) are known when a model enters service, and
no outcome from a test year enters a forecast for it. That is the sense in which the years are
held out. We also developed and chose the models on these years, so this is a rolling-origin
development evaluation, not a test on a period locked away in advance
(Section~\ref{sec:limits}). We repeat the comparisons of lifetime distributions in \emph{quarterly
blocks}, with training ending every three months from December 2017 to March 2026 and the
following quarter as the test: 34 folds with the same held-out failures.

We score a point forecast $\mu_c$ by the Poisson log score of the test year's cells,
$\sum_c[d_c\log\mu_c-\mu_c-\log d_c!]$. We score a model that gives a predictive distribution by
the logarithm of the joint predictive probability of all the test year's counts. Larger is
better, and differences are in natural log units. We compare two candidates by the sum over
folds of the differences of their scores, with a standard error from the fold-to-fold spread
of those differences. With nine folds a ratio of about 2.3 corresponds to a two-sided 5\%
test. We check the level of a forecast by the ratio of predicted to observed failures in each
test year.

\emph{Point and predictive.} A Poisson distribution at a fitted mean is itself a predictive
distribution. Throughout, a \emph{point} forecast means the plug-in Poisson forecast at the
fitted expected count, and a \emph{predictive} forecast means the distribution that also
integrates the uncertainty of the drive-model multiplier (and, for the Bayesian variants, of the
other parameters and the shocks). Both are scored on the same counts, so their scores are
comparable.

\textbf{Calibration.} A log score ranks predictive distributions but does not say whether an
interval quoted from one can be trusted. For each fold and drive model we place the observed
count of the test year in its predictive distribution. We report the coverage of central 50,
80 and 95\% intervals and the randomized probability integral transform, as described with its nonrandomized alternative by
Czado et al.~\cite{czado2009}. We
call a forecaster \emph{calibrated} for a stated group of drive-model-years when the coverage of
each of its central intervals is within the sampling error of that group's size from the nominal
rate and the two tails are near 2.5\% each. A single coverage figure on nine fleet totals does
not establish calibration, and we report it as a count of years.

\textbf{Three words used with a fixed meaning.} A parameter is \emph{identifiable} if distinct values imply distinct distributions
of the observations under the model. A finite estimated standard error does not establish
identifiability. We use \emph{weakly determined} for an estimate whose uncertainty is large
relative to its magnitude. We state structural non-identifiability separately where it
occurs (Section~\ref{sec:apc}). A model is \emph{validated} only in the sense above: scored on
the nine development years with no outcome from a test year entering its forecast.

\subsection{Scores for Per-Drive Forecasts}\label{sec:perdrive}

The prospective evaluation of Sections~\ref{sec:landmark} and~\ref{sec:boost} gives each drive
a probability $p_i$ of an outcome $y_i\in\{0,1\}$, and uses four scores. The \emph{log loss} is
$-\sum_i[y_i\log p_i+(1-y_i)\log(1-p_i)]$, summed over the drives of a cohort. Smaller is
better. It is the per-drive form of the log score above, so a difference of log losses is in
the same natural log units as a difference of log scores. The \emph{Brier score} is
$n^{-1}\sum_{i=1}^{n}(p_i-y_i)^2$ for a cohort of $n$ drives, smaller better. We multiply it
by 10{,}000 for reporting. The \emph{AUC} is the area
under the receiver operating characteristic curve, computed as the Mann--Whitney statistic: the
probability that a randomly chosen drive that failed received a higher $p$ than a randomly
chosen drive that did not, ties counting one half. It depends only on the ranking, so a
forecaster that gives every drive the same probability has AUC 0.5. A \emph{gain} in a table
is the reference forecaster's loss minus the candidate's, positive when the candidate is
better. We report AUC as is. We form \emph{deciles} within each cohort by sorting the drives
by $p$ into ten equal groups. A run of equal probabilities that straddles a boundary is divided
between the groups in proportion to its overlap (the expected allocation under random
tie-breaking), so a forecaster with no ranking puts a tenth of the failures in each decile.
The \emph{top tenth} is the tenth decile pooled over the cohorts of a kind.

\subsection{Notation}\label{sec:notation}

Table~\ref{tab:notation} collects the symbols that recur across sections. Distributions are
written in one convention: $\Gam(a,b)$ is the gamma distribution with shape $a$ and rate $b$,
density $b^a x^{a-1}e^{-bx}/\Gamma(a)$, mean $a/b$ and variance $a/b^2$. $\Normal(m,v)$ is the
normal distribution with mean $m$ and variance $v$; with a vector mean $m$ and a covariance
matrix $V$, $\Normal(m,V)$ is the multivariate normal distribution. The gamma function $\Gamma(a)=\int_0^\infty
u^{a-1}e^{-u}\,\mathrm{d}u$ is the normalizing constant of that density. The upper incomplete
gamma function $\Gamma(a,x)=\int_x^\infty u^{a-1}e^{-u}\,\mathrm{d}u$ appears once, in the survival
function of the generalized gamma distribution (Section~\ref{sec:families}).

\begin{table}[h]
\centering\small
\caption{Symbols Used in More Than One Section.}\label{tab:notation}
\footnotesize\setlength{\tabcolsep}{4pt}
\begin{tabular}{@{}lp{11.2cm}@{}}
\toprule
Symbol & Meaning (units; index) \\
\midrule
$i$, $g$, $y$, $b$, $c$, $s$ & a drive or grouped record; a drive model; a calendar year; an age bin; a cell of the Lexis table (drive model, month, bin); a SMART state \\
$a$, $a_0$, $a_1$ & power-on age; age at entry (first sighting, or the start of a training window); age at exit (last day, failure, or window end). Years, except in the Lexis table, where bins are in whole days \\
$\Delta t$ & one day, $1/365.25$ year \\
$\delta_i$, $w_i$ & failure indicator of record $i$; multiplicity of a grouped record \\
$h$, $H$, $S$ & hazard, cumulative hazard, and survival function of a lifetime distribution; $h_0,H_0$ the baseline distribution \\
$\beta_j$, $\eta_j$ & Weibull exponent of component $j$ (dimensionless) and scale (years); $\beta<1$, $=1$, $>1$: a hazard that falls, is constant, rises with age \\
$c_1$, $k_2$, $a_\star$ & $\eta_1^{-\beta_1}$; the wear-out cumulative hazard at the pivot age $a_\star=4$ yr \\
$Z_g$, $\mu_g$, $\gamma$ & hazard multiplier of drive model $g$; its prior mean; the coefficients of $\log\mu_g$ (vintage, log capacity, manufacturer, helium) \\
$\alpha$, $\phi=1/\alpha$ & gamma precision of $Z_g$ about $\mu_g$; $\sqrt\phi$ is the coefficient of variation of $Z_g$ about $\mu_g$ \\
$E$, $M_g$ & exposure in drive-years; exposure of model $g$ in units of baseline cumulative hazard (hazard-weighted exposure) \\
$D_g$, $d_c$, $\mu_c$, $m_c$ & failures of model $g$; observed count, expected count, and baseline expectation (exposure times mean baseline hazard) of cell $c$ \\
$Q_y$, $s_y=\log Q_y$ & fleet-wide hazard multiplier of calendar year $y$ ($Q_{2013}=1$ under the random-walk processes; random under the independent and AR(1) ones); $\sigma_Q$, $d$, $\rho$ its step standard deviation, drift, and AR(1) coefficient \\
$U_{gy}$, $\kappa_{gy}$, $\kappa_0$, $\xi$ & shock to model $g$ in year $y$ (mean 1); its gamma precision $\kappa_0 m_{gy}^{\xi}$, with $m_{gy}$ the model-year's expected failures \\
$\psi_s(g)$ & hazard multiplier of SMART state $s$ for drive model $g$ (Appendix~\ref{sec:ext}) \\
$\lambda_1$, $q$, $\lambda_2$ & rates of the two-state hidden Markov model (per year) \\
$\omega$ & share fraction of the fixed-share algorithm (Appendix~\ref{sec:track}) \\
$L$, $\tau$ & landmark date; forecast horizon in days (90 or 365) \\
$p_i$, $y_i$ & a drive's forecast probability of failing within $\tau$; its outcome (0 or 1; $y$ without a subscript is a calendar year) \\
\midrule
\multicolumn{2}{@{}p{15.4cm}@{}}{\footnotesize Individual lifetime distributions define their parameters locally (Section~\ref{sec:families}). The gamma posterior uses shape $a_g$ and rate $b_g$. The one-day cumulative-hazard increment $\Delta_i$ (Section~\ref{sec:hier}) differs from the SMART changes $\Delta_1,\Delta_3$ (Section~\ref{sec:boost}). A cell's baseline expectation $m_c$ excludes the multipliers $Q_y,Z_g$; a model-year's expected count $m_{gy}$ includes them.} \\
\bottomrule
\end{tabular}
\end{table}

\subsection{Sources of Method}

The left-truncated, right-censored likelihood is standard~\cite{turnbull1976,meeker2022smrd,lawless2003}.
Gamma frailty goes back to Vaupel, Manton, and Stallard~\cite{vaupel1979}, and its shared form
to Clayton~\cite{clayton1978}. See also Hougaard~\cite{hougaard2000} and Duchateau and
Janssen~\cite{duchateau2008}. Proschan~\cite{proschan1963} showed that a mixture of units with
different constant rates has a decreasing hazard, which is one reason a pooled fleet shows
apparent infant mortality. The equivalence between a survival likelihood with piecewise-constant
hazard and a Poisson likelihood on an age-by-period table is shown by Holford~\cite{holford1980}
and applied to log-linear models by Laird and Olivier~\cite{laird1981}, who credit its
theoretical basis to Holford and, independently, to Aitkin and Clayton. Age, calendar year, and a drive's service-start cohort are linearly dependent,
so their linear trends cannot be separated without assumptions~\cite{holford1983,clayton1987apc}.
This matters in Section~\ref{sec:apc}. Empirical-Bayes shrinkage of failure rates (each group's estimate pulled toward the population's, by an
amount the data set; Appendix~\ref{sec:primereb}) with a gamma-Poisson model
is in Gaver and O'Muircheartaigh~\cite{gaver1987}. We compare forecasts with the logarithmic
score, a proper scoring rule~\cite{gneiting2007,czado2009}, on a rolling origin~\cite{tashman2000}.
Our Bayesian fits use the No-U-Turn sampler~\cite{hoffman2014nuts} in NumPyro~\cite{phan2019numpyro},
with split-$\hat R$ and effective sample size as computed by NumPyro~\cite{gelman2013bda} ($\hat R$
compares the spread between chains with the spread within them and is 1 when they agree; the
effective sample size is the number of independent draws a correlated chain is worth;
Appendix~\ref{sec:primerbayes}). The minimum of
independent Weibull draws is the standard way to sample a competing-risk
distribution~\cite{devroye1986}.

\section{How Failure Risk Changes with Drive Age}\label{sec:families}

We first treat the fleet as one population, fit fourteen lifetime distributions to it and evaluate
the fixed three-component benchmark, pooled over all years and year by year. Section~\ref{sec:hier} then separates the drive models.

\subsection{The Lifetime Models We Compare}

We fit fourteen distributions to the whole fleet by maximizing \eqref{eq:lldisc}, with age in years,
and evaluate the fixed benchmark, which has no estimated parameters, by the same likelihood. Writing
$H_W(t;\beta,\eta) = (t/\eta)^\beta$:
\begin{itemize}[leftmargin=1.4em]
\item \emph{Exponential, Weibull, gamma, lognormal}.
\item \emph{Gompertz}: $h(t) = a e^{bt}$, $H(t) = a(e^{bt}-1)/b$.
\item \emph{Log-logistic}: $S(t) = [1+(t/\alpha)^{k}]^{-1}$.
\item \emph{Generalized gamma} (Stacy): $f(t) = \dfrac{p}{a^{d}\Gamma(d/p)}\,t^{d-1}e^{-(t/a)^{p}}$,
  $S(t) = \Gamma\bigl(d/p,(t/a)^{p}\bigr)/\Gamma(d/p)$, with $\Gamma(\cdot,\cdot)$ the upper
  incomplete gamma function and $\Gamma(\cdot)$ the gamma function (Section~\ref{sec:notation}).
\item \emph{Exponentiated Weibull}: $F(t) = \bigl[1-e^{-H_W(t;\beta,\eta)}\bigr]^{\theta}$.
\item \emph{$k$ competing Weibulls} ($k=2,3$): a drive fails at the first of $k$
  independent Weibull mechanisms, so hazards add:
  $H(t) = \sum_{j=1}^{k} H_W(t;\beta_j,\eta_j)$ with $\beta_1<\dots<\beta_k$.
\item \emph{Mixture of two Weibulls}: a fraction $p$ of drives follow one Weibull and the rest
  another, so survival functions average: $S(t) = p\,e^{-H_W(t;\beta_1,\eta_1)} + (1-p)\,e^{-H_W(t;\beta_2,\eta_2)}$.
\item \emph{Piecewise constant hazard} on age-years $0$--$1,\dots,9$--$10$, $10+$ (11 rates), as a
  flexible reference.
\item \emph{Three forms of the bathtub of Section~\ref{sec:elerathlit}}: its default constants with
  nothing estimated, the same with one common scale factor $s$ multiplying all three $\eta$,
  and its three exponents $(0.5, 1, 3)$ with all three scales free.
\end{itemize}
We start multi-parameter fits from three or four points and keep the best.

\subsection{Fitting One Lifetime Model to the Whole Fleet}

This comparison describes the whole fleet as recorded through 2025, with one interval per
drive and all years pooled (489{,}184 drives, 31{,}000 failures, 1{,}884{,}831 drive-years on the power-on scale; 489{,}633 drives, 31{,}021 failures on the first-seen scale
). It includes about 8{,}000
solid-state and small boot drives that the models from Section~\ref{sec:hier} on exclude. The
held-out comparisons of Section~\ref{sec:valid} and Appendix~\ref{sec:laws} use the data drives
only.

\begin{table}[h]
\centering\small
\caption{Fourteen Distributions Fitted to the Pooled Fleet, and the Fixed Benchmark. Rows are ordered by AIC on power-on age. $\Delta$AIC is relative to the best distribution on the same age scale.}\label{tab:cohort}
\begin{tabular}{@{}lrrrrr@{}}
\toprule
& & \multicolumn{2}{c}{Power-on age} & \multicolumn{2}{c}{First-seen age} \\
Distribution & $k$ & $\ell$ & $\Delta$AIC & $\ell$ & $\Delta$AIC \\
\midrule
Mixture of 2 Weibulls & 5 & $-339{,}878.7$ & 0.0 & $-340{,}756.3$ & 0.0 \\
2 competing Weibulls & 4 & $-339{,}917.7$ & 75.9 & $-340{,}759.1$ & 3.6 \\
3 competing Weibulls & 6 & $-339{,}917.7$ & 79.9 & $-340{,}759.1$ & 7.6 \\
Piecewise constant (age-years) & 11 & $-340{,}229.4$ & 713.3 & $-340{,}789.1$ & 77.5 \\
Gompertz & 2 & $-340{,}448.3$ & 1{,}133.2 & $-341{,}001.4$ & 484.1 \\
3 competing, exponents 0.5/1/3, free scales & 3 & $-340{,}466.4$ & 1{,}171.4 & $-340{,}904.2$ & 291.7 \\
Exponentiated Weibull & 3 & $-340{,}696.0$ & 1{,}630.6 & $-341{,}225.5$ & 934.3 \\
Generalized gamma & 3 & $-340{,}696.0$ & 1{,}630.6 & $-341{,}225.5$ & 934.3 \\
Weibull & 2 & $-340{,}780.8$ & 1{,}798.1 & $-341{,}324.4$ & 1{,}130.1 \\
Gamma & 2 & $-340{,}827.3$ & 1{,}891.2 & $-341{,}349.9$ & 1{,}181.1 \\
Log-logistic & 2 & $-340{,}879.2$ & 1{,}994.9 & $-341{,}435.4$ & 1{,}352.2 \\
Exponential & 1 & $-341{,}253.1$ & 2{,}740.8 & $-341{,}492.7$ & 1{,}464.8 \\
Bathtub constants, one free scale & 1 & $-341{,}429.5$ & 3{,}093.6 & $-342{,}305.5$ & 3{,}090.3 \\
Lognormal & 2 & $-342{,}651.2$ & 5{,}539.0 & $-343{,}274.1$ & 5{,}029.6 \\
Bathtub constants, fixed & 0 & $-350{,}282.9$ & 20{,}798.3 & $-351{,}554.9$ & 21{,}587.1 \\
\bottomrule
\end{tabular}
\end{table}

\begin{itemize}[leftmargin=1.4em]
\item Distributions with two components are far ahead. On power-on age the mixture is best: 28\% of
  drives have $\beta = 1.67$ and $\eta = 10.9$~yr, and the remainder a slowly decreasing hazard,
  $\beta = 0.30$. The two competing Weibulls are 76 AIC units behind, with $\beta_1 = 0.23$ for
  early life and $\beta_2 = 1.459 \pm 0.012$, $\eta_2 = 30.6 \pm 0.4$~yr for wear-out. Given the
  dependence documented in Section~\ref{sec:time}, a 76-unit difference does not separate the
  two distributions. Below, we also call two competing Weibulls the two-Weibull distribution; the failure time is the first of two competing times, not the end of successive stages. We do not carry
  the mixture further, because a mixture over drives is what the drive-model effect of
  Section~\ref{sec:hier} models directly.
\item A third competing Weibull does not improve the pooled likelihood at all: the optimizer
  drives the extra component to zero. Year by year a third component occasionally helps. It is
  the AIC-best distribution in 2018, by 8 units over two components.
\item The generalized gamma and the exponentiated Weibull reach the same likelihood because
  both run to the same boundary of their parameter space (a very large inner power). Their
  standard errors are not meaningful, and they are one fit, not two.
\item The single Weibull is 1{,}798 AIC units behind, and even the 11-rate piecewise model is
  713 behind the mixture of two Weibulls and 637 behind the competing pair.
\item The bathtub with default constants is last, 20{,}798 units behind and far worse than a
  constant hazard. Given one free scale it prefers all three characteristic lives 52\% longer
  and is still behind the single Weibull. Keeping only its exponents $(0.5,1,3)$ and freeing the
  three scales brings it to 1{,}171 behind, with a wear-out scale of 21.7 years
  instead of the default $120{,}000\ \text{h} = 13.7$ years.
\end{itemize}

\begin{figure}[h]
\centering
\includegraphics[width=\textwidth]{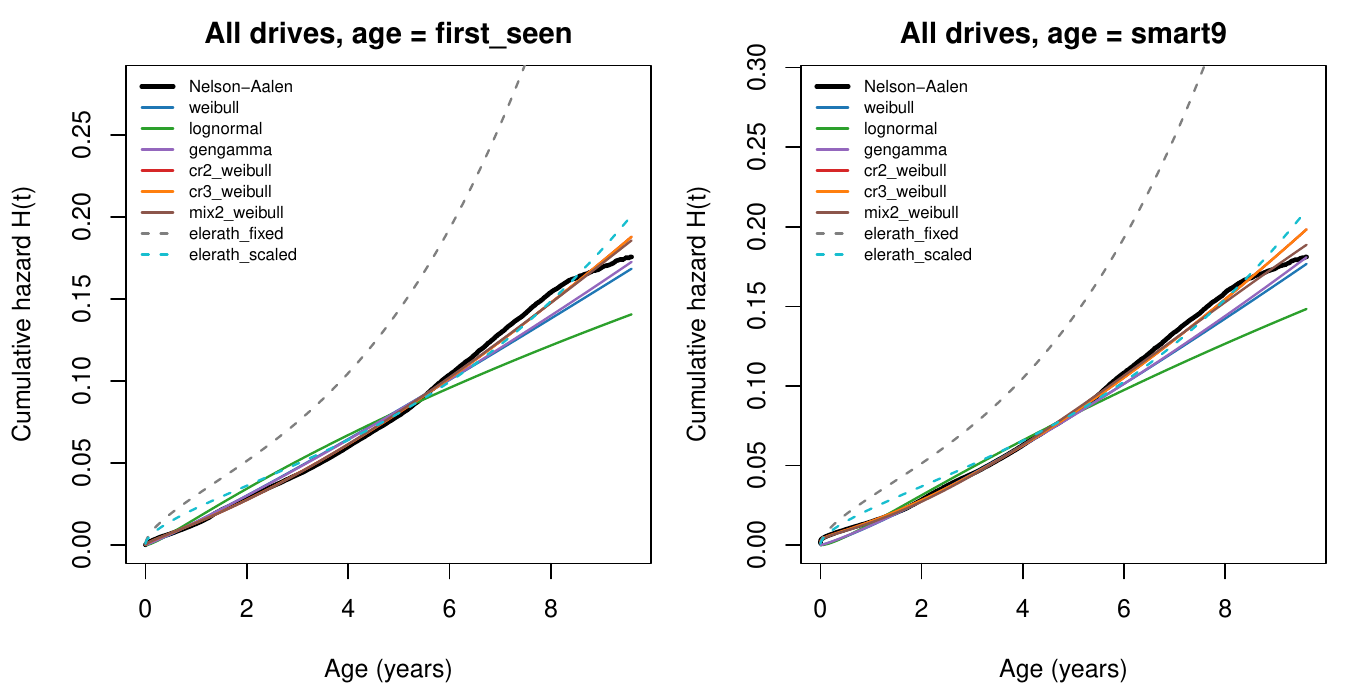}
\caption{Cumulative hazard of the pooled fleet: Nelson--Aalen estimate (black; ages with at least
1{,}000 drives at risk) and fitted distributions. Left: age since first sighting. Right: power-on age. The
dashed grey curve is the bathtub with default constants.}\label{fig:cumhaz}
\end{figure}

The nonparametric reference in Figure~\ref{fig:cumhaz} is the Nelson--Aalen estimator with left
truncation, $\hat H(t) = \sum_{u\le t} d_u / Y_u$, where $d_u$ is the number of failures at age
$u$ days and $Y_u = \#\{i: a_{0i} < u \le a_{1i}\}$ the number at risk.

\subsection{Year by Year}

Fitted to one calendar year at a time (Appendix~\ref{sec:fleetyear}), a single Weibull on
power-on age has exponent between about 1.0 and 1.6. The fits fall into three regimes. In 2014
the Seagate ST3000DM001 dominates the failures (exponent 1.42, scale under ten years). From 2015
to 2021 the fitted exponent is close to 1, a nearly constant hazard. From 2022 the exponents are 1.35 to 1.57 as the fleet
ages.
A single Weibull is never the AIC-best distribution in any year, and the default bathtub is worse than
it in every year but 2015: its constants describe drives of the mid-2010s tolerably and
current drives poorly.

\section{Accounting for Differences Between Drive Models}\label{sec:hier}

The fleet consists of about a hundred drive models, deployed at different times in different
numbers, with very different failure rates. A lifetime distribution fit to the pool describes the
mixture that Backblaze happened to own. That is why its parameters drift from year to year
(Table~\ref{tab:periodweibull} in Appendix~\ref{sec:fleetyear}). A model meant to be used elsewhere should separate
what is common (how hazard depends on age) from what is specific (the failure level of a given
drive model), and should say how uncertain the second part is for a drive model with little or no
history.

The population is the data drives of Section~\ref{sec:population}: 504{,}624 drives of 99 drive models, 31{,}881 failures, 2{,}023{,}457 drive-years
. We treat two drive models (five drives)
from manufacturers other than the four named as Seagate.

\subsection{Model}

Appendix~\ref{sec:primer} (\ref{sec:primerre} and~\ref{sec:primereb}) explains
random effects, shrinkage and empirical Bayes.

For a drive of model $g$ at power-on age $a$ (years),
\begin{equation}
\begin{aligned}
h(a \mid g) = Z_g\,h_0(a),\qquad
H_0(a) = \Bigl(\frac{a}{\eta_1}\Bigr)^{\beta_1} + \Bigl(\frac{a}{\eta_2}\Bigr)^{\beta_2},\\
Z_g \sim \Gam\bigl(\alpha,\ \alpha/\mu_g\bigr)\quad(\text{shape }\alpha,\ \text{rate }\alpha/\mu_g),
\end{aligned}
\label{eq:hier}
\end{equation}
so that $\E[Z_g] = \mu_g$ and $\mathrm{Var}(Z_g) = \phi\,\mu_g^2$ with $\phi = 1/\alpha$
(the shape--rate convention is fixed in Section~\ref{sec:notation}). The
baseline $h_0$ is the two-Weibull distribution (two competing Weibulls) of Section~\ref{sec:families}. The multiplier $Z_g$
scales that baseline hazard for drive model $g$. Its prior mean depends on what is known about a model before any
failure is seen:
\begin{equation}
\log\mu_g = \gamma_{\text{intro}}\,(y_g - 2018) + \gamma_{\text{cap}}\,\log\frac{C_g}{8\ \text{TB}}
          + \gamma_{\text{HGST}}\mathbf{1}_{\text{HGST}} + \gamma_{\text{Toshiba}}\mathbf{1}_{\text{Toshiba}} + \gamma_{\text{WDC}}\mathbf{1}_{\text{WDC}}
          + \gamma_{\text{He}}\mathbf{1}_{\text{He}},
\end{equation}
with $y_g$ the (fractional) year model $g$ first appears, $C_g$ its capacity, and Seagate the
reference manufacturer. HGST includes the WUH Ultrastars, and $\mathbf{1}_{\text{He}}$ marks helium
drives (Section~\ref{sec:population}). The reference drive model ($\mu = 1$) is thus an 8~TB
air-filled Seagate introduced in 2018.

\subsection{Likelihood, Posterior, and Predictive Distribution}

Condition on $Z_g = z$. With failures known to the day ($\Delta t$ one day, ages in years), a
failure at age $a_1$ after entry at $a_0$ has probability
\begin{align}
e^{-z[H_0(a_1-\Delta t)-H_0(a_0)]}\bigl(1-e^{-z\,\Delta_i}\bigr)
 &= z\,\Delta_i\;e^{-z[H_0(a_1-\Delta t)-H_0(a_0)+\Delta_i/2]}\,\bigl(1+O(z^2\Delta_i^2)\bigr), \\
\Delta_i &= H_0(a_1)-H_0(a_1-\Delta t), \notag
\end{align}
using $1-e^{-x} = x\,e^{-x/2}(1+x^2/24+\dots)$. In the fitted baseline the one-day increment
$\Delta_i$ is 0.0046 on the first day of life, 0.0010 on the second, and about $10^{-4}$ from
the second month on. At the largest posterior-mean multiplier ($z = 23.6$, a 1.5~TB Seagate
model) the relative correction on the first day is $5\times10^{-4}$. At the posterior-mean
multipliers, even placing every failure on its first day gives a total log-likelihood
correction below 0.2 units. This is a check at the fitted values, not a uniform bound over the
gamma posterior's unbounded support. We neglect the correction. A censored interval has probability
$e^{-z[H_0(a_1)-H_0(a_0)]}$. Hence for drive model $g$
\begin{equation}
L_g(z) = z^{D_g}\,e^{-z M_g}\prod_{i\in g}\Delta_i^{w_i\delta_i},\qquad
D_g = \sum_{i\in g} w_i\delta_i,\qquad
M_g = \sum_{i\in g} w_i\bigl[\tilde H_0(a_{1i}) - H_0(a_{0i})\bigr],
\end{equation}
where the product runs over the grouped records of model $g$ with their multiplicities $w_i$,
and $\tilde H_0(a_1)$ is $H_0(a_1)$ for a censored interval and
$\tfrac12[H_0(a_1-\Delta t)+H_0(a_1)]$ for a failure. $D_g$ is the number of failures and $M_g$ the
exposure of model $g$ measured in units of baseline cumulative hazard. This is the kernel of a
gamma density in $z$, so everything that follows is closed-form.

\emph{Marginal likelihood.} Integrating $z$ against the prior in \eqref{eq:hier},
\begin{equation}
\ell(\theta) = \sum_g\Bigl[\sum_{i\in g} w_i\delta_i\log\Delta_i + D_g\log\mu_g
   + \log\Gamma(D_g+\alpha) - \log\Gamma(\alpha) + \alpha\log\alpha - (D_g+\alpha)\log(\mu_g M_g+\alpha)\Bigr],
\label{eq:marg}
\end{equation}
which is maximized over $\theta = (\beta_1,\eta_1,\beta_2,\eta_2,\phi,\gamma)$: eleven parameters
with the six covariates, including helium.

\emph{Posterior for a drive model.} Given the data,
\begin{equation}
Z_g \mid \text{data} \sim \Gam\bigl(\alpha + D_g,\ \alpha/\mu_g + M_g\bigr),\qquad
\E[Z_g\mid\text{data}] = \frac{\alpha + D_g}{\alpha/\mu_g + M_g}.
\label{eq:post}
\end{equation}
A model with much exposure is estimated by its own failure rate $D_g/M_g$. A model with little
is pulled toward the prior mean $\mu_g$. A model never seen has the prior itself.

\emph{Predictive distribution.} Let future exposure of model $g$ be divided into cells $c$ with
baseline expectation $m_c$ (exposure times mean baseline hazard), and let $Z_g$ currently have a
$\Gam(a,b)$ distribution (posterior or prior). The future counts $d_c$ are jointly
\begin{equation}
p(d \mid a,b) = \prod_c\frac{m_c^{d_c}}{d_c!}\;\frac{\Gamma(a+D')}{\Gamma(a)}\;\frac{b^{a}}{(b+M')^{a+D'}},
\qquad D' = \sum_c d_c,\quad M' = \sum_c m_c,
\label{eq:pred}
\end{equation}
a negative-multinomial distribution. It differs from independent Poisson counts at the posterior mean by
carrying the uncertainty in $Z_g$, which is large for young or new models.

\emph{A caveat on truncation.} The conditional likelihood given $Z_g$ handles each drive's left
truncation correctly. Integrating over the prior of $Z_g$ assumes that which drive models
appear in the corpus, and in what numbers, does not depend on $Z_g$. That need not hold if
purchasing responds to early results. The likelihood does not model such a response
(Section~\ref{sec:time}).

\subsection{Estimates}

The model with the two-Weibull baseline and covariate-dependent gamma frailty (the per-model
hazard multiplier $Z_g$, a \emph{frailty} in the survival-analysis sense; Appendix~\ref{sec:primerre})
is the \emph{empirical-Bayes model} of the rest of the paper. In the validation tables we label it M2 + covariates. M0 and
M1 are the same model with a constant and a single-Weibull baseline.

\begin{table}[h]
\centering\small
\caption{Nested Models on the Same Data. The data are the HDD data drives through 30~June 2026. $\ell$ is the marginal log-likelihood \eqref{eq:marg}.}\label{tab:finalfits}
\begin{tabular}{@{}lrrr@{}}
\toprule
Model & Parameters & $\ell$ & AIC \\
\midrule
Constant hazard, pooled & 1 & $-352{,}320.5$ & 704{,}643.0 \\
Constant hazard + frailty & 2 & $-343{,}497.8$ & 686{,}999.7 \\
Constant hazard + covariate frailty & 8 & $-343{,}456.7$ & 686{,}929.4 \\
Weibull, pooled & 2 & $-351{,}765.4$ & 703{,}534.8 \\
Weibull + frailty & 3 & $-343{,}067.7$ & 686{,}141.3 \\
Weibull + covariate frailty & 9 & $-343{,}027.1$ & 686{,}072.2 \\
Two competing Weibulls, pooled & 4 & $-350{,}900.6$ & 701{,}809.2 \\
Two competing Weibulls + frailty & 5 & $-342{,}090.4$ & 684{,}190.8 \\
Two competing Weibulls + covariate frailty & 11 & $-342{,}048.0$ & 684{,}118.0 \\
\bottomrule
\end{tabular}
\end{table}

Adding frailty improves log likelihood by about 8{,}700 to 8{,}820 units across these baselines.
With frailty included, replacing a constant hazard with a Weibull gains about 430 units, and
replacing that Weibull with two competing Weibulls gains about 980. Adding the six covariate
coefficients gains about 42. Each step improves AIC, which here assumes independence
(Section~\ref{sec:time}).

\begin{table}[h]
\centering\small
\caption{Estimates of the Empirical-Bayes Model. The model has a two-Weibull baseline (two competing Weibulls) and a gamma frailty with covariate-dependent mean, fitted to data through 30~June 2026.}\label{tab:finalpar}
\begin{tabular}{@{}lrrl@{}}
\toprule
Parameter & Estimate & Std.\ error & Meaning \\
\midrule
$\beta_1$ & 0.28439 & 0.0186 & early-life exponent \\
$\eta_1$ (yr) & 4.5662e+05 & 6.55e+05 & early-life scale (weakly identified) \\
$\beta_2$ & 1.5713 & 0.0161 & wear-out exponent \\
$\eta_2$ (yr) & 14.011 & 2.95 & wear-out scale, reference drive model \\
$\phi$ & 0.75943 & 0.138 & frailty variance relative to squared mean \\
$\gamma_{\text{intro}}$ & 0.079574 & 0.061 & per year of introduction after 2018 \\
$\gamma_{\text{cap}}$ & -1.4537 & 0.357 & per unit of $\log(\text{TB}/8)$ \\
$\gamma_{\text{HGST}}$ & -1.6169 & 0.266 & HGST line, including WUH, relative to Seagate \\
$\gamma_{\text{Toshiba}}$ & -1.0434 & 0.322 & Toshiba relative to Seagate \\
$\gamma_{\text{WDC}}$ & -0.77794 & 0.432 & WD-native drives relative to Seagate \\
$\gamma_{\text{He}}$ & 0.46364 & 0.462 & helium relative to air \\
\bottomrule
\end{tabular}
\end{table}

\begin{itemize}[leftmargin=1.4em]
\item \emph{Wear-out} has Weibull exponent $\beta_2=1.57$: hazard grows roughly as
  age$^{0.57}$, a much gentler wear-out than the bathtub's exponent of 3.
\item \emph{Early life} has Weibull exponent $\beta_1=0.28$: a hazard that falls steeply
  over the first weeks. These data determine its scale $\eta_1$ weakly (the standard error is
  larger than the estimate), though it is not structurally non-identifiable. The early-life
  cumulative hazard is $H_{\text{early}}(a) = c\,a^{\beta_1}$ with
  $c = \eta_1^{-\beta_1} = 0.025$. The early-life hazard
  exceeds the wear-out hazard until age 0.37 years (4.5 months); the cumulative terms cross at 1.4 years.
\item \emph{Capacity}: $\gamma_{\text{cap}} = -1.45$ means doubling capacity multiplies the prior
  mean hazard by $2^{-1.45} = 0.37$, holding the other covariates fixed.
\item \emph{Manufacturer}, relative to Seagate at equal capacity, vintage and helium status:
  HGST $0.20$, Toshiba $0.35$, WDC $0.46$.
\item \emph{Helium}: $\gamma_{\text{He}}=0.464$ corresponds to a multiplier of $1.59$ at
  fixed manufacturer, capacity and vintage. Its standard error is $0.462$; the coefficient
  is not distinguishable from zero.
\item \emph{Vintage}: $+8\%$ per year ($e^{0.080}$), not distinguishable from zero.
\item \emph{Residual spread}: the gamma precision is $\alpha=1/\phi$ with $\phi = 0.76$, a
  coefficient of variation of $\sqrt{\phi} = 0.87$ between drive models that share manufacturer,
  capacity, vintage and helium status. Without covariates
  $\phi = 1.89$ (CV 1.37), so the covariates reduce the residual dispersion $\phi$ of the
  multipliers about their predicted mean by 60\%. The total variance of $Z_g$ also includes the
  spread of $\mu_g$ across drive models.
\end{itemize}

\begin{table}[h]
\centering\small
\caption{Hazard Multipliers of the 22 Drive Models with the Most Failures. Prior is $\mu_g$ from the covariates, including helium. Posterior is the mean of \eqref{eq:post}, with its 95\% interval. The full table covers 99 harmonized drive models.}\label{tab:models}
\footnotesize\setlength{\tabcolsep}{4pt}\begin{tabular}{@{}llrrrrrrr@{}}
\toprule
Drive model & Maker & TB & Intro & Drives & Failures & Prior & Posterior & 95\% interval \\
\midrule
ST4000DM000 & Seagate & 4 & 2013.3 & 37{,}040 & 5{,}790 & 1.89 & 0.487 & 0.475--0.500 \\
ST12000NM0008 & Seagate & 12 & 2019.8 & 21{,}036 & 2{,}610 & 1.02 & 0.425 & 0.409--0.442 \\
ST8000NM0055 & Seagate & 8 & 2016.9 & 15{,}680 & 2{,}514 & 0.92 & 0.338 & 0.325--0.352 \\
ST12000NM0007 & Seagate & 12 & 2017.7 & 38{,}842 & 2{,}303 & 0.86 & 0.576 & 0.553--0.600 \\
TOSHIBA MG07ACA14TA & Toshiba & 14 & 2018.4 & 39{,}386 & 2{,}290 & 0.26 & 0.220 & 0.211--0.229 \\
HGST HUH721212ALN604 & HGST & 12 & 2018.5 & 11{,}438 & 1{,}857 & 0.18 & 0.458 & 0.438--0.479 \\
ST3000DM001 & Seagate & 3 & 2013.3 & 4{,}707 & 1{,}708 & 2.86 & 6.872 & 6.550--7.202 \\
ST8000DM002 & Seagate & 8 & 2016.4 & 10{,}319 & 1{,}415 & 0.88 & 0.251 & 0.239--0.265 \\
HGST HUH721212ALE604 & HGST & 12 & 2020.5 & 14{,}345 & 1{,}222 & 0.21 & 0.380 & 0.359--0.401 \\
TOSHIBA MG08ACA16TA & Toshiba & 16 & 2020.1 & 41{,}061 & 1{,}215 & 0.24 & 0.271 & 0.256--0.286 \\
ST16000NM001G & Seagate & 16 & 2019.8 & 36{,}022 & 870 & 0.67 & 0.156 & 0.146--0.167 \\
ST14000NM001G & Seagate & 14 & 2020.7 & 11{,}268 & 794 & 0.87 & 0.293 & 0.273--0.314 \\
ST12000NM001G & Seagate & 12 & 2020.4 & 13{,}873 & 717 & 1.07 & 0.208 & 0.193--0.223 \\
TOSHIBA MG08ACA16TEY & Toshiba & 16 & 2020.8 & 5{,}352 & 549 & 0.26 & 0.529 & 0.485--0.574 \\
WDC WUH721816ALE6L4 & HGST & 16 & 2022.1 & 27{,}308 & 474 & 0.16 & 0.141 & 0.129--0.154 \\
ST14000NM0138 & Seagate & 14 & 2020.9 & 1{,}690 & 463 & 0.89 & 1.254 & 1.142--1.371 \\
HGST HMS5C4040BLE640 & HGST & 4 & 2014.2 & 16{,}348 & 455 & 0.40 & 0.072 & 0.066--0.079 \\
WDC WUH722222ALE6L4 & HGST & 22 & 2023.3 & 46{,}039 & 421 & 0.11 & 0.166 & 0.151--0.182 \\
ST31500541AS & Seagate & 2 & 2013.3 & 2{,}188 & 397 & 7.82 & 1.476 & 1.335--1.625 \\
TOSHIBA MG08ACA16TE & Toshiba & 16 & 2021.7 & 6{,}664 & 340 & 0.27 & 0.275 & 0.246--0.304 \\
ST10000NM0086 & Seagate & 10 & 2017.7 & 1{,}304 & 328 & 1.12 & 0.569 & 0.509--0.632 \\
ST24000NM002H & Seagate & 24 & 2025.0 & 9{,}900 & 302 & 0.56 & 0.885 & 0.788--0.988 \\
\bottomrule
\end{tabular}
\end{table}

\begin{figure}[h]
\centering
\includegraphics[width=\textwidth]{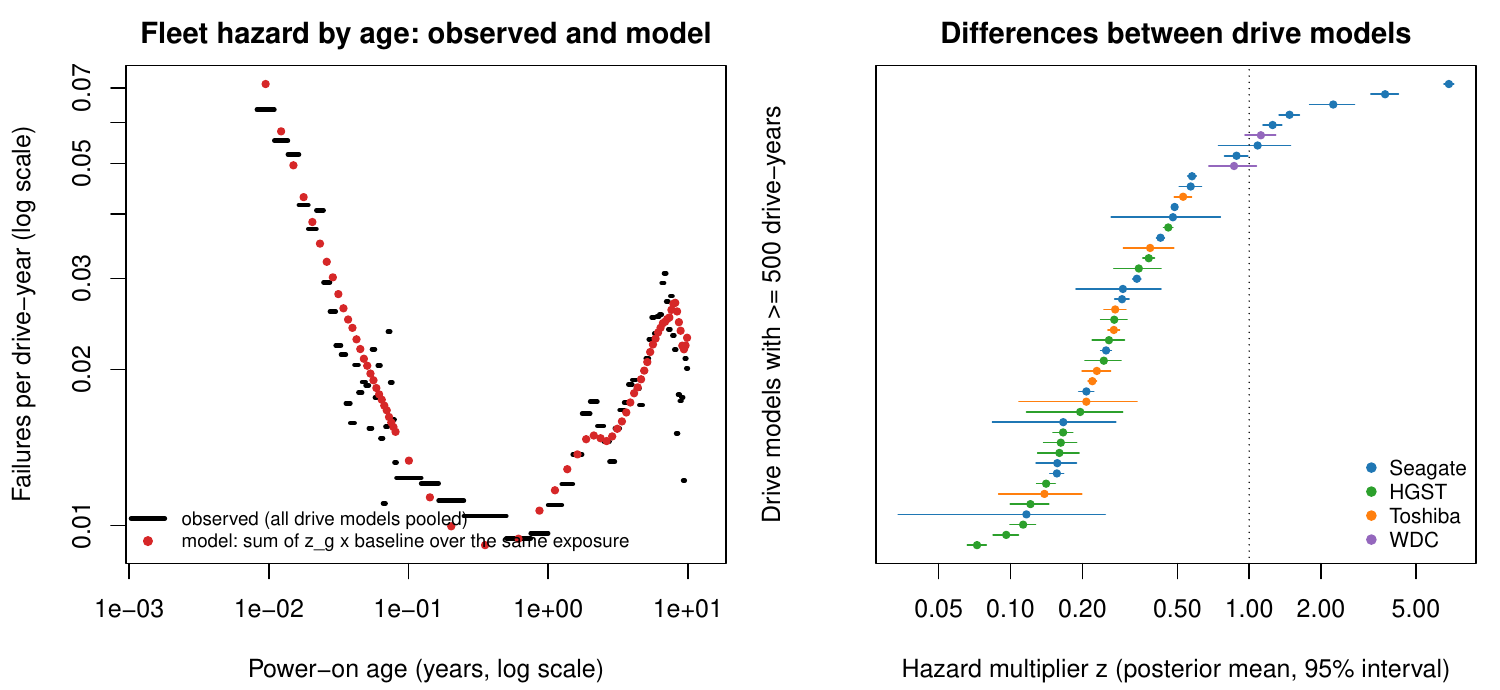}
\caption{Left: observed fleet hazard by power-on age (black bars, one per age bin; log scales)
and the empirical-Bayes model's expectation for the same exposure (red); bins with under 300
drive-years or no failures are not drawn. Table~\ref{tab:age} gives the same comparison by age
band. Right:
posterior hazard multipliers of drive models with at least 500 drive-years.}\label{fig:model}
\end{figure}

Among drive models with at least 500 drive-years the posterior multipliers run from $0.072$
(HGST HMS5C4040BLE640) to $6.9$ (ST3000DM001), a factor of about 95
(Figure~\ref{fig:model}, right). For scale, a multiplier
$z$ gives an effective wear-out scale $\eta_2 z^{-1/\beta_2}$: 74.5 years at $z = 0.0724$,
14.0 years at $z = 1$, and 4.1 years at $z = 6.87$. The fleet-year Weibull scales in
Table~\ref{tab:periodweibull} in Appendix~\ref{sec:fleetyear} (10 years in 2014, 25 to 86 years otherwise) are averages over
whichever models were in service that year.

\textbf{What is and is not identified.} The hazard $Z_g h_0(a)$ of a drive model is identified.
The split between the multiplier and the baseline scale is fixed only by the convention that
$Z$ has prior mean $\mu_g$. In an earlier simulation without the helium covariate, we generated
twelve replicates of 120 drive models and about 510{,}000 drives with known parameters and
refitted them. That test detected no
bias at its resolution of about a third of a standard error, and the reported standard errors
match the spread of estimates (standardized errors have standard deviation 0.8 to 1.2). The
95\% interval for the cumulative hazard $z_g H_0(5\,\text{yr})$ of a large drive model covers the
truth 95\% of the time. The interval for $z_g$ alone covers only 48\% of the time, because it is
conditional on the estimated baseline scale. The intervals in Table~\ref{tab:models} are therefore intervals
for a model's hazard relative to the fitted baseline, not for $Z_g$ in isolation.

Twelve of the sixteen drive models with at least 10{,}000 drives have a posterior mean below
their prior mean. A gamma distribution with this precision puts 62\% of its mass below its mean, so
that count is unremarkable.

We also fitted a single Weibull to each of the 21 drive models with at least 300 failures
separately (through 2025-12-31, power-on age). The exponents run from 0.57 to 4.04. For the ST3000DM001 the fit on its
full history (4{,}707 drives, 1{,}708 failures) is $\hat\beta = 4.04$, $\hat\eta = 2.93$~yr.

\subsection{Drive Models Do Not Share One Age Curve}\label{sec:permodel}

\begin{figure}[h]
\centering
\includegraphics[width=.98\textwidth]{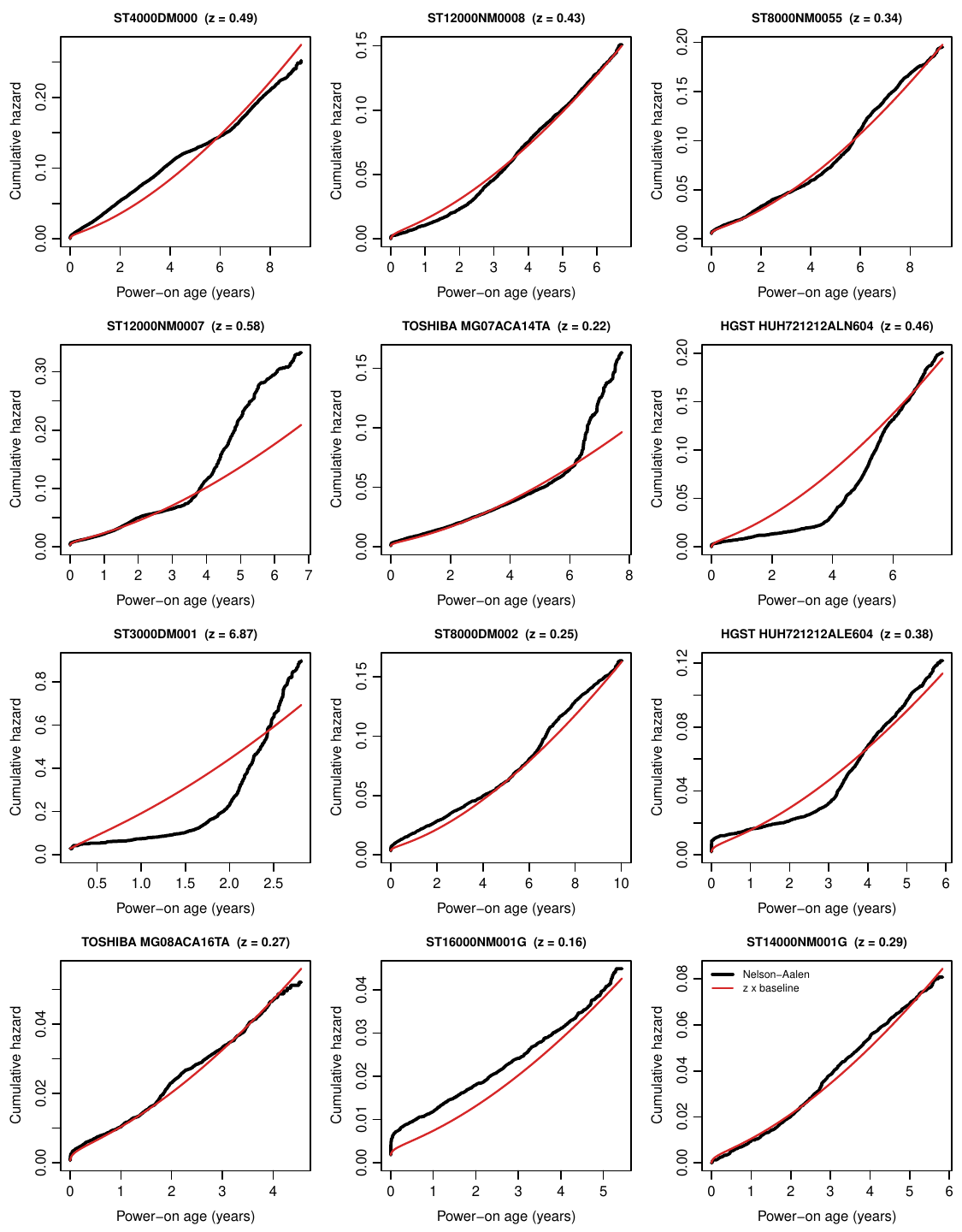}
\caption{The twelve drive models with the most failures: Nelson--Aalen cumulative hazard on
power-on age (black; ages with at least 200 drives at risk) and the empirical-Bayes model's
$z_g H_0(a)$ (red).}\label{fig:permodel}
\end{figure}

In Figure~\ref{fig:permodel} the common age curve times one number fits about half of the
large models well. For the others the cumulative hazard rises sharply after several years
of few failures (ST12000NM0007 near age 4, HGST HUH721212ALN604 near age 4,
Toshiba MG07ACA14TA near age 6, ST3000DM001 near age 2). Multiplying this common baseline by
a constant cannot reproduce these model-specific changes in slope.

Within one drive model, age and calendar time are nearly the same variable, because a model is
deployed over a year or two. The data therefore do not clearly separate wear-out specific
to that model (an age effect) from an episode that struck that model at some date (a period
effect).

Giving each drive model its own wear-out exponent and scale (Appendix~\ref{sec:perwear})
reproduces the changes in slope of the two HGST 12~TB models and of the ST3000DM001 but not those of the
ST12000NM0007 or the Toshiba MG07ACA14TA (Figure~\ref{fig:permodel3}). Its Laplace estimate
(a normal approximation about the mode) of the marginal log-likelihood is $-340{,}235$,
against $-342{,}048$ for the empirical-Bayes fit. Its fits reach their stopping rule but suffer a
severe held-out extrapolation failure (Section~\ref{sec:valid}). The extension does not
establish a predictive advantage.

\subsection{New Drive Models}\label{sec:newmodels}

For drive models with no history, the empirical-Bayes point forecast (the prior mean $\mu_g$)
summed over the nine test years predicts 769 failures against 126 observed. That total is
dominated by one test year and two drive models: the ST12000NM001G and ST14000NM001G in 2020,
for which 587 failures were predicted and 43 occurred. Across the other eight years the
forecast is 158 against 71, and in two years (2018, 2026) it is too low. Forty-seven drive models
entered across the nine folds, most with little exposure in their first year. This evidence
supports only two statements. The prior mean was a poor point forecast for two large Seagate
models. The wide prior predictive distribution still scored better on these cells than any
point forecast.

\begin{table}[h]
\centering\small
\caption{Observed and Expected Failures by Age Band. Expected is the empirical-Bayes model's in-sample expectation, including helium, on power-on age (HDD data drives, all data).}\label{tab:age}
\begin{tabular}{@{}lrrrr@{}}
\toprule
Age band & Drive-years & Observed & Expected & Obs/exp \\
\midrule
Days 1--3 & 367 & 107 & 76.0 & 1.41 \\
Days 4--30 & 24{,}347 & 539 & 579.2 & 0.93 \\
Days 31--91 & 73{,}664 & 860 & 818.6 & 1.05 \\
Day 92 to 1 yr & 332{,}479 & 3{,}275 & 3{,}243.8 & 1.01 \\
1--2 yr & 413{,}511 & 5{,}458 & 5{,}454.7 & 1.00 \\
2--3 yr & 347{,}730 & 5{,}314 & 5{,}137.6 & 1.03 \\
3--4 yr & 273{,}785 & 4{,}643 & 4{,}442.8 & 1.05 \\
4--5 yr & 217{,}375 & 4{,}041 & 4{,}093.6 & 0.99 \\
5--6 yr & 140{,}262 & 3{,}224 & 3{,}054.2 & 1.06 \\
6--7 yr & 79{,}850 & 2{,}195 & 1{,}932.8 & 1.14 \\
7--8 yr & 56{,}866 & 1{,}455 & 1{,}463.6 & 0.99 \\
8--9 yr & 34{,}432 & 632 & 883.3 & 0.72 \\
9--10 yr & 8{,}211 & 138 & 182.3 & 0.76 \\
Over 10 yr & 147 & 0 & 3.8 & 0.00 \\
\bottomrule
\end{tabular}
\end{table}

\section{Accounting for Changes over Time and Shared Risks}\label{sec:time}

The empirical-Bayes model treats a drive model's relative failure rate as fixed over the
years. A fleet can nevertheless have unusually good or bad years, and drives of one model
can fare differently from the rest. We measure this variation and fit a Bayesian model
with a fleet-wide annual change and an additional annual risk multiplier, or \emph{shock},
shared by drives of the same model. Its variants are scored on the held-out
years in Section~\ref{sec:bayesresults}, and what it implies about aging is in
Section~\ref{sec:apc}.

\subsection{What the Empirical-Bayes Model Leaves Out}\label{sec:limits2}

Around the fitted model (all data), let $O$ and $E$ be observed and expected failures in a group
of cells. If groups share a gamma-distributed multiplier with mean 1 and precision $\kappa$ (variance $1/\kappa$), counts
are negative binomial with size $\kappa$. We estimate $\kappa$ by maximum likelihood, separately
at each level of grouping.

\begin{table}[h]
\centering\small
\caption{Residual Clustering Around the Empirical-Bayes Model. Pearson dispersion is $\sum (O-E)^2/E$ per group, which is 1 for Poisson counts. CV is $1/\sqrt{\kappa}$, the coefficient of variation of the implied multiplier. In the last two rows each cell's expectation is first rescaled by its calendar year's fleet-wide $O/E$.}\label{tab:cluster}
\begin{tabular}{@{}lrrrr@{}}
\toprule
Grouping & Groups & Pearson dispersion & $\hat\kappa$ & CV \\
\midrule
fleet x year & 14 & 127.5 & 12.91 & 0.28 \\
fleet x month & 159 & 23.4 & 8.39 & 0.35 \\
model x year & 601 & 18.5 & 4.21 & 0.49 \\
model x month & 5{,}964 & 3.9 & 3.12 & 0.57 \\
model x year, net of fleet year & 601 & 17.2 & 4.65 & 0.46 \\
model x month, net of fleet year & 5{,}964 & 3.8 & 3.53 & 0.53 \\
\bottomrule
\end{tabular}
\end{table}

\begin{table}[h]
\centering\small
\caption{The Fleet by Calendar Year. Observed counts the failures of HDD data drives. Expected and the in-sample ratio come from the model fitted to all data. Obs/forecast comes from the rolling-origin validation of M2 + covariates and is out of sample.}\label{tab:year}
\begin{tabular}{@{}rrrrrr@{}}
\toprule
Year & Drive-years & Observed & Expected & Obs/exp (in-sample) & Obs/forecast \\
\midrule
2013 & 13{,}576 & 712 & 934 & 0.76 & -- \\
2014 & 33{,}181 & 2{,}134 & 1{,}493 & 1.43 & -- \\
2015 & 46{,}589 & 1{,}273 & 851 & 1.50 & -- \\
2016 & 65{,}515 & 1{,}280 & 924 & 1.38 & -- \\
2017 & 81{,}680 & 1{,}507 & 1{,}151 & 1.31 & -- \\
2018 & 98{,}520 & 1{,}240 & 1{,}492 & 0.83 & 0.63 \\
2019 & 109{,}434 & 2{,}048 & 1{,}832 & 1.12 & 1.14 \\
2020 & 140{,}502 & 1{,}307 & 2{,}396 & 0.55 & 0.44 \\
2021 & 180{,}801 & 1{,}804 & 2{,}606 & 0.69 & 0.87 \\
2022 & 215{,}846 & 2{,}996 & 3{,}071 & 0.98 & 1.33 \\
2023 & 246{,}522 & 4{,}192 & 3{,}666 & 1.14 & 1.44 \\
2024 & 284{,}213 & 4{,}528 & 3{,}900 & 1.16 & 1.25 \\
2025 & 317{,}286 & 4{,}308 & 4{,}536 & 0.95 & 0.92 \\
2026 & 169{,}272 & 2{,}552 & 2{,}516 & 1.01 & 1.03 \\
\bottomrule
\end{tabular}
\end{table}

Given the model, failures are far from independent. A drive model in a given month
fails at a rate that varies with a CV of about 0.57 around its expectation, and the whole fleet
in a given year with a CV of about 0.28 in-sample. Out of sample the standard deviation of
$\log(\text{observed}/\text{forecast})$ over the nine test years is 0.39. A simulator that draws
lifetimes independently omits shared variation of this kind. These annual and monthly counts
do not determine the probability of failures within a shorter repair window. Table~\ref{tab:cluster}'s
entries are residual summaries estimated one level at a time, not jointly and not within the
likelihood.

Standard errors and reference distributions that assume independence may therefore be
misleading; we have not estimated the effect on each parameter or comparison. The Pearson dispersions
grow with the size of the group (4 for a drive model in a month, 18 for a drive model in a year,
127 for the fleet in a year), so these summaries do not supply a single inflation factor.

\subsection{A Bayesian Model with Shocks}\label{sec:bayesmodel}

We define the model's hazard, the processes we compare for the year effect, the forms of the
drive-model-year shock, the likelihood, the priors and the sampler, and we check that the
sampler recovers known parameters from simulated data.

\subsubsection{Hazard}

Appendix~\ref{sec:primer} (\ref{sec:primerbayes} and~\ref{sec:primershock}) explains Bayesian inference, the
sampler and its diagnostics, and shared shocks.

For a drive of model $g$ at power-on age $a$ (years) in calendar year $y$,
\begin{equation}\label{eq:bayes}
h(a\mid g,y) = Q_y\,U_{gy}\,Z_g\,h_0(a),\qquad
H_0(a) = c_1 a^{\beta_1} + k_2\Bigl(\frac{a}{a_\star}\Bigr)^{\beta_2},\quad a_\star = 4\text{ yr}.
\end{equation}
The baseline is the two-Weibull distribution of Section~\ref{sec:hier} in a different parametrization:
$c_1=\eta_1^{-\beta_1}$, because $\eta_1$ is weakly determined on its own, and
$k_2=(a_\star/\eta_2)^{\beta_2}$, the wear-out cumulative hazard at the pivot age $a_\star$ of four
years, which is better determined than $\eta_2$. The three multipliers are (gamma distributions
in shape--rate form throughout, Section~\ref{sec:notation}):
\begin{itemize}[leftmargin=1.4em]
\item $Z_g\sim\Gam\bigl(1/\phi,\ 1/(\phi\mu_g)\bigr)$, one per drive model, so that
  $\E[Z_g]=\mu_g$ and $\mathrm{Var}(Z_g)=\phi\mu_g^2$ as in Section~\ref{sec:hier}, with
  $\log\mu_g$ linear in vintage, log capacity, manufacturer and helium as in \eqref{eq:hier};
\item $Q_y$, one per calendar year, shared by the whole fleet, with logarithm $s_y$.
  The random-walk variants set $Q_{2013}=1$; the independent and AR(1) variants do not;
\item $U_{gy}\sim\Gam(\kappa_{gy},\kappa_{gy})$, one per drive model per year, with mean 1.
\end{itemize}

\subsubsection{The year effect}

We compare five processes for $s_y=\log Q_y$, with $\varepsilon_y$ independent standard
normal innovations, one per year from 2014:
\begin{align*}
\text{none:}\quad & s_y = 0; \\
\text{independent:}\quad & s_y\sim\Normal(0,\sigma_Q^2); \\
\text{AR(1):}\quad & s_y = \rho s_{y-1} + \sigma_Q\varepsilon_y,\\
& s_{2013}\sim\Normal\bigl(0,\ \sigma_Q^2/(1-\rho^2)\bigr)\ \text{(the stationary distribution)}; \\
\text{random walk:}\quad & s_y = s_{y-1} + \sigma_Q\varepsilon_y,\quad s_{2013}=0; \\
\text{random walk with drift:}\quad & s_y = s_{y-1} + d + \sigma_Q\varepsilon_y,\quad s_{2013}=0.
\end{align*}
These processes give different forecasts on the log scale. Without a year effect, $s_y=0$;
the independent process has conditional mean zero, and AR(1) returns toward zero. A random
walk's conditional mean stays at the last $s_y$, or advances by $d$ per year with drift.
On the hazard scale these are median paths. For example, conditional on the parameters,
$\E[Q_{y+1}\mid Q_y]=Q_y\exp(d+\sigma_Q^2/2)$ under the random walk with drift.

\subsubsection{The drive-model-year shock}

We compare two forms. In the first (B1) the precision is one number, $\kappa_{gy}=\kappa$, so the
coefficient of variation of the shock is the same for a drive model with ten drives and one
with fifty thousand. In the second (B3),
\begin{equation}\label{eq:kappa}
\kappa_{gy} = \kappa_0\,m_{gy}^{\xi},\qquad m_{gy} = Q_y Z_g M_{gy},
\end{equation}
where $M_{gy}$ is the baseline expectation of the group (below) and $m_{gy}$ its expected number
of failures. The squared coefficient of variation of the shock is $1/\kappa_{gy}$, so
$\xi>0$ means large groups are relatively steadier. $\xi=0$ is B1. $\xi=1$ gives a
variance proportional to the mean, which is what would be seen if a group were a sum of many
independent sub-populations (pods, vaults, purchase batches) each with its own shock. The
Pearson dispersions of Table~\ref{tab:cluster} grow with group size, but more slowly than
proportionally, as $0<\xi<1$ describes.

Two cautions apply to the shock's precision. First, we estimated the relation \eqref{eq:kappa}
across drive models within one fleet, where group size is confounded with the identity and age
of the drive model. We never tested it on subsets of one drive model. Second, for $\xi<1$ it is
not consistent under subdivision. A group expected to have 1{,}000 failures has a shock with
coefficient of variation 0.27, while the same drives treated as 100 independent groups of ten
(0.63 each) have an aggregate coefficient of variation of 0.063. The clustering a simulation
shows therefore depends on what the caller declares a group to be. We estimated the relation
with a group defined as all drives of one drive model in the system, for one year.

\subsubsection{Likelihood}

The data are the Lexis table of Section~\ref{sec:lexis}: for each drive model, month, and age
bin $b=(\ell_b,u_b]$ (the convention of Section~\ref{sec:lexis}: a bin holds the ages above its
lower edge up to and including its upper edge, in whole days), the drive-years $E$ and the
failures $D$. Convert the bin endpoints to years before evaluating $H_0$. With a hazard constant within a
bin at its mean $\bar h_{b} = [H_0(u_b)-H_0(\ell_b)]/(u_b-\ell_b)$, the likelihood of a
survival model is proportional to that of independent Poisson counts with means $E\times$ hazard
\cite{holford1980,laird1981}. Summing a drive model's cells over a year,
\[
M_{gy} = \sum_{b} E_{gyb}\,\bar h_b,\qquad D_{gy} = \sum_b D_{gyb}.
\]
Given $Z_g$ and $Q_y$, integrating the gamma shock $U_{gy}$ gives a negative binomial for the
year's failures, and the log-likelihood is
\begin{multline}\label{eq:bayesll}
\ell = \sum_{g,b} D_{g\cdot b}\log\bar h_b
 + \sum_{g,y}\Bigl[ D_{gy}\log(Q_yZ_g) + \log\Gamma(\kappa_{gy}+D_{gy}) - \log\Gamma(\kappa_{gy})
 + \kappa_{gy}\log\kappa_{gy} \\ - (\kappa_{gy}+D_{gy})\log(\kappa_{gy}+Q_yZ_gM_{gy})\Bigr] + \text{const}.
\end{multline}
Without shocks (B0) the bracket is the Poisson term $D_{gy}\log Z_g - Z_gM_{gy}$. The likelihood
costs one $99\times81$ table of bin hazards and a few hundred terms, and a fit takes one to
seven minutes on four cores. This likelihood is on binned ages and observed drive-days, where
the empirical-Bayes fit used exact day intervals and calendar spans. B0 checks that the two
agree.

\subsubsection{Priors}

\begin{center}\small
\begin{tabular}{@{}ll@{}}
\toprule
Parameter & Prior \\
\midrule
$\log\beta_1$ & $\Normal(\log 0.4,\ 1)$ \\
$\log c_1$ & $\Normal(-3.5,\ 3^2)$ \\
$\log(\beta_2-\beta_1)$ & $\Normal(0,\ 1)$ \\
$\log k_2$ & $\Normal(\log 0.2,\ 3^2)$ \\
$\gamma_{\text{intro}}$ & $\Normal(0,\ 0.5^2)$ per year \\
$\gamma_{\text{cap}}$, manufacturer and helium effects & $\Normal(0,\ 2^2)$ \\
$\sqrt{\phi}$ (coefficient of variation of $Z_g$ about $\mu_g$) & half-normal, scale 2 \\
$1/\sqrt{\kappa}$ or $1/\sqrt{\kappa_0}$ & half-normal, scale 1 \\
$\xi$ & uniform on $(0,\ 1.5)$ \\
$\sigma_Q$ & half-normal, scale 0.5 \\
$\rho$ & uniform on $(-0.95,\ 0.95)$ \\
$d$ & $\Normal(0,\ 0.5^2)$ \\
$\sigma_\beta$ & half-normal, scale 0.3 \\
\bottomrule
\end{tabular}
\end{center}
These priors are weak relative to the data for every parameter except $\rho$, $d$, and
$\sigma_Q$, which are learned from thirteen yearly steps.

\subsubsection{Computation}

We sample the posterior with the No-U-Turn sampler \cite{hoffman2014nuts} in NumPyro
\cite{phan2019numpyro}: four chains, 1{,}500 warm-up and 1{,}000 retained iterations each, target acceptance 0.9. We sample $Z_g$ and the yearly innovations
explicitly and integrate $U_{gy}$ out as above. Table~\ref{tab:bayesdiag} gives, for each
variant, the worst $\hat R$ and smallest effective sample size over all parameters and all ten
fits (nine folds and the fit to all data), and the number of divergent transitions (steps at
which the sampler's numerical integration of its trajectory failed, a sign of a region of the
posterior it cannot explore).

\begin{table}[h]
\centering
\caption{Sampler Diagnostics by Variant. Each variant has ten fits. Max $\hat R$ is the largest split-$\hat R$ and Min ESS the smallest effective sample size over all sampled parameters. Divergences are totals over 40{,}000 retained draws per variant. Seconds is the longest sampling time of a fit, on four cores of a node shared with other fits.}\label{tab:bayesdiag}
\footnotesize\setlength{\tabcolsep}{4pt}\begin{tabular}{@{}llrrrrr@{}}
\toprule
Shocks & Year effect & Fits & Max $\hat R$ & Min ESS & Divergences & Seconds \\
\midrule
B0: none & --- & 10 & 1.017 & 358 & 0 & 638 \\
B1: constant precision & none & 10 & 1.007 & 573 & 0 & 105 \\
B1: constant precision & independent & 10 & 1.011 & 373 & 0 & 128 \\
B1: constant precision & AR(1) & 10 & 1.009 & 697 & 11 & 213 \\
B1: constant precision & random walk & 10 & 1.009 & 514 & 0 & 179 \\
B1: constant precision & random walk with drift & 10 & 1.011 & 536 & 25 & 257 \\
B3: size-dependent precision & none & 10 & 1.009 & 555 & 0 & 107 \\
B3: size-dependent precision & independent & 10 & 1.014 & 448 & 0 & 160 \\
B3: size-dependent precision & random walk & 10 & 1.012 & 642 & 0 & 217 \\
B3: size-dependent precision & random walk with drift & 10 & 1.011 & 523 & 12 & 312 \\
\bottomrule
\end{tabular}
\end{table}

These are the worst values over all parameters and ten fits, and most variants miss the
thresholds of Vehtari et al.~\cite{vehtari2021rhat} ($\hat R<1.01$, effective sample size
above 400; Appendix~\ref{sec:primer}). Four variants have $\hat R<1.01$: B1 with no year effect, with AR(1) shocks and with a
random walk, and B3 with no year effect. B3 with drift is at 1.011 and B0 reaches 1.017. Two of
the ten have a smallest effective sample size below 400. The standard errors of the
bridge-sampling estimates of the predictive scores (Section~\ref{sec:bayesvalid}) are computed from the spread between chains and so include some of
this Monte Carlo error. Longer chains and dispersed starting points are needed to assess
the remaining sampling uncertainty, particularly for the extensions in Section~\ref{sec:pms}.

\subsubsection{Parameter recovery}

To check that the sampler and the likelihood recover known parameters, we simulated failure
counts from B3 with drift on the real exposure table. The hyperparameters are at the posterior
means of Table~\ref{tab:bayespar}. Each drive model's multiplier is drawn from its gamma prior,
the year effect from the random walk, and a shock for each drive model and year. The counts are
Poisson given these. We generated eighteen data sets and fitted each as we fitted the real one.
Table~\ref{tab:recovery} compares the posteriors with the truth.

\begin{table}[h]
\centering\small
\caption{Parameter Recovery over 18 Simulated Data Sets. Mean of means and SD of means summarize the 18 posterior means. Covered counts the central 95\% intervals that contain the truth. The standardized error is the posterior mean minus the truth, over the posterior standard deviation, and its mean has a standard error of 0.24.}\label{tab:recovery}
\begin{tabular}{@{}lrrrrrr@{}}
\toprule
Parameter & Truth & Mean of means & SD of means & Mean post.\ SD & Covered & Mean std.\ error \\
\midrule
$\beta_1$ & 0.381 & 0.384 & 0.012 & 0.016 & 18/18 & $+0.18$ \\
$c_1$ & 0.059 & 0.070 & 0.019 & 0.021 & 17/18 & $+0.29$ \\
$\beta_2$ & 1.929 & 1.933 & 0.026 & 0.026 & 17/18 & $+0.16$ \\
$k_2$ & 0.352 & 0.416 & 0.121 & 0.127 & 17/18 & $+0.29$ \\
$\eta_2$ (yr) & 6.877 & 6.746 & 1.064 & 1.061 & 17/18 & $-0.32$ \\
$\phi$ & 0.577 & 0.582 & 0.106 & 0.117 & 18/18 & $-0.09$ \\
$\gamma_{\text{intro}}$ & 0.169 & 0.171 & 0.054 & 0.053 & 17/18 & $+0.00$ \\
$\gamma_{\text{cap}}$ & -1.229 & -1.161 & 0.307 & 0.303 & 16/18 & $+0.22$ \\
$\gamma_{\text{HGST}}$ & -1.510 & -1.544 & 0.182 & 0.230 & 18/18 & $-0.17$ \\
$\gamma_{\text{Toshiba}}$ & -0.790 & -0.787 & 0.239 & 0.288 & 18/18 & $-0.00$ \\
$\gamma_{\text{WDC}}$ & -0.448 & -0.386 & 0.320 & 0.410 & 18/18 & $+0.15$ \\
$\gamma_{\text{He}}$ & 0.623 & 0.487 & 0.351 & 0.392 & 17/18 & $-0.36$ \\
$\kappa_0$ & 1.054 & 1.220 & 0.361 & 0.294 & 16/18 & $+0.26$ \\
$\sigma_Q$ & 0.234 & 0.257 & 0.091 & 0.076 & 16/18 & $+0.03$ \\
$d$ & -0.134 & -0.140 & 0.062 & 0.076 & 18/18 & $-0.18$ \\
$\xi$ & 0.371 & 0.347 & 0.059 & 0.051 & 16/18 & $-0.42$ \\
\bottomrule
\end{tabular}
\end{table}

Of 288 intervals, 274 contain the truth (95.1\%). No parameter shows a bias beyond what 18
replicates allow, and the largest mean standardized error is for $\xi$. For
every parameter the spread of the posterior means across data sets is close to the posterior
standard deviation: 0.026 against 0.026 for $\beta_2$, and at the largest excess, for $\kappa_0$,
0.361 against 0.294.

The simulated count tables mostly contain more failures than the real table. Their totals run from 26{,}300
to 325{,}100 failures with a median of 69{,}700, against 31{,}881 (the failures in the Lexis
table) in the real data, and 17 of the 18 exceed it. The median is about 2.2 times the
observed count. These simulations hold the observed exposure fixed; they do not generate
drive lifetimes, deployments or removals. Two features may contribute to the excess.
First, the real fleet's exposure sits on drive models whose multipliers are below
their prior means: the exposure-weighted mean of $z_g/\mu_g$ is 0.76, for all 99 drive models
and for the twenty that carry 92\% of the exposure. The prior for $Z_g$ does not model the
association between deployment volume and fitted hazard; these data do not establish the
purchasing decisions behind it. Second, $Q_y$ is lognormal conditional on the random-walk
parameters, so its mean exceeds its
median by a factor that grows with time. The rest is not accounted for. The excess
resembles the prior mean's over-prediction for new drive models (Section~\ref{sec:newmodels}) at the scale
of the fleet, and it means that parameter recovery was tested mostly on data sets with more
failures than the real one.
The largest $\hat R$ over the 18 fits was 1.011. Thirteen fits had no divergent transitions, and
five had two to eight.

\section{What a Drive Reports About Itself}\label{sec:drive}

The models so far use only a drive's model and age. Each daily record also carries the
drive's SMART attributes. We define a warning state from five of them and
summarize how they and three other quantities the drive reports predict failure.
The regression is given in full in Appendix~\ref{sec:cov}, and the warning state enters the
Bayesian model in Appendix~\ref{sec:extsmart}.

\subsection{The SMART Warning State}\label{sec:smartstate}

The hidden Markov model of Appendix~\ref{sec:hmm} includes an unobserved degraded state. The
SMART record gives something close to an observed one. We call a drive \emph{warned} from the
first day on which any of SMART attributes 5, 187, 188, 197, or 198 has a non-zero raw value.
Before that it is \emph{clear}, meaning no recorded warning, not that every attribute was
measured and found to be zero.
These are the five that Backblaze uses to judge a drive~\cite{bb2014smart}. A
\emph{SMART warning} is this event; the dataset's \emph{failure flag} is the separate record
that the drive was removed as failed. For each of the
504{,}656 HDD data drives we extracted the first such day from the daily files and divided
exposure and failures at it. (This is every HDD data drive, with 31{,}899 failures. The
lifetime fits use the 504{,}624 with a usable power-on age, with 31{,}881.)

\begin{table}[h]
\centering
\caption{HDD Data Drives by SMART State and Calendar Year. A drive is warned from the first day any of SMART 5, 187, 188, 197 and 198 is non-zero. The warning rate is drives newly warned per clear drive-year. The failure rates are failures per drive-year in each state, and Ratio is their quotient. The last column is the share of the year's failures that were warned on or before the day of failure.}\label{tab:smartyear}
\footnotesize\setlength{\tabcolsep}{4pt}\begin{tabular}{@{}rrrrrrrrr@{}}
\toprule
& \multicolumn{2}{c}{Drive-years} & Warned & Warning & \multicolumn{2}{c}{Failure rate (\%/yr)} & & Failures \\
Year & clear & warned & share (\%) & rate (\%/yr) & clear & warned & Ratio & warned (\%) \\
\midrule
2013 & 15{,}640 & 1{,}578 & 9.2 & 8.4 & 1.52 & 30.1 & 20 & 67 \\
2014 & 28{,}915 & 5{,}355 & 15.6 & 20.6 & 0.90 & 35.0 & 39 & 88 \\
2015 & 43{,}344 & 3{,}835 & 8.1 & 4.4 & 0.78 & 24.3 & 31 & 73 \\
2016 & 62{,}815 & 3{,}436 & 5.2 & 2.1 & 0.62 & 25.9 & 42 & 70 \\
2017 & 80{,}148 & 3{,}586 & 4.3 & 2.6 & 0.78 & 24.7 & 32 & 59 \\
2018 & 96{,}926 & 3{,}870 & 3.8 & 2.3 & 0.49 & 19.7 & 40 & 62 \\
2019 & 105{,}331 & 6{,}237 & 5.6 & 7.1 & 0.51 & 24.3 & 48 & 74 \\
2020 & 131{,}352 & 10{,}770 & 7.6 & 4.5 & 0.35 & 7.9 & 22 & 65 \\
2021 & 170{,}325 & 11{,}762 & 6.5 & 3.0 & 0.36 & 10.2 & 28 & 66 \\
2022 & 201{,}909 & 15{,}287 & 7.0 & 3.3 & 0.35 & 14.9 & 42 & 76 \\
2023 & 229{,}206 & 19{,}062 & 7.7 & 3.5 & 0.35 & 17.9 & 52 & 81 \\
2024 & 262{,}517 & 22{,}222 & 7.8 & 3.4 & 0.28 & 17.0 & 60 & 84 \\
2025 & 291{,}406 & 26{,}732 & 8.4 & 3.5 & 0.31 & 12.7 & 41 & 79 \\
2026 & 154{,}168 & 15{,}724 & 9.3 & 4.1 & 0.32 & 13.1 & 41 & 81 \\
\midrule
All & 1{,}874{,}002 & 149{,}457 & 7.4 & 3.9 & 0.40 & 16.3 & 40 & 76 \\
\bottomrule
\end{tabular}
\end{table}

\begin{table}[h]
\centering\small
\caption{Warning Lead Time and the Five Attributes. Left: the 31{,}899 failures of HDD data drives by how long the drive had been warned. Right: for each attribute, the drives on which it was ever non-zero, and the shares of failed and of unfailed drives on which it was.}\label{tab:smartlead}
\begin{tabular}{@{}lrr@{}}
\toprule
At failure & Failures & \% \\
\midrule
No warning before or on the day of failure & 7{,}589 & 23.8 \\
First warned on the day of failure & 850 & 2.7 \\
Warned for 2 to 7 days & 1{,}938 & 6.1 \\
Warned for 8 to 30 days & 2{,}971 & 9.3 \\
Warned for more than 30 days & 18{,}551 & 58.2 \\
\bottomrule
\end{tabular}
\quad
\begin{tabular}{@{}lrrr@{}}
\toprule
Attribute & Drives & Failed (\%) & Unfailed (\%) \\
\midrule
5, reallocated sectors & 43{,}860 & 52.3 & 5.8 \\
187, reported uncorrectable errors & 23{,}585 & 40.8 & 2.2 \\
188, command timeout & 23{,}448 & 19.0 & 3.7 \\
197, current pending sectors & 35{,}516 & 54.8 & 3.8 \\
198, offline uncorrectable sectors & 24{,}871 & 42.2 & 2.4 \\
\bottomrule
\end{tabular}
\end{table}

\begin{table}[h]
\centering\small
\caption{SMART State Rates by Manufacturer. The rates are those of Table~\ref{tab:smartyear}, over all years.}\label{tab:smartmfr}
\begin{tabular}{@{}lrrrrr@{}}
\toprule
& & Warning & \multicolumn{2}{c}{Failure rate (\%/yr)} & Failures \\
Manufacturer & Drives & rate (\%/yr) & clear & warned & warned (\%) \\
\midrule
Seagate & 211{,}239 & 5.6 & 0.48 & 15.1 & 80 \\
HGST & 160{,}946 & 2.1 & 0.26 & 19.3 & 75 \\
Toshiba & 129{,}893 & 2.9 & 0.42 & 19.4 & 64 \\
WDC & 2{,}573 & 3.4 & 2.70 & 50.7 & 41 \\
\bottomrule
\end{tabular}
\end{table}

Over the whole record a clear drive fails at 0.40\% a year and a warned one at 16.3\% a
year, forty times the rate. Drives become warned at 3.9\% a year, and warned drives are
7.4\% of the exposure and 76\% of the failures (Table~\ref{tab:smartyear}). Of the failures, 23.8\%
came with no warning, and 58\% had been warned for more than 30 days
(Table~\ref{tab:smartlead}). Backblaze reported in 2016 that 76.7\% of failed drives had at
least one of the five attributes non-zero~\cite{bb2016smart}. Over the whole record from 2013
to mid-2026 we find 76.2\%.

\begin{table}[h]
\centering\small
\caption{Warning Attributes by Manufacturer, and Warnings Before Failure by Year. Left: HDD data drives on which each attribute was ever non-zero. Right: the share of each year's failures on drives already warned on 1~January, and the share warned at any time up to the failure.}\label{tab:smartmfrattr}
\begin{tabular}{@{}lrrrrrr@{}}
\toprule
Manufacturer & Drives & 5 & 187 & 188 & 197 & 198 \\
\midrule
Seagate & 211{,}239 & 27{,}362 & 23{,}359 & 23{,}148 & 18{,}847 & 18{,}204 \\
HGST & 160{,}946 & 8{,}808 & 211 & 287 & 9{,}984 & 4{,}389 \\
Toshiba & 129{,}893 & 7{,}659 & 11 & 13 & 6{,}332 & 2{,}189 \\
WDC & 2{,}573 & 31 & 0 & 0 & 349 & 89 \\
\bottomrule
\end{tabular}
\quad
\begin{tabular}{@{}rrrr@{}}
\toprule
Year & Failures & 1 Jan (\%) & Ever (\%) \\
\midrule
2014 & 2{,}134 & 33 & 88 \\
2015 & 1{,}273 & 37 & 73 \\
2016 & 1{,}280 & 23 & 70 \\
2017 & 1{,}507 & 17 & 59 \\
2018 & 1{,}241 & 19 & 62 \\
2019 & 2{,}048 & 20 & 74 \\
2020 & 1{,}307 & 29 & 65 \\
2021 & 1{,}814 & 31 & 66 \\
2022 & 2{,}998 & 38 & 76 \\
2023 & 4{,}194 & 44 & 81 \\
2024 & 4{,}530 & 45 & 84 \\
2025 & 4{,}309 & 48 & 79 \\
2026 & 2{,}552 & 63 & 81 \\
\midrule
2014--2026 & 31{,}187 & 39 & 76 \\
\bottomrule
\end{tabular}
\end{table}

\textbf{The warning does not mean the same thing for every manufacturer.} Attributes 187 and 188
are non-zero almost only on Seagate drives (Table~\ref{tab:smartmfrattr}): 99\% of the drives
ever non-zero on either are Seagate. For many other drives the warning rests on three
attributes. An unreported attribute contributes no positive flag to this binary warning;
that convention does not establish that its value was zero. The differences
between manufacturers in Table~\ref{tab:smartmfr}, and part of the movement by year in
Table~\ref{tab:smartyear}, therefore reflect what each manufacturer reports as well as how its
drives behave. With the warning defined by the three attributes that all report (5, 197, 198) the
pooled rates are 0.47\% a year for clear drives, 3.0\% a year for a first warning, and 22.8\% a year for warned drives, and
72\% of failures are warned.

\textbf{What a forecast can use.} The 76\% includes warnings raised during the year in which the
drive failed, 850 of them on the day of failure. A forecast for a calendar year made on
1~January, which is the forecast we score in Sections~\ref{sec:valid}
and~\ref{sec:bayesresults}, sees only warnings already raised. Over 2014--2026, 39\% of failures
were on drives warned by then, between 17\% and 48\% in full years
(Table~\ref{tab:smartmfrattr}). The other 61\% come from drives clear at the forecast date,
so the warning rate itself would have to be forecast.

Four further cautions apply. (i) These are constant-rate summaries that pool ages, drive
models, and years. Table~\ref{tab:smartyear} shows the clear rate falling five-fold from 2013 to 2024, like
the year effect of Section~\ref{sec:bayesfit}. (ii) The failure rate of warned drives is partly
policy. Backblaze removes drives on the evidence of these same attributes, and a removal is a
failure in the data, so the forty-fold ratio is not a property of the drives alone. (iii)
We treat a warning as permanent, though some attributes can return to zero. (iv) The raw value
of attribute 188 is a composite on some drive models, so we use it only as zero or non-zero.
We use the state as a predictor in Appendices~\ref{sec:cov} and~\ref{sec:ext}.
The observed rates differ from those of the hidden Markov model (Appendix~\ref{sec:cov}).

The raw-label cohorts of WUH721816ALE6L4 (Section~\ref{sec:population}) show an exception:
25 of the 624 drives without the WDC prefix have a positive attribute 188 before exit,
whereas none is recorded for the 26{,}685 WDC-prefixed drives. Excluding 188 reduces the
number of warned drives in the former cohort from 32 to 10 and failures while warned from
3 to 1. With state fixed on 1~January, it moves 1{,}492 days of calendar exposure and one
failure from warned to clear; the WDC-prefixed cohort is unchanged. This is a sensitivity
of the state assignment, without refitting the failure model.

\subsection{SMART Attributes, Temperature and Load as Predictors}\label{sec:covsum}

We put the warning state, the counts of the five warning attributes,
temperature (attribute 194), and load cycles (193) and power cycles (12) per year of power-on
time into a Poisson regression on monthly cells (Appendix~\ref{sec:cov}). The regression has age, manufacturer and a
random effect per drive model, but no year effect and no shocks. We score it on the nine
held-out years. The
warning alone gains 27{,}353 log units one month ahead and 11{,}128 one year ahead, in nine years
of nine. Given age and drive model, a drive warned at the end of the previous month fails at
27 times the rate of a clear one, and a drive warned on 1~January at 9.6 times the rate
over the year. Temperature, load cycles and power cycles add little beside the warning
attributes. On their own, temperature gains 424 log units one month ahead and 289 one year
ahead, and the two cycle counts together 851 and 436, against 36{,}010 and 14{,}383 for the
counts of the warning attributes. Added to those counts, relative temperature and both cycle
counts together gain a further 334 and 129. The regression's coefficients are associations,
not effects.

\emph{What this means for a simulator.} The lifetime models of Sections~\ref{sec:hier} and~\ref{sec:time} describe a
drive about which nothing is known but its model and age. For a fleet that exists, we observe
the state of each drive on the day the simulation starts. The warned drives, under a tenth of the fleet,
account for about two fifths of the coming year's failures (Table~\ref{tab:smartmfrattr}). A
simulator that starts from an observed fleet should therefore start each drive in its observed
state. That needs a model of the transition from clear to warned as well as the two
failure rates. The rates of Section~\ref{sec:smartstate} split exposure at the first warning
and pool ages, drive models, calendar years and durations since warning. They are descriptive
state-specific rates; the frozen 1~January state is a different predictor used in
Appendix~\ref{sec:extsmart}. We fit a transition model across the fleet in Section~\ref{sec:mstate}, following Oakley
et al.~\cite{oakley2023multistate}, who fitted one for a single drive model. The two-state
sampler built from these rates illustrates cost and is not that fitted model
(Appendix~\ref{sec:extspeed}). Its marginal lifetime distribution also differs from that of
Section~\ref{sec:hier}: putting the unconditional hazard in the clear state and a multiple of
it in the warned state raises the marginal hazard above the unconditional one wherever
warned survivors are present.

% Part 3: forecasting results
\section{How Well Do the Models Predict Failure Counts?}\label{sec:valid}

We compare the models of Sections~\ref{sec:families} and~\ref{sec:hier} with simple forecasts
and with flexible regressions on the nine held-out years of Section~\ref{sec:validdesign}. The
candidates are:
\begin{itemize}[leftmargin=1.4em]
\item \emph{Simple forecasts}: the last training year's annualized failure rate as a constant rate;
  the bathtub with its constants, on power-on age; and a per-year Weibull, fitted to the drives
  in service in the last training year on age since first sighting and applied to the next year
  (the fleet-year fits of Table~\ref{tab:periodweibull} in Appendix~\ref{sec:fleetyear} used as a forecast).
\item \emph{Lifetime-distribution baselines} (M0, M1, M2): constant, Weibull, and two-Weibull (two competing Weibulls) baselines on
  power-on age, each with no drive-model effect (``pooled''), with plain gamma frailty
  ($\mu_g\equiv1$), and with covariate frailty. Point forecasts use the posterior mean of $Z_g$,
  and the prior mean for unseen models. A variant multiplies the forecast by the fleet's
  observed-to-expected ratio in the last training year (``year calibration'').
\item \emph{Additive models} A1--A5 (\code{mgcv::bam}, Poisson, log link, offset $\log E$): a
  spline in log age with a random effect per drive model (A1); plus manufacturer, a capacity
  spline, a per-model random slope in log age, and a random year effect (A2); plus a vintage
  spline (A3); plus an age-by-vintage interaction (A4); and a separate shrunken age curve per
  drive model (A5). We carry the year effect of the last training year forward and drop model
  effects for unseen models.
\item \emph{Boosted trees}: four gradient-boosted hazard models with Poisson loss on the
  Lexis cells, summarized in Section~\ref{sec:boost} and described in
  Appendix~\ref{sec:boostlexis}. They appear in the tables here so that we score every candidate
  on the same folds.
\item \emph{Per-model wear-out}: the empirical-Bayes model with a wear-out exponent of its own for
  each drive model (Appendix~\ref{sec:perwear}).
\end{itemize}

\subsection*{Results}

\begin{table}[p]
\centering
\caption{Held-Out Results Summed over Nine Test Years. The test years run from 2018 to the first half of 2026. Rows are ordered by log score on all cells. Pred./obs.\ is total predicted over total observed failures. $\Delta$ is the deficit in log score relative to the best candidate in each column: all cells, cells of drive models present in the training data (24{,}849 failures), and cells of drive models new in the test year (126 failures). Predictive rows score the joint distribution \eqref{eq:pred}. All other rows score independent Poisson counts at a point forecast. Table~\ref{tab:rollyearls} compares the per-model wear-out model year by year.}\label{tab:rolling}
\footnotesize\setlength{\tabcolsep}{4pt}\begin{tabular}{@{}p{8.6cm}rrrrr@{}}
\toprule
Candidate & Pred./obs. & Log score & $\Delta$ all & $\Delta$ seen & $\Delta$ new \\
\midrule
M2 + covariates + year calibration, predictive & 1.00 & $-29{,}324$ & $0$ & $0$ & $-2$ \\
M2 + covariates, predictive & 0.99 & $-29{,}446$ & $-122$ & $-124$ & $0$ \\
M2, predictive & 1.04 & $-29{,}447$ & $-123$ & $-121$ & $-4$ \\
M0 + covariates + year calibration, predictive & 0.97 & $-29{,}623$ & $-299$ & $-292$ & $-9$ \\
M1 + covariates + year calibration, predictive & 0.99 & $-29{,}669$ & $-345$ & $-337$ & $-10$ \\
M1, predictive & 0.95 & $-29{,}972$ & $-648$ & $-642$ & $-8$ \\
M1 + covariates, predictive & 0.90 & $-29{,}974$ & $-651$ & $-643$ & $-10$ \\
M0, predictive & 0.89 & $-30{,}210$ & $-886$ & $-880$ & $-8$ \\
M0 + covariates, predictive & 0.84 & $-30{,}211$ & $-887$ & $-881$ & $-9$ \\
Additive A2: + covariates, per-model age slope, year effect & 1.13 & $-30{,}769$ & $-1{,}445$ & $-1{,}388$ & $-60$ \\
Additive A3: A2 + vintage spline & 1.14 & $-30{,}946$ & $-1{,}622$ & $-1{,}374$ & $-250$ \\
Boosting with drive-model identity and year, tuned rounds (Section~\ref{sec:boost}) & 1.13 & $-31{,}143$ & $-1{,}819$ & $-1{,}341$ & $-480$ \\
M2 + covariates + year calibration, point & 1.00 & $-31{,}436$ & $-2{,}112$ & $-1{,}637$ & $-476$ \\
M2 + covariates, point & 0.99 & $-31{,}491$ & $-2{,}168$ & $-1{,}671$ & $-498$ \\
Boosting with drive-model identity, tuned rounds (Section~\ref{sec:boost}) & 1.17 & $-31{,}696$ & $-2{,}372$ & $-1{,}898$ & $-476$ \\
Additive A1: age spline + drive-model effect & 0.97 & $-31{,}742$ & $-2{,}418$ & $-2{,}123$ & $-296$ \\
M0 + covariates + year calibration, point & 0.97 & $-31{,}756$ & $-2{,}432$ & $-1{,}842$ & $-592$ \\
M1 + covariates + year calibration, point & 0.99 & $-31{,}888$ & $-2{,}564$ & $-1{,}946$ & $-621$ \\
Boosting with drive-model identity & 1.09 & $-31{,}930$ & $-2{,}606$ & $-2{,}462$ & $-146$ \\
Boosting on covariates only & 1.08 & $-31{,}984$ & $-2{,}660$ & $-2{,}542$ & $-120$ \\
Additive A5: per-model age curves & 1.23 & $-32{,}080$ & $-2{,}756$ & $-2{,}426$ & $-332$ \\
Additive A4: A3 + age-by-vintage interaction & 1.22 & $-32{,}090$ & $-2{,}767$ & $-1{,}870$ & $-899$ \\
M1 + covariates, point & 0.90 & $-32{,}095$ & $-2{,}771$ & $-2{,}191$ & $-582$ \\
M0 + covariates, point & 0.84 & $-32{,}267$ & $-2{,}943$ & $-2{,}411$ & $-534$ \\
M2, point & 1.04 & $-32{,}527$ & $-3{,}204$ & $-1{,}900$ & $-1{,}306$ \\
M1, point & 0.95 & $-32{,}956$ & $-3{,}632$ & $-2{,}378$ & $-1{,}256$ \\
Per-year Weibull on first-seen age & 1.02 & $-33{,}236$ & $-3{,}912$ & $-3{,}864$ & $-49$ \\
M0, point & 0.89 & $-33{,}278$ & $-3{,}954$ & $-2{,}634$ & $-1{,}322$ \\
Two competing Weibulls, pooled & 1.36 & $-34{,}060$ & $-4{,}736$ & $-4{,}673$ & $-65$ \\
Constant rate: last year's AFR & 1.01 & $-34{,}168$ & $-4{,}845$ & $-4{,}798$ & $-48$ \\
Pooled Weibull, no drive-model effect & 1.33 & $-34{,}680$ & $-5{,}356$ & $-5{,}285$ & $-74$ \\
Pooled constant rate, all history & 1.28 & $-35{,}062$ & $-5{,}738$ & $-5{,}644$ & $-96$ \\
Bathtub with default constants & 2.39 & $-45{,}430$ & $-16{,}106$ & $-15{,}829$ & $-278$ \\
\bottomrule
\end{tabular}
\end{table}

Nine years are a small sample when the years differ this much. For the main comparisons,
Table~\ref{tab:pairs} gives the sum of the nine yearly score differences, the number of years
the first candidate won, and a standard error of the sum from the spread of the yearly
differences. The conventional 5\% threshold for their ratio is about 2.3 (Student's $t$, 8
degrees of freedom). We do not adjust the eighteen comparisons for multiplicity, so only
ratios well above 2.3 should be read as differences.

\begin{table}[h]
\centering
\caption{Paired Comparisons over the Nine Test Years. Sum is the sum of the yearly differences in log score, positive when the first-named candidate is better. Its standard error is $\sqrt{n}$ times the standard deviation of the yearly differences. Ratio is the sum over its standard error.}\label{tab:pairs}
\footnotesize\setlength{\tabcolsep}{4pt}\begin{tabular}{@{}p{10.2cm}rrrr@{}}
\toprule
Comparison & Sum & Years won & Std.\ error & Ratio \\
\midrule
M2 + covariates (point) minus bathtub & $13{,}938$ & 9 of 9 & 3{,}107 & $4.5$ \\
M2 + covariates (point) minus constant AFR & $2{,}677$ & 8 of 9 & 1{,}090 & $2.5$ \\
M2 + covariates (point) minus per-year Weibull & $1{,}744$ & 8 of 9 & 909 & $1.9$ \\
Per-year Weibull minus constant AFR & $933$ & 7 of 9 & 393 & $2.4$ \\
M2 + covariates (point) minus pooled two-Weibull distribution (value of drive model, given age) & $2{,}569$ & 8 of 9 & 660 & $3.9$ \\
M2 + covariates (point) minus M0 + covariates (value of age, given drive model) & $776$ & 6 of 9 & 651 & $1.2$ \\
M2 + covariates (point) minus M1 + covariates (value of the second component) & $603$ & 6 of 9 & 447 & $1.3$ \\
M2 + covariates (point) minus M2 (value of covariates, point) & $1{,}036$ & 8 of 9 & 326 & $3.2$ \\
M2 + covariates minus M2 (value of covariates, predictive) & $1$ & 3 of 9 & 11 & $0.1$ \\
M2 minus M1, both + covariates, predictive & $529$ & 6 of 9 & 418 & $1.3$ \\
M2 + covariates (point) minus additive A2 & $-722$ & 4 of 9 & 430 & $-1.7$ \\
M2 + covariates (point) minus additive A3 & $-546$ & 2 of 9 & 443 & $-1.2$ \\
M2 + covariates (point) minus additive A1 & $250$ & 6 of 9 & 501 & $0.5$ \\
M2 + covariates (point) minus boosting with model identity & $438$ & 4 of 9 & 499 & $0.9$ \\
M2 + covariates (point) minus boosting with model identity and year, tuned & $-348$ & 4 of 9 & 363 & $-1.0$ \\
Additive A2 minus boosting with model identity and year, tuned & $374$ & 7 of 9 & 213 & $1.8$ \\
Boosting with year minus boosting without year (both tuned) & $553$ & 7 of 9 & 193 & $2.9$ \\
Per-model wear-out minus M2 + covariates (point), 2019 excluded & $-531$ & 5 of 8 & 794 & $-0.7$ \\
\bottomrule
\end{tabular}
\end{table}

The comparison shows four patterns:
\begin{enumerate}[leftmargin=1.6em]
\item \textbf{The bathtub with default constants is last among the candidates in
  Table~\ref{tab:rolling}}, in total, and it loses to the
  empirical-Bayes model in each of the nine years. It predicts 2.39 times the failures observed.
\item \textbf{Drive-model effects improve the held-out score.} Adding them to the two-Weibull
  distribution gains 2{,}569 log units and wins eight of nine years. Every pooled candidate
  over-predicts by about 28--36\%. Fitted to all history, a pooled distribution is dominated by the
  high-failure drive models of the early fleet.
\item \textbf{The empirical-Bayes model is ahead of the simple forecasts.} It leads a constant
  AFR by 2{,}677 (eight of nine years, ratio 2.5) and the per-year Weibull by 1{,}744 (eight of
  nine years, ratio 1.9). The per-year Weibull is ahead of a constant AFR by 933 (ratio 2.4). The
  first and third ratios are near the threshold of 2.3; the second is below it.
\item \textbf{Covariates help the point forecast} (1{,}036, eight of nine years) but not the
  predictive score (1 unit). They give a better centre for drive models with little history,
  and the predictive distribution already allows for the uncertainty of those models'
  multipliers.
\end{enumerate}

It does not establish three others:
\begin{enumerate}[leftmargin=1.6em]
\item \textbf{That the form of the age curve matters, given the drive model.} The two-Weibull baseline beats a
  constant hazard by 776 and a single Weibull by 603, winning six of nine years each, with
  ratios near 1. In sample the evidence for two competing components is strong (Section~\ref{sec:families}).
  Out of sample, over nine years, it is suggestive only.
\item \textbf{A ranking among the strongest candidates.} Two additive models have better
  point-forecast scores than the empirical-Bayes model: A2 by 722 (winning five of nine years,
  ratio 1.7) and A3 by 546 (seven of nine, ratio 1.2). Of A2's 722, 439 come from cells of
  new drive models. The empirical-Bayes model is ahead of A1 by 250 and of the fixed boosted
  ensemble with drive-model identity by 438 (four of nine years). It is behind the tuned boosted ensemble with a year
  feature by 348 (five of nine, ratio 1.0), and A2 is ahead of that ensemble by 374 (seven
  of nine, ratio 1.8). None of these differences is clearly distinguished from the year-to-year
  variation of the scores. That does not make the candidates equivalent. The additive models
  A4 and A5 are behind A1.
\item \textbf{That the predictive rows are better models.} The empirical-Bayes model's predictive
  row is 1{,}323 units ahead of the best point forecast (A2) and 2{,}046 ahead of its own point
  forecast, because it is a distribution with a dispersion term scored against plug-in Poisson
  forecasts. Both are legitimate probabilities for the same counts, but no additive or boosted candidate was
  given a predictive distribution, so the comparison is not like for like. The margin shows
  that carrying the uncertainty of the drive-model multipliers gains more than separates any
  two of the leading point forecasts (A2, A3, the tuned boosted ensemble with a year feature and the
  empirical-Bayes model, which span 722 units).
\end{enumerate}

\begin{table}[h]
\centering
\caption{Predicted over Observed Failures by Test Year. Each forecast uses the data through the previous December. Boosting is the tuned ensemble with the drive-model identity and the calendar year (Section~\ref{sec:boost}).}\label{tab:rollyear}
\footnotesize\setlength{\tabcolsep}{4pt}\begin{tabular}{@{}rrrrrrrrr@{}}
\toprule
Test year & Observed & Bathtub & Const.\ AFR & Per-year Weibull & Boosting & Additive A2 & M2 + cov. & Per-model wear-out \\
\midrule
2018 & 1{,}240 & 2.05 & 1.47 & 1.43 & 1.39 & 1.50 & 1.60 & 1.56 \\
2019 & 2{,}048 & 1.40 & 0.67 & 0.66 & 0.78 & 0.84 & 0.88 & $>10^{3}$ \\
2020 & 1{,}307 & 3.15 & 2.01 & 1.97 & 2.62 & 2.46 & 2.35 & 2.75 \\
2021 & 1{,}804 & 3.14 & 0.93 & 0.92 & 0.93 & 1.00 & 1.14 & 1.14 \\
2022 & 2{,}996 & 2.36 & 0.72 & 0.74 & 0.75 & 0.84 & 0.74 & 0.80 \\
2023 & 4{,}192 & 2.12 & 0.82 & 0.87 & 0.93 & 0.94 & 0.69 & 0.90 \\
2024 & 4{,}528 & 2.22 & 1.07 & 1.04 & 1.19 & 1.10 & 0.80 & 1.10 \\
2025 & 4{,}308 & 2.72 & 1.17 & 1.21 & 1.31 & 1.31 & 1.09 & 1.54 \\
2026 & 2{,}552 & 2.64 & 0.90 & 0.94 & 1.01 & 0.98 & 0.97 & 1.20 \\
\bottomrule
\end{tabular}
\end{table}

\begin{figure}[h]
\centering
\includegraphics[width=\textwidth]{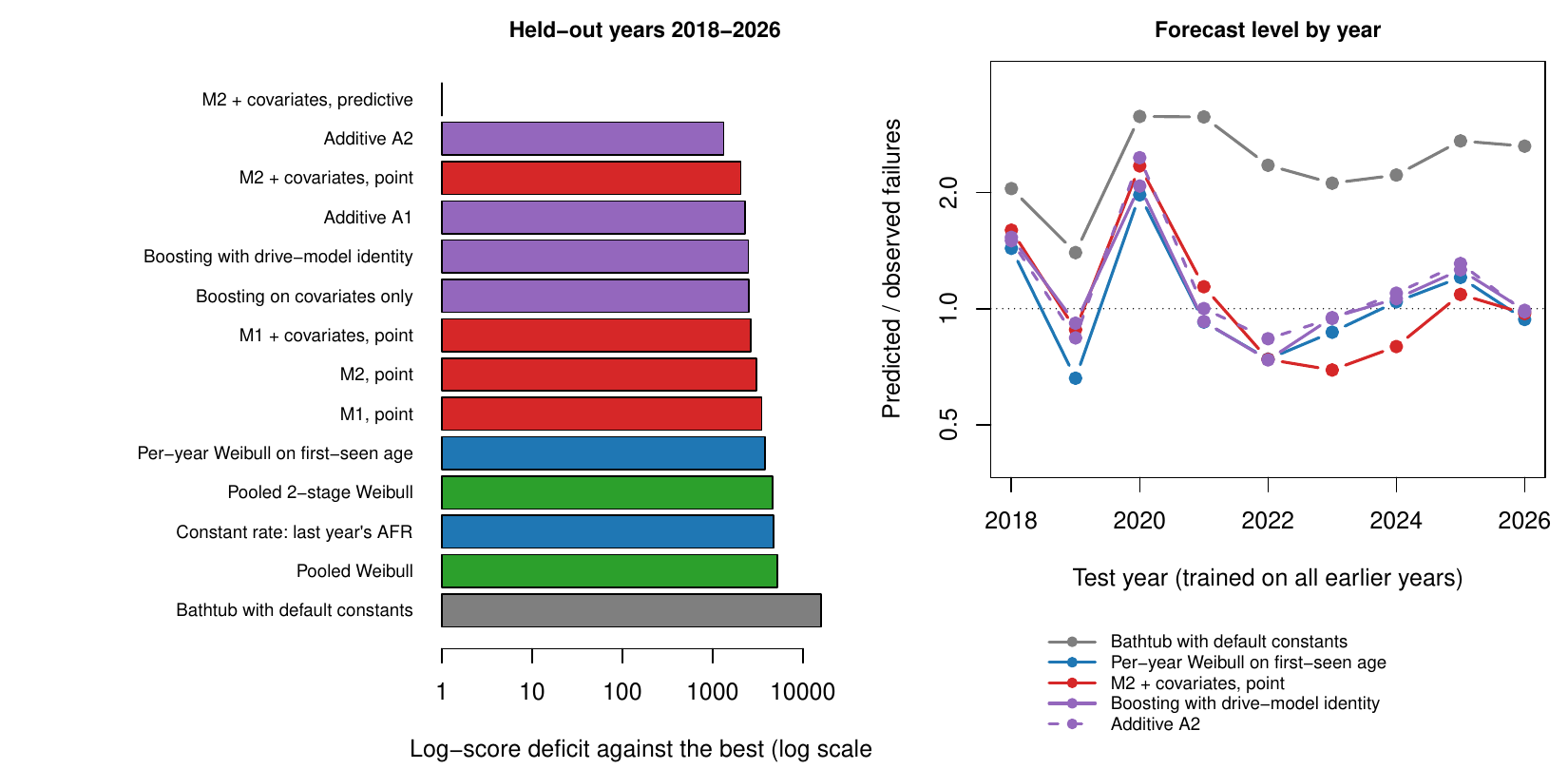}
\caption{Left: log-score deficit of selected candidates relative to the best of those shown, summed
over nine test years (log scale; grey: bathtub; blue: per-year Weibull and constant AFR; green: pooled distributions with no
drive-model effect; purple: additive models and boosting; red: models M1 and M2 of
Section~\ref{sec:hier}; the per-model wear-out model is omitted). Right: predicted over observed
failures by test year, in the same colours (additive A2 dashed).}\label{fig:rolling}
\end{figure}

Every candidate in Table~\ref{tab:rollyear} over-predicts 2018 by at least 39\% and 2020 by a
factor of at least 1.97 (Figure~\ref{fig:rolling}). The fleet did better in those years than any
method we tried forecast from its history.

\begin{table}[h]
\centering
\caption{Log Score by Test Year on All Cells.}\label{tab:rollyearls}
\footnotesize\setlength{\tabcolsep}{4pt}\begin{tabular}{@{}rrrrrrr@{}}
\toprule
Test year & M2 + cov.\ (point) & Per-model wear-out & Additive A2 & Boosting & Per-year Weibull & M2 + cov.\ (predictive) \\
\midrule
2018 & $-1{,}748$ & $-1{,}665$ & $-1{,}635$ & $-1{,}607$ & $-1{,}766$ & $-1{,}673$ \\
2019 & $-2{,}079$ & $-4.47\times 10^{6}$ & $-1{,}877$ & $-2{,}024$ & $-2{,}104$ & $-1{,}759$ \\
2020 & $-2{,}592$ & $-3{,}027$ & $-2{,}656$ & $-2{,}747$ & $-2{,}152$ & $-2{,}033$ \\
2021 & $-2{,}514$ & $-2{,}525$ & $-2{,}518$ & $-2{,}522$ & $-2{,}716$ & $-2{,}426$ \\
2022 & $-3{,}419$ & $-3{,}380$ & $-3{,}426$ & $-3{,}490$ & $-3{,}917$ & $-3{,}304$ \\
2023 & $-4{,}773$ & $-4{,}677$ & $-4{,}417$ & $-4{,}554$ & $-4{,}905$ & $-4{,}529$ \\
2024 & $-5{,}730$ & $-5{,}535$ & $-5{,}572$ & $-5{,}574$ & $-6{,}172$ & $-5{,}415$ \\
2025 & $-5{,}627$ & $-6{,}205$ & $-5{,}721$ & $-5{,}675$ & $-6{,}054$ & $-5{,}330$ \\
2026 & $-3{,}011$ & $-2{,}929$ & $-2{,}947$ & $-2{,}950$ & $-3{,}450$ & $-2{,}977$ \\
\bottomrule
\end{tabular}
\end{table}

The per-model wear-out model of Appendix~\ref{sec:perwear} fails as an extrapolator.
All ten fits (nine training
folds and the all-data fit) reached the stopping rule defined in the appendix after
266--900 iterations. Nevertheless, the model predicted 4{,}471{,}820 failures in 2019
against 2{,}048 observed. Toshiba MG07ACA14TA, introduced in 2018 with eight failures in
training, accounted for 4{,}470{,}180 predicted failures against nine observed. Its fitted
wear-out exponent was 11.99 and its scale was 0.630 years.

Even with 2019 set aside, the model gave a log score of $-29{,}944.0$ over the other
eight test years, against $-29{,}412.8$ for the empirical-Bayes point forecast with the
helium covariate: a deficit of 531.2 with a standard error of 793.7, which is inconclusive.
No additional prior was introduced to suppress the failure. Reaching the stopping rule
does not establish acceptable extrapolation or a predictive advantage for per-model
wear-out exponents.

\section{Do Shared Shocks Improve Failure Forecasts?}\label{sec:bayesresults}

Does allowing unusually good or bad years make forecasts more useful? We test the models
of Section~\ref{sec:time} against nine later years, asking both how well they anticipate
failure counts and whether their forecast ranges contain the observed counts as often as
intended. We also examine what the fitted parameters say about changes in the fleet.

\subsection{Held-Out Comparison}\label{sec:bayesvalid}

We first describe how we score a Bayesian variant on a held-out year, then compare the
variants with each other and with the empirical-Bayes model.

\subsubsection{Design}

We use the folds of Section~\ref{sec:valid}. We fit through December of $Y=2017,\dots,2025$
and score calendar year $Y+1$ on its Lexis cells given their exposure. We also score each
variant two and three years ahead, as far as the data allow.

The score is the logarithm of the joint posterior predictive probability of every failure count
of the test year,
\begin{equation}\label{eq:bayespred}
\log\int p(\text{test}\mid\theta,s_{Y+1},Z)\,p(s_{Y+1}\mid\theta,s_Y)\,p(Z,\theta,s_{\le Y}\mid\text{training})\,
\mathrm{d}s_{Y+1}\,\mathrm{d}Z\,\mathrm{d}\theta .
\end{equation}
All drive models share the test year's shock $s_{Y+1}$, which we integrate by Gauss--Hermite
quadrature (numerical integration against the shock's normal density at 24 fixed points). Given it, drive models are independent. We integrate each $Z_g$
numerically against its own training likelihood on a grid of 241 points in $\log Z$ (for B0
the integral is the closed form \eqref{eq:pred}). For the numerical variants with
common wear-out shape, drive models new in the test year take the prior of $Z$
at 200 fixed quantile nodes. The per-model-shape variants use 12{,}800 seeded
prior draws, the same set in every run, including shape deviations for B4; new
models in B5 use the common shape.
That leaves a function
$l(\theta)=p(\text{test}\mid\text{training},\theta)$ of the remaining parameters $\theta$ (the
hyperparameters and the year effects through $Y$) to be averaged over their posterior.

The plain average of $l$ over posterior draws is unbiased, but $\log l$ varies across draws
with a standard deviation of up to 8 for the bridged variants. The average is then dominated by one or two draws and
has a very large variance, and its logarithm, the score, is biased downward (Jensen's
inequality). Instead we use the fact that \eqref{eq:bayespred} is the ratio of the normalizing
constants of two posteriors over $\theta$, given the training data and given training and test
data. For an unchanged likelihood and forecast rule, the second is the next fold's fit.
B5 is an exception: its second posterior must use the training age cutoffs and the same
new-model rule as the forecast. Extending the ordinary fit changes those cutoffs and does
not produce the required target.

For the shock variants we estimate the ratio by bridge sampling~\cite{meng1996bridge}, which
estimates a ratio of normalizing constants from draws of both posteriors, from 1{,}000 draws of
each posterior. Repeating the estimate on each of the four chains gives a per-fold
error from the between-chain spread. The root sum of squares of these errors is
0.2 to 1.1 over nine years. The estimate differs from the plain average by
at most 4.8 on the sum. The estimate needs draws of both posteriors on both sides of it. Over the 117 fits of the nine shock variants that were bridged, the share of training-posterior draws with $l$ above the estimate is at least 0.1\% (median 11\%), and the share of draws from the second posterior with $l$ below it is at least 0.2\% (median 8\%); in 7 of them one of the two shares is under 1\%. The iteration converged in every case.

Writing the fold errors as $s_y$, the reported aggregate is
$s_{\mathrm{RSS}}=(\sum_y s_y^2)^{1/2}$. It omits the cross-fold covariance terms
in the finite-variance identity:
\[
 \operatorname{Var}\!\left(\sum_y\widehat L_y\right)
 =\sum_y\operatorname{Var}(\widehat L_y)
   +2\sum_{y<z}\operatorname{Cov}(\widehat L_y,\widehat L_z).
\]
Here the variance concerns numerical estimation of the log scores for fixed data.
Adjacent ordinary bridges reuse posterior draws, and folds reuse a random seed
within each run; independence of their numerical errors is not established.
We do not estimate the cross-fold covariances. The reported aggregate is therefore
a zero-covariance approximation, which also does not measure sampling bias or
quadrature error.
With 200 prior draws, changing only the seed moved the nine-fold totals of B4 with
drift and B5 by up to 0.36, on unchanged posterior draws, and one B5 fold by 0.33,
six times its reported error. At 12{,}800 draws, twelve draws at origin 2023 show a
mean change of 0.0013 between two seeds.

For B0 bridge sampling fails. In six of the nine folds no draw from either posterior has a
value of $l$ on the far side of the estimate, so the estimate extrapolates across a gap and
moves when we change the number of draws. Without shocks, however, $\theta$ has eleven
hyperparameters, and both the training likelihood (with the multipliers integrated by
conjugacy) and $l$ are closed-form. We therefore compute each normalizing constant directly by
importance sampling from a multivariate $t$ distribution fitted to that posterior's draws
(4{,}000 points, effective sample size 2{,}100 to 2{,}300 for every one of the eighteen integrals).
The standard error is 0.02 a fold. In the three folds where the two posteriors overlap by at
least 7\% of their draws, the three methods agree to within 0.3. Over
nine folds the importance-sampling total is $-29{,}153$, the bridge estimate $-29{,}196$, and
the plain average $-29{,}277$.

We did not implement either estimator for scores two and three years ahead. We do
report forecast levels at those horizons.

\begingroup\widowpenalty=10000
Like the ``predictive'' rows of Table~\ref{tab:rolling}, the score is a probability for all the
test counts, so it is comparable with those rows and with the Poisson scores of point
forecasts. The point columns report the numerical forecast used by the implementation.
For drive models absent from training, B4 averages a finite set of prior-shape draws.
Thus these point
columns must not be interpreted uniformly as posterior means.\par
\endgroup

\paragraph{Existence of the count mean.}\label{sec:newshapemean}
Conditional on the hyperparameters, an unseen model in B4 has a lognormal
wear-out exponent $B=\exp(m+\sigma e)$, with $e\sim N(0,1)$ and $\sigma>0$.
For an age bin $[l,u]$ with $u>4$ years, its wear-out contribution to the mean hazard is
\[
 \frac{k_2}{u-l}\left\{(u/4)^B-(l/4)^B\right\}.
\]
For sufficiently large $B$, the second term is at most half the first. The expectation
of the first term is infinite: with $t=\log(u/4)>0$, the exponent in its normal integral
is $t\exp(m+\sigma e)-e^2/2$, which tends to positive infinity. Positive exposure, a fixed
positive drive-model multiplier and the unit-mean future model-year shock preserve this
divergence; integrating over the remaining parameters cannot remove it. The unseen-model count
therefore has an infinite mean under B4, even though its predictive
probabilities are well defined.

The same conclusion holds for a \emph{seen} model if all its training age bins end at or
before four years. Holding the other parameters fixed, its training bin hazards approach
positive finite limits as $B\to\infty$: the early-life term remains positive, while each
wear-out increment is bounded. The likelihood of the finite training data therefore
approaches a positive finite limit. The conditional posterior retains the lognormal shape
tail up to an asymptotically constant factor, and its count mean in any exposed bin ending
above four years is again infinite. This is a sufficient condition; means outside it are
not certified. B5 changes the extrapolation beyond a model's training support and uses the
common exponent for unseen models, so this argument does not apply to B5.

Across the nine one-year folds, this condition covers
577 target cells: 6{,}149{,}275 drive-days and 249 recorded failures. Of these, 566 cells
belong to 43 seen models. The remaining eleven cells belong to HGST HUH728080ALE604,
absent through 2020 and observed above four years in 2021 (21 drive-days, no failures).
At least one affected model occurs in every fold. Since the count contributions are
nonnegative, the posterior predictive mean of the fleet total is infinite in all nine
folds under B4.
For example, TOSHIBA MG07ACA14TEY's training bins through 2022 end at 2.25 years; its
2024 and 2025 targets include bins above four years. The deposited finite point summaries
do not estimate the infinite means. This limitation is separate from sampling convergence
and B5's bridge-target correction.

\paragraph{Predictive medians.}\label{sec:pmsmedians}
We report medians of the simulated fleet count for B4, whose fleet mean is infinite in
these folds, and for B5, whose count mean has not been established to be finite
(Table~\ref{tab:pmsmedians}). Over the nine held-out periods, the sums of the predictive
medians for B4 with drift, B4 without a year effect and B5 are 1.73, 1.50 and 1.70 times
the total observed failures, against 0.96 for B3 with drift. These are sums of separate
rolling predictions, not medians of a joint aggregate count. B4 with drift and B5 predict
approximately twice the observed count in 2024 and more than twice in 2025. The three runs
agree to within 0.02 on each ratio. The mean absolute errors of the nine fleet medians are
564 for B3, 2{,}023 for B4 with drift, 1{,}377 for B4 without a year effect and 1{,}967 for
B5, including the partial 2026 period without annualization. The excess is not confined
to sparse models: in 2024 the median for TOSHIBA
MG07ACA14TA under B4 with drift is 2{,}018 failures against 415 observed, while
HGST HUH721212ALN604 goes the other way, 113 against 590. B5, whose unseen models take the
common exponent, shows the same pattern, so it does not come from the infinite mean alone.

The same excess appears at unit model-year shock in the training years. Trained through
2023 and evaluated for 2023, the median fleet intensity with the model-year shock set to
one is 7{,}312 for B4 with drift and 7{,}418 for B5, against 4{,}192 observed and 4{,}214
for B3 with drift. Conditioning on the shock posterior gives median fleet intensities of
4{,}187 and 4{,}188 in this check. At the model-year level, for a fixed parameter draw,
the conditional intensity is $M(\kappa+D)/(\kappa+M)$, where $M$ is the intensity at unit
shock, $D$ is the observed count and $\kappa$ is the shape of the unit-mean gamma shock.
This is a weighted average of $M$ and $D$; agreement with the same training counts does
not establish model adequacy. These fits rely on inferred shocks below their unit prior
mean in the highlighted years. The cause of the held-out excess is not established.
These variants assign more probability
to what occurred than B3 does, yet their central forecasts of the number of failures are
much worse, and we do not recommend them for forecasting failure counts.

\begin{table}[h]
\centering
\caption{Median Predictive Fleet Failures on the Nine Held-Out Periods, Replicate A. Each median is of the fleet count simulated jointly across drive models from the saved draws. Absolute errors and coverage include the partial 2026 period. The last row counts the periods whose observed total lies in the central 95\% predictive interval.}\label{tab:pmsmedians}
\footnotesize\setlength{\tabcolsep}{4pt}\begin{tabular}{@{}lrrrrr@{}}
\toprule
Year & Observed & B3 drift & B4 drift & B4 none & B5 \\
\midrule
2018 & 1{,}240 & 1{,}340 & 1{,}800 & 2{,}186 & 1{,}766 \\
2019 & 2{,}048 & 1{,}462 & 2{,}009 & 2{,}206 & 2{,}036 \\
2020 & 1{,}307 & 2{,}482 & 3{,}710 & 3{,}746 & 3{,}631 \\
2021 & 1{,}804 & 1{,}547 & 2{,}329 & 2{,}701 & 2{,}237 \\
2022 & 2{,}996 & 2{,}002 & 3{,}058 & 3{,}307 & 2{,}942 \\
2023 & 4{,}192 & 3{,}317 & 6{,}034 & 4{,}821 & 5{,}934 \\
2024 & 4{,}528 & 4{,}418 & 9{,}114 & 6{,}269 & 9{,}006 \\
2025 & 4{,}308 & 5{,}047 & 9{,}750 & 7{,}523 & 9{,}706 \\
2026 (half) & 2{,}552 & 2{,}312 & 5{,}301 & 4{,}604 & 5{,}291 \\
\midrule
Sum of medians / observed & & 0.96 & 1.73 & 1.50 & 1.70 \\
Mean absolute error & & 564 & 2{,}023 & 1{,}377 & 1{,}967 \\
Years in 95\% interval & & 9 & 5 & 4 & 5 \\
\bottomrule
\end{tabular}
\end{table}

\subsubsection{Results}

\begin{table}[h]
\centering
\caption{Bayesian Variants on the Nine Held-Out Years. Pred./obs.\ is the numerical point forecast of total failures over the 24{,}975 observed. $\Delta$ is the gain in predictive log score over the empirical-Bayes predictive baseline ($-29{,}446$). Won counts the years won. SE is the standard error from the spread of the nine yearly differences, and Ratio is $\Delta$ over SE. Point is the Poisson log score of that forecast; the point rules and mean-existence limitation are in Section~\ref{sec:newshapemean}.}\label{tab:bayesroll}
\footnotesize\setlength{\tabcolsep}{4pt}\begin{tabular}{@{}llrrrrrrr@{}}
\toprule
Shocks & Year effect & Pred./obs. & Predictive & $\Delta$ & Won & SE & Ratio & Point \\
\midrule
B0: none & --- & 1.00 & $-29{,}153$ & $+293$ & 9/9 & 84 & 3.5 & $-31{,}545$ \\
\addlinespace
B1: constant precision & none & 1.53 & $-27{,}143$ & $+2{,}303$ & 9/9 & 479 & 4.8 & $-34{,}695$ \\
B1: constant precision & independent & 2.84 & $-27{,}129$ & $+2{,}317$ & 9/9 & 478 & 4.8 & $-51{,}225$ \\
B1: constant precision & AR(1) & 1.59 & $-27{,}120$ & $+2{,}326$ & 9/9 & 478 & 4.9 & $-34{,}563$ \\
B1: constant precision & random walk & 1.33 & $-27{,}118$ & $+2{,}328$ & 9/9 & 478 & 4.9 & $-32{,}338$ \\
B1: constant precision & random walk with drift & 1.04 & $-27{,}118$ & $+2{,}328$ & 9/9 & 478 & 4.9 & $-31{,}240$ \\
\addlinespace
B3: size-dependent precision & none & 1.38 & $-27{,}129$ & $+2{,}316$ & 9/9 & 476 & 4.9 & $-32{,}902$ \\
B3: size-dependent precision & independent & 2.64 & $-27{,}115$ & $+2{,}331$ & 9/9 & 471 & 4.9 & $-47{,}490$ \\
B3: size-dependent precision & random walk & 1.24 & $-27{,}103$ & $+2{,}343$ & 9/9 & 471 & 5.0 & $-31{,}360$ \\
B3: size-dependent precision & random walk with drift & 1.01 & $-27{,}103$ & $+2{,}343$ & 9/9 & 471 & 5.0 & $-30{,}690$ \\
\addlinespace
\bottomrule
\end{tabular}
\end{table}

\begin{table}[h]
\centering
\caption{Paired Comparisons of the Bayesian Variants. Columns are as in Table~\ref{tab:pairs}. Rows above the rule compare predictive log scores, and rows below compare the Poisson log scores of point forecasts.}\label{tab:bayespairs}
\footnotesize\setlength{\tabcolsep}{4pt}\begin{tabular}{@{}p{10.2cm}rrrr@{}}
\toprule
Comparison & Sum & Years won & Std.\ error & Ratio \\
\midrule
B0 against the empirical-Bayes model & $+293$ & 9/9 & 84 & 3.5 \\
B3 with drift against B0 (shocks against none) & $+2{,}050$ & 9/9 & 438 & 4.7 \\
size-dependent against constant shock precision (drift) & $+15$ & 6/9 & 8 & 2.0 \\
size-dependent against constant shock precision (random walk) & $+15$ & 6/9 & 8 & 2.0 \\
drifting year effect against no year effect (B3) & $+26$ & 7/9 & 11 & 2.5 \\
drift against no drift (B3) & $-0$ & 6/9 & 2 & -0.0 \\
random walk against independent years (B3) & $+12$ & 9/9 & 1 & 11.4 \\
\midrule
point forecasts: B3 with drift against additive model A2 & $+79$ & 3/9 & 463 & 0.2 \\
point forecasts: B3 with drift against the empirical-Bayes model & $+802$ & 6/9 & 430 & 1.9 \\
point forecasts: B3 with drift against B0 & $+856$ & 6/9 & 456 & 1.9 \\
\bottomrule
\end{tabular}
\end{table}

\textbf{Shocks.} Every variant with shocks in Table~\ref{tab:bayesroll} is 2{,}303 to 2{,}343 log units ahead of the empirical-Bayes
model and 2{,}010 to 2{,}050 ahead of B0, in every one of the nine years
(Tables~\ref{tab:bayesroll} and~\ref{tab:bayespairs}). For B3 with drift the gain over B0 is
2{,}050, with a ratio to its standard error of 4.7 on eight degrees of freedom.

\textbf{Year process.} Among the B1 and B3 variants the predictive scores lie within 40 log units
of one another. For B3, the summed predictive scores with and without drift differ by less than one
log unit. The drifting year effect gains 26 over no year effect. The year process does change
the forecast level. For B3 the pooled predicted-to-observed ratio is 2.64 with independent
years, 1.38 with no year effect, 1.24 with a random walk and 1.01 with a drifting random walk.
These ratios pool nine years, so errors of opposite sign cancel. The yearly ratios
(Table~\ref{tab:bayesyear}) for B3 with drift run from 0.70 to 2.03. The mean of their
logarithms measures bias: $+0.24$ (standard error 0.11) for the random walk, $+0.02$ (0.11)
for the walk with drift and $+0.06$ (0.13) for B0. Three years ahead (Table~\ref{tab:bayeshz})
the random walk without drift is biased, with a mean log ratio of $+0.80$ (0.19). The walk with
drift, $+0.15$ (0.21), and B0, $+0.24$ (0.23), are distinguished neither from zero nor from
each other (their paired difference in absolute log ratio is 0.04 with a standard error of
0.09). The mean absolute log ratio at three years is 0.48 for B3 with drift and 0.52 for B0: a
typical single-year forecast is off by a factor of about 1.6 under either. These standard
errors treat the folds as independent, which they are not beyond one year, because their test
years overlap. The bias of the variants without drift is explained in Section~\ref{sec:apc}.

\textbf{Shock precision.} Letting the gamma shock's precision grow with the expected count gains
15 log units with drift (six of nine years, ratio 2.0) and 15 without drift (six of nine
years, ratio 2.0). The predicted-to-observed ratio changes from 1.04 to 1.01 with drift and
from 1.33 to 1.24 without.

\textbf{Point forecasts.} The posterior-mean forecast of B3 with drift scores $-30{,}690$, against $-30{,}769$ for the additive model A2, which had the best point score in Section~\ref{sec:valid}, and $-31{,}491$ for the empirical-Bayes model. Neither difference is established: 79 with a standard error of 463, and 802 with a standard error of 430.
\unskip\ B1 with drift ($-31{,}240$) and B3 with a random
walk ($-31{,}360$) are also slightly ahead of B0 ($-31{,}545$). The other six shock variants,
whose forecast levels are further off, are behind it.

\textbf{What shocks do not change.} Every variant over-predicts 2020 by a factor of about two or
more and 2018 by at least 1.18. The standard deviation of the log ratio across years falls
from 0.40 (B0) to 0.32 (B3 with drift). The shocks widen the predictive distribution, but they
do not remove these errors of level.

\textbf{Per-model wear-out.}\label{sec:pms} Mittman et al.~\cite{mittman2019hierarchical} let the wear-out
parameters vary by drive model, with partial pooling. We fitted two variants of B3 with drift, and one of them also without year effects (Table~\ref{tab:pms}). In B4 the logarithm of
each drive model's wear-out exponent is normal about a regression on vintage, log capacity,
manufacturer and helium, and the sampler is non-centred. In the score each drive model's multiplier and
exponent deviation are integrated on a two-dimensional grid, so that B4 is bridged like B3. B5
uses a drive model's own exponent only up to the oldest age at which the model was
observed, and continues its hazard beyond that age with the reference drive model's exponent
parameter. We designed B5 after seeing the forecast levels of B4. We also fitted a simpler
version with no regression on drive-model attributes, whose sampler did not mix, and B4 with a
tight prior on the spread of the exponents, which scored like B4; we report neither.

In B4, the baseline wear-out contribution on an age bin $0<l<u<a_\star$, with
$a_\star=4$~yr, is
\[
\bar h_{2g}(l,u;b)=\frac{k_2}{u-l}
 \left\{\left(\frac{u}{a_\star}\right)^b-
             \left(\frac{l}{a_\star}\right)^b\right\},\qquad b=\beta_{2g}.
\]
As a function of $b>0$, it increases from zero to a unique maximum at
\[
b_{\max}(l,u)=
 \frac{\log\!\left\{\log(l/a_\star)/\log(u/a_\star)\right\}}{\log(u/l)},
\]
then decreases to zero. Thus a single such bin cannot distinguish the two
exponents giving a positive wear-out rate below that maximum, with the other
parameters fixed. The maximum depends on the bin; observations across bins and
the hierarchical prior can distinguish the exponents. For $l=0$ and $u<a_\star$,
the wear-out contribution instead decreases monotonically with $b$.
A restricted age range can therefore leave the local shape weakly determined.
We check individual drive models' exponents and predictive counts, since
acceptable diagnostics for a fleet total do not imply well-mixed components.

The cutoff in B5 does not by itself identify the local exponent. Let $A_g>0$ be
the frozen training cutoff. Before the cutoff its baseline wear-out cumulative
hazard is $k_2(a/a_\star)^{\beta_{2g}}$, where $a_\star=4$~yr. Beyond the cutoff,
the wear-out component of the baseline hazard is
\[
h_{2g}(a)=\frac{k_2}{A_g}\,\beta_{2g}
 \left(\frac{A_g}{a_\star}\right)^{\beta_{2g}}
 \left(\frac{a}{A_g}\right)^{\beta_2-1},\qquad a\ge A_g.
\]
\begingroup\widowpenalty=10000
For fixed $A_g<a_\star$, $k_2$ and $\beta_2$, its dependence on the local exponent
$b=\beta_{2g}$ is through $b(A_g/a_\star)^b$. This amplitude increases up to
$b=1/\log(a_\star/A_g)$ and then decreases: two distinct positive exponents can
give the same hazard beyond the cutoff. Observations before the cutoff and the
hierarchical prior can distinguish them. The cutoff rule alone therefore does
not guarantee an identified wear-out exponent or a well-mixed posterior.\par
\endgroup

Individual exponents show this in the sampler. In B5 trained through 2022, the indicator of
the upper branch of TOSHIBA MG07ACA14TEY ($\beta_{2g}>1.739$) has rank-normalized split
$\hat R=1.078$ and effective sample size 75, while its next-year count has $\hat R=1.006$.
A 16-chain B4 fit without year effects (32{,}000 draws, no divergent transitions) finds a
separated region for the same drive model: a standardized exponent deviation above 0.5 holds
4.81\% of the draws, and its indicator has $\hat R=1.115$ and effective sample size 139, so
that share is not a converged probability. Shifting the multiplier with exposure-weighted age,
which preserves the posterior, did not help (indicator $\hat R=1.096$, 11 divergent
transitions). The data do not determine the wear-out exponent of a drive model with no exposure
past the four-year pivot, and no sampler can supply that information. For such models we report
only that there are not enough data; their individual exponents are not estimates.

The score of these variants integrates each seen model's shape deviation numerically, and our
first implementation, 17 nodes on $[-4,4]$, was wrong. For a data-rich model the conditional
posterior of the deviation is narrower than the node spacing and can lie outside that range. The
error in the fleet log probability at a posterior draw reached 3.7 log units, and on the default
B4-with-drift draws the nine-fold total rose by 5.34 at 49 nodes. The reported scores replace
that grid: for each posterior draw and drive model, a first pass of 65 nodes on $[-16,16]$ finds
the range within 30 log units of the profile's maximum, and 65 nodes integrate over it. No first
pass reached the edge of $[-16,16]$. At twelve draws from each of two origins (2021, 2023) and
every variant, the fleet log probability is within $1.5\times10^{-4}$ of a 3{,}201-node grid
on $[-16,16]$, and separate code agrees with that grid at four draws to $5.5\times10^{-12}$.
These checks cover two of the nine origins.

Overlap between the two posteriors of a bridge is weak at origin 2023: as few as 0 to 6 draws
per 1{,}000 cross the estimate, on each side, although the iteration converges. The three runs
bound the effect empirically. Their largest difference in any one fold is 0.74 (B5, origin 2020),
and at origin 2023 it is 0.41 for B4 with drift, 0.63 without a year effect and 0.69 for B5.
All three runs share the model and the sampler, so this agreement does not exclude a common bias.

Table~\ref{tab:pmsdiag} gives each variant's sampler diagnostics and scores from three runs:
the default settings (four chains of 1{,}500 warm-up and 1{,}000 sampling iterations, target
acceptance 0.9), replicate A (3{,}000 and 2{,}000 iterations, target acceptance 0.97, the same
seed) and replicate B (the settings of A with another seed and dispersed starting points). A
fold passes the reported $\hat R$ screen when its largest value is
at most 1.05; the per-fold diagnostics are deposited with the summaries (\code{pms\_folds.csv}).
This is a descriptive screen, not a convergence certificate. It uses classical split-chain
$\hat R$ for the reported latent parameters and does not replace checks of derived quantities,
effective sample size or divergent transitions~\cite{vehtari2021rhat}. At the default settings B4 with drift and B4 without
a year effect each have one fold fit, trained through
2022\unskip, that does not mix ($\hat R$ 1.140 and 1.055,
effective sample sizes 36 and 82), and B5 passes. In replicate A every fold fit of the three
variants has $\hat R$ at most 1.010\unskip\ and an effective
sample size of at least 325\unskip. In replicate B the same
fold fails for B4 without a year effect ($\hat R$ 1.117, effective sample size 40) and for B5
($\hat R$ 1.054, effective sample size 91), and B4 with drift passes. One of B5's nine
frozen-cutoff target fits fails in the default run ($\hat R$ 1.342, origin 2021) and one in
replicate B (1.087, origin 2017). Every run also records divergent transitions,
0 to 14\unskip\ over the nine folds of a variant, which we did not
remove. The fits to all data supply the posterior means of Table~\ref{tab:pms} and the second
posterior in the last fold's bridge calculation for B4; all pass, with $\hat R$ at most 1.01. Across the three
runs each variant's gain over B3 with drift moves
by at most 2\unskip\ log units and its forecast ratio by at
most 0.01\unskip, and the reported root-sum-of-squares
Monte Carlo error of a nine-fold sum is at most 1.1\unskip. For
the 4\unskip\ fold fits that did not mix, the bridge estimate of
the fold's score is within 0.1\unskip\ log units of the
other runs' estimates of the same fold, and the numerical point forecast within
2\unskip\%. B4 with
drift gains
223\unskip\ log units over B3 with drift (standard error
86\unskip, eight of nine years), and the three variants gain
191 to 224\unskip. In these fits, allowing wear-out to differ by
drive model assigns a higher estimated probability to the next year's counts. Every variant
also over-predicts the next year's total, by a factor of
1.53 to 1.84\unskip\ at the default settings and
1.53 to 1.85\unskip\ over the three runs, so their point scores are far
worse than B3's. The excess
lies in the drive models seen in training (B4 forecasts 1.82\unskip\
times their failures, B3 0.99\unskip). New drive models are
over-predicted by both (5.9\unskip\ and
5.4\unskip\ times, on 126 failures). Two fitted quantities change
with the per-model exponents. The drift of the year effect, $-0.134$ in B3, goes to zero,
and the wear-out exponent of the reference drive model (a Seagate of 8~TB introduced in
2018) falls from 1.93 to 0.92 to 0.94\unskip; in B4 and B5 this is one point of a
regression on drive-model attributes, not an exponent shared by all drive models. Within a drive
model age and calendar time are nearly the same variable (Section~\ref{sec:permodel}), so a
per-model wear-out term and the fleet-wide calendar trend can stand in for each other. That is
consistent with the drift vanishing, but we have not established how it produces the excess in
the forecast level, and we do not offer it as the explanation.
B5 changes two things against B4: beyond its oldest observed age a drive model continues with
the reference drive model's exponent, and a new drive model takes that exponent
with no regression term. It does not remove the excess in the saved estimates.
Lewis-Beck et al.~\cite{lewisbeck2022prediction} checked
per-model forecasts one week ahead, in each of 26 weeks; ours forecast the level a full year
ahead. B3 with drift is the variant we retain for the
simulation examples of this paper, because its aggregate forecasts were closer to the observed
totals in these development folds. That does not establish it as the best general simulator or
the best model of the future trend. B4 and B5 are not supplied as samplers. Their score gains
are similar across the three runs, including the run in which every training fold passes the
$\hat R$ screen. This does not rule out shared sampling bias. Until the forecast level and posterior sampling are resolved they do
not establish a benefit from per-model wear-out for simulation.

\textbf{Where the excess lies.} Summed over the nine test years, the numerical point forecast
of B4 with drift exceeds that of B3 with drift by 20{,}668\unskip\
failures. The deposited decomposition is three marginal tables, by power-on age band, by drive
model and by test year; we did not cross them. By age, the excess is at every age the fleet
has exposure: B4's forecast is 1.2\unskip\ to
2.3\unskip\ times the observed failures in each band
below eight years, 59\unskip\% of the excess falls at ages
three to six and 1\unskip\% at eight and over. The bands
are absolute ages, so the table does not say whether a drive model's excess lies beyond the
ages at which it was observed in training; in a one-year-ahead fold every drive model is
forecast up to a year past them. By drive model,
83\unskip\% of the excess falls on five drive models and
34\unskip\% on one, the Toshiba MG07ACA14TA, which B4
forecasts at 4.2\unskip\ times its observed failures where
B3 forecasts 1.2\unskip\ times. By test year the ratio of
B4's forecast to the observed total is 1.57, 1.05, 3.03, 1.35, 1.07, 1.53, 2.16, 2.41, 2.21\unskip\ for
2018\unskip\ to 2026\unskip: it
is not monotone, it exceeds two in 2020 and in every year from 2024, and
79\unskip\% of the excess is in the test years from 2023
on. The excess is thus concentrated in a few large drive models and in the recent test
years. We have not established the mechanism, and we do not offer one.

\textbf{A damped drift.} The drift of B3 is the same step every year, so projected forward
it continues the fall of the year effect indefinitely. In a variant with a damped drift the
step in year $t$ has mean $\text{drift}\times\delta^{t-1}$, with a uniform prior on $\delta$
in $(0,1)$, so that the drift can die away. Fitted at the settings of replicate A, the
posterior mean of $\delta$ in the fit to all data is 0.66\unskip\
with a 95\% interval of $(0.05,\ 0.97)$, little narrower than the prior, and the first step
is -0.189\unskip. On the nine folds the variant gains
+1.1\unskip\ log units over B3 with drift (standard error
1.4\unskip, 5\unskip\ of nine years), and
its forecast ratio is 1.19\unskip\ against
1.01\unskip\ for B3 with drift. The yearly steps do not determine
$\delta$, the score does not change, and the forecast level is higher.

\begin{table}[h]
\centering\small
\caption{Per-Model Wear-Out Variants of the Bayesian Model on the Nine Held-Out Years. Predictive is
the summed bridge estimate of the predictive log score. The gain is over B3 with drift, with
the standard error from the spread of the nine yearly differences and the years won.
Pred./obs.\ and Point are the numerical point forecast's ratio to the observed total and its
Poisson log score (see Section~\ref{sec:newshapemean}). $\beta_2$ is the wear-out exponent of the reference drive model (in B3
shared by all drive models), $\sigma$ the spread of the
per-model log exponents, and Drift the drift of the year effect, all posterior means
from the fit to all data. $\hat R$ is the largest over the nine fold fits.}\label{tab:pms}
\footnotesize\setlength{\tabcolsep}{3.5pt}
\begin{tabular}{@{}lrrrrrrrrrr@{}}
\toprule
Variant & Predictive & Gain & SE & Won & Pred./obs. & Point & $\beta_2$ & $\sigma$ & Drift & $\hat R$ \\
\midrule
B3 with drift (reference) & -27{,}103 & --- & --- & --- & 1.01 & -30{,}690 & 1.93 & --- & -0.134 & 1.011 \\
B4 with drift & -26{,}880 & +223 & 86 & 8 & 1.84 & -42{,}749 & 0.92 & 0.57 & +0.010 & 1.140 \\
B4, no year effect & -26{,}879 & +224 & 85 & 8 & 1.53 & -38{,}626 & 0.94 & 0.56 & --- & 1.055 \\
B5 with drift & -26{,}912 & +191 & 89 & 7 & 1.81 & -42{,}481 & 0.92 & 0.57 & +0.009 & 1.012 \\
\bottomrule
\end{tabular}
\end{table}

\begin{table}[h]
\centering\small
\caption{Sampler Diagnostics and Scores of the Per-Model Wear-Out Variants over Three Runs. Default: four chains of 1{,}500 warm-up and 1{,}000 sampling iterations, target acceptance 0.9. Replicate A: 3{,}000 and 2{,}000 iterations, target acceptance 0.97. Replicate B: as A, with another seed and dispersed starting points. $\hat R$, ESS and Div.\ are the largest $\hat R$, the smallest effective sample size and the total divergent transitions over the nine training-fold fits; $\hat R$ (all) is the largest $\hat R$ of the fit to all data. That fit supplies the posterior means of Table~\ref{tab:pms} and the second posterior for the last bridge estimate of B4. Gain and SE are as in Table~\ref{tab:pms}; MC SE is the root sum of squares of the per-fold chain errors, with cross-fold covariances omitted; Pred./obs.\ is the numerical point forecast over the observed total (Section~\ref{sec:newshapemean}). B5 bridge-based score entries use the frozen-cutoff correction described in Section~\ref{sec:bayesvalid}.}\label{tab:pmsdiag}
\footnotesize\setlength{\tabcolsep}{3.5pt}
\begin{tabular}{@{}llrrrrrrrr@{}}
\toprule
Variant & Run & $\hat R$ & ESS & Div. & $\hat R$ (all) & Gain & SE & MC SE & Pred./obs. \\
\midrule
B3 with drift (reference) & default & 1.011 & 569 & 12 & 1.01 & --- & --- & 0.3 & 1.01 \\
B4 with drift & default & 1.140 & 36 & 2 & 1.01 & +223 & 86 & 0.4 & 1.84 \\
B4, no year effect & default & 1.055 & 82 & 0 & 1.01 & +224 & 85 & 1.1 & 1.53 \\
B5 with drift & default & 1.012 & 317 & 2 & 1.01 & +191 & 89 & 0.7 & 1.81 \\
B3 with drift (reference) & replicate A & 1.005 & 1058 & 14 & 1.00 & --- & --- & 0.2 & 1.02 \\
B4 with drift & replicate A & 1.008 & 553 & 0 & 1.00 & +224 & 87 & 0.5 & 1.85 \\
B4, no year effect & replicate A & 1.010 & 325 & 0 & 1.01 & +225 & 85 & 0.6 & 1.54 \\
B5 with drift & replicate A & 1.007 & 723 & 1 & 1.00 & +190 & 89 & 0.7 & 1.82 \\
B3 with drift (reference) & replicate B & 1.004 & 827 & 7 & 1.00 & --- & --- & 0.2 & 1.02 \\
B4 with drift & replicate B & 1.005 & 853 & 4 & 1.00 & +224 & 87 & 0.4 & 1.85 \\
B4, no year effect & replicate B & 1.117 & 40 & 0 & 1.00 & +224 & 85 & 0.4 & 1.54 \\
B5 with drift & replicate B & 1.054 & 91 & 4 & 1.00 & +190 & 89 & 0.8 & 1.82 \\
\bottomrule
\end{tabular}
\end{table}

\begin{table}[h]
\centering
\caption{Predicted over Observed Failures by Test Year, Bayesian Variants. Each entry is the numerical point forecast, one year ahead (Section~\ref{sec:newshapemean}).}\label{tab:bayesyear}
\footnotesize\setlength{\tabcolsep}{4pt}\begin{tabular}{@{}rrrrrrrr@{}}
\toprule
Test year & Observed & B0 & B1, walk & B1, drift & B3, no year effect & B3, walk & B3, drift \\
\midrule
2018 & 1{,}240 & 1.59 & 1.63 & 1.18 & 2.07 & 1.58 & 1.20 \\
2019 & 2{,}048 & 0.89 & 1.16 & 0.84 & 1.22 & 0.99 & 0.76 \\
2020 & 1{,}307 & 2.39 & 2.99 & 2.08 & 3.09 & 2.72 & 2.03 \\
2021 & 1{,}804 & 1.15 & 1.24 & 0.90 & 1.88 & 1.19 & 0.90 \\
2022 & 2{,}996 & 0.74 & 0.97 & 0.69 & 1.23 & 0.92 & 0.70 \\
2023 & 4{,}192 & 0.69 & 1.05 & 0.83 & 1.07 & 1.02 & 0.84 \\
2024 & 4{,}528 & 0.80 & 1.27 & 1.07 & 1.08 & 1.19 & 1.02 \\
2025 & 4{,}308 & 1.09 & 1.51 & 1.29 & 1.34 & 1.43 & 1.22 \\
2026 & 2{,}552 & 0.97 & 1.18 & 1.01 & 1.20 & 1.10 & 0.95 \\
\midrule
mean of log ratio & & $+0.06$ & $+0.31$ & $+0.04$ & $+0.39$ & $+0.24$ & $+0.02$ \\
its standard error & & 0.13 & 0.11 & 0.11 & 0.12 & 0.11 & 0.11 \\
sd of log ratio & & 0.40 & 0.34 & 0.32 & 0.36 & 0.33 & 0.32 \\
\bottomrule
\end{tabular}
\end{table}

\begin{table}[h]
\centering
\caption{Forecasts One, Two and Three Years Ahead. There are nine, eight and seven folds. Pooled is total predicted over total observed failures. Mean log ratio is the mean over folds of $\log(\text{predicted}/\text{observed})$, with its standard error. Mean abs.\ is the mean of its absolute value.}\label{tab:bayeshz}
\footnotesize\setlength{\tabcolsep}{3pt}\begin{tabular}{@{}llrrrrrrrrr@{}}
\toprule
& & \multicolumn{3}{c}{Pooled} & \multicolumn{3}{c}{Mean log ratio (SE)} & \multicolumn{3}{c}{Mean abs.} \\
Shocks & Year effect & 1 yr & 2 yr & 3 yr & 1 yr & 2 yr & 3 yr & 1 yr & 2 yr & 3 yr \\
\midrule
B0: none & --- & 1.00 & 1.07 & 1.24 & $+0.06$ (0.13) & $+0.12$ (0.20) & $+0.24$ (0.23) & 0.29 & 0.38 & 0.52 \\
B1: constant precision & random walk & 1.33 & 1.76 & 2.50 & $+0.31$ (0.11) & $+0.60$ (0.19) & $+0.95$ (0.22) & 0.32 & 0.60 & 0.95 \\
B1: constant precision & random walk with drift & 1.04 & 1.08 & 1.16 & $+0.04$ (0.11) & $+0.09$ (0.19) & $+0.18$ (0.22) & 0.23 & 0.44 & 0.50 \\
B3: size-dependent precision & none & 1.38 & 1.53 & 1.81 & $+0.39$ (0.12) & $+0.48$ (0.16) & $+0.65$ (0.18) & 0.39 & 0.48 & 0.65 \\
B3: size-dependent precision & independent & 2.64 & 3.16 & 4.04 & $+1.01$ (0.11) & $+1.19$ (0.14) & $+1.44$ (0.14) & 1.01 & 1.19 & 1.44 \\
B3: size-dependent precision & random walk & 1.24 & 1.56 & 2.10 & $+0.24$ (0.11) & $+0.49$ (0.17) & $+0.80$ (0.19) & 0.26 & 0.49 & 0.80 \\
B3: size-dependent precision & random walk with drift & 1.01 & 1.03 & 1.11 & $+0.02$ (0.11) & $+0.05$ (0.17) & $+0.15$ (0.21) & 0.23 & 0.39 & 0.48 \\
\bottomrule
\end{tabular}
\end{table}

\subsection{The Fitted Model}\label{sec:bayesfit}

\begin{table}[h]
\centering
\caption{Posterior Means and 95\% Intervals of B0 and B3 with Drift. Both are fitted to all data through June 2026. B0 has no shocks. B3 has a drifting year effect and size-dependent drive-model-year shocks.}\label{tab:bayespar}
\small
\begin{tabular}{@{}lrrrrl@{}}
\toprule
& \multicolumn{2}{c}{B0} & \multicolumn{2}{c}{B3, drift} & \\
Parameter & Mean & 95\% interval & Mean & 95\% interval & Meaning \\
\midrule
$\beta_1$ & 0.294 & (0.258, 0.331) & 0.381 & (0.340, 0.423) & early-life exponent \\
$c_1$ & 0.0287 & (0.0145, 0.0545) & 0.0594 & (0.0293, 0.1090) & early-life cumulative hazard at 1 yr \\
$\beta_2$ & 1.573 & (1.542, 1.604) & 1.929 & (1.856, 2.007) & wear-out exponent \\
$k_2$ & 0.160 & (0.082, 0.304) & 0.352 & (0.173, 0.641) & wear-out cumulative hazard at 4 yr \\
$\eta_2$ (yr) & 13.59 & (8.53, 19.80) & 7.20 & (5.03, 10.00) & $a_\star k_2^{-1/\beta_2}$ \\
$\phi$ & 0.870 & (0.603, 1.226) & 0.577 & (0.366, 0.875) & frailty variance about $\mu_g$ \\
$\gamma_{\text{intro}}$ & 0.072 & (-0.060, 0.194) & 0.169 & (0.060, 0.278) & per year of first appearance \\
$\gamma_{\text{cap}}$ & -1.439 & (-2.118, -0.671) & -1.229 & (-1.864, -0.560) & per log capacity \\
$\gamma_{\text{HGST}}$ & -1.607 & (-2.154, -1.044) & -1.510 & (-1.993, -1.018) & HGST line, including WUH, relative to Seagate \\
$\gamma_{\text{Toshiba}}$ & -1.021 & (-1.667, -0.390) & -0.790 & (-1.386, -0.190) & Toshiba relative to Seagate \\
$\gamma_{\text{WDC}}$ & -0.669 & (-1.518, 0.291) & -0.448 & (-1.242, 0.419) & WD-native drives relative to Seagate \\
$\gamma_{\text{He}}$ & 0.415 & (-0.499, 1.281) & 0.623 & (-0.214, 1.439) & helium relative to air \\
$\kappa_0$ & &  & 1.054 & (0.699, 1.580) & shock precision at one expected failure \\
$\xi$ & &  & 0.371 & (0.263, 0.468) & growth of shock precision with group size \\
$\sigma_Q$ & &  & 0.234 & (0.132, 0.394) & yearly step of $\log Q_y$ \\
$d$ & &  & -0.134 & (-0.278, 0.004) & drift of $\log Q_y$ per year \\
\bottomrule
\end{tabular}
\end{table}

\begin{figure}[h]
\centering
\includegraphics[width=\textwidth]{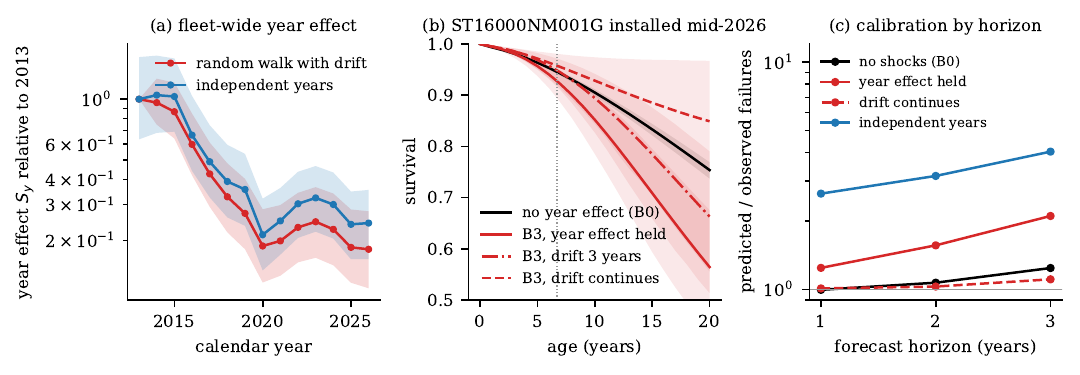}
\caption{(a) Posterior median and 95\% band of the fleet-wide year effect under two processes.
(b) Marginal survival of one drive model entering service after the data, under the four
combinations of model and future calendar trend defined in Section~\ref{sec:apc}
(Table~\ref{tab:implied}), with the 95\% band over posterior draws; the dotted line marks the
oldest age observed for that drive model. (c) Predicted over observed failures against
forecast horizon.}\label{fig:bayes}
\end{figure}

B0's 95\% intervals contain the empirical-Bayes estimates of Table~\ref{tab:finalpar}. They also
carry the uncertainty of the other parameters. For example, $\phi$ is 0.87 (0.60, 1.23) against
the empirical-Bayes point estimate of 0.76.

Adding shocks changes the fitted decomposition. The Weibull wear-out exponent rises from 1.57 to 1.93 and the
wear-out cumulative hazard at four years from 0.16 to 0.35, at the 2013 level of the year
effect. The year effect then falls: its posterior median in 2026 is 0.180 of its 2013 value
(0.116, 0.278), with most of the fall between 2015 and 2020 (Figure~\ref{fig:bayes}a). At the
2026 level the wear-out scale is $\eta_2 Q_{2026}^{-1/\beta_2} = 17.4$~yr. The drift is $-0.134$
per year with a 95\% interval of $(-0.278,\ 0.004)$. Thirteen yearly steps determine it
poorly.

The vintage coefficient is $+0.169$ per year $(0.060,\ 0.278)$: at the same age in the same
calendar year, a drive model introduced a year later fails about 18\% more often. Under B0 the
same coefficient is $+0.072$ with an interval that included zero.

For the drive-model-year shock, $\kappa_0=1.05$ and $\xi=0.37$ $(0.26,\ 0.47)$. The coefficient
of variation of the shock, $1/\sqrt{\kappa_0 m^{\xi}}$, is 0.97 for a group expected to have
one failure in the year, 0.63 for ten, 0.41 for a hundred, and 0.27 for a thousand. Both
$\xi=0$ (the same relative variability at every size) and $\xi=1$ (variance proportional
to the mean) are outside the interval.

Among the 33 drive models with at least 100 failures, the largest posterior-mean multiplier is
32 times the smallest (ST3000DM001 against HGST HMS5C4040BLE640). Under the empirical-Bayes model the
same ratio is 95: B3 attributes part of the ST3000DM001's excess to the years in which it
ran.

\begin{table}[h]
\centering\small
\caption{Numerical Convergence of the Predictive Scores. Default is the nine-year sum of the bridge estimates with the default approximations: 24 Gauss--Hermite nodes, a 241-point multiplier grid and 200 prior quantiles. Refined doubles all three. Max fold change is the largest change in a single fold. Bridge SE is the root sum of squares of the nine folds' bridge standard errors, with cross-fold covariances omitted.}\label{tab:finescores}
\begin{tabular}{@{}lrrrrr@{}}
\toprule
Variant & Default & Refined & Change & Max fold change & Bridge SE \\
\midrule
B1 with drift & -27,117.7 & -27,117.7 & +0.02 & 0.01 & 0.20 \\
B3, no year effect & -27,129.5 & -27,129.5 & +0.00 & 0.00 & 0.23 \\
B3, random walk & -27,103.0 & -27,102.9 & +0.04 & 0.03 & 0.27 \\
B3 with drift & -27,103.0 & -27,103.0 & +0.02 & 0.01 & 0.27 \\
\bottomrule
\end{tabular}
\end{table}

A separate check holds the posterior
draws fixed and increases the shock, multiplier and prior-node counts to
48, 481 and 400. Across all nine folds, the score sums change by
$+0.0245$ for B1 with drift, $+0.0034$ for B3 without year effects,
$+0.0385$ for B3 with a random walk and $+0.0161$ for B3 with drift.
The largest individual-fold change among these 36 comparisons is 0.0258.
This measures sensitivity to the combined refinement for B1 and B3; it does
not test posterior exploration.

\subsection{How Reliable Are the Prediction Intervals?}\label{sec:calib}

A log score ranks predictive distributions but does not say whether their intervals have
their nominal coverage. For each fold and each drive model, we compute the predictive
distribution of the number of failures in the test year and locate the observed number in
it.

Under the empirical-Bayes model the count is negative binomial: with $Z_g\sim\Gam(a_g,b_g)$ after
training and baseline expectation $e_g$ in the test year, the count has size $a_g$ and mean
$a_g e_g/b_g$. Under the Bayesian models we simulate the distribution. For each of 2{,}000
posterior draws, we draw the test year's year effect from its process, take each drive
model's multiplier from its posterior draw (a prior draw for a model not seen in training),
draw its shock, and draw a Poisson count. The fleet total is the sum over drive models within
a draw, so it carries the common year effect.

\begin{table}[h]
\centering
\caption{Calibration of the One-Year-Ahead Predictive Count of a Drive Model. The results are over nine test years. Coverage is the share of observed counts inside the central interval. The tails are the shares below the 2.5\% and above the 97.5\% points of the randomized probability integral transform, 2.5 each if calibrated. Fleet is the number of test years in which the observed fleet total lies in the 95\% interval.}\label{tab:calib}
\footnotesize\setlength{\tabcolsep}{4pt}\begin{tabular}{@{}lrrrrrrr@{}}
\toprule
& & \multicolumn{3}{c}{Coverage (\%) of interval} & \multicolumn{2}{c}{Tails (\%)} & \\
Model & $n$ & 50\% & 80\% & 95\% & below & above & Fleet \\
\midrule
\multicolumn{8}{@{}l}{\emph{Drive-model-years, five or more expected}} \\
Empirical Bayes & 189 & 19 & 32 & 47 & 27 & 29 & --- \\
B0: no shocks & 189 & 20 & 33 & 48 & 28 & 29 & 2 of 9 \\
B3, no year effect & 198 & 53 & 83 & 95 & 2 & 3 & 5 of 9 \\
B3, random walk & 194 & 56 & 87 & 97 & 1 & 3 & 8 of 9 \\
B1, random walk with drift & 190 & 58 & 82 & 94 & 0 & 6 & 9 of 9 \\
B3, random walk with drift & 189 & 56 & 83 & 94 & 0 & 6 & 8 of 9 \\
\addlinespace
\multicolumn{8}{@{}l}{\emph{Drive-model-years, fewer than five expected}} \\
Empirical Bayes & 174 & 75 & 89 & 97 & 3 & 5 & --- \\
B0: no shocks & 174 & 75 & 90 & 97 & 5 & 3 &  \\
B3, no year effect & 165 & 87 & 95 & 98 & 2 & 3 &  \\
B3, random walk & 169 & 85 & 95 & 98 & 2 & 3 &  \\
B1, random walk with drift & 173 & 82 & 95 & 97 & 2 & 3 &  \\
B3, random walk with drift & 174 & 84 & 94 & 98 & 4 & 3 &  \\
\addlinespace
\multicolumn{8}{@{}l}{\emph{Drive-model-years, new drive models}} \\
Empirical Bayes & 47 & 72 & 87 & 94 & 2 & 6 & --- \\
B0: no shocks & 47 & 74 & 91 & 96 & 0 & 4 &  \\
B3, no year effect & 47 & 77 & 91 & 96 & 4 & 4 &  \\
B3, random walk & 47 & 79 & 91 & 96 & 0 & 4 &  \\
B1, random walk with drift & 47 & 74 & 94 & 96 & 2 & 4 &  \\
B3, random walk with drift & 47 & 77 & 91 & 96 & 0 & 4 &  \\
\bottomrule
\end{tabular}
\end{table}

\textbf{One training-defined population.} Table~\ref{tab:calib} groups model-years by each
variant's own expected count, so the variants are not scored on the same model-years.
Table~\ref{tab:calibcommon} fixes the population without reference to any prediction. It
takes the 410 model-years of the nine test years in three strata defined from training
information alone: drive models unseen before the test year, seen models with fewer than 20
failures in the training years, and seen models with 20 or more. It reports both tails, the
median relative width of the 95\% interval, and a standard error of the 95\% coverage from a
bootstrap over test years, which treats the model-years of one year as dependent. On the
169\unskip\ heavily observed model-years, where the interval
widths are informative, B0 covers 44\unskip\% at nominal
95\%. It has 25\unskip\% of observations below and
30\unskip\% above its interval, and a median width of
0.58\unskip\ times the expected count. B3 with drift covers
92\unskip\% (standard error
1\unskip\ points) with a median width of
2.24\unskip, and B1 with drift
93\unskip\%. The variants without a declining year
effect reach the same or higher coverage (B3 without a year effect
94\unskip\%, the random walk
95\unskip\%) by over-predicting: their predicted-to-observed
ratios on this stratum are 1.37\unskip\ and 1.21
against 0.99\unskip\ for B3 with drift, with intervals up to
4\% wider. In this stratum the shocks supply the dispersion that B0 lacks and the drift
supplies the level. On the sparsely observed seen models every variant covers at or above
the nominal rate (95 to 99\unskip\% at nominal 95). Every
variant over-predicts the new drive models
(5.4 to 6.9\unskip\ times), with intervals wide enough to
cover anyway. This is the prior-predictive misfit of Section~\ref{sec:newmodels}.

\begin{table}[h]
\centering
\caption{Calibration on One Prespecified Population. The population is the 410 model-years of the nine test years, in three strata fixed from training information. Ratio is predicted over observed failures. Coverage is of the central 50, 80 and 95\% intervals, and SE is the bootstrap standard error of the 95\% coverage over the nine test years. The tails are the shares of observed counts below the 2.5\% and above the 97.5\% points. Width is the median width of the 95\% interval divided by the expected count (at least one).}\label{tab:calibcommon}
\footnotesize\setlength{\tabcolsep}{4pt}\begin{tabular}{@{}lrrrrrrrr@{}}
\toprule
& & \multicolumn{3}{c}{Coverage (\%)} & & \multicolumn{2}{c}{Tails (\%)} & \\
Model & Ratio & 50 & 80 & 95 & SE & below & above & Width \\
\midrule
\multicolumn{9}{@{}l}{\emph{Seen, $\ge 20$ training failures: 169 model-years, 22,960 failures}} \\
\quad B0 (no year effect, no shocks) & 0.97 & 21 & 32 & 44 & 5 & 25.4 & 30.2 & 0.58 \\
\quad B1 with drift & 1.02 & 60 & 83 & 93 & 2 & 0.0 & 7.1 & 2.51 \\
\quad B3, no year effect & 1.37 & 53 & 82 & 94 & 2 & 1.8 & 4.1 & 2.33 \\
\quad B3, random walk & 1.21 & 56 & 86 & 95 & 1 & 0.6 & 4.1 & 2.24 \\
\quad B3 with drift & 0.99 & 57 & 82 & 92 & 1 & 0.0 & 7.7 & 2.24 \\
\multicolumn{9}{@{}l}{\emph{Seen, $< 20$ training failures: 194 model-years, 1,889 failures}} \\
\quad B0 (no year effect, no shocks) & 0.90 & 68 & 85 & 95 & 1 & 1.5 & 3.6 & 2.02 \\
\quad B1 with drift & 1.02 & 77 & 92 & 98 & 1 & 0.0 & 2.1 & 3.09 \\
\quad B3, no year effect & 1.24 & 82 & 94 & 98 & 1 & 0.0 & 1.5 & 3.76 \\
\quad B3, random walk & 1.23 & 81 & 94 & 99 & 1 & 0.0 & 1.0 & 3.65 \\
\quad B3 with drift & 0.99 & 80 & 94 & 99 & 1 & 0.0 & 1.0 & 3.62 \\
\multicolumn{9}{@{}l}{\emph{New drive models: 47 model-years, 126 failures}} \\
\quad B0 (no year effect, no shocks) & 6.80 & 74 & 92 & 96 & 3 & 0.0 & 4.3 & 3.87 \\
\quad B1 with drift & 6.82 & 74 & 94 & 96 & 3 & 0.0 & 4.3 & 4.41 \\
\quad B3, no year effect & 5.51 & 77 & 92 & 96 & 3 & 0.0 & 4.3 & 4.15 \\
\quad B3, random walk & 6.93 & 79 & 92 & 96 & 3 & 0.0 & 4.3 & 4.00 \\
\quad B3 with drift & 5.37 & 77 & 92 & 96 & 3 & 0.0 & 4.3 & 4.06 \\
\multicolumn{9}{@{}l}{\emph{All: 410 model-years, 24,975 failures}} \\
\quad B0 (no year effect, no shocks) & 1.00 & 49 & 64 & 74 & 2 & 11.2 & 14.6 & 1.08 \\
\quad B1 with drift & 1.05 & 70 & 89 & 96 & 1 & 0.0 & 4.4 & 2.67 \\
\quad B3, no year effect & 1.38 & 69 & 89 & 96 & 1 & 0.7 & 2.9 & 2.84 \\
\quad B3, random walk & 1.24 & 70 & 90 & 97 & 1 & 0.2 & 2.7 & 2.83 \\
\quad B3 with drift & 1.01 & 70 & 88 & 96 & 1 & 0.0 & 4.1 & 2.62 \\
\bottomrule
\end{tabular}
\end{table}

\begin{table}[h]
\centering
\caption{Predictive Intervals for the Fleet Total. The table gives the observed failures in each test year and the one-year-ahead predictive median and 95\% interval of the fleet total. The 2026 row is the first half-year.}\label{tab:calibfleet}
\small\begin{tabular}{@{}lrrrrr@{}}
\toprule
& & \multicolumn{2}{c}{B0: no shocks} & \multicolumn{2}{c}{B3 with drift} \\
Year & Observed & Median & 95\% interval & Median & 95\% interval \\
\midrule
2018 & 1{,}240 & 1{,}965 & 1{,}802--2{,}147 & 1{,}366 & 540--3{,}147 \\
2019 & 2{,}048 & 1{,}780 & 1{,}621--2{,}239 & 1{,}441 & 720--3{,}112 \\
2020 & 1{,}307 & 2{,}918 & 2{,}464--5{,}017 & 2{,}451 & 1{,}330--5{,}289 \\
2021 & 1{,}804 & 2{,}059 & 1{,}894--2{,}306 & 1{,}564 & 800--2{,}893 \\
2022 & 2{,}996 & 2{,}225 & 2{,}083--2{,}386 & 1{,}990 & 1{,}168--3{,}635 \\
2023 & 4{,}192 & 2{,}912 & 2{,}765--3{,}076 & 3{,}328 & 1{,}714--6{,}842 \\
2024 & 4{,}528 & 3{,}616 & 3{,}440--3{,}802 & 4{,}434 & 2{,}276--8{,}392 \\
2025 & 4{,}308 & 4{,}680 & 4{,}467--4{,}889 & 5{,}005 & 2{,}729--9{,}488 \\
2026 & 2{,}552 & 2{,}478 & 2{,}364--2{,}596 & 2{,}302 & 1{,}261--4{,}161 \\
\bottomrule
\end{tabular}
\end{table}

Table~\ref{tab:calib} forms the group of drive models with five or more failures expected
from each variant's own expectation, so $n$ differs a little between rows. In this group the
models without shocks are badly overconfident. The 95\% interval of the empirical-Bayes model
contains 47\% of the observed counts and that of B0 48\%, with more than a quarter of the counts
beyond each end. The shock models' 95\% intervals contain 94 to 97\%. Their 50\% intervals contain 53
to 58\%, a little wide, and the two drift variants have a slightly heavy upper tail (6\% above
the 97.5\% point). The forecast level differs too. Over the model-years in this group, B3
without a year effect expects 33{,}523 failures where 24{,}721 occurred, and B3 with a random
walk 29{,}946. The drift variants and B0 are within about 4\%. For drive models with fewer
than five expected, and for new drive models, every model covers at or near the nominal rate,
which discreteness makes easy. These groups total about 221 model-years with 277 failures and
test little. For new drive models the level is also wrong. Every Bayesian variant expects
between 676 and 873 failures (Table~\ref{tab:calibcommon}) where 126 occurred, the
over-prediction of Section~\ref{sec:newmodels}.

For the fleet total the difference is larger (Table~\ref{tab:calibfleet}). Without shocks the
95\% interval has a half-width of 7\% of its median in the median year (from 4.5\% in 2025 to
44\% in 2020) and contains the observed total in two years of nine. The interval of B3 with
drift runs from about half to twice the median and contains the total in eight of nine (nine
in replicate A); 2020 falls just below it in the default run. That
of B3 without a year effect contains it in five, and that of B3 with a random walk and no drift
in eight. Nine years cannot tell a 95\% interval from a 90\% one. Its 50\% interval
contains five of the nine totals.

A simulator that draws lifetimes from a fitted distribution with no shocks produces, in any given year,
fleet failure counts within about $\pm 7\%$ of that model's median, whereas the observed total
differed from that median by more than 20\% in five of the nine years.

\subsection{Can Combining Models Improve Forecasts?}\label{sec:tracksum}

The ranking of the forecasters changes from one test period to the next. In
Appendix~\ref{sec:track} we combine the count forecasts of the point and predictive
forecasters for each drive model with the fixed-share algorithm of Herbster and
Warmuth~\cite{herbster1998tracking}. After each test period it reweights every forecaster by
the probability it gave the outcome, then redistributes a fraction $\omega$ of the weight
evenly. On the fleet's summed score, every
setting with $\omega\ge1/T$, for $T$ test periods, beats the best single forecaster in
hindsight (M2 + covariates, predictive) by 394 to 419 log units over 34 test quarters and by
518 to 527 over the nine test years. The Bayesian variants are not candidates in the
quarterly game and enter the yearly one only as point forecasts at their posterior means, so
the mixtures are not compared with the predictive distribution of B3. The uniform mixture ($\omega=1$), whose weights never
change, gains 402 of the 419 in the quarterly game and is the best setting in the yearly one
(527). Nearly all of the gain comes from mixing the forecasters, and little from reweighting
them. In the per-model games, the 95\% intervals of the fixed-share ($\omega=0.25$) and uniform
mixtures for a drive model's count cover 88 to 91\%, against 82\% (quarterly) and 73\% (yearly)
for the best single forecaster. They still under-cover at 95\% and over-cover at 50\%. These
coverages pool all drive models and are not comparable with those of Section~\ref{sec:calib}.

\section{Forecasting Failures in the Fleet Already in Service}\label{sec:landmark}

For the drives running today, how many will fail in the next 90 days or year, and which ones
are most at risk? These are the forecasts a fleet manager could use to plan replacements
and identify drives needing attention. The earlier comparisons
(Sections~\ref{sec:valid} and~\ref{sec:bayesresults}, Appendix~\ref{sec:track}) tested failure
rates using the operating time that actually occurred in each test year. That time includes
new deployments and ends when drives fail or leave service. It is useful for testing rates,
but is not known on 1~January. Here we forecast the drives already in service using only the
information available at the forecast date, subject to the metadata qualifications below.

\subsection{How We Forecast the Drives Already in Service}

A \emph{landmark} $L$ is a date at which we make a forecast from the information available by
that date. We also call it an \emph{origin} when it is one of the dates at which we form
training rows. The \emph{horizon} $\tau$ is the length of the future window, 90 or 365 days.
The cohort of a landmark is every HDD data drive with a row in the daily file of $L$ that had
not failed before $L$. A drive whose first and last dates bracket $L$ but which is absent from
that day's file is not in the cohort (252 to 2{,}553 such drives per yearly cohort; 0.7\% of the drives bracketed
\unskip). We fit each
forecaster to data before $L$ only, apart from the three fields of the drive table described
next. It gives every drive in the cohort a probability of
failing within $\tau$, from its drive model and its power-on age on $L$. Neither the cohort's
later exposure nor drives deployed after $L$ enter its forecast.

We take two fields of the shared drive table from the drive's full history rather than
reconstructing them at $L$: the drive-model name (the last observed) and the drive model's
capacity (its modal reported capacity, a per-model constant). The power-on age at first
sighting comes from the earliest usable SMART~9 reading, and a drive whose reading is in a
month after $L$ is not in the cohort at $L$, for scoring or for training: no drive in the
yearly cohorts, 255
\unskip\ drive-origin rows at
5
\unskip\ of the quarterly origins. Audits at every
yearly and quarterly origin (Appendix~\ref{sec:repro_landmark}) find that the two
full-history fields touch few rows. A capacity differing from the model's modal one by
more than 1\% occurs in 1\unskip\ of the
7,823,945\unskip\ comparable rows. We date a failure by the
drive's last row with a failure flag (Section~\ref{sec:data}). Table~\ref{tab:firstflag} rescores
every forecaster under the other convention, a failure at the drive's first flag: a drive whose
first flag precedes $L$ has already failed and leaves the cohort, and a drive fails within the
horizon if its first flag does. Over the eight January cohorts at one year that rule removes
44
\unskip\ of 1{,}487{,}544
\unskip\
drive-landmark observations and changes the outcome of
39
\unskip\ (20{,}552
\unskip\
failures become 20{,}556
\unskip). The forecasts are not
refitted: the training populations differ by the 57 drives with two flags in the whole record.
No log loss moves by more than 0.14
\unskip\% and no AUC by
more than 0.0009
\unskip, at either kind of origin, and the Poisson
score of the totals by at most 3.7
\unskip\ units. The
convention does not affect the rankings.

\begin{table}[h]
\centering\footnotesize\setlength{\tabcolsep}{4pt}
\caption{The Failure Convention and the Landmark Scores. Eight January cohorts, one-year horizon, all drives. Final is the paper's rule (a failure at the drive's final failure flag); $\Delta$ is the change under the first-flag rule, which removes the drive-landmark observations already failed at the origin under that rule. Log loss and the Poisson log score of the cohort totals are summed over the cohorts; AUC is the drive-weighted mean over cohorts.}\label{tab:firstflag}
\begin{tabular}{@{}lrrrrrr@{}}
\toprule
Forecaster & Log loss & $\Delta$ & AUC & $\Delta$ & Poisson & $\Delta$ \\
\midrule
Boosted trees, with the SMART record & 92{,}582 & +90 & 0.7438 & -0.0006 & -1{,}280 & +0.6 \\
Boosted trees, with the SMART warning & 99{,}027 & +65 & 0.7186 & -0.0007 & -1{,}184 & -1.2 \\
Boosted trees, drive model and age & 107{,}857 & +28 & 0.6055 & -0.0003 & -837 & -2.1 \\
The same, with a SMART warning before the landmark & 98{,}632 & +80 & 0.7350 & -0.0007 & -1{,}347 & -1.5 \\
Empirical-Bayes model, predictive & 107{,}247 & +38 & 0.6432 & -0.0004 & -1{,}373 & -1.6 \\
Empirical-Bayes model, posterior mean & 107{,}247 & +38 & 0.6432 & -0.0004 & -1{,}373 & -1.6 \\
B3 with drift & 106{,}282 & +36 & 0.6411 & -0.0003 & -865 & -2.4 \\
B3, random walk & 107{,}095 & +35 & 0.6407 & -0.0004 & -1{,}627 & -1.1 \\
B1 with drift & 106{,}522 & +36 & 0.6341 & -0.0003 & -928 & -2.6 \\
B3, no year effect & 109{,}282 & +36 & 0.6344 & -0.0004 & -3{,}018 & +1.8 \\
B0 (no shocks) & 107{,}231 & +38 & 0.6438 & -0.0004 & -1{,}355 & -1.6 \\
Two Weibulls, pooled & 109{,}816 & +20 & 0.5856 & -0.0004 & -2{,}542 & +1.7 \\
Per-year Weibull, first-seen age & 108{,}061 & +24 & 0.5686 & -0.0004 & -980 & -2.3 \\
Constant rate, last 12 months & 108{,}907 & +19 & 0.5000 & +0.0000 & -993 & -2.8 \\
Bathtub, default constants & 119{,}628 & +25 & 0.5820 & -0.0003 & -11{,}488 & +3.7 \\
\bottomrule
\end{tabular}
\end{table}

We call the models that convert fitted failure rates into probabilities \emph{hazard
forecasters}. For these forecasters, every drive in a cohort except those first seen on $L$ itself was in
the training data, and no cohort contained a drive model absent from training. The new-model
problem of Section~\ref{sec:newmodels} therefore does not arise for them. (The boosted
forecasters' training origins end a horizon before $L$, so for them the drive models deployed
in that last horizon are missing categories, Section~\ref{sec:boostlandmark}.) Forecasting
deployments that have not happened would need a deployment schedule, which the data do not
contain.

Replacement planning depends on both failure and retirement. A drive taken out of service
while still functioning can no longer contribute a recorded failure later in the forecast
window. Within that window a drive fails, is removed without a failure flag (retired,
Section~\ref{sec:retire}), or is still running at $L+\tau$. The \emph{primary target} of the
evaluation is the operator's recorded endpoint: whether the drive is recorded as failed within
$\tau$ days of $L$. A drive removed unfailed counts as a non-failure, and we score every drive
of the cohort. A hazard forecaster does not forecast that endpoint directly. Its probability
$1-\exp[-\int_0^\tau\lambda_F]$ is that of failing within $\tau$ under continued service. With
a retirement hazard $\lambda_R$, the probability of a recorded failure before retirement is
\begin{equation}\label{eq:cr}
\int_0^{\tau}\exp\Bigl[-\int_0^u\{\lambda_F(v)+\lambda_R(v)\}\,\mathrm{d}v\Bigr]\lambda_F(u)\,\mathrm{d}u,
\end{equation}
which is smaller. The hazard forecasters have no model of retirement and cannot compute it.
Scoring them on the primary target therefore penalizes them for omitting retirement.
For the hazard forecasters we report two further quantities as diagnostics, not as alternative
targets. The \emph{exposure-truncated} sum adds each drive's probability over the exposure it
actually had, from $L$ to its failure, removal or $L+\tau$. Its difference from the
full-horizon sum is not an estimate of the cost of retirement, because exposure also ends at
failure. With a constant hazard $\lambda$ and no retirement, the expected exposure-truncated
probability is $[1-e^{-2\lambda\tau}]/2$ against a failure probability of $1-e^{-\lambda\tau}$.
We therefore call the difference an exposure-truncation diagnostic, and
Table~\ref{tab:truncdecomp} measures its failure and removal components separately. The second
diagnostic is the per-drive score on the drives with \emph{complete follow-up} (failed, or
running at $L+\tau$). This checks how the choice of drives affects the results. It selects
on a future outcome and does not correct for retirement decisions related to a drive's
failure risk.

The yearly landmarks are 1~January of 2018 to 2026. The quarterly landmarks are 1~April,
1~July and 1~October from 2018 through the second quarter of 2026. January belongs to the
yearly group only. In both groups we drop horizons that run past the end of the data
(30~June 2026). The forecasters are those of Section~\ref{sec:valid} that can be evaluated for
a single drive:
\begin{itemize}[leftmargin=1.4em]
\item the constant rate of the twelve months before $L$;
\item the bathtub with its default constants;
\item the per-year Weibull on first-seen age. We fit it as a simple comparator with a
  continuous-time event density at the recorded exit age rather than the daily interval
  probability of \eqref{eq:lldisc}, a simplification that matters only where the hazard changes
  within a day;
\item two competing Weibulls pooled over the fleet;
\item the empirical-Bayes model, at its posterior-mean multiplier and with the multiplier
  integrated over its gamma posterior (the predictive, $1-[b_g/(b_g+\Delta H_0)]^{a_g}$);
\item at the yearly landmarks, where fits trained through the preceding December exist, the
  Bayesian variants B0, B1 with drift, and B3 without year effect, with random walk and with
  drift. Each is the average over 2{,}000 posterior draws of the failure probability. We step
  the year effect one year ahead and draw the drive-model-year shock with the precision the model
  gives a model-year of the cohort's annualized expected count. (The fitted precision also counted
  drives deployed during the year, which a prospective forecast cannot.)
\end{itemize}

One forecaster uses information the lifetime models do not: whether the drive had carried a
SMART warning (attributes 5, 187, 188, 197 or 198 non-zero) before $L$. It is the
empirical-Bayes predictive with the hazard of such drives multiplied by a manufacturer-specific
factor, which we estimate from the training years. The factor is the failure rate, in the
year after each 1~January from 2016 up to $L-365$ days, of drives warned by that date relative
to clear drives of the same manufacturer and age. We shrink it toward the pooled ratio when a
manufacturer has few warned failures. The factors range from 4.5 (Seagate, 2018) to 34 (HGST,
2018). One fleet-wide constant then rescales every drive's hazard so that the cohort's summed
posterior-mean cumulative hazard is unchanged. Because the predictive probability is concave
in the hazard, the cohort's expected failures fall by about 2\%. The warning thus reallocates
risk among drives and, by construction, from manufacturers with a lower warned share to those
with a higher one.

The three boosted forecasters of Section~\ref{sec:boostlandmark}, two of which read the
drive's SMART record, appear in the tables here so that we score every forecaster
on the same drives.

We score cohort totals by the ratio of predicted to observed failures and by the Poisson log
score of the observed total given the predicted one. We score individual drives by the log
loss, the Brier score and the AUC of Section~\ref{sec:perdrive}, against the primary target,
on all drives of the cohort. Lower log loss and Brier scores mean better probability
predictions; higher AUC means better ranking of drives that fail above those that do not.
We also give the AUC on the drives with complete follow-up. That
population excludes the removed drives, which bulk retirement makes mostly old ones. It
therefore flatters forecasters that rate old drives highly, and for some pairs of forecasters
the choice of population changes their order. To check calibration---whether predicted risks
match observed failure frequencies---we sort drives into ten equal-sized risk groups and
compare their predictions with their outcomes.

\subsection{What Helps Predict Failures in an Existing Fleet}

SMART warnings help identify vulnerable drives, but accurate replacement-demand forecasts
also depend on how the model handles retirement and changes in fleet-wide failure rates.
The comparisons below separate these questions: total failures, individual risk ranking,
and whether the predicted probabilities match observed frequencies.

\begin{table}[h]
\centering\small
\caption{The Yearly Landmark Cohorts and Their Outcomes over 365 Days. A cohort is the drives in service on 1~January. The ratios are predicted over observed failures: for the empirical-Bayes predictive forecast over the full horizon and exposure-truncated (the diagnostic of Section~\ref{sec:landmark}), and for B3 with drift over the full horizon. The AUC is on complete follow-up, for the empirical-Bayes forecast without and with a SMART warning before the landmark.}\label{tab:landmarkyears}
\footnotesize\setlength{\tabcolsep}{4pt}
\begin{tabular}{@{}lrrrrrrrr@{}}
\toprule
& & & & \multicolumn{2}{c}{EB predictive} & B3 drift & \multicolumn{2}{c}{AUC} \\
Landmark & Drives & Failures & Removed & full & trunc. & full & EB & EB + SMART \\
\midrule
2018 & 92{,}292 & 984 & 13{,}632 & 2.18 & 1.81 & 1.59 & 0.689 & 0.733 \\
2019 & 104{,}877 & 1{,}892 & 8{,}129 & 0.95 & 0.85 & 0.79 & 0.597 & 0.664 \\
2020 & 122{,}622 & 1{,}122 & 13{,}838 & 2.17 & 2.04 & 1.88 & 0.604 & 0.687 \\
2021 & 162{,}458 & 1{,}510 & 21{,}664 & 1.34 & 1.12 & 1.05 & 0.669 & 0.727 \\
2022 & 203{,}128 & 2{,}858 & 54 & 0.72 & 0.71 & 0.68 & 0.690 & 0.752 \\
2023 & 231{,}199 & 3{,}989 & 10{,}687 & 0.70 & 0.68 & 0.85 & 0.679 & 0.780 \\
2024 & 270{,}179 & 4{,}199 & 20{,}309 & 0.90 & 0.81 & 1.11 & 0.648 & 0.761 \\
2025 & 300{,}789 & 3{,}998 & 4{,}304 & 1.13 & 1.11 & 1.27 & 0.666 & 0.770 \\
\bottomrule
\end{tabular}
\end{table}

Table~\ref{tab:landmarkyears} gives the yearly cohorts. Over the eight 1~January landmarks
from 2018 to 2025 the cohorts hold 8 landmarks, 1{,}487{,}544 drive-cohort members (drive-landmark observations, not distinct drives), 20{,}552 failures, 92{,}617 removed without failure
. Removals are not
rare. Of the drives in service on a 1~January, 6.2\% were gone, unfailed, within the year,
and 15\% and 13\% in 2018 and 2021, when whole drive models were retired in bulk
(Section~\ref{sec:retire}). The exposure-truncation gap of the empirical-Bayes forecast is
largest in those two years, 17\% of the forecast in 2018 and 16\% in 2021 (ratios 2.18 against 1.81 and 1.34
against 1.12). It is 10\% in 2019 and 2024 and 1 to 6\% in the other years, and the
ratios are 1.05 against 0.97 over the eight years together. Table~\ref{tab:truncdecomp} splits
the gap by summing each drive's full-horizon minus truncated probability separately over the
drives that failed and over those removed without a failure flag. For the empirical-Bayes
predictive forecast over the eight January cohorts, the full-horizon sum is
21,522\unskip\ and the truncated sum
19,936\unskip, a gap of 7.4\unskip\%
of the full-horizon sum. Of that gap, 0.9\unskip\ percentage
points fall on the 20,552\unskip\ drives that failed and
6.4\unskip\ on the 92,617\unskip\
removed. The failures' share of the gap is 12\unskip\%
pooled and 3 to 99\unskip\% across the cohorts. When
individual risks differ, the fleet's average failure rate does not determine the failure
component. The measured share is what the forecast assigns under its own assumptions: how much
of its expected count fell on exposure that did not occur. It is not what the removed drives
would have done had they stayed.

\begin{table}[h]
\centering\small
\caption{Exposure-Truncation Decomposition, Eight January Cohorts at 365 Days. Full and Truncated are the sums of each hazard forecaster's probabilities over the full horizon and over the exposure that occurred. The gap is their difference as a percentage of the full-horizon sum. It is split between drives that failed within the horizon and drives removed without a failure flag, and the last column is the failures' share of the gap.}\label{tab:truncdecomp}
\footnotesize\setlength{\tabcolsep}{4pt}\begin{tabular}{@{}lrrrrrr@{}}
\toprule
& & & & \multicolumn{2}{c}{Gap on (\%)} & Failures' \\
\cmidrule(lr){5-6}
Forecaster & Full & Truncated & Gap (\%) & failures & removals & share (\%) \\
\midrule
Empirical Bayes, predictive & 21,522 & 19,936 & 7.4 & 0.9 & 6.4 & 12 \\
Empirical Bayes with SMART warning & 21,140 & 19,112 & 9.6 & 2.5 & 7.1 & 26 \\
B3 with drift & 21,848 & 20,481 & 6.3 & 0.9 & 5.3 & 15 \\
B0 & 21,481 & 19,903 & 7.3 & 0.9 & 6.4 & 13 \\
Two Weibulls, no multiplier & 29,631 & 28,027 & 5.4 & 0.7 & 4.7 & 13 \\
Per-year Weibull & 22,421 & 21,309 & 5.0 & 0.8 & 4.2 & 15 \\
Three-component benchmark & 49,764 & 46,597 & 6.4 & 0.8 & 5.6 & 13 \\
Constant rate & 21,079 & 20,201 & 4.2 & 0.7 & 3.5 & 17 \\
\bottomrule
\end{tabular}
\end{table}

\begin{table}[h]
\centering\small
\caption{Prospective Forecasts for the Yearly Landmarks, 365-Day Horizon. The eight landmarks are taken together. Pred./obs.\ is predicted over observed failures over the full horizon and, for the hazard forecasters, exposure-truncated. The truncated ratio is a diagnostic, and the classifiers have none because they forecast the recorded endpoint directly. $\Sigma|\log r|$ is the sum over landmarks of the absolute log ratio of the full-horizon forecast. The gains are differences from the empirical-Bayes predictive forecast on the primary target, positive when better: the Poisson log score of the cohort totals (which treats drives as independent), the log loss summed over all drives, and the Brier score per 10{,}000 drives. AUC is on all drives (the primary population) and on drives with complete follow-up, weighted over landmarks.}\label{tab:landmark365}
\footnotesize\setlength{\tabcolsep}{3.5pt}
\begin{tabular}{@{}lrrrrrrrr@{}}
\toprule
& \multicolumn{2}{c}{Pred./obs.} & & \multicolumn{3}{c}{Gain over EB predictive} & \multicolumn{2}{c}{AUC} \\
Forecaster & full & trunc. & $\Sigma|\log r|$ & totals & log loss & Brier & all & complete \\
\midrule
Boosted trees, with the SMART record & 1.19 & -- & 2.65 & +92 & +14{,}664 & +11.80 & 0.744 & 0.748 \\
Boosted trees, with the SMART warning & 1.15 & -- & 2.65 & +189 & +8{,}220 & +3.83 & 0.719 & 0.725 \\
Boosted trees, drive model and age (Section~\ref{sec:boost}) & 1.09 & -- & 2.17 & +535 & -610 & +0.03 & 0.606 & 0.615 \\
The same, with a SMART warning before the landmark & 1.03 & 0.93 & 2.80 & +26 & +8{,}615 & +3.15 & 0.735 & 0.747 \\
Empirical-Bayes model, predictive & 1.05 & 0.97 & 2.82 & +0 & +0 & +0.00 & 0.643 & 0.660 \\
Empirical-Bayes model, posterior mean & 1.05 & 0.97 & 2.82 & -0 & -0 & -0.00 & 0.643 & 0.660 \\
B3 with drift & 1.06 & 1.00 & 2.25 & +508 & +965 & +0.35 & 0.641 & 0.660 \\
B3, random walk & 1.30 & 1.22 & 2.78 & -254 & +152 & -0.03 & 0.641 & 0.660 \\
B1 with drift & 1.08 & 1.02 & 2.27 & +444 & +725 & +0.30 & 0.634 & 0.652 \\
B3, no year effect & 1.49 & 1.38 & 4.07 & -1{,}645 & -2{,}035 & -1.15 & 0.634 & 0.652 \\
B0 (no shocks) & 1.05 & 0.97 & 2.81 & +18 & +15 & +0.02 & 0.644 & 0.661 \\
Two Weibulls, pooled & 1.44 & 1.36 & 3.80 & -1{,}169 & -2{,}569 & -0.39 & 0.586 & 0.607 \\
Per-year Weibull, first-seen age & 1.09 & 1.04 & 2.42 & +393 & -814 & +0.11 & 0.569 & 0.594 \\
Constant rate, last 12 months & 1.03 & 0.98 & 2.44 & +380 & -1{,}660 & -0.04 & 0.500 & 0.500 \\
Bathtub, default constants & 2.42 & 2.27 & 7.04 & -10{,}115 & -12{,}381 & -5.83 & 0.582 & 0.601 \\
\bottomrule
\end{tabular}
\end{table}

\begin{table}[h]
\centering\small
\caption{Prospective Forecasts for the Yearly Landmarks, 90-Day Horizon. Columns are as in Table~\ref{tab:landmark365}, over the nine yearly landmarks of 2018 to 2026.}\label{tab:landmark90}
\footnotesize\setlength{\tabcolsep}{3.5pt}
\begin{tabular}{@{}lrrrrrrrr@{}}
\toprule
& \multicolumn{2}{c}{Pred./obs.} & & \multicolumn{3}{c}{Gain over EB predictive} & \multicolumn{2}{c}{AUC} \\
Forecaster & full & trunc. & $\Sigma|\log r|$ & totals & log loss & Brier & all & complete \\
\midrule
Boosted trees, with the SMART record & 1.22 & -- & 2.70 & -72 & +8{,}862 & +2.94 & 0.828 & 0.828 \\
Boosted trees, with the SMART warning & 1.23 & -- & 2.59 & -54 & +5{,}185 & +0.53 & 0.816 & 0.816 \\
Boosted trees, drive model and age (Section~\ref{sec:boost}) & 1.25 & -- & 2.71 & -85 & +16 & +0.00 & 0.660 & 0.660 \\
The same, with a SMART warning before the landmark & 1.12 & 1.09 & 2.35 & +5 & +4{,}213 & +0.34 & 0.799 & 0.801 \\
Empirical-Bayes model, predictive & 1.13 & 1.11 & 2.37 & +0 & +0 & +0.00 & 0.663 & 0.667 \\
Empirical-Bayes model, posterior mean & 1.13 & 1.11 & 2.37 & -0 & -0 & +0.00 & 0.663 & 0.667 \\
B3 with drift & 1.10 & 1.08 & 1.79 & +104 & +145 & +0.01 & 0.661 & 0.666 \\
B3, random walk & 1.34 & 1.32 & 2.79 & -126 & -88 & -0.00 & 0.661 & 0.665 \\
B1 with drift & 1.11 & 1.09 & 1.80 & +85 & +76 & +0.01 & 0.658 & 0.662 \\
B3, no year effect & 1.52 & 1.50 & 4.44 & -515 & -633 & -0.06 & 0.654 & 0.660 \\
B0 (no shocks) & 1.13 & 1.11 & 2.36 & +1 & +1 & +0.00 & 0.663 & 0.668 \\
Two Weibulls, pooled & 1.52 & 1.50 & 4.47 & -498 & -1{,}119 & -0.04 & 0.600 & 0.608 \\
Per-year Weibull, first-seen age & 1.16 & 1.15 & 2.04 & +60 & -529 & -0.01 & 0.592 & 0.601 \\
Constant rate, last 12 months & 1.13 & 1.12 & 2.05 & +72 & -816 & -0.02 & 0.500 & 0.500 \\
Bathtub, default constants & 2.69 & 2.64 & 8.52 & -3{,}803 & -4{,}515 & -0.41 & 0.591 & 0.599 \\
\bottomrule
\end{tabular}
\end{table}

\textbf{Forecasting replacement demand.} A forecast can get the total over several years
nearly right while leaving an operator short of replacements in a particular year.
Pooled over eight years, the empirical-Bayes forecast of the cohort's
failures is 5\% high over the full horizon and 3\% low after truncating exposure at failure or
removal. B0, the same model fitted by sampling, agrees with it, and the constant rate of the
previous twelve months is as close (1.03). These totals hide the year-to-year error
described in Section~\ref{sec:time}. Because the fleet-wide failure rate changed and the model has no
year effect, the empirical-Bayes forecast is 2.18 and 2.17 times the observed failures for the
2018 and 2020 cohorts and 0.72 and 0.70 times for 2022 and 2023. Among the lifetime models the
variants with a drifting year effect forecast the total best. B3 with drift is closer than the
empirical-Bayes model in five of the eight years and further in 2019, 2022 and 2025
(Table~\ref{tab:landmarkyears}). Its sum of absolute log ratios over the eight landmarks is
2.25 against 2.82. The sum is 2.27 for B1 with drift, 2.44 for the previous year's rate and
2.42 for the per-year Weibull. The boosted ensemble without SMART information
(Section~\ref{sec:boost}) is closer still at one year, 2.17. The random walk without drift
(1.30) and B3 without a year effect (1.49) are worse than the simple forecasts. The Poisson
score of the totals orders the forecasters nearly the same way: the two drift variants are
444 and 508 log units above the empirical-Bayes model, and the two simple forecasts 380 and
393 above. That score treats drives as independent. It therefore credits none of the shock models
for the spread of their own predictive distributions; it scores their means under a Poisson
assumption, rather than evaluating those distributions.
The fleet-wide movement a year ahead is thus of the same size as the differences between the
models. The summed error of the best lifetime models (both drift variants) is 7 to 8\% smaller than
that of the previous year's rate, and that of the hierarchy without a year effect is a sixth
larger. At the 90-day horizon, where the previous year's rate is more recent, the forecasters
fall in nearly the same order with smaller errors (Table~\ref{tab:landmark90}). Its nine cohorts hold
9 landmarks, 1{,}822{,}172 drive-cohort members (drive-landmark observations, not distinct drives), 5{,}607 failures, 34{,}463 removed without failure
.

\textbf{Identifying vulnerable drives.} SMART warnings concentrate attention on drives
that account for a much larger share of subsequent failures. The tenth of each cohort that
the SMART-informed forecast rates most at risk has 44\% of the year's failures, against
19\% for the empirical-Bayes model and 18\% for B3 with drift
(Table~\ref{tab:landmarkdeciles}). We rank each cohort's drives by their forecast probability. From drive
model and age alone, the empirical-Bayes model reaches an AUC on all drives, the primary
population, of 0.64 at one year and 0.66 at 90 days (0.66 and 0.67 on the drives with
complete follow-up). The constant rate, which gives every drive the same probability, is 0.50
by construction, and the pooled two-Weibull distribution 0.59 (0.61 on complete follow-up). B0 and B3
with a random walk or drift rank drives as the empirical-Bayes model does (0.64). B1 with
drift and B3 without a year effect are a little below (0.63). These variants differ in both
their shocks and their multiplier posteriors, and we did not isolate which difference costs
the 0.01. A SMART warning before the landmark raises the AUC to 0.73 at one year and 0.80 at
90 days (0.75 and 0.80 on complete follow-up). It lowers the summed log loss by 8{,}615 of
107{,}247 at one year (8{,}877 of 104{,}749 on complete follow-up) and by 4{,}213 of 37{,}287
at 90 days. The best lifetime model, B3 with drift, gains 965 (535 on complete follow-up). The
quarterly landmarks (23 with a one-year horizon, 25
with 90 days) give the same AUCs, 0.65 and 0.74 on all drives at one year.

\begin{table}[h]
\centering\small
\caption{Calibration by Decile of Forecast Risk, Yearly Landmarks, 365-Day Horizon. Deciles are formed within each cohort and pooled over the eight cohorts. $\bar p$ is the mean forecast probability in percent.}\label{tab:landmarkdeciles}
\footnotesize\setlength{\tabcolsep}{3.5pt}
\begin{tabular}{@{}rrrrrrrrrr@{}}
\toprule
& \multicolumn{3}{c}{EB predictive} & \multicolumn{3}{c}{EB + SMART} & \multicolumn{3}{c}{B3 with drift} \\
\cmidrule(lr){2-4}\cmidrule(lr){5-7}\cmidrule(lr){8-10}
Decile & $\bar p$ & obs. & pred. & $\bar p$ & obs. & pred. & $\bar p$ & obs. & pred. \\
\midrule
1 & 0.38 & 676 & 564 & 0.24 & 538 & 357 & 0.39 & 690 & 586 \\
2 & 0.58 & 781 & 859 & 0.36 & 555 & 542 & 0.58 & 652 & 860 \\
3 & 0.77 & 1{,}641 & 1{,}145 & 0.49 & 1{,}100 & 727 & 0.80 & 1{,}425 & 1{,}190 \\
4 & 0.96 & 1{,}269 & 1{,}422 & 0.61 & 1{,}005 & 908 & 1.05 & 1{,}540 & 1{,}564 \\
5 & 1.10 & 1{,}413 & 1{,}640 & 0.70 & 1{,}091 & 1{,}036 & 1.27 & 1{,}581 & 1{,}883 \\
6 & 1.25 & 1{,}907 & 1{,}861 & 0.82 & 1{,}500 & 1{,}218 & 1.44 & 2{,}034 & 2{,}137 \\
7 & 1.54 & 2{,}516 & 2{,}288 & 1.03 & 1{,}971 & 1{,}530 & 1.66 & 2{,}493 & 2{,}468 \\
8 & 1.93 & 3{,}126 & 2{,}867 & 1.34 & 1{,}577 & 1{,}994 & 1.95 & 2{,}942 & 2{,}906 \\
9 & 2.46 & 3{,}253 & 3{,}666 & 1.86 & 2{,}088 & 2{,}764 & 2.38 & 3{,}508 & 3{,}538 \\
10 & 3.50 & 3{,}969 & 5{,}210 & 6.77 & 9{,}126 & 10{,}064 & 3.17 & 3{,}686 & 4{,}717 \\
\bottomrule
\end{tabular}
\end{table}

\begin{table}[h]
\centering\small
\caption{Observed and Predicted Failures Within 365 Days by Warning State and Manufacturer. The warning is a SMART warning before the landmark. The yearly landmarks are pooled.}\label{tab:landmarkgroups}
\footnotesize\setlength{\tabcolsep}{4pt}
\begin{tabular}{@{}lrrrrr@{}}
\toprule
& & & \multicolumn{3}{c}{Predicted} \\
Group & Drives & Observed & EB predictive & EB + SMART & B3 with drift \\
\midrule
clear & 1{,}384{,}197 & 11{,}891 & 19{,}427 & 12{,}398 & 19{,}694 \\
warned & 103{,}347 & 8{,}661 & 2{,}094 & 8{,}742 & 2{,}155 \\
Seagate & 768{,}984 & 13{,}211 & 15{,}120 & 15{,}212 & 14{,}819 \\
HGST & 409{,}332 & 3{,}811 & 2{,}807 & 3{,}094 & 2{,}795 \\
Toshiba & 308{,}168 & 3{,}516 & 3{,}523 & 2{,}774 & 4{,}188 \\
WDC & 1{,}060 & 14 & 71 & 60 & 45 \\
\bottomrule
\end{tabular}
\end{table}

Table~\ref{tab:landmarkgroups} splits the forecasts by warning and manufacturer. Of the
20{,}552 failures, 8{,}661 were among the 103{,}347 drives (6.9\%) that had carried a warning
before the landmark. The empirical-Bayes model, which cannot see the warning, expects 2{,}094 of
them, and the SMART-informed forecast 8{,}742. By decile (Table~\ref{tab:landmarkdeciles}),
the empirical-Bayes forecast is too high in its top two deciles (5{,}210 predicted against
3{,}969 observed in the tenth), in the second, and in the fourth and fifth. It is too low in
the first, the third and the sixth to eighth. The SMART-informed forecast concentrates the
risk in the tenth decile (10{,}064 predicted, 9{,}126 observed) and is closer than the
empirical-Bayes forecast in the second, fourth and fifth. It predicts too few failures in the
first, sixth and seventh and too many in the eighth and ninth, each by a wider margin than the
empirical-Bayes forecast. Its reallocation by manufacturer follows from the single fleet-wide
rescaling. Every clear drive's hazard is multiplied by the same constant, so a manufacturer
whose drives are warned less often than the fleet's loses expected failures. Toshiba and WDC
lose (2{,}774 and 60 predicted against 3{,}523 and 71 without the warning, 3{,}516 and 14 observed)
and HGST gains (3{,}094 against 2{,}807, 3{,}811 observed).

\textbf{What changes when forecasting before the year begins.} The earlier failure-rate
comparisons and these forecasts of an existing fleet give a similar ordering of predicted
totals. The bathtub
forecasts 2.4 times the failures, the pooled two Weibulls 1.44, the hierarchy, B0 and the two
variants with a drifting year effect 1.05 to 1.08, the random walk 1.30 and B3 without a year
effect 1.49. Truncating exposure lowers the hazard forecasters' expected counts by about a
sixth in the two years of bulk retirement. A forecast of the recorded endpoint for a fleet
needs a retirement assumption or a retirement model, which these forecasters lack. By log loss
the SMART warning's gain (8{,}615) is five times the hierarchy's gain over a constant rate
(1{,}660). This is the drive-month result of Appendix~\ref{sec:cov} seen prospectively.

\subsection{Modeling Warnings, Failure, and Retirement Together}\label{sec:mstate}

A warning helps assess a drive today, but a simulation also needs to represent what happens
next: a clear drive may develop a warning, its risk may change as time passes, and it may
fail or be retired. The SMART-informed forecaster above uses a fixed multiplier for an
existing warning. Here we model those changes explicitly. Oakley et al.~\cite{oakley2023multistate}
fitted an illness-death model to one drive model: the drive can move from a clear state to a
warned state and then fail, or fail without a prior warning. We extend their idea to the whole
fleet and test it prospectively. Our warning is the one used throughout this paper, any of the five
attributes Backblaze uses. Oakley et al.\ dropped attribute 188 for their drive model, on which
it did not appear to change the failure rate.

\textbf{Model.} A drive is clear (C) or warned (F). It moves from C to F when the first of SMART
5, 187, 188, 197 and 198 becomes non-zero, and it leaves the record from either state by a
recorded failure or by removal without a failure flag. Each of the five transition hazards
(C to F, C to failure, F to failure, C to removal, F to removal) is constant within age bins and
calendar years and has a drive-model adjustment drawn from a normal distribution. The hazards out of F also
depend on the time since the first warning, in five bins (under 30 days, 30 to 90, 90 to 365, one to
two years, over two years), with drives already warned when first seen as a sixth level. A warning
raised on a drive's last observed day gives no warning, so such a drive stays in C to its exit.
We fit each hazard as a Poisson model on cells of drive model, calendar quarter, age bin and
state. The exposure is the days on which the drive appears in the daily files, the same day
masks as the Lexis table, split at its first warning; a drive absent from the files contributes
no exposure while it is absent. Two of the 104{,}458 warnings and failures fall in cells with no
observed exposure and are dropped. A removal is a drive without a failure flag last seen more than 30
days before a cutoff, as in Section~\ref{sec:retire}. For each landmark $L$ we refit all five
hazards to the cells before $L$. The removals are those known on $L$: drives last seen more
than 30 days before $L$, including drives then in an absence of more than 30 days, whatever
happened to them later. We hold the calendar-year term at the last complete year.

The forecast is the probability of a recorded failure before removal within $\tau$, the
primary target of the landmark evaluation, in continuous time. For a warned drive it is exact: within each
stretch on which the age bin and the bin of the time since the warning are both constant, the
competing exits are exponential. For a clear drive the path after a new warning is exact, and
the time of the warning is integrated on a grid of a quarter of a day. Two closed-form cases
check the computation (\code{bb\_multistate\_tests.R}). For the
122{,}622\unskip\ drives of the 2020-01-01\unskip\ cohort,
the summed forecasts of every variant at 90 and 365 days change by at most
0.006\unskip\ failures between steps of a quarter and a sixteenth of a day
(0.21\unskip\ with a step of one day).

Over the whole record a warned drive fails at 86\unskip\% a year in
the first 30 days after its first warning and at 5.7\unskip\% a year
once it has been warned for two years. Drives warned when first seen fail at
12.2\unskip\% a year. These are observed conditional rates. They do not
show that elapsed time lowers the risk, since the drives that fail soon after a warning leave
the warned population. A fixed factor for the warning averages over them.

\Needspace{3\baselineskip}
\textbf{What the warning history adds.} Accounting for time since the warning improves
individual probability forecasts beyond simply recording whether a warning occurred.
The transition model also provides the evolving drive state needed for simulation.
Table~\ref{tab:mstate} compares its forecasts with the alternatives on the same drives.
Relative to the empirical-Bayes predictive baseline, at the 8\unskip\ yearly landmarks it lowers
the summed log loss by 9{,}994\unskip\ at one year (standard error
3{,}175\unskip, treating the cohorts as independent, which they are not,
since they share drives), against 8{,}615\unskip\ for the
empirical-Bayes forecaster with the warning and 8{,}220\unskip\ for the
boosted trees that read the warning. Against these two directly, the paired difference is
1{,}379\unskip\ (standard error 1{,}090\unskip) and
1{,}774\unskip\ (781\unskip), positive when the
transition model has the lower loss. At 90 days the gains are
5{,}333\unskip, 4{,}213\unskip\ and
5{,}185\unskip. The quarterly landmarks overlap, and we give no standard
errors for them: one-year gains of 27{,}229\unskip,
26{,}974\unskip\ and 26{,}065\unskip, and 90-day
gains of 16{,}031\unskip, 13{,}141\unskip\ and
17{,}079\unskip. The boosted trees that read the full monthly SMART record
gain 14{,}664\unskip\ at one year; they use far more information about
each drive. Pooled over the yearly landmarks at one year, the model forecasts
1.13\unskip\ times the observed failures, against
1.03\unskip\ and 1.05\unskip\ for the two
empirical-Bayes forecasters. The pooled ratios hide errors of opposite sign: for the 2020 cohort
the model forecasts 2.74\unskip\ times the observed failures and the
empirical-Bayes forecasters 2.14 and 2.17\unskip.

Time since the warning accounts for more of this score improvement than modeling retirement
at these horizons. We assess their contributions by switching off each part in turn.
With the failure and removal rates after a warning not depending on the time
since the warning, the summed log loss rises by 2{,}378\unskip\ at the
yearly landmarks at one year and 591\unskip\ at 90 days. With the
removal hazards set to zero, so that the model forecasts failure under continued service, it
rises by 138\unskip\ and 5\unskip.

An earlier version of this model, which the previous draft reported, propagated the state day
by day and drew its training exposure from each drive's first to its last sighting, so that a
drive absent across a gap before $L$ and seen again after it counted as at risk through the
gap. With the observed-day exposure, the removals known on $L$ and the continuous-time forecast,
the one-year gain at the yearly landmarks is 9{,}994\unskip\ before and
9{,}994\unskip\ after, the 90-day gain 5{,}339\unskip\
and 5{,}333\unskip, and the pooled one-year ratio of predicted to observed
failures 1.11\unskip\ and 1.13\unskip. The
forecasts themselves changed (the summed one-year forecast rose by about 2\%), but the summed
log losses hardly moved.

\textbf{Samplers.} The fit to all data, with the calendar year held at 2025, has two samplers
that draw the state path, failure and removal of a drive exactly for these piecewise-constant
hazards: one in R (\code{bb\_ms\_query.R}) and one in the C++ library (\code{ms:MODEL},
Table~\ref{tab:crosswalk}). We checked both against the forecast's probabilities for
24\unskip\ drives (four drive models, three ages, clear and warned) at 90
and 365 days, with $2\times10^5$ draws each, with and without removal. The largest $|z|$ is
3.4\unskip\ with removal and 2.6 without it for the R sampler, and 2.9\unskip\ over
the 96\unskip\ C++ checks, about what that many independent
comparisons give by chance. In the oldest age bin, ten years and more,
the fitted failure rates are close to zero and rest on almost no observed failures. Neither
sampler should therefore be used to extrapolate lifetimes beyond ten years of age.

\textbf{Variants tried.} We specified the first row of Table~\ref{tab:mstate} before seeing any
result. After seeing its totals we tried four other treatments of the calendar year, all
reported in the table, none chosen in advance. Without a year effect the pooled one-year ratio
at the yearly landmarks is 1.05\unskip; a choice among the variants
would be made on the same landmarks that score them.

\begin{table}[h]
\centering\small
\caption{The Fitted Transition Model Among the Landmark Forecasters. The cohorts are the drives in service at each yearly or quarterly landmark, and the endpoint is a recorded failure before removal. Gains are reductions in the log loss summed over all drives, relative to the empirical-Bayes predictive forecast, at the two horizons. SE is the standard error of the yearly one-year gain from the spread of the per-landmark differences, which treats cohorts that share drives as independent; the quarterly cohorts overlap and have none. Pred./obs., $\Sigma|\log r|$ and AUC (on all drives, weighted over landmarks) are for the yearly landmarks at 365 days. B3 with drift was run at the yearly landmarks only.}\label{tab:mstate}
\footnotesize\setlength{\tabcolsep}{4pt}
\begin{tabular}{@{}lrrrrrrrr@{}}
\toprule
& \multicolumn{3}{c}{Yearly gain} & \multicolumn{2}{c}{Quarterly gain} & & & \\
\cmidrule(lr){2-4}\cmidrule(lr){5-6}
Forecaster & 365 days & SE & 90 days & 365 days & 90 days & Pred./obs. & $\Sigma|\log r|$ & AUC \\
\midrule
Transition model, last year's level & 9{,}994 & 3{,}175 & 5{,}333 & 27{,}229 & 16{,}031 & 1.13 & 2.88 & 0.739 \\
Transition model, year as a random effect & 10{,}000 & 3{,}173 & 5{,}334 & 27{,}277 & 16{,}045 & 1.13 & 2.87 & 0.739 \\
Transition model, no year effect & 9{,}632 & 2{,}950 & 5{,}234 & 28{,}872 & 16{,}603 & 1.05 & 2.81 & 0.735 \\
Transition model, calendar drift & 9{,}771 & 3{,}222 & 5{,}341 & 25{,}828 & 15{,}877 & 1.11 & 3.24 & 0.739 \\
Transition model, mean level of three years & 9{,}298 & 3{,}296 & 4{,}972 & 25{,}411 & 15{,}117 & 1.15 & 3.05 & 0.742 \\
Transition model, no time since the warning & 7{,}616 & 2{,}762 & 4{,}742 & 21{,}709 & 13{,}965 & 1.13 & 2.81 & 0.738 \\
Transition model, removal hazards set to zero & 9{,}856 & 3{,}261 & 5{,}329 & 26{,}357 & 15{,}976 & 1.17 & 2.79 & 0.739 \\
Empirical Bayes with the SMART warning & 8{,}615 & 2{,}615 & 4{,}213 & 26{,}974 & 13{,}141 & 1.03 & 2.80 & 0.735 \\
Boosted trees, SMART warning & 8{,}220 & 2{,}659 & 5{,}185 & 26{,}065 & 17{,}079 & 1.15 & 2.65 & 0.719 \\
Boosted trees, full SMART record & 14{,}664 & 4{,}135 & 8{,}862 & 46{,}608 & 28{,}415 & 1.19 & 2.65 & 0.744 \\
B3 with drift & 965 & 438 & 145 & --- & --- & 1.06 & 2.25 & 0.641 \\
Empirical Bayes, predictive & 0 & --- & 0 & 0 & 0 & 1.05 & 2.82 & 0.643 \\
\bottomrule
\end{tabular}
\end{table}

\section{Predicting Drive Failures with Boosted Trees}\label{sec:boost}

The drive's recent SMART history gives the tree models their main advantage. With that
history they produce the best individual probability forecasts in this study; without it,
their greater modeling capacity does not reliably improve on the lifetime models.
We test gradient-boosted decision trees on the drives
already in service at each forecast date (Section~\ref{sec:landmark}), both with and without
their SMART records. These models combine many small trees, each dividing drives into
groups using rules about their attributes. Successive trees improve the predictions of
the existing combination. They can learn relationships among age, drive model and condition
without prescribing a lifetime distribution.

We also use trees to predict failure counts in the Lexis table, which groups observations
by age, drive model and calendar time. This checks whether a more flexible treatment of
those same inputs improves on the fitted lifetime models. That comparison is summarized
in Section~\ref{sec:boostlexissum} and given in full in Appendix~\ref{sec:boostlexis}.

\subsection{How the Tree Models Learn}\label{sec:boosttheory}

Training uses a loss function to measure how poorly the current predictions match the
observations; each new tree seeks to reduce that loss. We want a probability of failure,
rather than only a yes-or-no classification. The chosen log loss rewards assigning high
probability to the outcome that occurs and penalizes confident mistakes. Gradient
boosting~\cite{friedman2001greedy} fits an additive model $F(x)=\sum_m f_m(x)$ one
term at a time, each term a small regression tree. At round $m$ it fits a tree $h_m$ to the
negative gradient of a differentiable loss $\ell(y,F)$ at the current fit,
$-\partial\ell(y_i,F)/\partial F$ at $F=F_{m-1}(x_i)$, and $F_m=F_{m-1}+\nu h_m$ with a
learning rate $\nu<1$. (Its origin in AdaBoost, and its relation to the expert tracking of
Appendix~\ref{sec:track}, are outlined in Appendix~\ref{sec:boostbg}.) The landmark forecasters use the logistic loss. For a
binary outcome with $F$ the \emph{logit} of $p$, $F=\log[p/(1-p)]$, the loss is
$-y\log p-(1-y)\log(1-p)$, the log loss of Section~\ref{sec:perdrive}, and the negative
gradient is $y-p$. The Poisson loss of the boosted hazard on the Lexis cells is in
Appendix~\ref{sec:boostlexis}. We do not use AdaBoost's exponential loss. It is a surrogate for the
misclassification rate whose fit is dominated by the examples it gets most wrong, and it is
less robust than the binomial deviance~\cite{friedman2000additive}. With one or two failures
per hundred drive-years, and a probability rather than a classification wanted, the loss is
the negative log-likelihood. We use two implementations: LightGBM~\cite{ke2017lightgbm} and,
for two fixed ensembles, scikit-learn's histogram gradient boosting. Both bin features into
histograms~\cite{ke2017lightgbm}. As implemented, both set each leaf's value by a Newton step,
the negative sum of loss gradients divided by the sum of second derivatives plus the
$L_2$ penalty. Without the penalty this is the usual Newton update for the
logistic loss~\cite{friedman2001greedy}. Two further
terms recur below. The \emph{split gain} of a feature is the total reduction in the training
loss achieved by the splits made on it, summed over all trees. A model refitted
\emph{without} a predictor is the same ensemble fitted again with that feature removed.

A tree is piecewise constant. An ensemble can fit interactions within the restrictions on
its splits, depth and leaf sizes. It cannot extrapolate a trend:
outside the range of a feature in the training data it returns the value of the nearest leaf.
A calendar-year feature is held at the leaf its last training years fall in. This need not
equal the fleet-wide rescaling called ``year calibration'' in Section~\ref{sec:valid}. A drive model
absent from training is a missing value of a categorical feature. LightGBM sends it down the
side of each split that holds the categories not selected for the other side. A new drive
model is therefore predicted whatever those categories were predicted, and the fit does not
control it.
To limit fitting to noise in the training data, we control complexity through the size of
the trees, the minimum number of examples in a leaf, an $L_2$ penalty on the leaf values,
the learning rate and, above all, the number of rounds. Except
for two fixed-round reference ensembles on the Lexis cells (Appendix~\ref{sec:boostlexis}), we choose the
number of rounds by early stopping---stopping when performance ceases to improve on the
latest part of the training data (the last training
year, or the last year of training origins)---and then refit the ensemble to all the training
data with that number of rounds.

\subsection{Predicting Failure Counts from Age, Drive Model, and Year}\label{sec:boostlexissum}

For aggregate failure counts, the trees did not establish a clear advantage over the
drive-model hierarchy with calendar adjustment, given the variation in scores between test years.
We fitted four boosted Poisson ensembles to the HDD cells of the Lexis table on the
rolling-origin folds of Section~\ref{sec:valid} (Appendix~\ref{sec:boostlexis},
Table~\ref{tab:rolling}). The best has the number of rounds chosen by early stopping and the
calendar year as a feature. On the annual data it scores $-31{,}143$, which is
348 log units ahead of the empirical-Bayes point forecast (five of nine years, 1.0 standard
errors of the sum) and 374 behind additive model A2 (which wins seven of nine years,
1.8 standard errors). It is 1{,}697 behind the empirical-Bayes predictive. That gap is of the size of the hierarchy's own
point-to-predictive gap and, like it, is not a like-for-like comparison
(Section~\ref{sec:valid}). On the cells of drive models already seen in training the ensemble
is 47 log units ahead of A2. Its overall deficit to A2 comes from the cells of new drive
models, where it is 421 behind. Given the
drive model, its age and the year, the scores of the parametric hierarchy with a year
calibration and of a few hundred trees that could form interactions among those variables
were not clearly distinguished, against the year-to-year variation of the scores. That does
not make them equivalent. The ensemble supplies fitted expected counts without a distribution
representing uncertainty in its fitted parameters, so the table reports only its point-forecast score.

\subsection{Predicting Failures Among Drives Already in Service}\label{sec:boostlandmark}

For every drive in service on $L$, the question of the landmark design
(Section~\ref{sec:landmark}) is whether it fails within $\tau$ days. We train one classifier
across earlier forecast dates, with one row per drive and date and an outcome of failure
or non-failure within the chosen horizon. Its predictors are information available at that
date, including the calendar year. The
pooling of landmarks follows van Houwelingen's supermodel~\cite{vanhouwelingen2007dynamic},
which uses proportional-hazards models and a pseudo-partial likelihood. A 2026 preprint by
Sandqvist~\cite{sandqvist2026landmark} combines landmark supermodels with boosted trees through
a Poisson formulation of the conditional hazard. It bridges the two boosted approaches used
here, the Lexis-cell hazard of Appendix~\ref{sec:boostlexis} and this classifier. We cite it
as a preprint and have not implemented it. Our classifier fits the binary $\tau$-day outcome
directly.

The training rows come from earlier origins: the first day of every quarter from January 2014
to $L-\tau$, so that each row's $\tau$-day window closed before $L$. Every HDD data drive in
service and not yet failed on an origin is a row, labelled by whether it failed within $\tau$
days. We label a drive removed unfailed inside the window a non-failure, which is what the
operator saw. For the 2025 landmark at one year that is 41 origins and 5.0 million rows. The
features are what the monthly extraction of Appendix~\ref{sec:cov} knew about the drive on the
origin:
\begin{itemize}[leftmargin=1.4em]
\item from the drive's last monthly record before the origin: reallocated sectors, reported
  uncorrectable errors, command timeouts, pending and offline uncorrectable sectors (attributes
  5, 187, 188, 197 and 198, with the low word of 188 as a separate feature). Also the changes
  of 5, 187, 197 and 198 over the previous one and three months, power-on hours and their
  change over one and three months, and start/stop, power-cycle, power-off-retract and
  load-cycle counts per power-on year (attributes 4, 12, 192, 193). Also the month's mean and
  maximum temperature, the mean's deviation from the drive model's fleet that month (attribute
  194), and the days the record covered.
\item the observation history: the months the drive had been observed and the gap since its
  last record.
\item the drive model as a categorical feature, with its capacity, manufacturer and vintage,
  the drive's power-on age and age since first sighting on the origin, and the origin's
  calendar year.
\end{itemize}
We compare three feature sets. \emph{Base} has the drive model, its capacity, manufacturer and
vintage, the two ages, the calendar year and the observation history: what the lifetime models
see, plus the observation history, and nothing from the SMART attributes. \emph{Warning} is base
plus the SMART warning of Section~\ref{sec:landmark}, as two features (ever before the
origin, and on the last record). \emph{SMART} is base plus the full record.

The ensembles have logistic loss, learning rate 0.05, 63 leaves, at least 500 rows per leaf,
$L_2$ penalty 10, and feature and row subsampling of 0.8. We choose the number of rounds by
early stopping on an inner temporal split of the training origins. The validation origins are
those in the last year before $L-\tau$. The fitting origins are those whose $\tau$-day windows
had closed before the first validation origin, so that no failure labels both a fitting row
and a validation row. We then refit the ensemble to all training origins with the chosen
number of rounds. A second set of ensembles stops on the validation AUC instead of the log
loss (Table~\ref{tab:boostrounds} in Appendix~\ref{sec:boostdiag}). The classifier's output is
a probability of the operator's recorded endpoint, and we score it exactly as the other
landmark forecasters. It has no exposure-truncated version: a probability that already treats
removal as a non-failure has no hazard to condition on the exposure observed. We fit each
cohort on its own, so nothing after its landmark enters it apart from the full-history
metadata quantified in Section~\ref{sec:landmark}.
Rounds chosen by early stopping, yearly landmarks: base 1 to 117 (one year), 7 to 66 (90 days); SMART 19 to 152 and 15 to 427. Training rows 0.6 to 5.0 million (one year). Fitting time of a SMART ensemble, including the early-stopping fit, 3 to 17~s on 5 threads. Stopped on AUC instead: base 1 to 487 rounds (one year), SMART 1 to 310.

\begin{table}[h]
\centering\small
\caption{The Boosted Forecasters Among the Landmark Forecasters, 365-Day Horizon. All are scored on the recorded endpoint at the yearly landmarks, as in Section~\ref{sec:landmark}. $\Sigma|\log r|$ is the sum over landmarks of $|\log(\text{predicted}/\text{observed})|$. $\Delta$ log loss is the gain over the empirical-Bayes predictive forecast, positive when better. AUC is on all drives and on drives with complete follow-up. Top 10\% is the share of the cohorts' failures in the tenth of drives each forecaster rates most at risk.}\label{tab:boostlandmark}
\footnotesize\setlength{\tabcolsep}{4pt}
\begin{tabular}{@{}lrrrrrr@{}}
\toprule
Forecaster & Pred./obs. & $\Sigma|\log r|$ & $\Delta$ log loss & AUC all & AUC complete & Top 10\% \\
\midrule
Boosted, SMART record & 1.19 & 2.65 & +14{,}664 & 0.744 & 0.748 & 47 \\
Boosted, SMART warning & 1.15 & 2.65 & +8{,}220 & 0.719 & 0.725 & 44 \\
Boosted, base & 1.09 & 2.17 & -610 & 0.606 & 0.615 & 13 \\
EB predictive + SMART warning & 1.03 & 2.80 & +8{,}615 & 0.735 & 0.747 & 44 \\
EB predictive & 1.05 & 2.82 & +0 & 0.643 & 0.660 & 19 \\
B3 with drift & 1.06 & 2.25 & +965 & 0.641 & 0.660 & 18 \\
Constant rate & 1.03 & 2.44 & -1{,}660 & 0.500 & 0.500 & 10 \\
\bottomrule
\end{tabular}
\end{table}

\begin{table}[h]
\centering\small
\caption{The Boosted Forecasters, 90-Day Horizon. Columns are as in Table~\ref{tab:boostlandmark}, over the nine yearly landmarks.}\label{tab:boostlandmark90}
\footnotesize\setlength{\tabcolsep}{4pt}
\begin{tabular}{@{}lrrrrrr@{}}
\toprule
Forecaster & Pred./obs. & $\Sigma|\log r|$ & $\Delta$ log loss & AUC all & AUC complete & Top 10\% \\
\midrule
Boosted, SMART record & 1.22 & 2.70 & +8{,}862 & 0.828 & 0.828 & 64 \\
Boosted, SMART warning & 1.23 & 2.59 & +5{,}185 & 0.816 & 0.816 & 62 \\
Boosted, base & 1.25 & 2.71 & +16 & 0.660 & 0.660 & 21 \\
EB predictive + SMART warning & 1.12 & 2.35 & +4{,}213 & 0.799 & 0.801 & 61 \\
EB predictive & 1.13 & 2.37 & +0 & 0.663 & 0.667 & 23 \\
B3 with drift & 1.10 & 1.79 & +145 & 0.661 & 0.666 & 22 \\
Constant rate & 1.13 & 2.05 & -816 & 0.500 & 0.500 & 10 \\
\bottomrule
\end{tabular}
\end{table}

\begin{table}[h]
\centering\small
\caption{The Boosted Forecasters by Yearly Landmark, 365-Day Horizon. Pred./obs.\ is given for the three feature sets. The AUC is on all drives. Without SMART it is given for the empirical-Bayes predictive forecast and the base ensemble. With SMART it is given for the empirical-Bayes forecast with the SMART warning and for the warning and SMART ensembles. Table~\ref{tab:boostrounds} in Appendix~\ref{sec:boostdiag} gives the rounds, the AUC-stopped variants and the AUC on complete follow-up.}\label{tab:boostyears}
\footnotesize\setlength{\tabcolsep}{4pt}
\begin{tabular}{@{}rrrrrrrrrr@{}}
\toprule
& & \multicolumn{3}{c}{Pred./obs.} & \multicolumn{2}{c}{AUC, no SMART} & \multicolumn{3}{c}{AUC, with SMART} \\
\cmidrule(lr){3-5}\cmidrule(lr){6-7}\cmidrule(lr){8-10}
Year & Obs. & base & warning & SMART & EB & base & EB+warning & warning & SMART \\
\midrule
2018 & 984 & 1.88 & 2.06 & 2.32 & 0.638 & 0.650 & 0.688 & 0.701 & 0.715 \\
2019 & 1{,}892 & 0.62 & 0.62 & 0.82 & 0.585 & 0.606 & 0.653 & 0.660 & 0.664 \\
2020 & 1{,}122 & 1.83 & 2.08 & 2.15 & 0.580 & 0.534 & 0.671 & 0.630 & 0.655 \\
2021 & 1{,}510 & 0.98 & 1.12 & 1.25 & 0.614 & 0.507 & 0.689 & 0.672 & 0.734 \\
2022 & 2{,}858 & 1.09 & 0.88 & 0.89 & 0.689 & 0.617 & 0.752 & 0.710 & 0.750 \\
2023 & 3{,}989 & 0.90 & 0.94 & 0.91 & 0.666 & 0.619 & 0.773 & 0.732 & 0.759 \\
2024 & 4{,}199 & 1.06 & 1.17 & 1.12 & 0.630 & 0.620 & 0.752 & 0.748 & 0.760 \\
2025 & 3{,}998 & 1.19 & 1.29 & 1.35 & 0.670 & 0.643 & 0.773 & 0.775 & 0.791 \\
\bottomrule
\end{tabular}
\end{table}

Table~\ref{tab:boostlandmark} places the ensembles among the landmark forecasters and
Table~\ref{tab:boostyears} follows them year by year. We ask three things of them. Does a
learner with no lifetime distribution rank drives by model and age as well as the hierarchy does? What
does the drive's own record add? Does the sum of a classifier's probabilities forecast the
total?

\paragraph{Ranking by drive model and age.} More modeling capacity alone does not resolve
which drives are most vulnerable. At one year the base ensemble ranks drives worse
than the hierarchy. Its AUC is 0.606 against 0.643 on all drives and 0.615 against 0.660 on
complete follow-up, its log loss is 610 worse, and its top tenth holds 13\% of the failures
against 19\%. At 90 days the two are level: AUC 0.660 against 0.663 on all drives at the
yearly landmarks and 0.674 against 0.664 at the quarterly ones, with the ensemble's log loss
16 and 953 better. At the quarterly landmarks with a one-year horizon the gap is as at
the yearly ones (0.602 against 0.646). By year, on complete follow-up, the ensemble is within
0.015 of the hierarchy from 2018 to 2020 (0.684 against 0.689, 0.607 against 0.597, 0.589
against 0.604), and ahead on all drives in 2018 and 2019. It falls to 0.556 in 2021 and about
0.62 in 2022 and 2023, while the hierarchy stays near 0.68. It is 0.04 and 0.03 below the
hierarchy in 2024 and 2025. Power-on age alone, with no drive-model information, orders four of the five
cohorts of 2021 to 2025 better than the base ensemble does (0.568 to 0.658
against 0.556 to 0.638). It orders those of 2019 and 2020 worse than chance (0.429 and 0.489),
when the fleet's young drives failed more than its old ones.

To look for the cause, we refitted the 2021 to 2023 cohorts several times
(Appendix~\ref{sec:boostdiag}), changing one thing in each refit: more rounds, stopping on AUC, without the calendar year or the vintage, with the
drive model's identity replaced by its failure rate, and on the rows with complete follow-up
only. No single change removed the AUC difference in those three cohorts. The two forecasters
also differ in the information they were given. The ensemble's training origins end a horizon
before the landmark, while the hierarchy's multipliers use every drive-day to its eve. The two
also treat removal differently: the ensemble labels a removed drive a non-failure, and the
hierarchy treats removal as censoring. The diagnostics do
not distinguish those differences from restrictions of the model's form. Within the large
drive models both forecasters are near chance (from 2021 on,
12 to 22 drive models with at least 20 failures per cohort; mean within-model AUC 0.52 to 0.56 for the hierarchy, 0.52 to 0.55 for age alone and 0.50 to 0.52 for the base ensemble
\unskip). The AUC the hierarchy has and the ensemble lacks
therefore lies mostly in ordering drives of different models and ages across the cohort.

\paragraph{Forecasting the total number of failures.} Adding the predicted probability of
failure for every drive gives the expected replacement demand from failures. As forecasts of
the cohort's failures, the ensembles' sums and the hierarchy's are of similar quality, with
neither consistently ahead. By the summed absolute log ratio, all three ensembles are ahead
of the hierarchy at the yearly one-year landmarks (2.17, 2.65 and 2.65 against 2.82). There
the base ensemble's Poisson score of the totals is 535 above the hierarchy's
(Table~\ref{tab:landmark365}). All three ensembles are behind at the yearly 90-day landmarks
(2.59 to 2.71 against 2.37). At the 25 quarterly 90-day landmarks all three are ahead (4.06 to
4.84 against 6.20). At the 23 quarterly one-year landmarks the base and SMART ensembles are
ahead (4.85 and 5.84 against 6.02) and the warning ensemble behind (6.09). There the base
ensemble's ratio is 1.10 and the hierarchy's 1.00. At the yearly landmarks the constant rate
of the previous twelve months is ahead of the ensembles at 90 days (2.05) and between them at
one year (2.44). The Bayesian variants with a drifting year effect, where available, are the
best at 90 days (1.79 and 1.80) and behind the base ensemble at one year
(Section~\ref{sec:landmark}: 2.25 against 2.17). The two kinds of forecaster do not
have the same target. The ensembles' probabilities sum to an expectation of the recorded
endpoint, failures before removal, which is what the observed total counts. The hazard
models' full-horizon sum is their expectation under continued service
(Section~\ref{sec:landmark}). The hazard models' ratios to the observed total therefore carry
that difference as well as error.

\begin{table}[h]
\centering\small
\caption{Where the SMART Ensemble's Fit Comes From. Each entry is a feature's share of the total split gain, averaged over the yearly landmarks. The sixteen features with the largest share at either horizon are shown. SMART attribute numbers are in parentheses, and $\Delta_1$ and $\Delta_3$ are changes over the previous one and three months. Split gain favours a categorical feature with many levels, so the drive model's share is likely inflated.}\label{tab:boostimportance}
\footnotesize
\begin{tabular}{@{}lrr@{}}
\toprule
Feature & 365 days (\%) & 90 days (\%) \\
\midrule
warning on last record & 16.6 & 19.9 \\
drive model & 19.7 & 7.7 \\
warning ever & 16.6 & 15.1 \\
pending sectors (197) & 8.4 & 12.1 \\
$\Delta_1$ 5 & 3.7 & 9.4 \\
$\Delta_1$ 187 & 0.5 & 8.0 \\
reported uncorrectable (187) & 4.3 & 4.1 \\
calendar year & 3.5 & 2.7 \\
$\Delta_3$ 5 & 2.9 & 1.9 \\
manufacturer & 2.5 & 0.8 \\
$\Delta_3$ 187 & 2.3 & 0.7 \\
load cycles per year (193) & 2.0 & 1.6 \\
power-on age & 2.0 & 1.2 \\
power-off retracts per year (192) & 1.7 & 1.3 \\
vintage & 1.7 & 0.8 \\
reallocated sectors (5) & 1.6 & 1.0 \\
all other features (20) & 10.1 & 11.7 \\
\bottomrule
\end{tabular}
\end{table}

\paragraph{What the drive's record adds.} SMART records provide the largest improvement in
the individual probability forecasts, especially in distinguishing risks among drives that
have already raised a warning. The tenth of each cohort that the full-record ensemble rates
most at risk holds 47\% of the year's failures and 64\% of the quarter's, against 44\% and
62\% for the warning ensemble and 19\% and 23\% for the lifetime model.
At one year the warning ensemble gains
8{,}220 log units over the empirical-Bayes predictive forecast, and the hierarchy with its
manufacturer-specific warning factor gains 8{,}615. At 90 days the gains are 5{,}185 and
4{,}213 (the ensemble also uses the warning on the last record). The ensemble's AUC on all drives
is lower at one year (0.719 against 0.735) and higher at 90 days (0.816 against 0.799). Given
the full record, the ensemble has the lowest log loss of any per-drive forecaster in this
paper at one year, 14{,}664 below the empirical-Bayes predictive. At 90 days it is 8{,}862
below, and its AUC-stopped variant 8{,}483 below (Appendix~\ref{sec:boostdiag}). These are 1.8
and 1.7 times what the same ensemble gains from the warning. Its AUC gain over the warning ensemble on all drives is 0.025 at
one year and 0.012 at 90 days.

The two measures move by different amounts because they weigh the two parts of the cohort
differently: among the drives warned before the landmark (6.7\% of the drives with complete follow-up at one year, 42\% of their failures) the SMART ensemble's AUC is 0.83, the warning ensemble's 0.62 and the hierarchy's with the warning 0.59; among the clear drives the three are 0.61, 0.58 and 0.62 (failure-weighted means over the yearly cohorts); of the SMART ensemble's log-loss gain over the warning ensemble on these drives, 5{,}767 units come from the warned drives and 846 from the clear ones
\unskip. The AUC counts pairs of one failure and
one survivor, and pairs with both drives warned are 0.6 to 3.8\% of the pairs the AUC counts in a cohort, pairs of clear drives 45 to 76\% and mixed pairs 23 to 51\%; of the pooled AUC gain summed over the eight cohorts, 22\% comes from the warned pairs, 71\% from the clear pairs and 7\% from the mixed pairs
\unskip. By rank, then, the record's gain is
mostly a small reordering of the many clear drives. The log loss weighs drives by what
happens to them. The warned drives hold 42\% of the failures and 87\% of the log-loss gain,
so by probability the record's gain is in saying which of the drives the warning has already
singled out are the ones failing. We made this decomposition for the one-year cohorts only.

Table~\ref{tab:boostimportance} shows what the ensemble splits on, averaged over the yearly
landmarks: the SMART warning and error-count features together carry 59\% of the split gain at one year and 76\% at 90 days; drive model, year, manufacturer, vintage, capacity, age and observation history 32\% and 15\%; temperature, the start/stop, power-cycle, power-off-retract and load-cycle rates, power-on hours and the days covered by the last record the remaining 9\% and 9\%
\unskip. The one-month change in reallocated
sectors accounts for 9.4\% of training split gain at 90 days and 3.7\% at one year. The
one-month change in reported uncorrectable errors accounts for 8.0\%
against 0.5\%. The drive model has the largest split-gain share at one year (20\%) and falls
to 8\% at 90 days. For a failure in the next year the drive's state still carries more than
half the split gain, but drive model, cohort, age and observation history carry a third, twice
their share at 90 days. The regression of Appendix~\ref{sec:cov} found temperature to add
little given the error counts, and power cycling to keep a smaller effect.

\paragraph{What these forecasts can be used for.} SMART measurements help identify
vulnerable drives and improve their predicted failure probabilities. A forecast of who is
most at risk and a forecast of how many replacements to stock are distinctly different
tasks: the full-record ensemble still predicts about 19\% too many failures at one year
and 22\% too many at 90 days (Tables~\ref{tab:boostlandmark} and~\ref{tab:boostlandmark90}).
Its predictions concern recorded failures, including precautionary removals. These data
cannot separate physical deterioration from the
operator's response to the same SMART measurements (Section~\ref{sec:smartstate}). A
simulator that needs both a fleet failure rate and forecasts from a drive's current condition
must model how that condition changes and how failure risk changes with it. Neither a hazard
with removal as censoring nor a fixed-horizon classifier supplies one. Oakley et
al.~\cite{oakley2023multistate} fit such a model for one drive model, and
we fit one across the fleet in Section~\ref{sec:mstate}. Other SMART-based predictors on this
corpus~\cite{botezatu2016kdd,aussel2017icmla} classify drives from windows of daily SMART records, and Mahdisoltani et al.~\cite{mahdisoltani2017atc} predict sector
errors from them. Their designs and measures (Aussel et al., for example, report precision and
recall on daily snapshots) are not comparable with ours. Our ensembles
read one monthly record and its changes and forecast a quarter or a year ahead for every drive
in service. We score them only on cohorts whose training windows closed before the landmark,
the design under which they can be compared with the lifetime models. For the 2025 one-year
ensembles, evaluating an ensemble costs \frag{0.11 to 5.2}~$\mu$s per drive in a batch,
depending on its number of trees, and \frag{16 to 28}~$\mu$s for one drive per call.
The number of trees varies by two orders of magnitude between cohorts. We measured these costs
through LightGBM's Python binding, so they include the binding's overhead and are not compared
with the compiled sampler timings (Appendix~\ref{sec:boostspeed}).

% Part 4: aging and extrapolation; further structure in the data
\section{What Can We Say About Long-Term Drive Survival?}\label{sec:apc}

Reliable short-term forecasts do not settle how long a drive installed today will last.
As drives age, calendar time also advances, and new hardware enters a changing fleet.
Different accounts of aging and changes in operating conditions can therefore fit the
observed years similarly while predicting very different survival over ten or fifteen years.
The key difficulty is separating these effects. This matters most when a simulation keeps
drives running well beyond the ages for which there is substantial observation.

The forecasts illustrate the practical scale of the uncertainty
(Table~\ref{tab:implied}, Figure~\ref{fig:bayes}b). At five years the four scenarios differ by
at most 3.9 percentage points in survival probability for each of the eight drive models.
At fifteen years the range within a drive model is 15 to 27 percentage points. For the
ST12000NM0008, survival is 0.52 if the fleet-wide rate varies without a continuing downward
trend, 0.61 under the model without calendar effects (B0), 0.63 if the downward trend
continues for three more years and stops, and 0.79 if it never stops. Short-term agreement
can therefore coexist with substantially different predictions for a long-serving drive.

To see why, consider when a drive entered service. The age--period--cohort identity relates
a drive's age to calendar time and to its own
start of service, not to the first appearance of its hardware model
\cite{holford1983,clayton1987apc}. If drive $i$ first appears at calendar time $t_{0i}$ with
power-on age $a_{0i}$, its age at time $t$ is $a_i(t)=a_{0i}+t-t_{0i}$, in years. Putting the
drive-model vintage in place of the service start $t_{0i}-a_{0i}$ adds the differences in
deployment date and age at entry, so age, calendar time and drive-model vintage can be
strongly associated without being interchangeable. Here, vintage means the hardware
model's first appearance in the corpus; an individual drive's service start can be earlier
or later.

In this corpus the drive model's first appearance explains a share
0.87\unskip\ of the variance of the service starts $t_{0i}-a_{0i}$
(the model means alone, 0.96\unskip). Most drives are
deployed for years after their model first appears, and
23\unskip\ models have drives in service before it. A
drive's service start differs from its model's vintage by a median of
1.3\unskip\ years (3.3\unskip\
years at the 90th percentile), and 42\unskip\% of drives
are within a year of it. Within a drive model the interquartile range of service starts is
0.34\unskip\ years at the median and
0.2 to 1.6\unskip\ years across the
16\unskip\ models with at least 10{,}000 drives.
16\unskip\% of drives entered observation already in
service, at a median age of 0.2\unskip\ years. The
drive-model multipliers absorb the per-model offset. What can separate age from calendar time
in the variants with a period term is the within-model spread of service starts, and it is
modest. The fitted decompositions differ:
\begin{itemize}[leftmargin=1.4em]
\item B0 has no year effect. Its fitted Weibull wear-out exponent is 1.57, with
  no clear vintage trend.
\item B3 has Weibull wear-out exponent 1.93, a fleet-wide rate falling at 13\%
  a year, and newer drive models worse by 18\% a year at fixed age and date. For successive
  vintages observed at the same age, the last two nearly cancel.
\end{itemize}
The forecasts for later years differ under these assumptions (Table~\ref{tab:bayeshz}). In
a random walk, the fleet-wide rate changes from one year to the next; drift adds an average
annual direction to those changes. B3 with a driftless random walk combines steep wear-out
with no average downward trend in the log year effect.
It over-predicts failures by an amount that grows with the horizon: a mean log ratio of
$+0.24$, $+0.49$ and $+0.80$ at one, two and three years. The last is 4.1 nominal standard
errors from zero (approximate, because the test windows overlap). B3 with continued drift
and B0 have mean log ratios within the reported uncertainty of zero at every horizon we
examined. Their forecast levels are not clearly distinguished. B3 has the better predictive
log score.

Two results weaken the case for extrapolating the fitted drift. Its own 95\% interval,
$(-0.278,\ 0.004)$, includes zero. And the fitted year effect is nearly flat from 2020 (0.187)
to 2026 (0.180): the decline that the drift summarizes happened before 2020. The drifting
variant's forecasts stay level over 2021--2026 because two fitted quantities compensate. The
wear-out exponent $\beta_2=1.93$ raises the forecast hazard as the fleet ages, and the drift
$d=-0.134$ per year lowers it by about the same amount over those years. The sum fits the
level, but we have not tested either term separately.

Continuing an apparent improvement indefinitely can produce unrealistic lifetime
predictions. A drift that never stops makes the wear-out hazard, proportional to $e^{-0.134a}a^{0.93}$,
peak at seven years and decline afterwards. That is a property of the extrapolation, not an
observation: the oldest of these drive models is under seven years old, and of the two
million drive-years in the corpus 2\% are beyond eight years of age and 147 are beyond ten
(Table~\ref{tab:age}). With the hazard proportional to $e^{-ra}a^{\beta-1}$ and $r>0$ the
cumulative hazard also converges, $\int_0^\infty e^{-ra}a^{\beta-1}\,da = \Gamma(\beta)/r^{\beta}$,
so a drive has a positive probability of never failing. We compute that probability with the
year effect on its deterministic drift path and the yearly steps and drive-model-year shocks at
their means. Under the full stochastic forecast the random steps make it a random variable
with a different distribution over paths. For a drive entering service in 2026, when the year
effect is at 0.18 of its 2013 level, the probability of never failing over the posterior draws
of B3 with drift has median 0.67 at multiplier $Z=1$ (2.5th to 97.5th percentile 0.00 to
0.91) and 0.74 for the ST12000NM0008 ($z=0.75$). Of the draws, 2.0\% have a non-negative drift
and no such probability. The lifetime distribution under indefinite drift is therefore
\emph{defective}, the technical term for a distribution that assigns some probability to
never failing.
Most drives simulated under it reach the samplers' cap (Section~\ref{sec:sim}), and the column
is a sensitivity scenario, not a forecast.

For simulation, the useful response is to compare plausible future trends and report
whether the system-level conclusion changes. The models we compared do not resolve the
separation of aging from calendar-time effects. The
information that would do most to resolve it, the same drive model deployed in well-separated
years, is scarce in this corpus, because a fleet operator buys a model over a year or two. A
simulation that depends on drive lifetimes beyond about eight years should be run under at
least two of the scenarios in Table~\ref{tab:implied}. If the system-level results disagree,
the conclusion depends on the assumed future trend and should be reported that way.

Where the spread of service starts exists, the record does separate an event in calendar time
from wear-out. The Seagate ST3000DM001 is the case Backblaze described: 2.7\% of its drives
failed in 2012, 5.4\% in 2013 and 47.2\% in 2014~\cite{bb2015csi}. In our record
97\unskip\% of its 4{,}707\unskip\
drives are present in the record's first month, and their service starts, read from their
power-on hours, run from February 2012\unskip\ to
January 2013\unskip\ (5th to 95th percentile). That spread is enough to
cross-tabulate calendar quarter against power-on age (Table~\ref{tab:st3000qa}). Among drives
one to two years old the failure rate was 6\unskip\ per
100 drive-years in the second quarter of 2013, 24\unskip\
in the fourth, 61\unskip\ in the third quarter of 2014 and
28\unskip\ in the first quarter of 2015: the same age, a
tenfold swing by date. Within the third quarter of 2014 the rate was
32\unskip\ for drives under a year old,
61\unskip\ at one to two years and
178\unskip\ at two to three: the same date, a steep rise
with age. The record shows an event in time, peaking in the summer of 2014 and hitting the
older drives hardest, and not a wear-out curve alone. It is what a drive-model-year shock
(Section~\ref{sec:bayesmodel}) stands for, and the record has no lot numbers or firmware
versions that would say what caused it. Of the data drives that reappear in the files after a
failure flag (Section~\ref{sec:population}), 348\unskip\
are this model, consistent with Backblaze's account of pulling, bench-testing and reusing
them~\cite{bb2015csi}.

\begin{table}[h]
\centering\footnotesize\setlength{\tabcolsep}{5pt}
\caption{The ST3000DM001 by Calendar Quarter and Power-On Age. Failures per 100 drive-years in each cell of the Lexis table (drive-years in parentheses), for the ages with at least 20 drive-years in the quarter. Rows are calendar quarters, columns power-on age in years. Blank cells have under 20 drive-years.}\label{tab:st3000qa}
\begin{tabular}{@{}lrrr@{}}
\toprule
Quarter & Age 0--1 & Age 1--2 & Age 2--3 \\
\midrule
2013~Q2 & 6 (935) & 6 (81) & -- \\
2013~Q3 & 3 (461) & 10 (165) & -- \\
2013~Q4 & 4 (312) & 24 (617) & -- \\
2014~Q1 & 6 (160) & 26 (866) & -- \\
2014~Q2 & 10 (48) & 31 (873) & 51 (51) \\
2014~Q3 & 32 (25) & 61 (592) & 178 (158) \\
2014~Q4 & -- & 28 (227) & 86 (107) \\
2015~Q1 & -- & 28 (79) & 25 (95) \\
2015~Q2 & -- & -- & 39 (82) \\
2015~Q3 & -- & -- & 35 (49) \\
\bottomrule
\end{tabular}
\end{table}

\begin{table}[h]
\centering
\caption{Probability That a Drive Entering Service in Mid-2026 Survives 5, 10 and 15 Years. The drive models are the eight introduced since 2019 with the most exposure. Each column is one combination of fitted model and future trend. Each entry is a marginal probability, averaged over the posterior. For B3 it is also averaged over the future steps of the year effect (64 simulated paths per posterior draw) and over the drive-model-year shocks, for a group expected to have 100 failures a year. Held lets the year effect walk without drift from its 2026 level. 3~yr continues the fitted drift for three years first. Drift continues it for the life of the drive.}\label{tab:implied}
\footnotesize\setlength{\tabcolsep}{3.5pt}\begin{tabular}{@{}lrrrrrrrrrrrr@{}}
\toprule
& \multicolumn{4}{c}{5 years} & \multicolumn{4}{c}{10 years} & \multicolumn{4}{c}{15 years} \\
\cmidrule(lr){2-5}\cmidrule(lr){6-9}\cmidrule(lr){10-13}
& & \multicolumn{3}{c}{B3} & & \multicolumn{3}{c}{B3} & & \multicolumn{3}{c}{B3} \\
Drive model & B0 & held & 3 yr & drift & B0 & held & 3 yr & drift & B0 & held & 3 yr & drift \\
\midrule
ST16000NM001G & 0.964 & 0.957 & 0.969 & 0.972 & 0.905 & 0.852 & 0.895 & 0.928 & 0.833 & 0.709 & 0.785 & 0.885 \\
ST12000NM0008 & 0.904 & 0.914 & 0.938 & 0.943 & 0.763 & 0.728 & 0.801 & 0.862 & 0.609 & 0.522 & 0.629 & 0.791 \\
TOSHIBA MG08ACA16TA & 0.938 & 0.928 & 0.948 & 0.952 & 0.842 & 0.767 & 0.831 & 0.883 & 0.730 & 0.577 & 0.676 & 0.820 \\
WDC WUH721816ALE6L4 & 0.967 & 0.964 & 0.974 & 0.977 & 0.914 & 0.876 & 0.912 & 0.941 & 0.849 & 0.751 & 0.817 & 0.904 \\
WDC WUH722222ALE6L4 & 0.961 & 0.951 & 0.965 & 0.968 & 0.900 & 0.834 & 0.881 & 0.919 & 0.825 & 0.680 & 0.761 & 0.872 \\
ST12000NM001G & 0.952 & 0.951 & 0.965 & 0.968 & 0.876 & 0.836 & 0.882 & 0.920 & 0.785 & 0.682 & 0.763 & 0.873 \\
HGST HUH721212ALE604 & 0.914 & 0.927 & 0.947 & 0.952 & 0.785 & 0.765 & 0.829 & 0.882 & 0.642 & 0.574 & 0.674 & 0.819 \\
ST14000NM001G & 0.933 & 0.933 & 0.952 & 0.956 & 0.830 & 0.781 & 0.841 & 0.891 & 0.711 & 0.597 & 0.694 & 0.831 \\
\bottomrule
\end{tabular}
\end{table}

% Part 5: simulation; limitations and conclusions
\section{Using the Models in Simulators}\label{sec:sim}

A simulator draws one lifetime per drive installed or replaced. A rare-event simulation of
data loss may draw billions, so the cost of a draw matters. The simulation also needs the
remaining life of drives already in service and a way to share elevated risks across drives.
We distribute the lifetime models listed in Table~\ref{tab:crosswalk} as small tables of fitted
parameters, with one row per drive model and, for the Bayesian models, 1{,}000 plausible
parameter sets drawn from their fitted distribution (posterior draws). Functions in R and
samplers in a C++
library use these tables to calculate risks and generate lifetimes (Appendix~\ref{sec:raid}).
A separate Rust program reimplements the samplers so that
we can time them (Section~\ref{sec:speedsum}, Appendix~\ref{sec:speed}).

\subsection{Generating Lifetimes and Sharing Risks}\label{sec:query}

Using a fitted model in a simulator involves two choices: how to generate each drive's
remaining lifetime, and which risks that drive shares with the rest of the system. The
functions below support both. Table~\ref{tab:crosswalk} connects the statistical model names
used in the paper to their result files and sampler names; a hardware model is the type of
drive to which those functions are applied.

\begin{table}[h]
\centering\footnotesize\setlength{\tabcolsep}{4pt}
\caption{Names of the Models in the Text, the Result Files and the Sampler Library. Sampler specs are those of the C++ library's \code{make\_sampler}. A hardware model is a drive model such as ST4000DM000, and the per-model samplers take its name after the colon.}\label{tab:crosswalk}
\begin{tabular}{@{}>{\raggedright\arraybackslash}p{2.7cm}>{\raggedright\arraybackslash}p{3.3cm}>{\raggedright\arraybackslash}p{3.9cm}>{\raggedright\arraybackslash}p{5.3cm}@{}}
\toprule
Family & Name in the text & Fitted variant (results) & Sampler spec \\
\midrule
Single distributions on power-on age & exponential, Weibull, gamma, \dots & \code{models/\allowbreak thru-2025-12/\allowbreak fits.csv} & \code{exponential}, \code{weibull}, \code{gamma}, \code{lognormal}, \code{loglogistic}, \code{gompertz}, \code{gengamma}, \code{expweibull}, \code{piecewise} \\
Two and three competing Weibulls & two-Weibull distribution; three Weibulls & \code{cr2\_weibull}, \code{cr3\_weibull} & \code{cr2}, \code{cr3} \\
Mixture, GLFP, Weibull--Gompertz & by name & \code{mix2\_weibull}; \code{hier2\_laws} \code{glfp}, \code{wg} & \code{mix2}, \code{glfp}, \code{wg} \\
Two-state hidden Markov & two-state model & \code{hier2\_hmm\_fix2} \code{hmm2} & \code{hmm2}; \code{eb-hmm2:MODEL} \\
Simulator baselines & fixed three-component benchmark; per-year Weibull & constants; \code{landmark\_params.csv} & \code{elerath}, \code{elerath-scaled}, \code{april2025}, \code{yearweib:YYYY} \\
Empirical-Bayes hierarchy & the empirical-Bayes model (M2) & \code{final/bb\_model.toml}, \code{frailty\_\allowbreak posterior\_\allowbreak *.csv} & \code{eb-cr2:MODEL[:predictive]}, \code{eb-weibull}, \code{eb-exponential}, \code{eb-glfp}, \code{eb-hmm2} \\
Per-model wear-out & hier3 & \code{hier3/\allowbreak hier3\_\allowbreak drive\_\allowbreak models.csv} & \code{hier3:MODEL} \\
SMART state & observed-state model; EB with SMART & \code{smart/\allowbreak smart\_\allowbreak states\_\allowbreak by\_\allowbreak mfr.csv}; \code{bayes2} smart & \code{smart-obs:MFR}; \code{eb-smart:MODEL} \\
Bayesian, no year effect & B0 & \code{b0} & \code{b0-mean}, \code{b0-draw} \\
Bayesian with shocks & B3 with drift & \code{b3-rwd} & \code{b3-mean}, \code{b3-draw}, \code{b3-yearly}, \code{Context} (\code{raid-sim b3-context}) \\
Other Bayesian variants & B1, B3 with a random walk or no year effect, B4 and B5 & \code{b1-rwd}, \code{b3-rw}, \code{b3-none}, \code{b4-*}, \code{b5-rwd} & none \\
Transition model & fitted transition model (Section~\ref{sec:mstate}) & \code{multistate\_final/\allowbreak ms\_final\_rates.csv} & \code{ms:MODEL[:retire]}; R sampler \code{bb\_ms\_query.R} \\
Boosted forecasters & boosted hazard; landmark ensembles & \code{hier2/gbm\_*}; \code{landmark/gbm} & none (LightGBM models, not lifetime distributions) \\
Expert tracking & fixed-share mixture & \code{track/} & none (count forecasts) \\
\bottomrule
\end{tabular}
\end{table}

For the empirical-Bayes model the functions are:
\begin{itemize}[leftmargin=1.4em]
\item \code{bb\_z}: the multiplier for a known drive model (posterior mean) or for a
  manufacturer, helium status, capacity, and vintage (prior mean);
\item \code{bb\_hazard}, \code{bb\_survival}: $z\,h_0(a)$ and
  $P(T>a_1\mid T>a_0) = \exp\{-z[H_0(a_1)-H_0(a_0)]\}$;
\item \code{bb\_sample}: remaining life of a drive of age $a$. Because the baseline is two
  competing Weibulls and $z$ scales both, the draw is the minimum of two closed-form conditional
  Weibull draws,
  \begin{equation}
  T_j = \eta_j\Bigl[\Bigl(\frac{a}{\eta_j}\Bigr)^{\beta_j} + \frac{E_j}{z}\Bigr]^{1/\beta_j},\quad E_j\sim\mathrm{Exp}(1),\qquad
  R = \min(T_1,T_2) - a;
  \end{equation}
\item \code{bb\_draw\_z}: a draw of the multiplier from its posterior (known model) or prior
  (new model), to carry parameter uncertainty into a simulation.
\end{itemize}

\begin{table}[h]
\centering\small
\caption{Cost of the Query Functions and a Check of the Sampler. Left: the cost of the R query functions on a performance core of a GB10 machine, in nanoseconds per element of a vectorized call over $10^6$ random drive models and ages (five calls per round, each round a fresh process with its own inputs). The two Cortex-X925 cores used (\frag{7 and 17}) differ by \frag{0.9 to 4.0}\% for these calls, more than sessions on one core vary, so each core's cost is the mean of \frag{eight} pilot-bench sessions on \frag{two} machines and the table gives the mean of the two cores; the $\pm$ column is the half-width of its 95\% interval. Right: five-year survival from age $a_0$ at multiplier $z$, from the formula and from $4\times10^5$ sampled lifetimes.}\label{tab:bench}
\begin{tabular}{@{}lrrr@{}}
\toprule
Operation & ns & $\pm$ & Per second \\
\midrule
multiplier lookup & 45.4 & 0.14 & 22 million \\
hazard & 22.7 & 0.035 & 44 million \\
5-year conditional survival & 43.3 & 0.089 & 23 million \\
remaining-life sample & 78.0 & 0.10 & 13 million \\
\bottomrule
\end{tabular}
\qquad
\begin{tabular}{@{}rrrr@{}}
\toprule
$a_0$ (yr) & $z$ & Formula & Sampled \\
\midrule
0.01 & 1 & 0.7382 & 0.7387 \\
3 & 1 & 0.6423 & 0.6424 \\
6 & 0.3 & 0.8447 & 0.8447 \\
1 & 4 & 0.2490 & 0.2485 \\
\bottomrule
\end{tabular}
\end{table}

To represent good and bad years, the simulator can raise or lower failure rates for a group
of drives together. It can add this shared variation in two ways. One is a single $\Gam(\kappa,\kappa)$
multiplier per drive model per year with $\kappa\approx4.2$, which already contains the
fleet-wide year effect. The other is a fleet-wide yearly multiplier with $\kappa\approx13$
together with a drive-model-year multiplier with $\kappa\approx4.7$ (the ``net of fleet year''
row of Table~\ref{tab:cluster}). The two unadjusted rows should not be stacked. These are
descriptive calibrations around the fitted empirical-Bayes model. The Bayesian model estimates
the same structure jointly.

The Bayesian R functions, with prefix \code{bbb\_}, support B1 and B3 with no year effect,
a random walk, or a random walk with drift (\code{none}, \code{rw}, \code{rwd}). The loader
rejects other calendar processes and per-model wear-out shapes, which these samplers do not
implement. For the supported variants a draw is a table lookup followed by closed-form
arithmetic, at one of three levels.
\begin{itemize}[leftmargin=1.4em]
\item \emph{Plug-in.} Parameters and multipliers at their posterior means: the minimum of two
  conditional Weibull draws, as for the empirical-Bayes model.
\item \emph{One draw of everything, held for the whole life.} Pick one of the 1{,}000 posterior
  rows, draw one year effect and one drive-model-year shock, then draw as above. This carries
  parameter uncertainty and over-dispersion into a drive's marginal lifetime. It is not the
  fitted model, in which the multipliers change every calendar year, and it gives a different
  lifetime distribution.
\item \emph{Shocks by year.} As the second, with the year effect stepping and the
  drive-model-year shock redrawn for each year of the drive's life. This is the fitted model.
  The shocks multiply both competing Weibulls, so each year is still the minimum of two
  closed-form draws. The drive retains two random failure thresholds while its cumulative
  risk increases from year to year; these are the exponential budgets used by the algorithm. The cost
  is per simulated year. The shared year effect and shocks are drawn once per system and year.
  A simulator that caches them computes only the two Weibull steps per drive and year.
\end{itemize}
Drives of the same model share its multiplier $Z_g$. All drives in the same calendar year
share the fleet-wide multiplier $Q_y$, while drives of the same model in that year also
share $U_{gy}$. To reproduce clustering, a simulator must draw these at their shared level and
reuse them for every drive in the group. \code{bbb\_sample} shares them only within one call
(\code{shared = TRUE}). A \emph{context} stores one scenario for the whole simulated system,
so successive calls for different drives and their replacements reuse the same shared risks.
\code{bbb\_context}
draws four things:
\begin{itemize}[leftmargin=1.4em]
\item one posterior row;
\item the multiplier of every drive model the system will contain (a prior draw, from
  manufacturer, helium status, capacity and vintage, for a model not in the table);
\item the calendar path of the year effect from a stated first simulated year for a stated
  number of years;
\item the shock of every listed drive model in every year of that path.
\end{itemize}
\code{bbb\_sample\_ctx} then draws lifetimes for any number of drives of a model against that
context, installed in any calendar year of the path at any age. Drives of one model installed
in different years, and their replacements, thus see the same $Z_g$, $Q_y$ and $U_{gy}$. It
returns the remaining life, the calendar year of failure, and an indicator that is false for a
drive still running at the end of the path, whose remaining life is then a censoring time. The
precision of $U_{gy}$ \eqref{eq:kappa} depends on the expected number of failures of that drive
model in that year \emph{in the system being simulated}. A ten-drive array and a
ten-thousand-drive pool have different relative variability. The caller supplies that number per drive model
when the context is made. The checks of the accompanying
fleet example's replacement scheduler are in Appendix~\ref{sec:reproE}.

How long the drift continues is an argument (\code{years\_drift}) in both implementations. Its
default of three years is the longest horizon we examined in Section~\ref{sec:bayesvalid}. It
is a convention, and the data do not determine it (Section~\ref{sec:apc}).

For a drive model that is not in the table, the samplers draw the multiplier from the prior at
its manufacturer, helium status, capacity and year of introduction. With the fitted vintage trend that prior
is too high: in the held-out years its mean over-predicted the first-year failures of new
drive models five- to seven-fold (Sections~\ref{sec:newmodels} and~\ref{sec:calib}).

Table~\ref{tab:bayesquery} checks the R sampler against the survival formula.

\begin{table}[h]
\centering\small
\caption{Five-Year Survival of an ST16000NM001G from Age $a_0$, Formula Against Sampler. Each sampled value uses $4\times10^5$ lifetimes, for a group expected to have 10 failures. In the first four rows both shocks are switched off, so the two should agree. In the last two the formula uses the median path of the shocks and the sampler draws them, so the difference is the effect of the shocks on the average.}\label{tab:bayesquery}
\begin{tabular}{@{}lrrrr@{}}
\toprule
Check & $a_0$ (yr) & Years of drift & Formula & Sampled \\
\midrule
sampler vs formula, shocks off & 0 & 0 & 0.9608 & 0.9610 \\
sampler vs formula, shocks off & 0 & 5 & 0.9741 & 0.9747 \\
sampler vs formula, shocks off & 3 & 0 & 0.9302 & 0.9305 \\
sampler vs formula, shocks off & 3 & 5 & 0.9529 & 0.9528 \\
shocks on their median path vs shocks drawn & 0 & 0 & 0.9608 & 0.9569 \\
shocks on their median path vs shocks drawn & 3 & 0 & 0.9302 & 0.9240 \\
\bottomrule
\end{tabular}
\end{table}

\subsection{How Much Does Lifetime Sampling Cost?}\label{sec:speedsum}

The implementations make it inexpensive to generate lifetimes while allowing drives to
differ and failure risk to change over time. To measure that cost, we timed the lifetime
generation calculations separately from the rest of a storage simulation. The benchmark
and every measurement are in Appendix~\ref{sec:speed} (Table~\ref{tab:benchall}).
We ran a Rust program on the performance cores of \frag{three} GB10
machines and, as a second architecture, an Intel i5. On the GB10 a draw from the empirical-Bayes model or
from B3 at posterior means takes \frag{30 to 31}~ns for a new drive and
\frag{93 to 97}~ns for an aged one, about \frag{10} million
aged-drive lifetimes per second on one core. A posterior draw with one year effect and one
drive-model-year shock takes \frag{147}~ns for an aged drive. Redrawing the
shocks each year takes about \frag{154}~nanoseconds of computation for
each year of a drive's simulated service, or \frag{3.8}~$\mu$s per lifetime with
the drift continued for three years. The per-lifetime figure depends on the 60-year cap and on
the drift assumption (Appendix~\ref{sec:speed}). In the C++ library the three distributions without a closed-form
conditional draw cost 0.8 to 5.3\unskip~$\mu$s by bisection and
30 to 56\unskip~ns from tabulated inverses, against
12 to 128\unskip~ns for the distributions with a closed form.
The extensions (a vault multiplier, the SMART two-state drive
and the hidden Markov model) are timed in Appendix~\ref{sec:extspeed}. Samplers within about \frag{57}\% of each other on one machine can change
places on the other. The large differences hold on both.

The simulation's time limit must not be mistaken for a drive failure. A drive that reaches
the cap is still running; its eventual failure time is unknown. The R marginal sampler, the benchmark program and
\code{bbb\_sample} cap remaining life at 60 years and return the cap value, which the caller
must treat as a censoring time. The documented C++ lifetime interface caps total age at 60
years and returns an infinite value. A shared context ends at its declared calendar horizon,
and \code{bbb\_sample\_ctx} returns an explicit indicator. Means computed from capped
draws, including the last column of Table~\ref{tab:benchall}, are restricted means: they
average only time up to the cap, rather than the full lifetime.

The C++ library also implements the shared context of
\code{bbb\_context}. When the caller gives the system's drive count, the context's shock size
for a drive model is the expected number of failures of that model in the first simulated
year, which for an array of ten drives is a fraction of one. Otherwise the interface's
default is 10. The effect of that choice on the array's loss probability is in Appendix~\ref{sec:raid}.

\subsection{Choosing a Sampler for the Simulation}\label{sec:simrec}

For the Bayesian simulation examples, we use one posterior draw
per simulated system, one path of year effects for the system, and one shock per drive model
and year shared by the drives of that model, with the multipliers advanced every calendar year.
This preserves the shared variation in the fitted model. In the benchmark, generating
lifetimes with new shocks each year takes about \frag{154}~nanoseconds
per drive for each simulated year on the GB10, or
3.3 to 3.8\unskip~$\mu$s per lifetime in the C++ library. Those timings
include drawing shocks separately for each drive. The shared-context implementation reuses
them across drives and was not timed separately. A vault
multiplier adds at most \frag{0.6}~ns. Holding the first year's multipliers for a drive's whole life costs
\frag{74 to 147}~ns per lifetime (Appendix~\ref{sec:speed}) and is a different distribution. It is
adequate only when the simulation is short compared with a year, and should be labelled as
such.

To simulate a fleet that exists, start each drive in its observed SMART state. That needs
a transition model, which we fit in Section~\ref{sec:mstate}. The R sampler
\code{bb\_ms\_query.R} and the C++ sampler \code{ms:MODEL} draw from its fit to all data;
the array example uses the C++ sampler for new drives under continued service. Its fitted
hazards beyond ten years should not be used for lifetime extrapolation. The two-state sampler timed in
Appendix~\ref{sec:extspeed} (\frag{80 to 136}~ns) shows the cost of such a model, not its
predictions. The hidden Markov model costs about the same when sampled by thinning and is not
more accurate.

Sampling need not involve an expensive numerical search. In the Rust benchmark, bisection
on the cumulative hazard costs
\frag{2.4}~$\mu$s for an aged drive where a closed form costs
\frag{103}~ns. Use the closed-form implementation when one is available;
the tabulated alternatives and their checks are described in
Appendices~\ref{sec:speed} and~\ref{sec:extspeed}.

The empirical-Bayes per-model wear-out sampler (\code{hier3:MODEL}) is supplied for
comparison, but its underlying fit did not converge and it should not be used as it stands
(Section~\ref{sec:valid}). The Bayesian per-model wear-out variants B4 and B5 have no
sampler implementation in the library (Table~\ref{tab:crosswalk}).

\section{Practical Implications and Study Limitations}\label{sec:practical}\label{sec:limits}

The results suggest where effort in storage modeling is most likely to pay off: estimate
rates for the drive models actually installed, allow for good and bad years, and use the
current condition of drives already in service. The choice of lifetime distribution still
matters, but its practical importance depends on the service ages and decisions being
studied. We consider replacement planning, identifying vulnerable drives and estimating
data-loss risk in turn, then explain the limits of the available evidence.

\subsection{Choosing a Useful Baseline}\label{sec:practical_baseline}

A useful starting point is a simple lifetime distribution with a failure rate estimated for
the drive model and deployment context, combined with an explicit replacement policy.
Differences between drive models deserve attention before fine distinctions among the
leading lifetime distributions. To compare forecasts, we ask how much probability each
assigns to the failure counts that actually occur. A forecast scores better when those
outcomes are more plausible under its predictions; a confident forecast that misses them
is penalized heavily. We add the natural logarithms of these probabilities over the
age, drive-model and calendar-time groups in nine test years. For this comparison, each
forecast's expected count defines a Poisson distribution. Higher log scores mean the
observed failure counts were better anticipated.

Keeping the two-Weibull lifetime distribution, adding drive-model multipliers and their
predictors (manufacturer, helium status, capacity and vintage) improves that score by
2{,}568.6. Keeping that drive-model treatment, replacing a single Weibull with two competing
Weibulls improves it by 603.4 (Table~\ref{tab:pairs}). The ratio is $2{,}568.6/603.4\approx4.26$,
or about 4.3. The latter is the largest improvement among the six nonconstant distributions
in Appendix~\ref{sec:laws}. For comparison, replacing
a constant-rate distribution with two Weibulls, with the same drive-model treatment, gains
775.6 and gives a ratio of about 3.3 instead. These ratios compare gains in log score;
Table~\ref{tab:pairs} also reports how those gains vary across test years.

The practical lesson is to estimate rates for the drives being simulated before refining
the shape of an otherwise generic lifetime curve. The fixed three-component benchmark's 2.4-fold
over-prediction also shows the cost of using prescribed parameters without checking their
fit to the deployment.

The replacement policy defines the ages over which a lifetime model needs to be useful.
If functioning drives are retired after five to seven years, differences between lifetime
distributions beyond that age have no direct effect on failures during service under an
enforced age limit. Predictions within the service window still matter, as do failures
among replacement drives and any dependence between them. Five to seven years is an
illustrative operating policy here. Under such a policy, a constant-rate model with a
drive-specific rate is a useful baseline to test against the age-dependent alternatives;
this study has not evaluated whether it is sufficient for that service window.

An age-dependent model becomes useful when its predictions within the service window
change the expected replacement workload or the comparison of policies. A simulator should
therefore compare models under the same retirement, repair, and replacement rules. Agreement
under those rules is useful evidence that a result is insensitive to the lifetime
distribution; disagreement identifies an assumption that needs further investigation.

\subsection{Planning for Average Failures and Bad Periods}\label{sec:practical_shared}

The expected number of failures is useful for planning replacement stock and maintenance
work. Redundancy and repair also depend on how much failures vary around that expectation
and whether elevated risks are shared. Several failures during a repair window can defeat
protection that handles isolated failures well. Predicting the mean accurately is therefore
only one requirement for estimating data-loss risk.

The interval comparison makes this distinction visible. On the common population of
169 combinations of drive model and year with substantial observation, selected using
earlier records, forecast ranges intended to contain 95\% of failure counts contain only
44\% without shared annual variation, compared with 92\% when that variation and changes
in the fleet-wide failure rate are included (Table~\ref{tab:calibcommon}). The simpler
model is too confident about annual counts. Allowing shared variation provides a more
credible account of unusually good or bad years. These effects can be added to a simple
lifetime distribution; they do not require a particular age-curve family.

For simulation, a bad year for one drive model must affect its drives together. Drawing a
separate annual shock for every drive would average away variation that a common shock
preserves. Here, a shock means a shared change in failure risk, rather than a requirement
that the drives fail at the same instant. The
context interface supplies one plausible set of fitted parameters, a path of fleet-wide
annual rates, and an annual risk multiplier shared by drives of the same model within a
simulated system (Section~\ref{sec:query}). The matched experiment in
Appendix~\ref{sec:raid} shows that this choice can substantially change simulated data-loss
probabilities. It is a sensitivity result: the fleet records do not establish the appropriate
shock scale for a particular array or dependence within its repair window. The annual
coverage improvement supports modeling shared variation, but does not by itself validate
an array-level data-loss estimate.

\subsection{Using SMART Information in an Existing Fleet}\label{sec:practical_smart}

A drive already in service has an observed condition as well as an age and model.
For the January cohorts, a SMART warning raises one-year AUC from 0.64 to 0.73
(Section~\ref{sec:landmark}). This improvement means that the forecast is better at
distinguishing drives that subsequently fail from those that do not. It makes current
condition useful when deciding which drives warrant closer attention, although the study
does not evaluate a maintenance rule based on that ranking.

Identifying high-risk drives and estimating replacement demand are distinctly different
tasks. The boosted forecasters using monthly SMART records assign better probabilities to
individual outcomes, as measured by log loss, yet predict about 19\% too many failures
at one year and 22\% too many at 90 days (Section~\ref{sec:boost}). Using those totals
directly for procurement or staffing would inherit the excess. A deployment should check
ranking, whether predicted probabilities match observed failure frequencies, and total
predicted failures against the quantity it actually needs. For example, a ranking can guide
which drives receive attention, while the summed probabilities determine expected stock
requirements from failures.

Simulation adds another requirement: the condition of a surviving drive changes over time.
A warning known at the forecast date can inform a fixed-horizon prediction, but a multi-year
simulation needs a process for entering the warned state, failing, and retiring. The fitted
transition model of Section~\ref{sec:mstate} supplies such a process. It also distinguishes
failure from retirement, which is necessary when the operational outcome is failure before
removal. Future deployments require a separate deployment schedule; forecasts for today's
fleet do not supply one.

\subsection{What Better Predictions Mean for Decisions}\label{sec:practical_decisions}

The comparisons give reasons to improve models in specific ways: use drive-specific rates,
represent uncertainty and shared variation, and use observed drive condition when it is
available. They also show where further complexity has uncertain value. Several leading
lifetime distributions have similar predictive scores, and some more flexible models
improve those scores while substantially over-predicting totals. The Bayesian variants with
wear-out that varies by drive model make this concrete: the sums of their median fleet
forecasts are 1.50 to 1.73 times the observed failures, compared with 0.96 for the model with
a common wear-out curve, a changing fleet-wide rate and shared annual variation (B3 with drift).
Their mean absolute errors over the nine periods are 1{,}377 to 2{,}023 failures, compared
with 564 for B3 (Table~\ref{tab:pmsmedians}). These medians provide a finite comparison even
where a predictive mean does not exist. A model can assign more probability to the observed
counts while giving a less useful central forecast of replacement demand. Model choice
should follow the quantity being forecast and the decision it informs.

The supplied implementations let these alternatives be tested in a common simulation.
Allowing failure risk to change from year to year is computationally inexpensive: in our
benchmark, generating a drive's lifetime takes about \frag{154}~nanoseconds
of computation for each year of simulated service (Section~\ref{sec:speedsum}). This measures
the lifetime generation step; a complete simulator also performs other work. A useful
study can vary the lifetime
distribution, shared-risk assumptions, and available drive-state information while holding
the system and its policies fixed.

The RAID experiments show large differences in estimated loss probabilities under their
chosen policies. They replace failed drives and allow functioning drives to remain in
service. For an operator who retires drives after five to seven years, the next useful
comparison is to apply that same retirement rule across the candidate models. Comparing
replacement, maintenance or redundancy decisions also requires the same deployment and
repair assumptions, explicit action costs and consequences, and an evaluation of the
outcomes each policy produces. That connects the forecasting improvements demonstrated
here to the operational benefit the operator needs to assess.

\subsection{Limitations of the Data}\label{sec:limits_data}

The records describe one operator's drives, purchasing decisions, environment, workload,
and replacement practices. A recorded failure includes removal because a drive stopped
working or was judged likely to fail (Section~\ref{sec:reconcile}). It is an operational
endpoint, rather than a measurement of physical breakdown alone. Other removals are also
informative: drive models that performed badly could be retired early, changing which
drives remain available to reveal late-life behavior (Section~\ref{sec:retire}). A fitted
rate therefore describes a drive model in this deployment, not an intrinsic property that
can be transferred unchanged to another operator.

Some distinctions must be inferred from the records. We identify bulk removals by a
threshold on same-day removals of one drive model, and assign drive classes from model
names and capacity because the record does not give each drive's role. Reconciling spacing
differences and equivalent part labels leaves 99 drive models. Those labels can still conceal
different observed cohorts. Helium status is inferred from sampled SMART-attribute presence
and model-name rules. Helium drives also tend to have greater capacity, making the two
associations difficult to separate
(Section~\ref{sec:population}). Power-on age depends
on the interpretation of SMART attribute~9. The analyses consistently exclude rows after
the final failure flag, but the reason some drives reappear is unknown. The sensitivity
checks limit the impact of several recording choices on the reported results; they do not
recover missing reasons for removal or unobserved service history. Detailed qualifications,
exposure conventions, and provenance checks are retained in
Appendix~\ref{sec:qualifications} and Appendix~\ref{sec:repro}.

\subsection{Limitations of the Models and Long-Term Predictions}\label{sec:limits_models}

The models describe the observed range much better than they determine the far tail of a
drive's lifetime. Only 2\% of observed operating time is beyond eight years of age, and we
check predicted failure counts no further than three years ahead. Drives age as calendar time
advances, while newer hardware enters the fleet; these closely linked changes are difficult
to separate. Different fitted accounts of aging and calendar time differ by up to 18
percentage points in fifteen-year survival probability, and assumptions about future
calendar trends by up to 27 points
(Section~\ref{sec:apc}). A continuing decline in the calendar effect is consequently a
scenario to examine, not an established long-term improvement in reliability. Even within
the observed range, the first three days and ages eight to ten show departures from the
fitted age curve (Table~\ref{tab:age}).

Predictions for newly introduced drive models are particularly uncertain. The
empirical-Bayes forecast made before any model-specific failure history predicted 769
failures where 126 occurred, and the Bayesian variants
between 676 and 874; much of the excess comes from two drive models in one year
(Section~\ref{sec:newmodels}). Nor does greater flexibility ensure better forecast levels:
the Bayesian per-model wear-out variants have summed median fleet forecasts 1.50 to 1.73
times the observed failures, despite higher predictive scores (Section~\ref{sec:pmsmedians}).
For B4, the predicted distribution of fleet failure counts has such a heavy upper tail that
its mean is infinite in all nine test periods. This is a property of the statistical model,
not a physical prediction of infinitely many failures. Finite averages of simulated
intensities cannot estimate that mean.
The argument does not apply to B5, whose count mean has not been established to be finite
(Section~\ref{sec:newshapemean}). The empirical-Bayes per-model wear-out fit did not converge.
B5 uses each drive model's own wear-out exponent only up to its oldest age in training.
The reported scores now keep that cutoff fixed when evaluating the model, correcting the
earlier calculation. Some fits still fail checks that the sampling chains
explore the fitted distribution consistently, and repeated runs could share the same bias
(Section~\ref{sec:pms}). These results limit claims that one modeling
hierarchy or lifetime distribution is uniformly preferable.

The models also make different simplifying assumptions. Empirical Bayes estimates the
parameters shared across drive models once, then treats them as known when forecasting.
The count likelihood averages failure rates within
age bins, and some extensions share an age-curve shape across drive models. The covariate
regression, SMART-state extensions, and transition model use different information and
different assumptions about how condition evolves. Their coefficients describe associations;
they do not identify causal effects of warnings, temperature, or load. Comparisons with a
predictive version of the additive model, a hierarchical limited-failure-population model,
alternative multiplier distributions, and a plateau calendar trend remain open
(Appendix~\ref{sec:qualifications}).

\subsection{Limitations of the Evaluation}\label{sec:limits_evaluation}

Each held-out forecast is fitted using observations before its test period, but the same
nine test years informed model development. No final period remained untouched. In
particular, the preference for B3 with drift followed comparison of ten variants and
inspection of their forecast levels. The improvement from shocks does not depend on that
choice; a claim that drift forecasts the level better than B0 is not established. The
quarterly comparisons reuse years and drives, and their reported standard errors treat
overlapping test periods as independent. The amount of independent evidence is smaller than the
number of forecasts might suggest.

We check whether forecast ranges contain observed counts as often as intended one year
ahead, for five Bayesian variants, over nine test years.
The common-population comparison uses groups selected from training data and reports interval
widths and both tails, but uncertainty is estimated by resampling only nine years. Forecasts of
the fleet in service also depend on how removal is represented: the hazard forecasters
without retirement and the classifiers predicting failure before removal do not target
identical processes. The fitted transition model includes removal explicitly. None of these
prospective cohorts evaluates drives deployed after the forecast date.

The monthly SMART features omit the daily trajectory, and overlapping forecast origins
produce repeated observations of the same drives. Some ensemble diagnostics were designed
after the main results were known. Metadata audits check what was available at the forecast
origin at monthly resolution; they do not establish the source dates within that month.
Failure-flag sensitivity scores were recomputed, but the corresponding training-side
refits have not been run (Appendix~\ref{sec:repro_landmark}). Numerical calculations also
introduce uncertainty. More precise integration changes the B1 and B3 scores little.
The per-model wear-out variants require a separate integration method, whose checks cover
selected parameter draws at two of the nine forecast dates. Some sampling runs still
explore the fitted distribution poorly, and the distributions used together to calculate
the scores overlap weakly, making that calculation difficult (Section~\ref{sec:pms}).
The reported numerical error estimates for the aggregate scores also omit dependence between test
periods. These checks do not establish convergence or remove possible shared bias.
The numerical qualifications are preserved in Appendix~\ref{sec:qualifications}.

\subsection{Limitations of the Simulation Evidence}\label{sec:limits_simulation}

Annual variation in counts for a drive model does not determine dependence among a small
array's drives over hours or days of repair. Model-year shocks are independent between
years, contain no within-year timing structure, and use an arbitrary calendar boundary.
Splitting a group into smaller groups changes the aggregate variation implied by the fitted
rule for shock size. The
vault-style rule is consistent only under a joint construction with a shared common factor,
which neither the fitted marginal gamma model nor the samplers implement
(Appendix~\ref{sec:extvault}). The array-derived shock size in the RAID example is therefore
a sensitivity assumption, rather than a quantity identified by the likelihood.

The illustrative RAID simulation isolates consequences of modeling assumptions under its
specified repair and replacement rules; it does not estimate Backblaze's actual data-loss
probability. The transition model's hazards beyond ten years rest on almost no failures.
Different deployment, retirement, repair, and redundancy policies could change the
importance of the lifetime tail or of shared variation, so these policies should accompany
any reported simulation result.

The timing results measure the calculations that generate drive lifetimes on the stated
processors and compiler. They exclude the work of scheduling events, repairing arrays, and
maintaining the rest of a storage simulation. The Rust benchmark does not implement the shared
context interface, whose cost per drive and year was not separately timed. The boosted
forecasters were measured through a Python binding, including per-call overhead, so the
timings do not establish a cost ratio between those forecasters and the compiled lifetime
samplers. These measurements establish implementation costs under the benchmark conditions;
they do not establish the total cost of using a model in another system.

\section{Conclusions}\label{sec:conclusions}

Storage forecasts benefit most from knowing which drive models are installed, allowing
for good and bad years, and using the condition of drives already in service. The clearest
comparison concerns drive-model differences. We score forecasts by how much probability
they assign to the failure counts that actually occur. Adding drive-model differences and
attributes to a two-Weibull baseline improves this score about 4.3 times as much as
replacing a single Weibull with two competing Weibulls while retaining the same drive-model
treatment. This is the largest improvement from changing the lifetime distribution among
the six nonconstant families compared. For a simulator designer, the priority is therefore
to estimate rates for the drives being deployed before refining the lifetime curve.

Across the fleet, failure risk falls in early life and rises at older ages. The pattern
resembles the familiar bathtub curve, but the operational failure records do not identify
physical mechanisms or determine a universal curve. The fixed bathtub distribution we tested predicts
about 2.4 times as many failures as were recorded. Capturing the general shape is not enough;
the rates and the timing of changes must fit the drives being modeled.

Planning for bad periods matters as well as getting the average right. On a common
population selected using earlier records, forecast ranges intended to contain 95\% of
annual failure counts contain only 44\unskip\% without
shared annual variation and changes in the fleet-wide rate, and
92\unskip\% with them (Table~\ref{tab:calibcommon}).
Treating drives as independent under a fixed rate can therefore give planners too little
allowance for unusually demanding years. Shared changes in risk can be represented even
with a simple lifetime distribution; they need not wait for a more elaborate age curve.

For an existing fleet, SMART records help identify vulnerable drives and improve their
predicted failure probabilities. These gains make drive condition useful when deciding
where to direct attention. Estimating total replacement demand and identifying individual
vulnerable drives remain distinctly different tasks: the full SMART classifier gives the
best per-drive probability forecasts in our comparison, yet predicts about 19\% too many
failures at one year and 22\% at 90 days. A predictor suited to prioritizing inspection may
still need adjustment before its probabilities are added up to plan spare-drive stocks.

The intended service life also changes which uncertainties matter. Forecasts with different
accounts of aging and calendar time give failure counts consistent with the observations
at horizons up to three years, yet their fifteen-year survival probabilities differ by up
to 18 percentage points. Alternative assumptions about future calendar trends widen the
spread to 27 points. Only 2\% of observed operating time is beyond eight years of age, and
retirement further limits what we can learn about late failures. For a fleet that replaces
drives after five to seven years, the priority is performance over that service period.
For a simulation that keeps working drives much longer, the lifetime tail becomes a major
assumption to test (Table~\ref{tab:implied}).

The supplied samplers make these choices practical to explore. In our benchmark, the
calendar-time sampler takes about \frag{154}~nanoseconds of computation
for each year of a drive's simulated service; this measures lifetime calculations rather
than the running time of a complete storage simulation. The implementations can also
share annual risks across drives and represent the arrival of SMART warnings, failure,
and retirement. The RAID example shows why the choice matters: simulated loss probabilities
differ greatly, and a matched comparison attributes most of the difference to annual
risk shared by drives of the same model. The strength of that dependence within a small
array remains an assumption. The practical next step is to compare these models under the
same deployment, replacement, and repair policies, asking whether the differences change
capacity, maintenance, or data-loss decisions.

\clearpage
\bibliographystyle{ssl-tr-style/IEEEtran}
\bibliography{report}

@phdthesis{elerath2007diss,
  author = {Elerath, Jon Garrett},
  title  = {Reliability Model and Assessment of Redundant Arrays of Inexpensive Disks ({RAID}) Incorporating Latent Defects and Non-Homogeneous {P}oisson Process Events},
  school = {University of Maryland, College Park},
  year   = {2007},
  url    = {https://drum.lib.umd.edu/items/3175b1d2-9ea8-4073-8747-bca4ae60df59}
}

@inproceedings{elerath2007dsn,
  author    = {Elerath, Jon G. and Pecht, Michael},
  title     = {Enhanced Reliability Modeling of {RAID} Storage Systems},
  booktitle = {37th Annual IEEE/IFIP International Conference on Dependable Systems and Networks (DSN'07)},
  year      = {2007},
  pages     = {175--184},
  doi       = {10.1109/DSN.2007.41}
}

@article{elerath2009tc,
  author  = {Elerath, Jon G. and Pecht, Michael},
  title   = {A Highly Accurate Method for Assessing Reliability of Redundant Arrays of Inexpensive Disks ({RAID})},
  journal = {IEEE Transactions on Computers},
  year    = {2009},
  volume  = {58},
  number  = {3},
  pages   = {289--299},
  doi     = {10.1109/TC.2008.163}
}

@article{elerath2014tos,
  author  = {Elerath, Jon G. and Schindler, Jiri},
  title   = {Beyond {MTTDL}: A Closed-Form {RAID} 6 Reliability Equation},
  journal = {ACM Transactions on Storage},
  year    = {2014},
  volume  = {10},
  number  = {2},
  pages   = {1--21},
  doi     = {10.1145/2577386}
}

@article{elerath2007queue,
  author  = {Elerath, Jon},
  title   = {Hard Disk Drives: The Good, the Bad and the Ugly!},
  journal = {ACM Queue},
  year    = {2007},
  volume  = {5},
  number  = {6},
  pages   = {28--37},
  doi     = {10.1145/1317394.1317403}
}

@misc{karmakar2015scalable,
  author       = {Karmakar, Prasenjit and Gopinath, K.},
  title        = {Scalable Reliability Modelling of {RAID} Storage Subsystems},
  howpublished = {arXiv:1508.02055},
  year         = {2015},
  url          = {https://arxiv.org/abs/1508.02055}
}

@article{kishani2018human,
  author  = {Kishani, Mostafa and Asadi, Hossein},
  title   = {Modeling Impact of Human Errors on the Data Unavailability and Data Loss of Storage Systems},
  journal = {IEEE Transactions on Reliability},
  year    = {2018},
  volume  = {67},
  number  = {3},
  pages   = {1111--1127},
  doi     = {10.1109/TR.2018.2846675}
}

@article{hall2016tools,
  author  = {Hall, Robert J.},
  title   = {Tools for Predicting the Reliability of Large-Scale Storage Systems},
  journal = {ACM Transactions on Storage},
  year    = {2016},
  volume  = {12},
  number  = {4},
  pages   = {1--30},
  doi     = {10.1145/2911987}
}

@inproceedings{schroeder2007fast,
  author    = {Schroeder, Bianca and Gibson, Garth A.},
  title     = {Disk Failures in the Real World: What Does an {MTTF} of 1,000,000 Hours Mean to You?},
  booktitle = {5th USENIX Conference on File and Storage Technologies (FAST '07)},
  year      = {2007},
  url       = {https://www.usenix.org/conference/fast-07/disk-failures-real-world-what-does-mttf-1000000-hours-mean-you}
}

@inproceedings{pinheiro2007fast,
  author    = {Pinheiro, Eduardo and Weber, Wolf-Dietrich and Barroso, Luiz Andr{\'e}},
  title     = {Failure Trends in a Large Disk Drive Population},
  booktitle = {5th USENIX Conference on File and Storage Technologies (FAST '07)},
  year      = {2007},
  url       = {https://www.usenix.org/conference/fast-07/failure-trends-large-disk-drive-population}
}

@inproceedings{jiang2008fast,
  author    = {Jiang, Weihang and Hu, Chongfeng and Zhou, Yuanyuan and Kanevsky, Arkady},
  title     = {Are Disks the Dominant Contributor for Storage Failures? {A} Comprehensive Study of Storage Subsystem Failure Characteristics},
  booktitle = {6th USENIX Conference on File and Storage Technologies (FAST '08)},
  year      = {2008},
  url       = {https://www.usenix.org/conference/fast-08/are-disks-dominant-contributor-storage-failures-comprehensive-study-storage}
}

@inproceedings{greenan2010mttm,
  author    = {Greenan, Kevin M. and Plank, James S. and Wylie, Jay J.},
  title     = {Mean Time to Meaningless: {MTTDL}, {M}arkov Models, and Storage System Reliability},
  booktitle = {2nd USENIX Workshop on Hot Topics in Storage and File Systems (HotStorage '10)},
  year      = {2010},
  url       = {https://www.usenix.org/conference/hotstorage-10/mean-time-meaningless-mttdl-markov-models-and-storage-system-reliability}
}

@inproceedings{ford2010availability,
  author    = {Ford, Daniel and Labelle, Fran{\c{c}}ois and Popovici, Florentina I. and Stokely, Murray and Truong, Van-Anh and Barroso, Luiz and Grimes, Carrie and Quinlan, Sean},
  title     = {Availability in Globally Distributed Storage Systems},
  booktitle = {9th USENIX Symposium on Operating Systems Design and Implementation (OSDI '10)},
  year      = {2010},
  url       = {https://www.usenix.org/conference/osdi10/availability-globally-distributed-storage-systems}
}

@misc{bb2013howlong,
  author       = {Beach, Brian},
  title        = {How Long Do Disk Drives Last?},
  howpublished = {Backblaze blog, 12 November 2013. Archived copy},
  year         = {2013},
  url          = {https://web.archive.org/web/20141221110805/https://www.backblaze.com/blog/how-long-do-disk-drives-last/}
}

@misc{bb2021bathtub,
  author       = {{Backblaze}},
  title        = {Drive Failure Over Time: The Bathtub Curve Is Leaking},
  howpublished = {Backblaze blog, 26 October 2021},
  year         = {2021},
  url          = {https://www.backblaze.com/blog/drive-failure-over-time-the-bathtub-curve-is-leaking/}
}

@misc{bb2025bathtub,
  author       = {{Backblaze}},
  title        = {Are Hard Drives Getting Better? {L}et's Revisit the Bathtub Curve},
  howpublished = {Backblaze blog, 15 October 2025},
  year         = {2025},
  url          = {https://www.backblaze.com/blog/are-hard-drives-getting-better-lets-revisit-the-bathtub-curve/}
}

@misc{bb2014smart,
  author       = {{Backblaze}},
  title        = {Hard Drive {SMART} Stats},
  howpublished = {Backblaze blog, 12 November 2014},
  year         = {2014},
  url          = {https://www.backblaze.com/blog/hard-drive-smart-stats/}
}

@misc{bb2016smart,
  author       = {{Backblaze}},
  title        = {What {SMART} Stats Tell Us About Hard Drives},
  howpublished = {Backblaze blog, 6 October 2016},
  year         = {2016},
  url          = {https://www.backblaze.com/blog/what-smart-stats-indicate-hard-drive-failures/}
}

@misc{bb2015csi,
  author       = {{Backblaze}},
  title        = {{CSI}: {B}ackblaze -- Dissecting {3TB} Drive Failure},
  howpublished = {Backblaze blog, 15 April 2015},
  year         = {2015},
  url          = {https://www.backblaze.com/blog/3tb-hard-drive-failure/}
}

@misc{bb2018durability,
  author       = {{Backblaze}},
  title        = {Backblaze Durability Calculates at 99.999999999\% -- And Why It Doesn't Matter},
  howpublished = {Backblaze blog, 17 July 2018},
  year         = {2018},
  url          = {https://www.backblaze.com/blog/cloud-storage-durability/}
}

@misc{bb2025q3,
  author       = {{Backblaze}},
  title        = {Backblaze Drive Stats for {Q3} 2025},
  howpublished = {Backblaze blog, 11 November 2025},
  year         = {2025},
  url          = {https://www.backblaze.com/blog/backblaze-drive-stats-for-q3-2025/}
}

@misc{bb2026testdata,
  author       = {{Backblaze}},
  title        = {Hard Drive Reliability and Test Data (Drive Stats dataset page)},
  howpublished = {Web page},
  note         = {Accessed 16 October 2026},
  year         = {2026},
  url          = {https://www.backblaze.com/cloud-storage/resources/hard-drive-test-data}
}

@article{mittman2019hierarchical,
  author  = {Mittman, Eric and Lewis-Beck, Colin and Meeker, William Q.},
  title   = {A Hierarchical Model for Heterogenous Reliability Field Data},
  journal = {Technometrics},
  year    = {2019},
  volume  = {61},
  number  = {3},
  pages   = {354--368},
  doi     = {10.1080/00401706.2018.1518273}
}

@phdthesis{mittman2017diss,
  author = {Mittman, Eric T.},
  title  = {Applications of {B}ayesian Hierarchical Models in Gene Expression and Product Reliability},
  school = {Iowa State University},
  year   = {2018},
  doi    = {10.31274/etd-180810-6046},
  url    = {https://dr.lib.iastate.edu/handle/20.500.12876/30599}
}

@article{lewisbeck2022prediction,
  author  = {Lewis-Beck, Colin and Tian, Qinglong and Meeker, William Q.},
  title   = {Prediction of Future Failures for Heterogeneous Reliability Field Data},
  journal = {Technometrics},
  year    = {2022},
  volume  = {64},
  number  = {1},
  pages   = {125--138},
  doi     = {10.1080/00401706.2021.1921036}
}

@article{chan1999glfp,
  author  = {Chan, Victor and Meeker, William Q.},
  title   = {A Failure-Time Model for Infant-Mortality and Wearout Failure Modes},
  journal = {IEEE Transactions on Reliability},
  year    = {1999},
  volume  = {48},
  number  = {4},
  pages   = {377--387},
  doi     = {10.1109/24.814520}
}

@article{ruan2025subsampling,
  author  = {Ruan, Yixiao and Li, Zan and Li, Zhaohui and Lin, Dennis K. J. and Hu, Qingpei and Yu, Dan},
  title   = {Novel Subsampling Strategies for Heavily Censored Reliability Data},
  journal = {Statistics and Its Interface},
  year    = {2025},
  volume  = {18},
  number  = {4},
  pages   = {511--538},
  doi     = {10.4310/SII.250522054738}
}

@article{oakley2023multistate,
  author  = {Oakley, Jordan L. and Forshaw, Matthew and Philipson, Pete and Wilson, Kevin J.},
  title   = {Examining the Impact of Critical Attributes on Hard Drive Failure Times: Multi-State Models for Left-Truncated and Right-Censored Semi-Competing Risks Data},
  journal = {Applied Stochastic Models in Business and Industry},
  year    = {2024},
  volume  = {40},
  number  = {3},
  pages   = {684--709},
  doi     = {10.1002/asmb.2829}
}

@article{siemroth2026manufacturer,
  author  = {Siemroth, Christoph and Park, Yeomyung},
  title   = {Are There Manufacturer Differences in Hard-Drive Reliability?},
  journal = {IEEE Transactions on Cloud Computing},
  year    = {2026},
  volume  = {14},
  number  = {2},
  pages   = {1015--1024},
  doi     = {10.1109/TCC.2026.3679404}
}

@article{ahmed2024survival,
  author  = {Ahmed, Jishan and Green II, Robert C.},
  title   = {Leveraging Survival Analysis in Cost-Aware Deepnet for Efficient Hard Drive Failure Prediction},
  journal = {Neural Computing and Applications},
  year    = {2025},
  volume  = {37},
  number  = {2},
  pages   = {1089--1104},
  doi     = {10.1007/s00521-024-10479-6}
}

@article{chen2024grey,
  author  = {Chen, Rongxing and Xiao, Xinping},
  title   = {New {W}eibull Log-Logistic Grey Forecasting Model for a Hard Disk Drive Failures},
  journal = {Applied Mathematical Modelling},
  year    = {2024},
  volume  = {131},
  pages   = {669--690},
  doi     = {10.1016/j.apm.2024.04.025}
}

@inproceedings{mohapatra2023eol,
  author    = {Mohapatra, Rohan and Coursey, Austin and Sengupta, Saptarshi},
  title     = {Large-Scale End-of-Life Prediction of Hard Disks in Distributed Datacenters},
  booktitle = {2023 IEEE International Conference on Smart Computing (SMARTCOMP)},
  year      = {2023},
  pages     = {261--266},
  doi       = {10.1109/SMARTCOMP58114.2023.00069}
}

@inproceedings{botezatu2016kdd,
  author    = {Botezatu, Mirela Madalina and Giurgiu, Ioana and Bogojeska, Jasmina and Wiesmann, Dorothea},
  title     = {Predicting Disk Replacement towards Reliable Data Centers},
  booktitle = {Proceedings of the 22nd ACM SIGKDD International Conference on Knowledge Discovery and Data Mining},
  year      = {2016},
  pages     = {39--48},
  doi       = {10.1145/2939672.2939699}
}

@inproceedings{aussel2017icmla,
  author    = {Aussel, Nicolas and Jaulin, Samuel and Gandon, Guillaume and Petetin, Yohan and Fazli, Eriza and Chabridon, Sophie},
  title     = {Predictive Models of Hard Drive Failures Based on Operational Data},
  booktitle = {2017 16th IEEE International Conference on Machine Learning and Applications (ICMLA)},
  year      = {2017},
  pages     = {619--625},
  doi       = {10.1109/ICMLA.2017.00-92}
}

@inproceedings{mahdisoltani2017atc,
  author    = {Mahdisoltani, Farzaneh and Stefanovici, Ioan and Schroeder, Bianca},
  title     = {Proactive Error Prediction to Improve Storage System Reliability},
  booktitle = {2017 USENIX Annual Technical Conference (USENIX ATC '17)},
  year      = {2017},
  url       = {https://www.usenix.org/conference/atc17/technical-sessions/presentation/mahdisoltani}
}

@inproceedings{kadekodi2019heart,
  author    = {Kadekodi, Saurabh and Rashmi, K. V. and Ganger, Gregory R.},
  title     = {Cluster Storage Systems Gotta Have {HeART}: Improving Storage Efficiency by Exploiting Disk-Reliability Heterogeneity},
  booktitle = {17th USENIX Conference on File and Storage Technologies (FAST '19)},
  year      = {2019},
  url       = {https://www.usenix.org/conference/fast19/presentation/kadekodi},
  pages     = {345--358}
}

@inproceedings{kadekodi2020pacemaker,
  author    = {Kadekodi, Saurabh and Maturana, Francisco and Subramanya, Suhas Jayaram and Yang, Juncheng and Rashmi, K. V. and Ganger, Gregory R.},
  title     = {{PACEMAKER}: Avoiding {HeART} Attacks in Storage Clusters with Disk-Adaptive Redundancy},
  booktitle = {14th USENIX Symposium on Operating Systems Design and Implementation (OSDI '20)},
  year      = {2020},
  url       = {https://www.usenix.org/conference/osdi20/presentation/kadekodi},
  pages     = {369--385}
}

@inproceedings{kadekodi2022tiger,
  author    = {Kadekodi, Saurabh and Maturana, Francisco and Athlur, Sanjith and Merchant, Arif and Rashmi, K. V. and Ganger, Gregory R.},
  title     = {Tiger: Disk-Adaptive Redundancy Without Placement Restrictions},
  booktitle = {16th USENIX Symposium on Operating Systems Design and Implementation (OSDI '22)},
  year      = {2022},
  url       = {https://www.usenix.org/conference/osdi22/presentation/kadekodi},
  pages     = {413--429}
}

@article{turnbull1976,
  author  = {Turnbull, Bruce W.},
  title   = {The Empirical Distribution Function with Arbitrarily Grouped, Censored and Truncated Data},
  journal = {Journal of the Royal Statistical Society: Series B (Methodological)},
  year    = {1976},
  volume  = {38},
  number  = {3},
  pages   = {290--295},
  doi     = {10.1111/j.2517-6161.1976.tb01597.x}
}

@book{meeker2022smrd,
  author    = {Meeker, William Q. and Escobar, Luis A. and Pascual, Francis G.},
  title     = {Statistical Methods for Reliability Data},
  edition   = {2},
  publisher = {Wiley},
  year      = {2022}
}

@book{lawless2003,
  author    = {Lawless, Jerald F.},
  title     = {Statistical Models and Methods for Lifetime Data},
  edition   = {2},
  publisher = {Wiley},
  year      = {2003},
  doi       = {10.1002/9781118033005}
}

@article{berger1993polyweibull,
  author  = {Berger, James O. and Sun, Dongchu},
  title   = {Bayesian Analysis for the Poly-{W}eibull Distribution},
  journal = {Journal of the American Statistical Association},
  year    = {1993},
  volume  = {88},
  number  = {424},
  pages   = {1412--1418},
  doi     = {10.1080/01621459.1993.10476426}
}

@article{xie1996additive,
  author  = {Xie, M. and Lai, C. D.},
  title   = {Reliability Analysis Using an Additive {W}eibull Model with Bathtub-Shaped Failure Rate Function},
  journal = {Reliability Engineering \& System Safety},
  year    = {1996},
  volume  = {52},
  number  = {1},
  pages   = {87--93},
  doi     = {10.1016/0951-8320(95)00149-2}
}

@article{vaupel1979,
  author  = {Vaupel, James W. and Manton, Kenneth G. and Stallard, Eric},
  title   = {The Impact of Heterogeneity in Individual Frailty on the Dynamics of Mortality},
  journal = {Demography},
  year    = {1979},
  volume  = {16},
  number  = {3},
  pages   = {439--454},
  doi     = {10.2307/2061224}
}

@article{clayton1978,
  author  = {Clayton, D. G.},
  title   = {A Model for Association in Bivariate Life Tables and Its Application in Epidemiological Studies of Familial Tendency in Chronic Disease Incidence},
  journal = {Biometrika},
  year    = {1978},
  volume  = {65},
  number  = {1},
  pages   = {141--151},
  doi     = {10.1093/biomet/65.1.141}
}

@book{hougaard2000,
  author    = {Hougaard, Philip},
  title     = {Analysis of Multivariate Survival Data},
  publisher = {Springer},
  year      = {2000},
  doi       = {10.1007/978-1-4612-1304-8}
}

@book{duchateau2008,
  author    = {Duchateau, Luc and Janssen, Paul},
  title     = {The Frailty Model},
  publisher = {Springer},
  year      = {2008},
  doi       = {10.1007/978-0-387-72835-3}
}

@book{aalen2008,
  author    = {Aalen, Odd O. and Borgan, {\O}rnulf and Gjessing, H{\aa}kon K.},
  title     = {Survival and Event History Analysis: A Process Point of View},
  publisher = {Springer},
  year      = {2008},
  doi       = {10.1007/978-0-387-68560-1}
}

@article{proschan1963,
  author  = {Proschan, Frank},
  title   = {Theoretical Explanation of Observed Decreasing Failure Rate},
  journal = {Technometrics},
  year    = {1963},
  volume  = {5},
  number  = {3},
  pages   = {375--383},
  doi     = {10.1080/00401706.1963.10490105}
}

@article{holford1980,
  author  = {Holford, Theodore R.},
  title   = {The Analysis of Rates and of Survivorship Using Log-Linear Models},
  journal = {Biometrics},
  year    = {1980},
  volume  = {36},
  number  = {2},
  pages   = {299--305},
  doi     = {10.2307/2529982}
}

@article{laird1981,
  author  = {Laird, Nan and Olivier, Donald},
  title   = {Covariance Analysis of Censored Survival Data Using Log-Linear Analysis Techniques},
  journal = {Journal of the American Statistical Association},
  year    = {1981},
  volume  = {76},
  number  = {374},
  pages   = {231--240},
  doi     = {10.1080/01621459.1981.10477634}
}

@article{keiding1990,
  author  = {Keiding, Niels},
  title   = {Statistical Inference in the {L}exis Diagram},
  journal = {Philosophical Transactions of the Royal Society of London. Series A},
  year    = {1990},
  volume  = {332},
  number  = {1627},
  pages   = {487--509},
  doi     = {10.1098/rsta.1990.0128}
}

@article{carstensen2007,
  author  = {Carstensen, Bendix},
  title   = {Age-Period-Cohort Models for the {L}exis Diagram},
  journal = {Statistics in Medicine},
  year    = {2007},
  volume  = {26},
  number  = {15},
  pages   = {3018--3045},
  doi     = {10.1002/sim.2764}
}

@article{holford1983,
  author  = {Holford, Theodore R.},
  title   = {The Estimation of Age, Period and Cohort Effects for Vital Rates},
  journal = {Biometrics},
  year    = {1983},
  volume  = {39},
  number  = {2},
  pages   = {311--324},
  doi     = {10.2307/2531004}
}

@article{clayton1987apc,
  author  = {Clayton, D. and Schifflers, E.},
  title   = {Models for Temporal Variation in Cancer Rates. {II}: Age-Period-Cohort Models},
  journal = {Statistics in Medicine},
  year    = {1987},
  volume  = {6},
  number  = {4},
  pages   = {469--481},
  doi     = {10.1002/sim.4780060406}
}

@article{morris1983,
  author  = {Morris, Carl N.},
  title   = {Parametric Empirical {B}ayes Inference: Theory and Applications},
  journal = {Journal of the American Statistical Association},
  year    = {1983},
  volume  = {78},
  number  = {381},
  pages   = {47--55},
  doi     = {10.1080/01621459.1983.10477920}
}

@article{gaver1987,
  author  = {Gaver, Donald P. and O'Muircheartaigh, I. G.},
  title   = {Robust Empirical {B}ayes Analyses of Event Rates},
  journal = {Technometrics},
  year    = {1987},
  volume  = {29},
  number  = {1},
  pages   = {1--15},
  doi     = {10.1080/00401706.1987.10488178}
}

@article{gneiting2007,
  author  = {Gneiting, Tilmann and Raftery, Adrian E.},
  title   = {Strictly Proper Scoring Rules, Prediction, and Estimation},
  journal = {Journal of the American Statistical Association},
  year    = {2007},
  volume  = {102},
  number  = {477},
  pages   = {359--378},
  doi     = {10.1198/016214506000001437}
}

@article{czado2009,
  author  = {Czado, Claudia and Gneiting, Tilmann and Held, Leonhard},
  title   = {Predictive Model Assessment for Count Data},
  journal = {Biometrics},
  year    = {2009},
  volume  = {65},
  number  = {4},
  pages   = {1254--1261},
  doi     = {10.1111/j.1541-0420.2009.01191.x}
}

@article{tashman2000,
  author  = {Tashman, Leonard J.},
  title   = {Out-of-Sample Tests of Forecasting Accuracy: An Analysis and Review},
  journal = {International Journal of Forecasting},
  year    = {2000},
  volume  = {16},
  number  = {4},
  pages   = {437--450},
  doi     = {10.1016/S0169-2070(00)00065-0}
}

@article{hoffman2014nuts,
  author  = {Hoffman, Matthew D. and Gelman, Andrew},
  title   = {The {No-U-Turn} Sampler: Adaptively Setting Path Lengths in {H}amiltonian {M}onte {C}arlo},
  journal = {Journal of Machine Learning Research},
  year    = {2014},
  volume  = {15},
  url     = {https://jmlr.org/papers/v15/hoffman14a.html},
  number    = {47},
  pages     = {1593--1623}
}

@misc{phan2019numpyro,
  author       = {Phan, Du and Pradhan, Neeraj and Jankowiak, Martin},
  title        = {Composable Effects for Flexible and Accelerated Probabilistic Programming in {NumPyro}},
  howpublished = {arXiv:1912.11554},
  year         = {2019},
  url          = {https://arxiv.org/abs/1912.11554}
}

@misc{betancourt2017,
  author       = {Betancourt, Michael},
  title        = {A Conceptual Introduction to {H}amiltonian {M}onte {C}arlo},
  howpublished = {arXiv:1701.02434},
  year         = {2017},
  url          = {https://arxiv.org/abs/1701.02434}
}

@article{vehtari2021rhat,
  author  = {Vehtari, Aki and Gelman, Andrew and Simpson, Daniel and Carpenter, Bob and B{\"u}rkner, Paul-Christian},
  title   = {Rank-Normalization, Folding, and Localization: An Improved {$\widehat{R}$} for Assessing Convergence of {MCMC} (with Discussion)},
  journal = {Bayesian Analysis},
  year    = {2021},
  volume  = {16},
  number  = {2},
  pages   = {667--718},
  doi     = {10.1214/20-BA1221}
}

@book{gelman2013bda,
  author    = {Gelman, Andrew and Carlin, John B. and Stern, Hal S. and Dunson, David B. and Vehtari, Aki and Rubin, Donald B.},
  title     = {Bayesian Data Analysis},
  edition   = {3},
  publisher = {Chapman and Hall/CRC},
  year      = {2013},
  doi       = {10.1201/b16018}
}

@article{lewis1979thinning,
  author  = {Lewis, P. A. W. and Shedler, G. S.},
  title   = {Simulation of Nonhomogeneous {P}oisson Processes by Thinning},
  journal = {Naval Research Logistics Quarterly},
  year    = {1979},
  volume  = {26},
  number  = {3},
  pages   = {403--413},
  doi     = {10.1002/nav.3800260304}
}

@book{devroye1986,
  author    = {Devroye, Luc},
  title     = {Non-Uniform Random Variate Generation},
  publisher = {Springer},
  year      = {1986},
  doi       = {10.1007/978-1-4613-8643-8}
}

@article{meng1996bridge,
  author  = {Meng, Xiao-Li and Wong, Wing Hung},
  title   = {Simulating Ratios of Normalizing Constants via a Simple Identity: A Theoretical Exploration},
  journal = {Statistica Sinica},
  year    = {1996},
  volume  = {6},
  number  = {4},
  pages   = {831--860}
}

@incollection{neal2011hmc,
  author    = {Neal, Radford M.},
  title     = {{MCMC} Using {H}amiltonian Dynamics},
  booktitle = {Handbook of Markov Chain Monte Carlo},
  editor    = {Brooks, Steve and Gelman, Andrew and Jones, Galin L. and Meng, Xiao-Li},
  publisher = {Chapman and Hall/CRC},
  year      = {2011},
  pages     = {113--162}
}

@book{neuts1981,
  author    = {Neuts, Marcel F.},
  title     = {Matrix-Geometric Solutions in Stochastic Models: An Algorithmic Approach},
  publisher = {Johns Hopkins University Press},
  address   = {Baltimore},
  year      = {1981}
}

@article{monaco2018frailtysurv,
  author  = {Monaco, John V. and Gorfine, Malka and Hsu, Li},
  title   = {General Semiparametric Shared Frailty Model: Estimation and Simulation with {frailtySurv}},
  journal = {Journal of Statistical Software},
  year    = {2018},
  volume  = {86},
  number  = {4},
  pages   = {1--42},
  doi     = {10.18637/jss.v086.i04}
}

@article{herbster1998tracking,
  author  = {Herbster, Mark and Warmuth, Manfred K.},
  title   = {Tracking the Best Expert},
  journal = {Machine Learning},
  year    = {1998},
  volume  = {32},
  number  = {2},
  pages   = {151--178},
  doi     = {10.1023/A:1007424614876}
}

@article{littlestone1994weighted,
  author  = {Littlestone, Nick and Warmuth, Manfred K.},
  title   = {The Weighted Majority Algorithm},
  journal = {Information and Computation},
  year    = {1994},
  volume  = {108},
  number  = {2},
  pages   = {212--261},
  doi     = {10.1006/inco.1994.1009}
}

@article{freund1997decision,
  author  = {Freund, Yoav and Schapire, Robert E.},
  title   = {A Decision-Theoretic Generalization of On-Line Learning and an Application to Boosting},
  journal = {Journal of Computer and System Sciences},
  year    = {1997},
  volume  = {55},
  number  = {1},
  pages   = {119--139},
  doi     = {10.1006/jcss.1997.1504},
}

@article{friedman2000additive,
  author  = {Friedman, Jerome and Hastie, Trevor and Tibshirani, Robert},
  title   = {Additive Logistic Regression: A Statistical View of Boosting},
  journal = {The Annals of Statistics},
  year    = {2000},
  volume  = {28},
  number  = {2},
  pages   = {337--407},
  doi     = {10.1214/aos/1016218223},
}

@article{friedman2001greedy,
  author  = {Friedman, Jerome H.},
  title   = {Greedy Function Approximation: A Gradient Boosting Machine},
  journal = {The Annals of Statistics},
  year    = {2001},
  volume  = {29},
  number  = {5},
  pages   = {1189--1232},
  doi     = {10.1214/aos/1013203451},
}

@inproceedings{ke2017lightgbm,
  author    = {Ke, Guolin and Meng, Qi and Finley, Thomas and Wang, Taifeng and Chen, Wei and Ma, Weidong and Ye, Qiwei and Liu, Tie-Yan},
  title     = {{LightGBM}: A Highly Efficient Gradient Boosting Decision Tree},
  booktitle = {Advances in Neural Information Processing Systems 30},
  year      = {2017},
  pages     = {3146--3154},
}

@article{vanhouwelingen2007dynamic,
  author  = {van Houwelingen, Hans C.},
  title   = {Dynamic Prediction by Landmarking in Event History Analysis},
  journal = {Scandinavian Journal of Statistics},
  year    = {2007},
  volume  = {34},
  number  = {1},
  pages   = {70--85},
  doi     = {10.1111/j.1467-9469.2006.00529.x},
}

@misc{sandqvist2026landmark,
  author = {Sandqvist, Oliver Lunding},
  title  = {Event History Analysis with Time-Dependent Covariates via Landmarking Supermodels and Boosted Trees},
  year   = {2026},
  note   = {Preprint, arXiv:2601.17605},
  eprint = {2601.17605},
  archivePrefix = {arXiv},
  url    = {https://arxiv.org/abs/2601.17605}
}

@article{arslan2021heavytailed,
  author  = {Arslan, {\c{S}}uayb {\c{S}}efik and Zeydan, Engin},
  title   = {On the Distribution Modeling of Heavy-Tailed Disk Failure Lifetime in Big Data Centers},
  journal = {IEEE Transactions on Reliability},
  year    = {2021},
  volume  = {70},
  number  = {2},
  pages   = {507--524},
  doi     = {10.1109/TR.2020.3007127}
}

@article{pinciroli2023lifespan,
  author  = {Pinciroli, Riccardo and Yang, Lishan and Alter, Jacob and Smirni, Evgenia},
  title   = {Lifespan and Failures of {SSDs} and {HDDs}: Similarities, Differences, and Prediction Models},
  journal = {IEEE Transactions on Dependable and Secure Computing},
  year    = {2023},
  volume  = {20},
  number  = {1},
  pages   = {256--272},
  doi     = {10.1109/TDSC.2021.3131571}
}

@article{amram2021interpretable,
  author  = {Amram, Maxime and Dunn, Jack and Toledano, Jeremy J. and Zhuo, Ying Daisy},
  title   = {Interpretable Predictive Maintenance for Hard Drives},
  journal = {Machine Learning with Applications},
  year    = {2021},
  volume  = {5},
  pages   = {100042},
  doi     = {10.1016/j.mlwa.2021.100042}
}

@inproceedings{gonthier2025drex,
  author    = {Gonthier, Maxime and Sanchez-Gallegos, Dante D. and Pan, Haochen and Nicolae, Bogdan and Zhou, Sicheng and Nguyen, Hai Duc and Hayot-Sasson, Valerie and Pauloski, J. Gregory and Carretero, Jesus and Chard, Kyle and Foster, Ian},
  title     = {{D-Rex}: Heterogeneity-Aware Reliability Framework and Adaptive Algorithms for Distributed Storage},
  booktitle = {Proceedings of the 39th ACM International Conference on Supercomputing (ICS '25)},
  year      = {2025},
  pages     = {853--867},
  doi       = {10.1145/3721145.3730412}
}

@article{christoffersen2021dynamichazard,
  author  = {Christoffersen, Benjamin},
  title   = {dynamichazard: Dynamic Hazard Models Using State Space Models},
  journal = {Journal of Statistical Software},
  year    = {2021},
  volume  = {99},
  number  = {7},
  doi     = {10.18637/jss.v099.i07}
}

@article{hu2024sequential,
  author  = {Hu, Jun and So, Hon Yiu and Zhuang, Yan},
  title   = {Estimation of the Mean Exponential Survival Time under a Sequential Censoring Scheme},
  journal = {Journal of Applied Statistics},
  year    = {2025},
  volume  = {52},
  number  = {3},
  pages   = {669--686},
  doi     = {10.1080/02664763.2024.2386609}
}

@inproceedings{yang2025censored,
  author    = {Yang, Yuqi and Wang, Chengjie and Hu, Qingpei},
  title     = {Highly Censored Reliability Big Data Analysis: Hard Disk Case Revisited},
  booktitle = {2025 IEEE International Conference on Industrial Engineering and Engineering Management (IEEM)},
  year      = {2025},
  pages     = {430--434},
  doi       = {10.1109/ieem63636.2025.11357750}
}

@article{nasyrov2023distribution,
  author  = {Nasyrov, I. N. and Nasyrov, I. I. and Nasyrov, R. I.},
  title   = {Distribution of Failed Data Storage Devices from Operation Time in Data Centers},
  journal = {Vestnik NSUEM},
  year    = {2023},
  number  = {3},
  pages   = {132--143},
  note    = {In Russian},
  doi     = {10.34020/2073-6495-2023-3-132-143}
}

@article{nasyrov2022bigdata,
  author  = {Nasyrov, I. N. and Nasyrov, I. I. and Nasyrov, R. I.},
  title   = {Big Data on Storage Devices Reliability in Data Centers},
  journal = {Tsifrovaya Ekonomika (Digital Economy)},
  year    = {2022},
  number  = {2(18)},
  note    = {In Russian},
  doi     = {10.34706/de-2022-02-04}
}

@article{vasilev2024stratified,
  author  = {Vasilev, I. A. and Filimonova, I. O. and Petrovskiy, M. I. and Mashechkin, I. V.},
  title   = {Stratified Statistical Models in Hardware Reliability Analysis},
  journal = {Doklady Mathematics},
  year    = {2024},
  volume  = {110},
  number  = {S1},
  pages   = {S103--S109},
  doi     = {10.1134/S1064562424601963}
}

@inproceedings{vasilev2025censoring,
  author    = {Vasilev, Yu. A.},
  title     = {Vliyanie informativnosti tsenzurirovaniya v zadachakh analiza degradatsii diskovykh nakopitelei [The Influence of Informative Censoring in Disk-Drive Degradation Analysis]},
  booktitle = {Programmirovanie i vychislitel'naya matematika: sbornik materialov konferentsii, posvyashchennoi 100-letiyu so dnya rozhdeniya N. P. Trifonova},
  publisher = {Moscow University Press},
  address   = {Moscow},
  year      = {2025},
  pages     = {37--40},
  note      = {In Russian},
  doi       = {10.55959/MSU012316-6-2025-37-40}
}

@misc{dasilva2026cindex,
  author        = {da Silva, Rafael and Alvares, Danilo},
  title         = {The {C}-Index Illusion: Discrimination Without Calibration in Published Survival Models},
  howpublished  = {arXiv:2607.19526 (v3, 11 September 2026)},
  year          = {2026},
  doi           = {10.48550/arXiv.2607.19526}
}

@inproceedings{frank2025uncertain,
  author    = {Frank, Reginald and Lomeli, Octavio and Giridharan, Neil and Ponnapalli, Soujanya and Aguilera, Marcos K. and Crooks, Natacha},
  title     = {Real Life Is Uncertain. {C}onsensus Should Be Too!},
  booktitle = {Proceedings of the Workshop on Hot Topics in Operating Systems (HotOS '25)},
  year      = {2025},
  pages     = {80--87},
  doi       = {10.1145/3713082.3730374}
}

@misc{wang2026slot,
  author        = {Wang, Yi},
  title         = {Slot-Adjacent Drive Failures Cluster Within Servers: Code, Derived Data and Results},
  howpublished  = {Zenodo, version 1.0.0, 19 October 2026},
  year          = {2026},
  doi           = {10.5281/zenodo.23478229}
}

@article{pecht2020circuitworld,
  author  = {Pecht, Michael and Elburn, Edmond},
  title   = {Commercial Hard Drive Failures in a Data Center Application and the Role of {SMART} Attribute Information},
  journal = {Circuit World},
  year    = {2021},
  volume  = {47},
  number  = {4},
  pages   = {408--426},
  doi     = {10.1108/CW-07-2020-0127}
}

@misc{seagate2024openseachest,
  author = {{Seagate Technology}},
  title = {Drive Health and {SMART}, {openSeaChest} wiki},
  howpublished = {\url{https://github.com/Seagate/openSeaChest/wiki/Drive-Health-and-SMART}},
  note = {Accessed 2026-10-31}
}

\clearpage
\appendix
\section*{Supplementary Material}
\addcontentsline{toc}{section}{Supplementary Material}
The supplement holds the catalogue of other lifetime distributions with the per-model wear-out extension and the fleet-year
fits, the profiles of individual drive models,
the location and secondary-class results, the model extensions, the covariate regression,
the fixed-share tracking of the best forecaster, the statistical background, the boosted Lexis-cell hazard
and the diagnostics of the boosted ensemble that forecasts without the SMART record, the
benchmarks of sampling and prediction cost, the worked RAID example that runs every model through one simulation, and notes on
provenance and on corrections to earlier rounds of the analysis, together with detailed
qualifications of the data, models, evaluation, and benchmarks.
\section{Other Lifetime Distributions, Per-Model Wear-Out and Fleet-Year Fits}\label{sec:laws}

We give the fits that the main text summarizes: three further lifetime distributions, a
two-state hidden Markov model, a per-model wear-out extension of the hierarchy, fits to each
calendar year of the fleet (a single Weibull, and the best distribution of each year), and the comparison of distributions on quarterly test blocks.

The hidden
Markov profile limitations and earlier additional-start checks are identified below and should not be
combined numerically with the current tables.

\subsection{Limited Failure Population, Gompertz Wear-Out, and Three Weibulls}\label{sec:laws3}

Three further distributions are fitted alone and with the gamma drive-model effect and its covariates
(Section~\ref{sec:hier}), and scored on the same nine held-out years (Section~\ref{sec:valid})
and on the 34 quarterly folds.

\emph{General limited failure population (GLFP).} Chan and Meeker's distribution, fitted to Backblaze
data in hierarchical form by Mittman et al.~\cite{mittman2019hierarchical}: a fraction $\pi$ of
units carries a defect that fails by a Weibull $F_1$, and every unit wears out by a Weibull
$F_2$, so
\begin{equation}
1-F(t)=\bigl[1-\pi F_1(t)\bigr]\bigl[1-F_2(t)\bigr],\qquad
H(t)=-\log\bigl[1-\pi F_1(t)\bigr]+\Bigl(\frac{t}{\eta_2}\Bigr)^{\beta_2}.
\end{equation}
It has five parameters. Unlike two competing Weibulls, its early hazard integrates to a
finite total, $-\log(1-\pi)$.

\emph{Weibull and Gompertz.} The early Weibull term with a wear-out hazard that rises
exponentially, $H(t)=(t/\eta_1)^{\beta_1}+(b/c)(e^{ct}-1)$; four parameters.

\emph{Three competing Weibulls.} The form of the bathtub of Section~\ref{sec:elerathlit} with all six parameters
free and the exponents ordered, started from the bathtub's own constants and from three other
points.

For every distribution in this subsection the drive-model effect multiplies the cumulative hazard; Mittman
et al.\ instead let the parameters of the distribution vary by drive model, which was not done here for
these distributions.

Among the six families with the drive-model effect and covariates, the largest difference
in yearly point scores is 603 log units (two competing Weibulls over one Weibull;
Table~\ref{tab:laws}), against 2{,}569 for adding the drive-model effect and covariates to two competing
Weibulls (Table~\ref{tab:pairs}). This is a ratio of 4.26 between descriptive score gains,
not a statistical test of their difference.

\begin{table}[h]
\centering
\caption{Regenerated Fits and Scores for Six Distributions on Power-On Age, Each Alone and with the Drive-Model Effect. In-sample columns give the log-likelihood on all data. Held-out yearly columns give the log score summed over nine yearly folds: alone, and with the drive-model effect as a point forecast and as a predictive distribution. The last column is the predictive score over 34 quarterly folds. The rows for the Weibull, the hidden Markov model and two competing Weibulls repeat Table~\ref{tab:hmm}, and the two Weibull rows also appear in Table~\ref{tab:rolling}.}\label{tab:laws}
\footnotesize\setlength{\tabcolsep}{4pt}\begin{tabular}{@{}lrrrrrrr@{}}
\toprule
& & \multicolumn{2}{c}{In-sample} & \multicolumn{3}{c}{Held out, yearly} & Quarterly \\
Distribution & Par. & alone & with model & alone & point & predictive & predictive \\
\midrule
Weibull & 2 & -351{,}765 & -343{,}027 & -34{,}680 & -32{,}095 & -29{,}974 & -29{,}848 \\
Two-state hidden Markov & 3 & -351{,}357 & -342{,}570 & -34{,}528 & -31{,}681 & -29{,}696 & -29{,}500 \\
Two competing Weibulls & 4 & -350{,}901 & -342{,}048 & -34{,}060 & -31{,}491 & -29{,}446 & -29{,}247 \\
Weibull and Gompertz & 4 & -351{,}099 & -342{,}269 & -34{,}157 & -31{,}591 & -29{,}585 & -29{,}390 \\
Limited failure population (GLFP) & 5 & -350{,}896 & -342{,}046 & -34{,}084 & -31{,}571 & -29{,}473 & -29{,}262 \\
Three competing Weibulls & 6 & -350{,}901 & -342{,}048 & -34{,}060 & -31{,}507 & -29{,}448 & --- \\
\bottomrule
\end{tabular}
\end{table}

\begin{table}[h]
\centering
\caption{Paired Comparisons of the Distributions over the Nine Test Years. Columns are as in Table~\ref{tab:pairs}.}\label{tab:lawspairs}
\footnotesize\setlength{\tabcolsep}{4pt}\begin{tabular}{@{}p{9.6cm}rrrr@{}}
\toprule
Comparison & Sum & Years won & Std.\ error & Ratio \\
\midrule
Limited failure population over two Weibulls, predictive & $-27$ & 3 of 9 & 24 & $-1.1$ \\
the same, point forecast & $-80$ & 1 of 9 & 33 & $-2.4$ \\
the same, without the drive-model effect & $-24$ & 3 of 9 & 39 & $-0.6$ \\
Weibull and Gompertz over two Weibulls, predictive & $-139$ & 2 of 9 & 108 & $-1.3$ \\
the same, point forecast & $-100$ & 3 of 9 & 149 & $-0.7$ \\
the same, without the drive-model effect & $-97$ & 4 of 9 & 89 & $-1.1$ \\
Three Weibulls over two, without the drive-model effect & $+0.005$ & 3 of 9 & 0.004 & $+1.2$ \\
Three Weibulls, fitted, over the bathtub's constants & $+11{,}370$ & 8 of 9 & 2{,}812 & $+4.0$ \\
\bottomrule
\end{tabular}
\end{table}

\begin{table}[h]
\centering
\caption{Fitted Values of the Limited-Failure-Population and Gompertz Distributions. Both include the drive-model effect and covariates and are fitted to all data.}\label{tab:lawsparams}
\footnotesize\begin{tabular}{@{}lp{11.5cm}@{}}
\toprule
Distribution & Parameters \\
\midrule
Limited failure population & $\pi=0.0329$, early mode $\beta_1=0.366$, $\eta_1=0.818$ yr; wear-out $\beta_2=1.529$, $\eta_2=13.92$ yr; $\phi=0.761$ \\
Weibull and Gompertz & early $\beta_1=0.140$; wear-out hazard $0.0276\,e^{0.146\,a}$ per year; $\phi=0.735$ \\
\bottomrule
\end{tabular}
\end{table}

The limited-failure-population model fits the data 2.3 log-likelihood units better than two
competing Weibulls with the drive-model effect, for one parameter more, and is 27 behind on
the held-out years (standard error 24) as a predictive distribution and 80 behind as a point
forecast. As predictive distributions the two are not distinguished. As point forecasts the
limited-failure-population model loses eight years of nine, at a ratio of $-2.4$, past the
threshold of 2.3 used in Section~\ref{sec:valid}; the margin is small and favours two competing
Weibulls.

At a drive-model multiplier of one the distribution's susceptible
fraction is $\pi=3.3\%$, with an early-failure scale of 0.82 years. The multiplier acts on
the cumulative hazard, so the fraction for drive model $g$ is $1-(1-\pi)^{z_g}$.
Using each model's posterior mean multiplier, then averaging these fractions over
drives, gives 1.24\%; the drive-weighted mean multiplier is 0.38. Integrating that
nonlinear fraction over each gamma posterior also gives 1.24\% to this precision.
That fraction is a parameter of the distribution, the weight of its early mode, not a
count of physically defective drives: the data identify no drive as susceptible. The
wear-out Weibull has exponent 1.53 and scale 13.9 years, against 1.57 and 14.0 for the two
competing Weibulls.

The Gompertz wear-out term is worse in sample by 221 and behind on held-out years by 139
(standard error 108). Its hazard grows by 15.7\% per year of age. The data prefer a wear-out
hazard that rises as a power of age to one that rises exponentially, but do not exclude the
latter.

On all data, three competing Weibulls reach the same likelihood as two to the first
decimal. The third term is negligible over the observed ages: its exponent is at least
8 and its scale lies beyond the data. Without the drive-model effect the held-out
scores differ by only 0.005. With covariates, the three-Weibull model is 16 point-score
units and 1.94 predictive units behind two Weibulls (predictive standard error 2.36).
Thus the all-data fit effectively reduces to two competing Weibulls; the separately
refitted forecasting sequences are close, not identical. Without the drive-model
effect, freeing the bathtub's parameters improves its score by 11{,}370 log units over
the default constants (eight years of nine).

\subsection{A Two-State Hidden Markov Model}\label{sec:hmm}

A two-state hidden Markov model, a Coxian phase-type model (Appendix~\ref{sec:primer},
Appendix~\ref{sec:primerhmm}), has one parameter fewer than two competing Weibulls: a drive starts in a first state with a constant
failure rate $\lambda_1$, moves to a second state at rate $q$, and there fails at a constant
rate $\lambda_2$. The state is not observed. The lifetime then has
\begin{equation}
S(t) = e^{-(\lambda_1+q)t} + \frac{q}{\lambda_1+q-\lambda_2}\Bigl(e^{-\lambda_2 t}-e^{-(\lambda_1+q)t}\Bigr),
\end{equation}
which we evaluate as $S(t)=e^{-\lambda_2t}[e^{-x}+qt\,g(x)]$ with $x=(\lambda_1+q-\lambda_2)t$ and
$g(x)=(1-e^{-x})/x$, a form without a singularity at $\lambda_2=\lambda_1+q$. Unlike a
Weibull, its hazard is bounded: it moves from $\lambda_1$ towards $\min(\lambda_2,\lambda_1+q)$.

\begin{samepage}
The positive rates are not globally identifiable from lifetimes alone. When
$\lambda_2>\lambda_1$, the transformation
\begin{equation}\label{eq:hmmsymmetry}
 (\lambda_1,q,\lambda_2)\longmapsto
 (\lambda_1,\lambda_2-\lambda_1,\lambda_1+q)
\end{equation}
leaves the lifetime distribution unchanged.
\end{samepage}

To see this, put $a=\lambda_1+q$
and $b=\lambda_2$. The Laplace transform of the lifetime density is
\[
 \mathcal L_f(s)=\frac{\lambda_1}{s+a}
   +\frac{q}{s+a}\frac{b}{s+b}
   =\frac{\lambda_1s+ab}{(s+a)(s+b)},
\]
which is symmetric in $a$ and $b$. The transformation exchanges them and is its
own inverse; the two rate triples differ unless $a=b$. For example,
$(0.01,0.05,0.09)$ and $(0.01,0.08,0.06)$ per year give exactly the same distribution.
When $\lambda_2=\lambda_1$, the distribution is exponential at rate $\lambda_1$ for every
positive $q$. Equality of the survival functions also preserves cumulative-hazard
multipliers and common scaling of all three rates. These are likelihood
equivalences; a prior on the rates need not preserve the symmetry.

It is fitted here in two ways: with the gamma drive-model effect on the cumulative hazard, as
for the other distributions, and with constants of its own for each drive model that scale its rates.
We fit it in hierarchical form inside the Bayesian model in Appendix~\ref{sec:exthmm}.

\subsubsection{With the drive-model effect}

The model goes through the procedure of Sections~\ref{sec:hier} and~\ref{sec:valid} unchanged:
the same interval likelihood, alone and with the gamma drive-model effect and its covariates,
and the same nine held-out years (Tables~\ref{tab:hmm} and~\ref{tab:hmmpairs}). With the effect
on the cumulative hazard the result is no longer a two-state chain with scaled rates;
the rates themselves vary in Appendix~\ref{sec:hmmper}.

\begin{table}[h]
\centering
\caption{Three Baselines on Power-On Age. In-sample columns give the log-likelihood on all data without and with the drive-model effect. Held-out columns give log scores summed over nine years: a point forecast without the drive-model effect, and a point forecast and a predictive distribution with it. Pred./obs.\ is for the point forecast with the drive-model effect.}\label{tab:hmm}
\footnotesize\setlength{\tabcolsep}{4pt}\begin{tabular}{@{}lrrrrrrr@{}}
\toprule
& & \multicolumn{2}{c}{In-sample log-likelihood} & \multicolumn{3}{c}{Held-out log score} & \\
Baseline & Parameters & alone & with drive model & alone & with, point & with, predictive & Pred./obs. \\
\midrule
Weibull & 2 & $-351{,}765.4$ & $-343{,}027.1$ & $-34{,}680$ & $-32{,}095$ & $-29{,}974$ & 0.90 \\
Two-state hidden Markov & 3 & $-351{,}356.8$ & $-342{,}569.7$ & $-34{,}528$ & $-31{,}681$ & $-29{,}696$ & 0.92 \\
Two competing Weibulls & 4 & $-350{,}900.6$ & $-342{,}048.0$ & $-34{,}060$ & $-31{,}491$ & $-29{,}446$ & 0.99 \\
\bottomrule
\end{tabular}
\end{table}

\begin{table}[h]
\centering
\caption{Paired Comparisons for the Hidden Markov Baseline over the Nine Test Years. Columns are as in Table~\ref{tab:pairs}.}\label{tab:hmmpairs}
\footnotesize\setlength{\tabcolsep}{4pt}\begin{tabular}{@{}p{10.2cm}rrrr@{}}
\toprule
Comparison & Sum & Years won & Std.\ error & Ratio \\
\midrule
hidden Markov against Weibull, no drive-model effect & $+152$ & 7/9 & 215 & 0.7 \\
hidden Markov against two-Weibull distribution, no drive-model effect & $-468$ & 2/9 & 308 & -1.5 \\
hidden Markov against Weibull, with drive-model effect (point) & $+414$ & 6/9 & 356 & 1.2 \\
hidden Markov against two-Weibull distribution, with drive-model effect (point) & $-190$ & 4/9 & 540 & -0.4 \\
the same, predictive & $-250$ & 4/9 & 389 & -0.6 \\
drive-model effect, given the hidden Markov baseline & $+2{,}847$ & 9/9 & 594 & 4.8 \\
\bottomrule
\end{tabular}
\end{table}

The hidden Markov model falls between the single Weibull and the two-Weibull model on every measure in
Table~\ref{tab:hmm}. In
sample it is 409 log-likelihood units above the Weibull and 456 below the two-Weibull model
without the drive-model effect, and 457 above and 522 below with it. Out of sample, with the
drive-model effect, it is 190 log units behind the two-Weibull model as a point forecast and
250 behind as a predictive distribution, with ratios to the standard error of $-0.4$ and $-0.6$:
nine years do not separate them. Without the drive-model effect it is behind by 468 (ratio
$-1.5$).

The drive-model effect gains more than the change of baseline: 2{,}847 log units given the
hidden Markov baseline, in nine of nine years, against 2{,}569 given the two-Weibull model.

The pooled fit lies close to $\lambda_2 = \lambda_1 + q$ ($\lambda_1=0.0098$, $q=0.0502$, $\lambda_2=0.0600$ per
year), as do the two fits with the drive-model effect. The starting points also cover a regime in which
the first state is fragile and left quickly ($q$ of tens per year) for a state with a low
constant rate: early failures, then a flat hazard. The fits without and with the gamma
effect use seven starting points, with training likelihood selecting the result in each fold; the
fit with covariates starts from the latter's optimum, and an earlier check of all seven starts
gave the same optimum in every fold. That check has not been repeated on the current data.

The deposited auxiliary profile uses a pooled Lexis Poisson objective, not the
interval objective used for the fits above. Its direct subtraction of exponentials
also loses precision near coincident rates: stable reevaluation lowers its recorded
maximum by 0.107 log units at the same parameter point. No refit is claimed here,
and this profile is not used to calibrate uncertainty. The surface
$\lambda_2=\lambda_1+q$ is interior to the parameter space of three positive rates, not a
boundary of it; it is special because of the model's form. On it the two exponential rates
coincide, $S(t)=e^{-\lambda_2 t}(1+qt)$, and the hazard is $\lambda_2 - q/(1+qt)$: it rises from
1.0\% a year to 2.7\% a year at age ten and can never exceed 6.0\%. Off the surface the hazard
tends to $\min(\lambda_2,\lambda_1+q)$. The coincident-rate surface does not maximize the
initial rise: $h'(0)=q(\lambda_2-\lambda_1)$, which continues to increase with $\lambda_2$
at fixed $\lambda_1$ and $q$. Nor does a fitted point on that surface by itself reduce the
three-parameter model to two free parameters. Writing $m=(a+b)/2$ and
$d=(a-b)/2$, exchanging the rates changes $d$ to $-d$ at fixed $m$ and
$\lambda_1$. The survival function is smooth and even in $d$, so its first
derivative in this direction vanishes at $d=0$. Ordinary regular likelihood
approximations cannot be assumed there. A profile can describe the likelihood's
shape, but does not remove the rate ambiguity or by itself justify confidence
calibration. The model extrapolates to a hazard
that levels off, where the Weibull's rises without limit. Neither is supported by observation
beyond about eight years of age (Section~\ref{sec:apc}).

\subsubsection{With rates of its own for each drive model}\label{sec:hmmper}

With the drive-model effect on the cumulative hazard, the model above is a hybrid of the
two-state model and a frailty model. Here the rates themselves vary by drive model, in the
simplest form: each drive model $g$ gets a constant $z_g$ that multiplies time, $S_g(t)=S(z_g t)$, which for the hidden Markov model
multiplies $\lambda_1$, $q$ and $\lambda_2$ alike and leaves each drive model a two-state
chain. The same is done for two competing Weibulls, and both distributions are also fitted with the
constant on the hazard. The constants are free, with no prior, and are profiled out: in closed
form on the hazard, by a one-dimensional search on time.
Each training fold includes drive models with at least twenty failures. On all data,
the 46 selected models hold 31{,}670 of the 31{,}881 failures. Scoring uses a continuous failure-time density per year, with a
failure placed at the end of its recorded final day. Conditional on entry at age $a_0$,
a drive ending at age $a_1$ contributes
$d\log h_g(a_1)-[H_g(a_1)-H_g(a_0)]$, where $d$ is the failure indicator. This is not the
failure probability integrated over the recorded day, and these density scores are not
on the scale of the earlier count-score tables.

Expressing the same density per day subtracts $D\log(365.25)$ from a fold's score, where
$D$ is its failure count; this unit change cancels in paired contrasts on the same drives.
Integrating the event probability over the last day is a different change: a failed
drive then contributes
$-H_g(a_1-\delta)+H_g(a_0)+\log\{1-\exp[-H_g(a_1)+H_g(a_1-\delta)]\}$,
with $\delta=1/365.25$. That change need not cancel. The table retains the continuous-density
convention used for fitting and scoring; no daily-event refit is claimed.

\begin{table}[h]
\centering
\caption{Two Distributions with a Free Constant per Drive Model on the Hazard or on Time. Difference is the held-out log score over nine annual test folds minus that of the first row, with its standard error.}\label{tab:hmmper}
\footnotesize\begin{tabular}{@{}llrrrrr@{}}
\toprule
Model & Constant on & Shared par. & In-sample & Held-out & Difference & Pred./obs. \\
\midrule
Two competing Weibulls & hazard & 3 & -152{,}946 & -117{,}435 & --- & 0.99 \\
Two competing Weibulls & time & 3 & -152{,}717 & -118{,}003 & $-567$ (503) & 0.92 \\
Two-state hidden Markov & hazard & 3 & -153{,}339 & -117{,}877 & $-442$ (360) & 0.92 \\
Two-state hidden Markov & time & 2 & -153{,}252 & -117{,}781 & $-345$ (430) & 0.91 \\
\bottomrule
\end{tabular}
\end{table}

Letting the rates vary by drive model raises the hidden Markov model's likelihood by 87 over the
same model with the constant on the hazard (Table~\ref{tab:hmmper}). It remains 306 below two
competing Weibulls with a constant per drive model on the hazard, and 535 below two competing
Weibulls with a constant per drive model on time. On held-out years no row is separated from the first: the hidden Markov model is
345 behind with its own rates (standard error 430) and 442 behind with the constant on the
hazard (360). Constants on time do 97 better than constants on the hazard for the hidden
Markov model (standard error 113) and 567 worse for two competing Weibulls (503).

Notes on the fits. Each hidden Markov fit is the best of five or six starting points, which include both
basins of the hidden Markov likelihood (Appendix~\ref{sec:hmm}); in the folds ending in 2020,
2021 and 2022, and with the constant on the hazard also in 2018, the fragile-first-state basin
has the higher training likelihood, and in the others and on all data the wear-out basin does.
The time scale of each drive model is found by a grid search over its logarithm followed by
refinement, because its profile can have two modes. With the constant on the hazard the all-data parameters are near a boundary at which
both failure rates $\lambda_1$ and $\lambda_2$ tend to zero, while the transition rate
$q$ remains finite and the constants grow inversely with the failure-rate scale.
To first order the baseline hazard is
$\lambda_1 e^{-qa}+\lambda_2(1-e^{-qa})$; up to an overall multiplier its shape is
$\lambda_1/(\lambda_2-\lambda_1)+1-e^{-qa}$, approximately
$0.058+1-e^{-0.01417a}$ at the fitted parameters.
The comparison ``with the constant on the hazard'' is this fixed-effects fit on 46
drive models, not the gamma-frailty fit of Appendix~\ref{sec:hmm}.
When every drive model with at least one failure is given a constant (83 on all data),
the hidden Markov model is compared with two competing Weibulls with a constant on
the hazard. It is 564 behind with its own rates and 270 behind with the constant on
the hazard (standard errors 440 and 486).
The survival function and hazard are evaluated in numerically stable forms.
The independent check reconstructs the constants at the deposited shared parameters;
it validates the scores to the precision shown in the table, not global optimality
of the fits.

\subsection{A Per-Model Wear-Out Model: Failed Extrapolation}\label{sec:perwear}

To let each drive model have its own wear-out, replace the hierarchy \eqref{eq:hier} of
Section~\ref{sec:hier} by
\begin{equation}
H_g(a) = c_g\,a^{\beta_1} + \Bigl(\frac{a}{\eta_g}\Bigr)^{\beta_{2g}},\qquad
\theta_g = (\log c_g,\ \log\beta_{2g},\ \log\eta_g)^\top \sim \Normal(B x_g,\ \Sigma),
\end{equation}
with a shared early-life exponent $\beta_1$ fixed at its pooled estimate, covariates $x_g$ of Section~\ref{sec:hier} plus an intercept, a
$3\times7$ coefficient matrix $B$ and a $3\times3$ covariance matrix $\Sigma$, so $\theta_g$ is
trivariate normal. There is no closed form.
It is fitted by an iteration of the EM type: given $(B,\Sigma)$, each drive model's
$\theta_g$ is set to its posterior mode $\hat\theta_g$ (maximizing its log-likelihood plus the
log prior), with posterior covariance $V_g$ from the Hessian at the mode (Laplace approximation);
then $B$ is updated by least squares of $\hat\theta_g$ on $x_g$ and
$\Sigma \leftarrow \frac1G\sum_g[(\hat\theta_g - Bx_g)(\hat\theta_g - Bx_g)^{\top} + V_g]$. The
original fit omitted the helium indicator (using a $3\times6$ matrix $B$), and its
iteration was stopped at 25 rounds without having converged: its Laplace objective peaked at
round 7 and the parameter change per round was not decreasing at the end. Its results are
therefore those of an unconverged fit.

The subsequent check in Section~\ref{sec:valid} uses the helium covariate defined above
and removes the iteration cap. It stops when the sum of the
largest absolute changes in the entries of $B$ and $\Sigma$ is below $10^{-4}$.
All ten fits met this rule, but the 2019 forecast still failed. The all-data fit used
below stopped after 348 iterations, with $\beta_1=0.22865$, 99 drive models,
504{,}624 drives and 31{,}881 failures. Independent reconstruction of its saved
likelihood, posterior curvature and empirical cumulative-hazard curves agrees with
the reported values to the precision shown. This checks the saved fit, not global
optimality or forecast calibration.

\begin{figure}[h]
\centering
\includegraphics[width=0.97\textwidth]{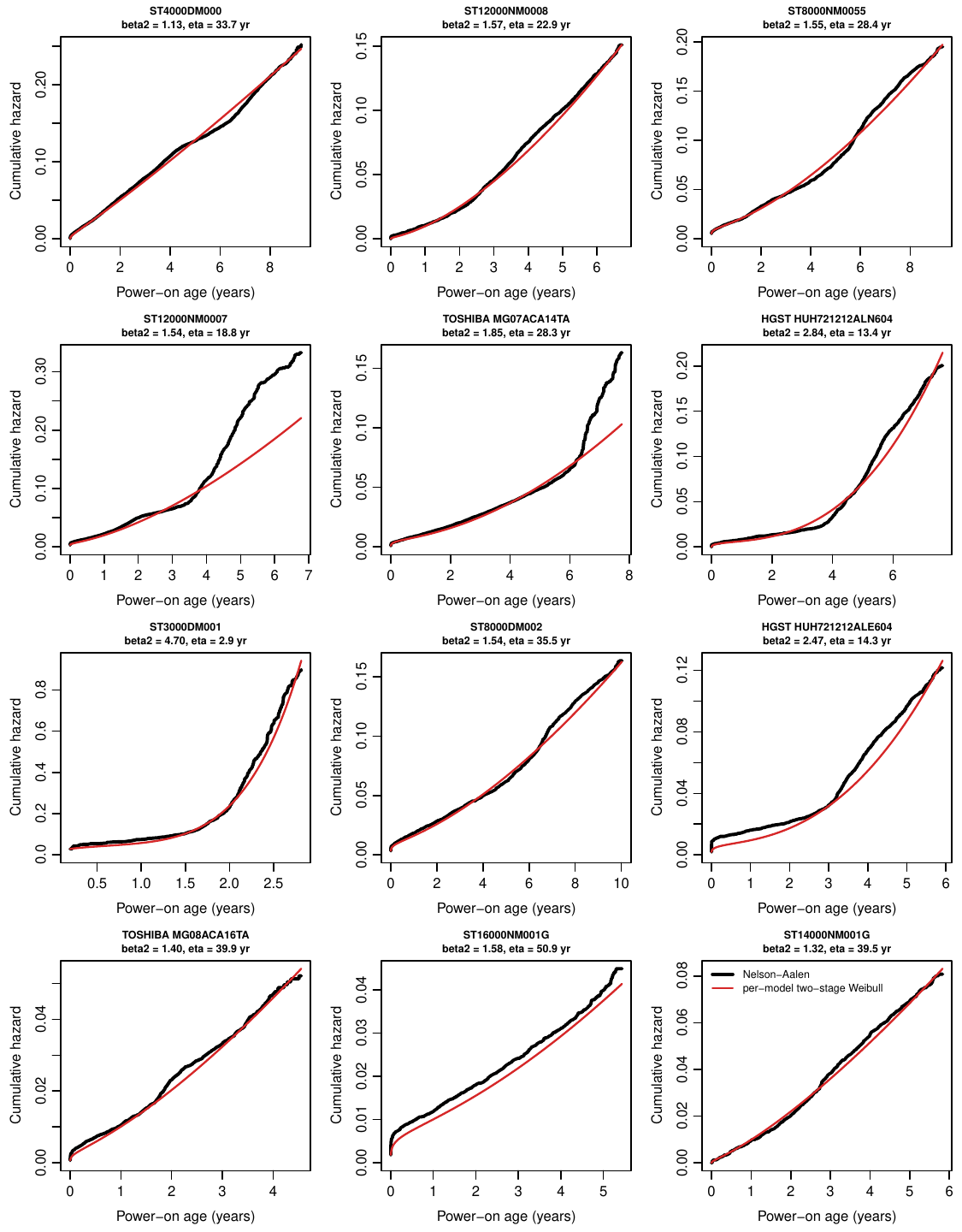}
\caption{As Figure~\ref{fig:permodel}, with each drive model's own two-Weibull distribution from the
per-model wear-out model (red), using the current fit with the helium covariate.
Each fitted curve is aligned to the empirical curve at the first age with at least
200 drives at risk; only ages meeting that risk-set threshold are shown.}\label{fig:permodel3}
\end{figure}

With three parameters per drive model it follows the data more closely in sample. It
reproduces the changes in slope (Section~\ref{sec:permodel}) of the two HGST 12~TB models and of the ST3000DM001 (fitted
wear-out exponents 2.5 to 4.7) but not those of the ST12000NM0007 or the Toshiba MG07ACA14TA
(Figure~\ref{fig:permodel3}). A Laplace estimate of its marginal log-likelihood is $-340{,}235$
against $-342{,}048$ for the empirical-Bayes model. The better in-sample fit does not resolve
the failed extrapolation in Section~\ref{sec:valid}.

\subsection{Fleet-Year Fits}\label{sec:fleetyear}

A fleet-year fit takes the drives in service in one calendar year $y$ and fits a distribution
to that year alone. These descriptive fits use all drive classes through December 2025,
including small or unclassified HDDs and SSDs, as in Section~\ref{sec:families}. They use
the absorbing-failure rule. Power-on-age fits exclude
drives without a usable age; first-sighting fits include them. A drive enters year $y$ at its age on 1~January (or on its first day, if later),
which makes it left-truncated at that age, and leaves at failure or, censored, at its last
day in the year; the likelihood is \eqref{eq:lldisc} restricted to those intervals. Each fit is a
period estimate for the mixture of drive models in service that year, not a distribution for any one
drive model. Table~\ref{tab:periodweibull} gives the single-Weibull fits on two age scales:
power-on age, and age since the drive's first appearance in the record.

\begin{table}[h]
\centering\small
\caption{Single Weibull by Fleet Year, All Drive Classes Through 2025. Drives, failures and drive-years are on the power-on scale. $\hat\beta$ and $\hat\eta$ (years) are given on power-on age and on age since first sighting. Standard errors, in parentheses, assume independent drives.}\label{tab:periodweibull}
\footnotesize\setlength{\tabcolsep}{4pt}
\begin{tabular}{@{}rrrrrrrr@{}}
\toprule
& & & & \multicolumn{2}{c}{Power-on age} & \multicolumn{2}{c}{Age since first sighting} \\
Year & Drives & Failures & Drive-years & $\hat\beta$ & $\hat\eta$ (yr) & $\hat\beta$ & $\hat\eta$ (yr) \\
\midrule
2013 & 29{,}071 & 724 & 17{,}730 & 0.990 (0.037) & 25.0 (2.2) & 0.919 (0.033) & 33.6 (4.9) \\
2014 & 47{,}831 & 2{,}205 & 35{,}639 & 1.423 (0.023) & 9.9 (0.2) & 1.546 (0.035) & 7.4 (0.3) \\
2015 & 62{,}902 & 1{,}413 & 48{,}745 & 1.083 (0.026) & 28.8 (1.7) & 1.011 (0.027) & 33.5 (2.4) \\
2016 & 81{,}093 & 1{,}427 & 67{,}910 & 1.110 (0.027) & 36.6 (2.3) & 1.080 (0.026) & 38.9 (2.6) \\
2017 & 108{,}193 & 1{,}545 & 85{,}432 & 1.088 (0.026) & 44.3 (2.8) & 1.069 (0.024) & 46.2 (2.9) \\
2018 & 125{,}147 & 1{,}357 & 102{,}696 & 1.011 (0.028) & 73.2 (6.5) & 0.990 (0.026) & 78.1 (6.7) \\
2019 & 137{,}144 & 2{,}226 & 113{,}561 & 0.991 (0.022) & 52.2 (3.1) & 0.994 (0.021) & 51.8 (3.0) \\
2020 & 182{,}146 & 1{,}495 & 144{,}596 & 1.039 (0.024) & 86.2 (6.2) & 1.039 (0.022) & 86.0 (5.9) \\
2021 & 231{,}298 & 2{,}139 & 185{,}244 & 1.084 (0.019) & 68.3 (3.7) & 1.043 (0.018) & 75.8 (4.1) \\
2022 & 238{,}584 & 3{,}155 & 220{,}860 & 1.465 (0.023) & 31.2 (0.9) & 1.428 (0.022) & 32.4 (1.0) \\
2023 & 289{,}765 & 4{,}350 & 252{,}255 & 1.571 (0.022) & 26.2 (0.6) & 1.496 (0.020) & 27.8 (0.6) \\
2024 & 330{,}431 & 4{,}655 & 288{,}524 & 1.355 (0.019) & 34.1 (0.9) & 1.293 (0.017) & 36.5 (0.9) \\
2025 & 348{,}850 & 4{,}309 & 321{,}640 & 1.405 (0.021) & 37.5 (1.0) & 1.383 (0.020) & 38.3 (1.0) \\
\bottomrule
\end{tabular}
\end{table}

\begin{figure}[h]
\centering
\includegraphics[width=0.7\textwidth]{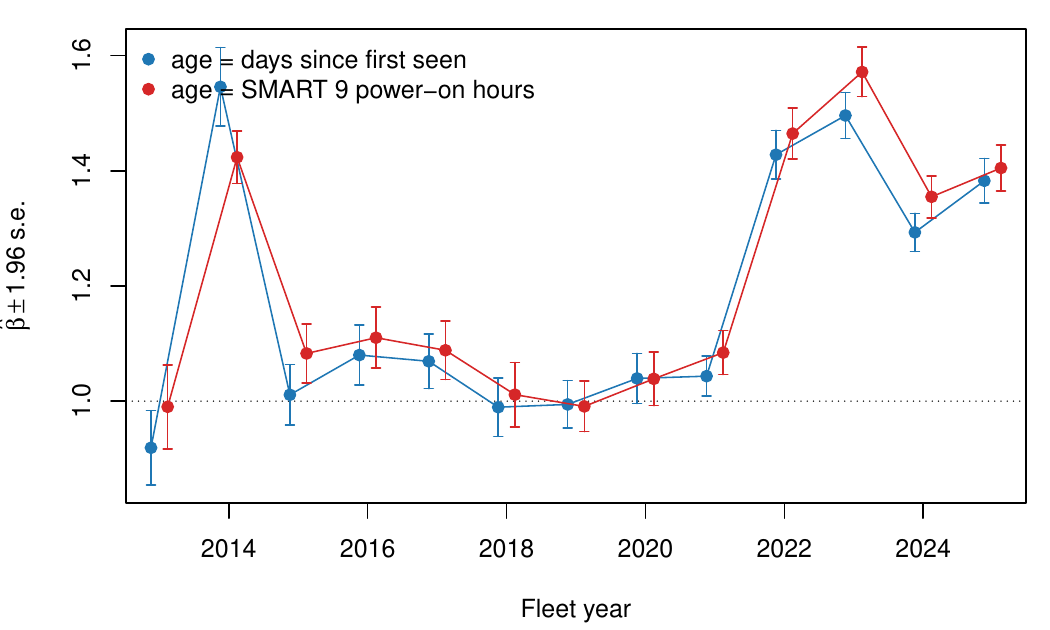}
\caption{Weibull exponent by fleet year for all drive classes through 2025, on power-on age
and on age since first sighting. The nominal 95\% Wald intervals assume independent drives.}\label{fig:beta}
\end{figure}

On power-on age the exponent lies between about 1.0 and 1.6 (Figure~\ref{fig:beta}), with
standard errors of 0.02 to 0.04. After 2013 (a small fleet; exponent 0.99, scale 25 years) there
are three regimes. In 2014 about 60\% of the fleet's failures are Seagate ST3000DM001s, whose
annualized failure rate that year was 43\%; the exponent is 1.42 and the scale under ten years.
From 2015 to 2021 the exponent is within 0.11 of 1 and the scale 29 to 86 years, close to a
constant hazard. From 2022 the exponent is 1.35 to 1.57 and the scale 26 to 38 years, as the
fleet ages. Age since first sighting counts drives that entered the record in service as new.
It changes the exponent most in the first years of the record: 0.92 against $0.99\pm0.04$ in
2013, 1.55 against 1.42 in 2014 and 1.01 against 1.08 in 2015. From 2016 the two age scales
give exponents within about two standard errors of each other, except in 2023 and 2024, where
first-sighting exponents are 0.08 and 0.06 lower.

\begin{table}[h]
\centering\small
\caption{Smallest AIC and AIC Deficits of Selected Distributions by Fleet Year. Fits use all drive classes with usable power-on age. Scale is the fitted common scale factor for the bathtub constants.}\label{tab:periodfam}
\footnotesize\setlength{\tabcolsep}{4pt}\begin{tabular}{@{}rlrrrrrr@{}}
\toprule
& & \multicolumn{6}{c}{$\Delta$AIC relative to the best distribution} \\
Year & Best distribution & 2 comp. & Weibull & Expon. & Bathtub, scaled & Scale & Bathtub, default \\
\midrule
2013 & Piecewise constant (age-years) & 157.1 & 215.2 & 213.3 & 164.5 & 0.72 & 300.5 \\
2014 & Mixture of 2 Weibulls & 1{,}912.8 & 1{,}962.1 & 2{,}359.2 & 2{,}621.1 & 0.58 & 3{,}924.1 \\
2015 & Piecewise constant (age-years) & 27.3 & 103.6 & 111.9 & 47.9 & 0.95 & 54.1 \\
2016 & Mixture of 2 Weibulls & 20.2 & 142.6 & 158.0 & 194.2 & 1.18 & 261.3 \\
2017 & Mixture of 2 Weibulls & 18.8 & 177.1 & 187.2 & 206.2 & 1.29 & 381.4 \\
2018 & 3 competing Weibulls & 7.7 & 121.4 & 119.6 & 135.7 & 1.74 & 791.5 \\
2019 & Mixture of 2 Weibulls & 362.6 & 358.6 & 356.8 & 609.5 & 1.30 & 896.1 \\
2020 & Mixture of 2 Weibulls & 16.3 & 63.0 & 63.7 & 256.2 & 2.37 & 2{,}553.3 \\
2021 & Mixture of 2 Weibulls & 30.5 & 215.2 & 232.5 & 384.9 & 2.10 & 3{,}277.2 \\
2022 & Piecewise constant (age-years) & 41.1 & 257.8 & 725.8 & 612.2 & 1.61 & 3{,}228.4 \\
2023 & Piecewise constant (age-years) & 468.5 & 625.3 & 1{,}429.1 & 1{,}248.9 & 1.47 & 4{,}032.4 \\
2024 & Mixture of 2 Weibulls & 342.6 & 925.6 & 1{,}350.3 & 1{,}142.6 & 1.53 & 4{,}571.4 \\
2025 & Piecewise constant (age-years) & 232.2 & 657.8 & 1{,}116.8 & 930.9 & 1.74 & 6{,}801.6 \\
\bottomrule
\end{tabular}
\end{table}

A single Weibull is never the best distribution in any year among the families tested. (In a few
years the winning mixture has one component with a scale of about a day, concentrating
failures at very young ages, plus a Weibull.) The default bathtub is worse than the single Weibull
in every year except 2015 (a deficit of 54 against 104). Its deficit is a few hundred units in
2013, 2016 and 2017, and thousands in 2014 and from 2020 on. Its fitted scale factor is 0.95 to
1.29 for the 2015--2017 fleet and 1.3 to 2.4 afterwards. The bathtub's constants describe
drives of the mid-2010s tolerably and current drives poorly.

\subsection{Quarterly Test Blocks}\label{sec:quarter}

The quarterly comparison uses the models and covariates of Section~\ref{sec:valid}. Training ends every three months from December 2017 to March
2026, with the following quarter as the test. These 34 nonoverlapping test blocks partition
the same 185{,}453 monthly Lexis cells and 24{,}975 failures as the annual comparison.
Table~\ref{tab:laws} gives the quarterly predictive scores with the matching current fits;
Table~\ref{tab:quarterscores} adds the point scores for both schedules.

\begin{table}[h]
\centering
\caption{Current Annual and Quarterly Scores with the Drive-Model Effect and Covariates.
Both schedules score the same failure counts. Point columns score plug-in Poisson forecasts;
predictive columns integrate the gamma multiplier of each drive model. Annual forecasts
are refitted at nine origins, quarterly forecasts at 34. No additional year calibration is
used in these rows.}\label{tab:quarterscores}
\footnotesize\setlength{\tabcolsep}{4pt}\begin{tabular}{@{}lrrrr@{}}
\toprule
& \multicolumn{2}{c}{Annual} & \multicolumn{2}{c}{Quarterly} \\
Distribution & Point & Predictive & Point & Predictive \\
\midrule
Weibull & -32{,}095 & -29{,}974 & -30{,}561 & -29{,}848 \\
Two-state hidden Markov & -31{,}681 & -29{,}696 & -30{,}141 & -29{,}500 \\
Two competing Weibulls & -31{,}491 & -29{,}446 & -29{,}943 & -29{,}247 \\
Weibull and Gompertz & -31{,}591 & -29{,}585 & -30{,}053 & -29{,}390 \\
Limited failure population (GLFP) & -31{,}571 & -29{,}473 & -30{,}001 & -29{,}262 \\
\bottomrule
\end{tabular}
\end{table}

\begin{table}[h]
\centering
\caption{Paired Comparisons over 34 Quarterly Folds on the Regenerated Data. Sum is the
difference in held-out log score summed over the folds. The standard error is from the
fold-to-fold spread, treating folds as independent, and Ratio is the sum over that standard
error. Dependence between quarters is not accounted for.}\label{tab:quarterpairs}
\footnotesize\setlength{\tabcolsep}{4pt}\begin{tabular}{@{}p{9.6cm}rrrr@{}}
\toprule
Comparison & Sum & Quarters won & Std.\ error & Ratio \\
\midrule
Two Weibulls over hidden Markov, predictive & $+253$ & 20 of 34 & 212 & $+1.2$ \\
the same, point forecast & $+198$ & 20 of 34 & 230 & $+0.9$ \\
the same, without the drive-model effect & $+460$ & 24 of 34 & 176 & $+2.6$ \\
Two Weibulls over one, predictive & $+602$ & 20 of 34 & 231 & $+2.6$ \\
Hidden Markov over one Weibull, predictive & $+348$ & 25 of 34 & 116 & $+3.0$ \\
Drive-model effect plus predictive averaging (two Weibulls) & $+4{,}395$ & 34 of 34 & 275 & $+16.0$ \\
Empirical-Bayes predictive over the per-year Weibull & $+3{,}502$ & 31 of 34 & 481 & $+7.3$ \\
Empirical-Bayes predictive over the bathtub & $+16{,}183$ & 34 of 34 & 1{,}798 & $+9.0$ \\
\bottomrule
\end{tabular}
\end{table}

The ordering of the five distributions in Table~\ref{tab:quarterscores} is the same under
both refit schedules. With the drive-model effect, two Weibulls lead the hidden Markov
model by 253 predictive log units (standard error 212, ratio 1.2) and 198 point-score units
(230, ratio 0.9). Both have higher predictive scores than one Weibull: the gains are 602
and 348, with ratios 2.6 and 3.0. The limited-failure-population model is 16 predictive log
units behind two Weibulls (standard error 14). These ratios summarize the quarterly
variation; they are not adjusted for dependence or for multiple comparisons.

The row labelled ``drive-model effect plus predictive averaging'' combines two changes.
Adding the drive-model effect and its covariates to the two-Weibull point forecast
gains 3{,}699 log units;
integrating the fitted gamma multiplier adds another 697, giving the reported combined
gain of 4{,}395. The full gain must not be attributed solely to the change in point forecast.

Every fitted predictor scored under both schedules improves on the quarterly folds.
The empirical-Bayes point score improves from $-31{,}491$ to $-29{,}943$, and its predictive
score from $-29{,}446$ to $-29{,}247$. The fixed bathtub has exactly the same total score,
because it uses the same rate on each of the same cells. Quarterly refitting changes the
information available: the first quarter has the same cutoff as the annual forecast, then
the next three fits incorporate three, six and nine additional months. The comparison is
between refit schedules, not between independent new sets of failures. It does not isolate
one source of improvement, such as tracking the calendar level, from the other parameter
updates.

For fixed baseline hazards and hyperparameters, the gamma--Poisson joint predictive
probability factorizes into successive quarterly conditional probabilities. Updating
only the gamma multiplier therefore leaves the total joint score unchanged: splitting
the test period alone does not explain the predictive gains above.

% Moved from the hierarchical-model section to the supplement: the per-drive-model profiles.
\section{Profiles of Drive Models}\label{sec:kiviat}

\begin{figure}[p]
\centering
\includegraphics[width=.98\textwidth]{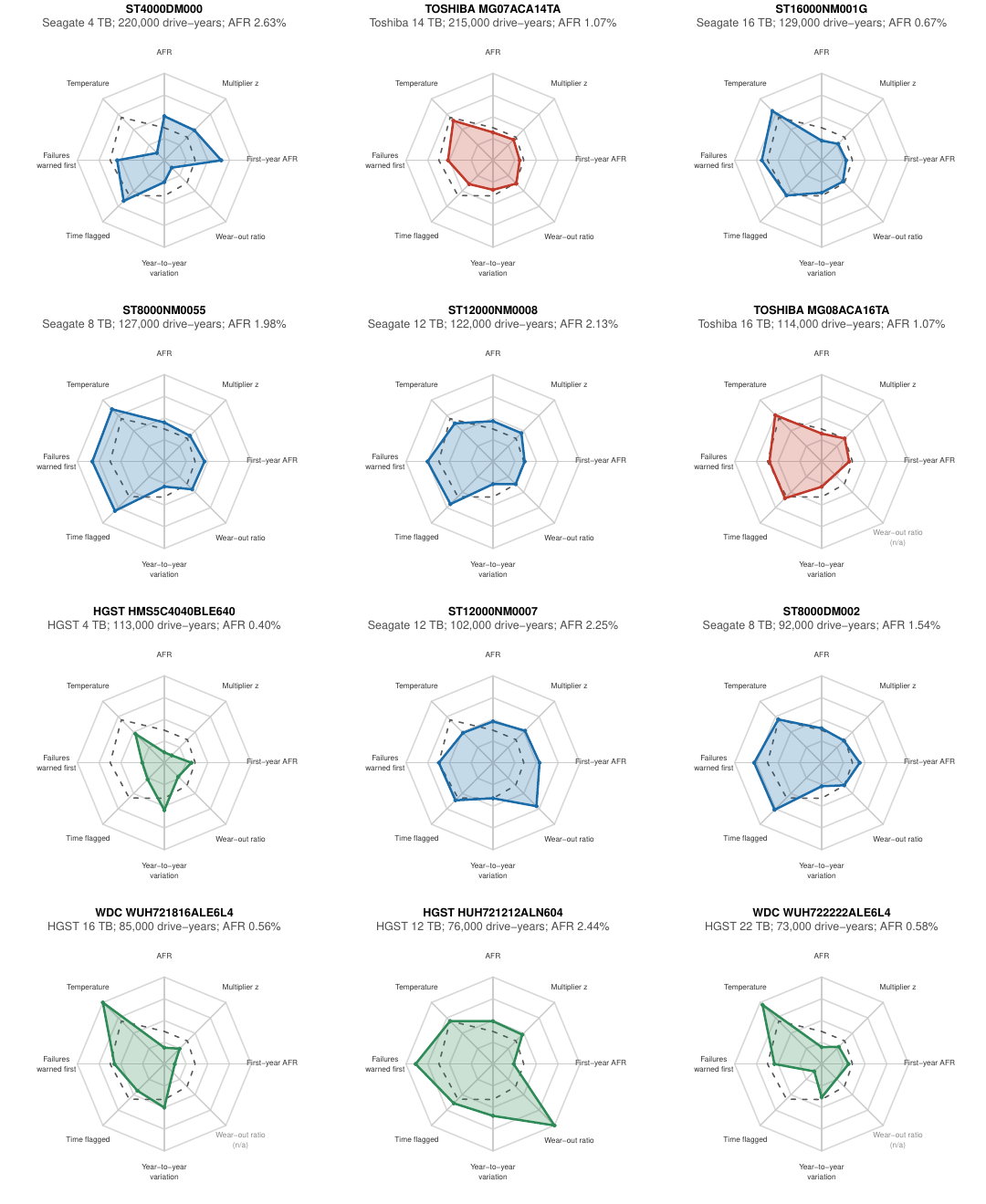}
\caption{Kiviat diagrams of the twelve drive models with the most drive-years, coloured by
manufacturer. Each axis runs from the smallest to the largest value among the drive models
measured on it (Table~\ref{tab:kiviataxes}); the dashed polygon is the median drive model, axis by axis. ``n/a'': fewer
than twenty failures at ages 4--6, so no wear-out ratio; the polygon then joins the two
neighbouring axes.}\label{fig:kiviat}
\end{figure}

Figure~\ref{fig:kiviat} draws, for the twelve drive models with the most drive-years, a Kiviat
diagram of eight measurements
(Table~\ref{tab:kiviataxes}): three of failure level (the raw annualized failure rate, the
empirical-Bayes multiplier, which adjusts for age, and the rate in the first year of power-on
age), the wear-out rate ratio defined in Table~\ref{tab:kiviataxes}, variation from year to
year, two SMART warning measures, and temperature. All but the multiplier are measured directly from the per-drive
monthly records. Each axis is scaled from the smallest to the largest value among the 36 drive
models with at least 2{,}000 drive-years and 20 failures, on a log scale where the values span
more than a factor of ten; outward always means more, and the dashed polygon is the median of
the 36. An axis that would rest on too few failures for a drive model is left out for that
model: the first-year rate needs 1{,}000 drive-years and five failures in the first year, the
wear-out ratio twenty failures in each of its two age ranges, and the year-to-year variation
four calendar years with at least 50 drive-years each.

The year-to-year variation is the coefficient of variation $c$ of a multiplier of the drive
model's failure rate that changes from year to year. The expected count $E_{y}$ of drive
model $g$ in year $y$ comes from a Poisson regression of failures on drive model, age bin and
calendar year (main effects, on drive-model$\times$age$\times$year cells), so that the fleet's
year effect is estimated net of its changing mix of drive models; with $O_y$ the observed count,
$c^2=\max\bigl(0,\ \sum_y[(O_y-E_y)^2-O_y]/\sum_y E_y^2\bigr)$. Subtracting $O_y$ removes the
Poisson part of the variance, so $c$ does not grow with the number of failures. It includes the
model-year shocks of Section~\ref{sec:time} and any departure of the drive model's own age
curve from the common one.

\begin{table}[h]
\centering\small
\caption{Axes of the Drive-Model Profiles. For each axis the table gives its definition, its scale, the number $n$ of drive models measured, and the smallest, median and largest value among them. The drive models are the 36 with at least 2{,}000 drive-years and 20 failures (HDD data drives, power-on age). The median is the lower median, the value of one drive model.}\label{tab:kiviataxes}
\footnotesize\setlength{\tabcolsep}{4pt}
\begin{tabular}{@{}lp{6.0cm}lrrrr@{}}
\toprule
Axis & Definition & Scale & $n$ & Smallest & Median & Largest \\
\midrule
AFR & failures per drive-year, all ages & log & 36 & 0.404\% & 1.38\% & 28.4\% \\
Multiplier $z$ & posterior mean multiplier of the empirical-Bayes model & log & 36 & 0.0724 & 0.271 & 6.87 \\
First-year AFR & failures per drive-year in the first year of power-on age & log & 31 & 0.398\% & 0.881\% & 7.88\% \\
Wear-out ratio & failure rate at ages 4--6 over the rate at ages 1--3 & log & 23 & 0.741 & 1.51 & 9.28 \\
Year-to-year variation & coefficient of variation of a yearly multiplier of the rate, about a Poisson regression on drive model, age and calendar year & linear & 31 & 0 & 0.261 & 0.798 \\
Time warned & share of drive-years after the first SMART warning & log & 36 & 0.432\% & 5.07\% & 52.4\% \\
Failures warned first & share of failures warned at the end of an earlier month & linear & 36 & 12.9\% & 59.4\% & 94.2\% \\
Temperature & mean drive temperature, $^\circ$C & linear & 36 & 24.4 & 31.7 & 35.6 \\
\bottomrule
\end{tabular}
\end{table}

Drive models with similar failure rates can differ on the other axes. The ST4000DM000 has a high raw rate (2.63\%) and the highest first-year rate of the twelve, but the
smallest wear-out ratio of the 23 measured, 0.74 (its rate fell between ages 1--3 and 4--6),
one of the steadiest rates from year to year ($c=0.12$), and the lowest mean temperature of the
twelve, 24.4$^\circ$C. The HGST HUH721212ALN604 has a similar raw rate (2.44\%) with the
opposite course over age: few early failures, a wear-out ratio of 9.3, the largest measured, a
year-to-year variation of 0.43, and 84\% of its failures warned by SMART in an earlier month.
The HGST HMS5C4040BLE640 has the smallest multiplier and rate of all 36 (0.40\%) and gave a
SMART warning before only a quarter of its failures.

Between drive models, temperature and failure are associated in the opposite direction to
within one drive model. Across the 36, cooler drive models failed more often (correlation $-0.44$ between mean temperature
and log failure rate, $-0.35$ with the log multiplier), and the coolest include the
ST3000DM001 (28\% a year). Mean temperature is correlated 0.84 with the year a drive model was
introduced: older drive models ran cooler and failed more. Within a drive model
the opposite holds (Appendix~\ref{sec:cov}): a drive more than 4$^\circ$C warmer than its model's
mean fails 1.32 times as often as one more than 4$^\circ$C cooler, and given age and drive model
the rate at 38$^\circ$C and above is 1.76 times that below 22$^\circ$C (1.06 one month ahead and
1.18 one year ahead once the SMART counts are in the model). The between-model association is
not evidence about the effect of temperature.

The axes are descriptive. The raw rate, the first-year rate, and the wear-out ratio mix age with
calendar year in the way described in Section~\ref{sec:apc}, and the share of failures warned
first depends on how a manufacturer reports the five SMART attributes.

\subsection*{Manufacturer Contrasts}\phantomsection\label{sec:mfrcontrast}

Siemroth and Park~\cite{siemroth2026manufacturer} report manufacturer hazard ratios from Cox and
Weibull regressions with installation-year, capacity and form-factor indicators and mean drive
temperature. Their Toshiba ratio, 1.07, differs from our prior-mean ratio $\exp(\gamma)$, 0.35.
We fitted their specification to our population, without the form factor, and then changed it
one step at a time toward ours (Table~\ref{tab:mfrcontrast}). We report every step we fitted. On
our data their Cox model gives Toshiba 0.85\unskip\ and a hazard ratio of
1.012\unskip\ per degree Celsius of mean temperature (theirs is 1.021).
Dropping temperature gives 0.84\unskip, and moving to power-on age with
left truncation changes little. Replacing the installation-year and capacity indicators with
vintage and log capacity moves Toshiba to 0.78, and adding the installation-year indicators back
moves it to 0.82\unskip. We also fitted Cox models with a gamma frailty per drive
model, the structure of our hierarchy. They returned incomplete manufacturer estimates in the
software we used (none in one fit, and all but Toshiba in the other), so the hierarchy's own multipliers
stand in for that step.

The hierarchy's contrasts are of a different kind, and we keep three of them apart. Its
$\exp(\gamma)$ is the ratio of the prior means of two drive models of equal capacity and
vintage. It is a regression coefficient, not an average over drive models: each drive model
enters through its own random effect, whose information and shrinkage depend on its exposure.
It describes the prior for a drive model, not a typical drive. The other two are summaries of
the posterior multipliers: their geometric mean with the drive models weighted equally, and
their mean weighted by exposure. Seagate has
49\unskip\ drive models in the data, many of them small and
with high multipliers, while most of its drives belong to a few large models with low ones.
With the drive models weighted equally, the empirical-Bayes posterior multipliers give Toshiba
0.29\unskip. Weighted by exposure they give
0.59\unskip. Unlike the regressions, these exposure-weighted contrasts
do not hold capacity and vintage fixed. Of Toshiba's drive-years,
98\unskip\% come from drives first seen in 2019 or later
(Seagate's: 43\unskip\%), when the fleet-wide hazard was at its
lowest, and a model without a year effect credits that low level to the manufacturer. Under B3
with drift, which adds a year effect and model-year shocks, the exposure-weighted Toshiba ratio
is 0.86\unskip, and adding installation-year indicators to the Cox
regression moves Toshiba the same way. Both are consistent with calendar time accounting for
much of the remaining difference, but B3 differs from the empirical-Bayes model in more than
the year effect, so neither step isolates it. The two papers' numbers answer different
questions. A simulator that draws a drive model from a manufacturer's catalogue should use the
prior-mean contrast. One that describes the drives a fleet operator actually ran should use an
exposure-weighted contrast with calendar effects.

\begin{table}[h]
\centering\small
\caption{Manufacturer Hazard Ratios Relative to Seagate. Rows run from the specification of Siemroth and Park~\cite{siemroth2026manufacturer} to the hierarchy's prior-mean contrast. Rows 3 to 8 are Cox or Weibull regressions on the drives of this paper (HDD data drives with a usable age, through June 2026). Exposure-weighted rows average each manufacturer's posterior-mean multipliers over its drive models with weights equal to their exposure. The row with drive models weighted equally is the geometric mean over drive models. The two small drive models of other makers (five drives) are counted with Seagate in the regressions and left out of the multiplier rows.}\label{tab:mfrcontrast}
\begin{tabular}{@{}lrrr@{}}
\toprule
& HGST & Toshiba & WDC \\
\midrule
Siemroth and Park, published, Cox (data to mid-2025) & 0.40 & 1.07 & 0.51 \\
Siemroth and Park, published, Weibull & 0.41 & 1.07 & 0.52 \\
Their specification on these data, Cox & 0.35 & 0.85 & 0.68 \\
Their specification on these data, Weibull & 0.35 & 0.86 & 0.66 \\
Without temperature, all drives & 0.35 & 0.84 & 0.67 \\
Power-on age with left truncation & 0.35 & 0.83 & 0.70 \\
Vintage and log capacity in place of the indicators & 0.42 & 0.78 & 1.50 \\
The same, with installation-year indicators added & 0.38 & 0.82 & 1.31 \\
Empirical Bayes, exposure-weighted multipliers & 0.46 & 0.59 & 2.45 \\
B3 with drift, exposure-weighted multipliers & 0.52 & 0.86 & 1.21 \\
Empirical Bayes, drive models weighted equally & 0.24 & 0.29 & 1.28 \\
Empirical Bayes, prior-mean ratio $\exp(\gamma)$ & 0.20 & 0.35 & 0.46 \\
B3 with drift, prior-mean ratio $\exp(\gamma)$ & 0.22 & 0.45 & 0.64 \\
\bottomrule
\end{tabular}
\end{table}

\section{Where Failures Happen: Vaults and Pods}\label{sec:vaults}

From July 2023 the daily files give each drive's datacenter, cluster, vault, and pod. Over the
36 months to June 2026 that covers 404{,}526 HDD data drives, 882{,}851 drive-years, and
13{,}224 failures in 6 datacenters, 9 clusters, 352 vaults, and 7{,}000 pods: about twenty pods
to a vault. (Drive-months without a location, 18{,}128 drive-years with 128 failures, about 1\%
of the failures in the window, are left out.) Backblaze spreads each stored object across a
vault as 17 data and 3 parity shards~\cite{bb2018durability}. Two separate questions bear on
durability: whether vaults differ in their failure rate, and whether failures within a vault
arrive together in time.

\textbf{Make-up of a vault.} In the last month of the data the median vault holds 1{,}200 drives
of which 98.9\% are one drive model, with drives of one age. Over the whole window, 227 of the
352 vaults are a \emph{single batch}: one drive model supplies at least 95\% of the drive-days
and the interquartile range of drive age at any one time is under a quarter of a year. The
other 125 are \emph{mixed}, and older.

\textbf{Expectation.} For each drive and month the expected number of failures was taken as
drive-days times $h_0(\text{age})$ from the empirical-Bayes model, rescaled so that expectations sum
to the observed failures of the drive model in that month. This allows for drive model, for
age through one age curve common to all drive models, and for anything that moves a drive
model's rate everywhere at once.

\begin{table}[h]
\centering\small
\caption{Clustering of Failures by Location, July 2023 to June 2026. Expectations are given the drive model's failures in each month. Pearson is $\sum(O-E)^2/E$ per group, 1 if failures fall at random, first without and then with the allowance for drive age. CV is the coefficient of variation of a gamma multiplier per group fitted by maximum likelihood (age-adjusted). Max is the largest observed-to-expected ratio among groups expecting at least five failures.}\label{tab:podcluster}
\begin{tabular}{@{}lrrrrrr@{}}
\toprule
Group & Groups & Mean expected & Pearson, plain & Pearson, age & CV & Max \\
\midrule
Pod, month & 187{,}795 & 0.07 & 1.16 & 1.10 & 1.17 & --- \\
Vault, month & 9{,}483 & 1.39 & 1.64 & 1.55 & 0.60 & 3.4 \\
Cluster, month & 306 & 43.22 & 1.59 & 1.65 & 0.12 & 2.4 \\
Datacenter, month & 198 & 66.79 & 1.56 & 1.62 & 0.10 & 1.8 \\
\addlinespace
Pod, whole window & 7{,}000 & 1.89 & 1.98 & 1.90 & 0.63 & 3.9 \\
Vault, whole window & 352 & 37.57 & 13.49 & 12.25 & 0.54 & 5.5 \\
Cluster, whole window & 9 & 1469.33 & 20.90 & 20.80 & 0.15 & 1.2 \\
Datacenter, whole window & 6 & 2204.00 & 18.93 & 18.72 & 0.08 & 1.1 \\
\bottomrule
\end{tabular}
\end{table}

\begin{table}[h]
\centering\small
\caption{Further Breakdowns of the Clustering, Age-Adjusted. Columns are as in Table~\ref{tab:podcluster}. In the rows given the vault's ratio, each group's expectation is multiplied by its vault's observed-to-expected ratio over the whole window, so what remains is variation within a vault.}\label{tab:podextra}
\begin{tabular}{@{}lrrrrr@{}}
\toprule
Group & Groups & Mean expected & Pearson & CV & Max \\
\midrule
All vaults, whole window & 352 & 37.57 & 12.25 & 0.54 & 5.5 \\
Single-batch vaults, whole window & 227 & 24.90 & 8.34 & 0.58 & 3.2 \\
Mixed vaults, whole window & 125 & 60.58 & 19.36 & 0.48 & 5.5 \\
Pods, whole window & 7{,}000 & 1.89 & 1.90 & 0.63 & 3.9 \\
Pods, given their vault's ratio & 6{,}888 & 1.92 & 1.23 & 0.32 & 3.6 \\
Vault-months, given the vault's ratio & 9{,}454 & 1.40 & 1.13 & 0.25 & 2.8 \\
Vault-weeks, given the vault's ratio & 41{,}590 & 0.32 & 1.16 & 0.59 & 2.7 \\
Vault-days, given the vault's ratio & 287{,}729 & 0.05 & 1.13 & 1.29 & --- \\
\bottomrule
\end{tabular}
\end{table}

\begin{table}[h]
\centering\small
\caption{Datacenters and the Vaults Furthest Above Expectation. Left: datacenters over the window, with the rate in \% a year. Obs/exp allows for drive model and age. Right: the eight vaults furthest above expectation among those expecting at least twenty failures, with the share of the vault's drive-days from its commonest drive model and its median drive age.}\label{tab:poddc}
\footnotesize\setlength{\tabcolsep}{3.5pt}
\begin{tabular}{@{}lrrrrrrr@{}}
\toprule
Site & Drive-yr & Pods & Models & Age & Fail. & Rate & Obs/exp \\
\midrule
phx1 & 284{,}675 & 2000 & 33 & 3.1 & 3{,}262 & 1.15 & 1.01 \\
sac0 & 238{,}477 & 1960 & 53 & 4.7 & 3{,}902 & 1.64 & 0.88 \\
sac2 & 194{,}827 & 1260 & 38 & 3.8 & 4{,}015 & 2.06 & 1.11 \\
iad1 & 99{,}005 & 940 & 10 & 1.2 & 1{,}006 & 1.02 & 1.06 \\
ams5 & 60{,}489 & 780 & 14 & 2.2 & 1{,}020 & 1.69 & 1.05 \\
yyz1 & 5{,}378 & 60 & 2 & 0.8 & 19 & 0.35 & 0.65 \\
\bottomrule
\end{tabular}
\quad
\begin{tabular}{@{}lrrrrr@{}}
\toprule
Vault & Top (\%) & Age & Fail. & Exp. & Ratio \\
\midrule
phx1 000 1063 & 92 & 4.7 & 244 & 44.2 & 5.5 \\
phx1 000 1066 & 94 & 4.6 & 177 & 44.1 & 4.0 \\
sac2 000 1152 & 99 & 3.9 & 76 & 25.9 & 2.9 \\
sac2 000 1133 & 96 & 4.9 & 130 & 45.7 & 2.8 \\
sac2 000 1111 & 95 & 6.0 & 126 & 50.3 & 2.5 \\
phx1 000 1081 & 100 & 3.7 & 70 & 28.8 & 2.4 \\
phx1 000 1087 & 100 & 3.4 & 68 & 28.1 & 2.4 \\
ams5 031 2011 & 96 & 3.0 & 60 & 25.8 & 2.3 \\
\bottomrule
\end{tabular}
\end{table}

\textbf{Vaults differ from one another, persistently over the window.} Summed over the window, the Pearson
statistic for vaults is 12.3 (1 under independence), and the fitted multiplier has a
coefficient of variation of 0.54 (Table~\ref{tab:podcluster}). Among the 220 vaults expecting at
least twenty failures, the ratio of observed to expected runs from 0.29 at the 5th percentile
to 2.01 at the 95th. The largest ratios belong to mixed, older vaults: the worst, with 244
failures against 44 expected, is 92\% one drive model with a median age of 4.7 years
(Table~\ref{tab:poddc}). Among the 227 single-batch vaults the coefficient of variation is 0.58
and the largest ratio is 3.2 (Table~\ref{tab:podextra}). Clusters and datacenters differ little
once drive model and age are allowed for: coefficients of variation of 0.15 and 0.08, and
observed-to-expected ratios from 0.88 to 1.11 for the five large datacenters.

\textbf{Ageing that the common curve misses.} Among single-batch vaults of the fifteen drive
models with at least five such vaults, giving each drive model its own slope in log age reduces
the dispersion among vaults from 8.6 to 6.7. Four drive models have
slopes more than 2.5 standard errors above zero (ST16000NM001G $1.60\pm0.49$, ST24000NM002H $1.54\pm0.58$, Toshiba
MG08ACA16TA $0.85\pm0.28$, WDC WUH721816ALE6L4 $1.82\pm0.58$): within those drive models, older
vaults fail more than the common age curve predicts. About a fifth of the between-vault
dispersion in these vaults is therefore drive-model-specific ageing or vintage, the confound of
Section~\ref{sec:apc}; ``allowing for age'' above means allowing for one common curve only.

\textbf{Pods.} Pods taken singly have a coefficient of variation of 0.63, most of which is their
vault's: with each pod's expectation multiplied by its vault's ratio, the Pearson statistic
falls from 1.90 to 1.23 and the coefficient of variation to 0.32, the size of the pod-level
effect that remains. That is a fixed difference between pods, not contagion within one. A
preregistered analysis of the same files, deposited as code and derived data without a
manuscript when we searched~\cite{wang2026slot} (later versions of the deposit retitle the
manuscript ``Local clustering of recorded drive failures within servers of one storage
operator''), asks whether a recorded failure raises the failure rate of the other drives of
its server beyond what their model, age and vault predict, and whether the excess depends on
the distance between slots. Its summary reports two ratios on a 2026 hold-out, about 1.3 and
about 1.9 with intervals above one, says that its verdicts were the same under its original
and its corrected definition of a server, and does not state which hypotheses those ratios
test or which it counts as confirmed. We did not test short-range contagion.

\textbf{Within a vault, in time.} With each vault's expectation multiplied by its own ratio over
the window, the vault-month Pearson statistic is 1.13 and the coefficient of variation 0.25,
against 1.55 and 0.60 before. Most of the month-level clustering in
Table~\ref{tab:podcluster} is the fixed difference between vaults; given a vault's own rate,
its monthly counts are close to Poisson. At shorter scales, spreading each vault-month's
expectation evenly over its days, the Pearson statistic is 1.16 for vault-weeks and 1.13 for
vault-days. The tail is heavier than these averages suggest: 1{,}222 vault-days have two or more recorded
failures where independence gives 617, and 306 vault-weeks have four or more where it gives
197; the largest counts are 8 in a day and 18 in a week. Part of this excess is the calendar:
fleet-wide, fewer failures are recorded at weekends. Spreading each vault-month's expectation
over its days in proportion to the rest of the fleet's failures on each day, independence gives
729 and 214, so the excess is a factor of 1.7 for days and 1.4 for weeks. The date of a failure
in this corpus is the date the drive was recorded as removed, so same-day coincidences may
reflect how replacements are scheduled as well as when drives fail.

\textbf{Implications.} A fixed multiplier per vault with a coefficient of variation of about
a half does not by itself make failures within a vault coincide. Because the probability of
loss is approximately proportional to a power of the failure rate in rare-loss calculations,
such variation can raise loss probability through the vaults with high multipliers. This
approximation is not a general convexity result for all rates and repair policies.
The evidence that failures within a vault arrive together is the excess of
multi-failure days and weeks, a factor of 1.4 to 1.7 over independence once the fleet's daily
pattern is allowed for, subject to the caveat about recorded dates. Nothing here measures
coincidence within a rebuild time shorter than a day. These data cannot identify what a
vault's multiplier represents: a single-batch vault shares a manufacturing lot, a firmware
level, a chassis generation and a place in a building, and the mixed vaults that head
Table~\ref{tab:poddc} share only the last two. For simulation, drive model and age do not
exhaust what is known about a drive's failure rate. The drive-model-year shock of
Section~\ref{sec:bayesmodel} is probably better broken down by vault, which bears on the inconsistency under
subdivision noted there; this has not been tried. The window is three years, so persistence
beyond that is untested.

\section{Extensions of the Bayesian Model}\label{sec:ext}

Three extensions of the Bayesian model share one table and one scoring rule: the SMART
state (Appendix~\ref{sec:extsmart}), a shock rule motivated by vaults (Appendix~\ref{sec:extvault}),
and the hidden Markov model in hierarchical form (Appendix~\ref{sec:exthmm}). Their scores are
comparable with each other but not with those of Section~\ref{sec:bayesvalid}, because both
the table and the score differ.

\subsection{Common Structure and Scoring}

The Lexis table is rebuilt with one more dimension, the drive's state on 1~January of each
year: \emph{clear}, \emph{warned} (one of SMART attributes
5, 187, 188, 197, 198 non-zero before that day), or \emph{new} (first seen during the year).
Exposure is the drive's span from first sighting to exit (Section~\ref{sec:population}),
2{,}023{,}456 drive-years with 31{,}881 failures: 1{,}665{,}165 drive-years clear with 16{,}455
failures (0.99\% a year), 117{,}937 warned with 12{,}076 (10.2\%), and 240{,}354 new with 3{,}350
(1.39\%). In every
variant the count in cell (drive model $g$, year $y$, age bin $b$, state $s$) is
\begin{equation}
D_{gybs}\sim\mathrm{Poisson}\bigl(E_{gybs}\,\bar h_g(b)\,\psi_{s}(g)\,Z_g\,Q_y\,U_{gy}\bigr),\qquad
U_{gy}\sim\Gam(\kappa_{gy},\kappa_{gy}),
\end{equation}
with $\log Q_y$ a random walk with drift, as in B3 with drift. Integrating $U_{gy}$ leaves a
negative binomial for the model-year total and a multinomial for its split over age bins and
states. The \emph{base} variant has $\psi\equiv1$, two competing Weibulls for $\bar h$, the
gamma $Z_g$ with covariate mean, and $\kappa_{gy}=\kappa_0 m_{gy}^{\xi}$: it is B3 with drift
on the new table.

For each test year and drive model, the score is the log of the posterior mean of the
probability of that drive model's observed cell counts, with the year effect integrated by
Gauss--Hermite quadrature within each draw; the drive models' scores are added. This is a sum
of marginal scores over drive models, not the joint score of Section~\ref{sec:bayesvalid}, and
it includes the split of failures over states. It was chosen because it is computed the same
way for a distribution with any number of parameters per drive model. Its weakness is the Monte Carlo
average over draws for each drive model: of 411 drive-model-years, between one and five per
variant have an effective sample size below 20 of 1{,}000 draws, and those carry $-348$ to
$-806$ of the totals in Table~\ref{tab:extscores}. Differences of a few units between
variants are therefore unreliable. Larger differences also need a Monte Carlo error estimate;
their magnitude alone does not establish that they are resolved.

\begin{table}[h]
\centering
\caption{Six Variants on the State-Split Table over Nine Test Years. Score is the sum over drive models and years of the log marginal predictive probability. Difference is from the base variant, except for the two rows that add to the SMART variant, which are compared with it. Its standard error is from the year-to-year spread, and Won counts the years won. Log ratio is the mean over the nine years of the log of predicted over observed failures, with its standard error. Cov.\ is the coverage of the 95\% interval for drive models with five or more failures expected. Fleet is the number of years in which the fleet total lies in its 95\% interval. $\hat R$ is the largest over the nine fold fits.}\label{tab:extscores}
\footnotesize\setlength{\tabcolsep}{3pt}\begin{tabular}{@{}p{5.4cm}rrrrrrrr@{}}
\toprule
Variant & Score & Difference & Won & Pred./obs. & Log ratio & Cov. & Fleet & $\hat R$ \\
\midrule
Base (B3 with drift) & -22{,}839 & --- & --- & 1.05 & $+0.06$ (0.11) & 96 & 9 & 1.01 \\
Common shock plus size-averaged shock & -22{,}841 & $-2.6$ (4.5) & 2 & 1.04 & $+0.03$ (0.10) & 94 & 9 & 1.01 \\
SMART state on 1 January & -12{,}053 & $+10{,}786$ (2{,}734) & 9 & 0.99 & $-0.02$ (0.12) & 95 & 7 & 1.21 \\
SMART state and the vault-style shock & -12{,}054 & $-1.3$ (7.3) & 5 & 0.98 & $-0.04$ (0.11) & 94 & 8 & 1.01 \\
Hierarchical hidden Markov model & -23{,}114 & $-275$ (116) & 1 & 1.31 & $+0.30$ (0.13) & 96 & 9 & 1.01 \\
Hierarchical hidden Markov model with SMART state & -12{,}215 & $-163$ (96) & 4 & 1.23 & $+0.22$ (0.15) & 96 & 8 & 1.01 \\
\bottomrule
\end{tabular}
\end{table}

\begin{table}[h]
\centering
\caption{Posterior Means and 95\% Intervals of the Extensions' Year Effect and Shock Rule. All fits are to all data. $Q_{2026}/Q_{2013}$ is the year effect in 2026 relative to 2013. $\sigma_Q$ and Drift are the standard deviation and drift of its yearly steps. The shock rule's two parameters are $\mathrm{cv}_u=\kappa_0^{-1/2}$ and $\xi$, and for the vault-style rule the common coefficient of variation and the one per thousand drive-years.}\label{tab:exthyper}
\footnotesize\setlength{\tabcolsep}{3pt}\begin{tabular}{@{}lrrrr@{}}
\toprule
Variant & $Q_{2026}/Q_{2013}$ & $\sigma_Q$ & Drift & Shock rule \\
\midrule
Base & $0.21$ ($0.13$ to $0.32$) & $0.25$ ($0.14$ to $0.42$) & $-0.12$ ($-0.27$ to $0.02$) & $0.96$ ($0.77$ to $1.17$); $0.36$ ($0.25$ to $0.46$) \\
SMART state & $0.17$ ($0.11$ to $0.25$) & $0.23$ ($0.13$ to $0.39$) & $-0.14$ ($-0.27$ to $-0.00$) & $0.94$ ($0.75$ to $1.15$); $0.37$ ($0.26$ to $0.47$) \\
Vault-style shock & $0.20$ ($0.12$ to $0.30$) & $0.24$ ($0.14$ to $0.40$) & $-0.13$ ($-0.27$ to $0.01$) & $0.45$ ($0.39$ to $0.50$); $0.11$ ($0.07$ to $0.16$) \\
Hidden Markov & $1.08$ ($0.72$ to $1.58$) & $0.34$ ($0.20$ to $0.54$) & $0.01$ ($-0.18$ to $0.20$) & $1.02$ ($0.81$ to $1.23$); $0.36$ ($0.24$ to $0.46$) \\
Hidden Markov, SMART & $0.39$ ($0.27$ to $0.57$) & $0.26$ ($0.14$ to $0.43$) & $-0.07$ ($-0.22$ to $0.08$) & $0.94$ ($0.73$ to $1.14$); $0.36$ ($0.24$ to $0.46$) \\
\bottomrule
\end{tabular}
\end{table}

\subsection{The SMART State}\label{sec:extsmart}

The SMART variant gives the warned state a multiplier per manufacturer and the new state one
multiplier; the clear state is the baseline. The shock $U_{gy}$ is shared by the three states
of a model-year.

\begin{table}[h]
\centering
\caption{State Multipliers Relative to a Clear Drive of the Same Model and Age. Entries are posterior means with 95\% intervals.}\label{tab:extrho}
\small\begin{tabular}{@{}lrr@{}}
\toprule
& Two competing Weibulls & Hidden Markov model \\
\midrule
Warned-rate multiplier, Seagate & $6.5$ ($6.3$ to $6.7$) & $6.5$ ($6.3$ to $6.7$) \\
Warned-rate multiplier, HGST & $23.4$ ($22.0$ to $24.9$) & $23.9$ ($22.4$ to $25.4$) \\
Warned-rate multiplier, Toshiba & $12.3$ ($11.4$ to $13.2$) & $12.4$ ($11.5$ to $13.3$) \\
Warned-rate multiplier, WDC & $9.5$ ($5.7$ to $14.4$) & $9.3$ ($5.5$ to $14.3$) \\
Multiplier for drives arriving in the year & $1.25$ ($1.16$ to $1.35$) & $1.55$ ($1.46$ to $1.66$) \\
\bottomrule
\end{tabular}
\end{table}

Knowing each drive's state on 1~January raises the score by 10{,}786 over the base variant,
in nine years of nine (Table~\ref{tab:extscores}). Leaving out the nine drive-model-years in
which either estimate has an effective sample size below 50, the gain is 9{,}571. The same step
on the hidden Markov model raises the score by 10{,}898. The warning gained 11{,}128 in the
regression of Appendix~\ref{sec:cov}. The multipliers (Table~\ref{tab:extrho}) are 6.5 for Seagate, 12 for
Toshiba, 23 for HGST and 9.5 for WDC; the regression gave 6.8, 11.6, 36 and 15, with HGST the
largest in both and WDC, on few warned drives, poorly determined. A drive that arrives during the year fails at 1.25 times the
rate of a clear drive of the same age. Part of this is an artefact of the first year of data:
every drive in 2013 is ``new'', so this multiplier and the step of the year effect from 2013 to
2014 are not separately determined, and that step accounts for the lower $Q_{2026}/Q_{2013}$ of
this variant (0.17 against 0.21).

The state does not explain the shocks. The model-year shock keeps its size
(Table~\ref{tab:extcv}: a coefficient of variation of 0.31 for the largest groups against
0.33) and the year effect keeps its fall (to 0.17 of its 2013 level, against 0.21 in the base
variant and 0.18 for B3 with drift in Section~\ref{sec:bayesfit}). Calibration for drive
models is nearly unchanged (95\% coverage for those with five or more expected, against 96\%). For the fleet total
it is a little worse: the 95\% interval is narrower (1.36 times its median against 1.57) and
misses two years of nine (2020, 1{,}307 observed against 1{,}358 to 5{,}426; 2022, 2{,}996 against
1{,}130 to 2{,}993), and the 50\% interval
contains four of the nine totals where the base variant's contains six. The state adds
information about which drives fail; the shocks still carry the uncertainty about how many.
The wear-out exponent does change: 1.49 (1.43 to 1.56) for a clear drive, against 1.91 in the base
variant, so part of the base variant's ageing is the growth of the warned share with age.

The forecast level is not distinguishable from the base variant's (predicted over observed
0.99; mean log ratio $-0.02$, standard error 0.12). One of its nine fold fits, trained through 2021, did not mix
($\hat R=1.21$, effective sample size 15, 307 divergent transitions); the other eight have 13 divergent transitions
and $\hat R\le1.008$, and the base variant has 5 over its nine.

\begin{table}[h]
\centering
\caption{Median Coefficient of Variation of the Model-Year Shock. Rows are bins of the drive-years of the model-year, in the fits to all data.}\label{tab:extcv}
\small\begin{tabular}{@{}lrrrrr@{}}
\toprule
Drive-years & Model-years & Base & SMART & Vault-style & Hidden Markov \\
\midrule
below 100 & 289 & 1.25 & 1.21 & 1.21 & 1.34 \\
100 to 1{,}000 & 79 & 0.67 & 0.65 & 0.49 & 0.75 \\
1{,}000 to 5{,}000 & 95 & 0.55 & 0.52 & 0.45 & 0.58 \\
5{,}000 to 20{,}000 & 93 & 0.40 & 0.39 & 0.45 & 0.43 \\
above 20{,}000 & 23 & 0.33 & 0.31 & 0.45 & 0.34 \\
\bottomrule
\end{tabular}
\end{table}

\subsection{A Shock Rule with a Common Part and a Part That Averages Out}\label{sec:extvault}

The rule $\kappa_{gy}=\kappa_0m_{gy}^\xi$ of Section~\ref{sec:bayesmodel} lets the shock shrink as a group
grows, but is not consistent when a group is split (Section~\ref{sec:bayesmodel}). A shared-factor construction supplies a second-moment rule that is consistent
under subdivision. Let $C_{gy}$ be a multiplier
common to all drives of drive model $g$ in year $y$, with mean 1 and coefficient of variation
$c$, and let the model-year's exposure be divided into $n_{gy}$ units of a fixed size, such as
vaults, each with its own independent multiplier $V_k$ of mean 1 and coefficient of variation
$v$. A drive in unit $k$ has multiplier $C_{gy}V_k$, and the model-year's effective multiplier,
the exposure-weighted mean over its units, is $U_{gy}=C_{gy}\sum_k r_kV_k$, where
$r_k$ is unit $k$'s fraction of the total exposure and $\sum_k r_k=1$. Assume that $C_{gy}$
is independent of all $V_k$, and that the $V_k$ are mutually independent. Then
$\operatorname{CV}^2(U_{gy})=c^2+(1+c^2)v^2\sum_k r_k^2$. For equal exposures,
$r_k=1/n_{gy}$ and this becomes $c^2+(1+c^2)v^2/n_{gy}$. The rule fitted here drops the cross term,
\begin{equation}
\frac{1}{\kappa_{gy}}=c^2+\frac{v^2}{n_{gy}},
\end{equation}
which is the squared coefficient of variation of the sum of a common and an averaged local
component rather than of their product; the two agree when $v$ is read as the local
coefficient of variation times $\sqrt{1+c^2}$, so the local parameter of the product
construction is $v/\sqrt{1+c^2}$, and with the fitted values below that is 0.10 against 0.11.
In the shared-factor construction, the common multiplier is retained when the group
is divided, and the local unit has a fixed meaning. Matching this second moment does not
make the product gamma-distributed or supply a joint distribution for separately fitted
gamma shocks. Here $n_{gy}$ is the model-year's exposure in thousands
of drive-years (a vault holds about 1{,}200 drives). Vault identifiers exist only from July
2023, but the rule needs only the size of the group, so it can be fitted to all years.

Its held-out score is not distinguished from that of the rule of
Section~\ref{sec:bayesmodel}: the difference is $-2.6$ (standard error 4.5). The fit puts most of the shock in the common part: $c=0.45$ (0.40 to
0.51) and $v=0.11$ (0.07 to 0.16). Under this rule the coefficient of variation is 0.57 at a
hundred drive-years, 0.49 at three hundred and between 0.45 and 0.46 above a thousand
(Table~\ref{tab:extcv}), where the rule of Section~\ref{sec:bayesmodel} falls from 0.67 to 0.33 over the range of
the table. For the largest groups the two rules differ by a factor of 1.8 in variance, and the
held-out years do not choose between a floor and a slow decline.

No vault enters this likelihood. The rule is a function of a model-year's exposure, and $v$
is determined mostly by the 368 model-years below a thousand drive-years, which are smaller than
one vault and hold 2\% of the exposure, so the fit is not evidence about vaults. It shows what
$\xi$ (0.36 in the base variant, 0.37 in the model of Section~\ref{sec:bayesfit}) already shows:
the yearly shock does not shrink as one over the size of the group, as it would if it were the
sum of independent shocks to equal sub-groups. The direct evidence on
vaults remains that of Appendix~\ref{sec:vaults}: vaults differ by a coefficient of variation
of 0.54 over three years, and given a vault's own level its failures are close to independent
from month to month.

Combining the two gives a candidate structure for a simulator,
\begin{equation}
\text{hazard multiplier of a drive}=Q_y\times U_{gy}\times Z_g\times V_{\text{vault}},
\end{equation}
with $U_{gy}$ common to the drive model in the year and $V_{\text{vault}}$ a multiplier drawn
once for each vault and kept. A common shock is consistent under subdivision, and a vault
multiplier matters for redundancy because the drives of a stripe share a vault. This is a
proposal, not a fitted model, and it leaves three things open:
\begin{itemize}[leftmargin=1.4em]
\item the size of the common shock for a large drive model, 0.33 or 0.45 depending on the rule;
\item the vault coefficient of variation of 0.54, which was measured against the drive model's
  prediction over 36 months and so contains whatever model-year shocks fell in those months;
  multiplying it by $U_{gy}$ counts some variation twice and may overstate the total dispersion;
\item whether a vault's multiplier is persistent or slowly varying, which 36 months cannot
  distinguish.
\end{itemize}
The combination of this rule with the SMART state samples
($\hat R\le1.013$, 22 divergent transitions over the nine fold fits), and its score is not
distinguished from the SMART variant's ($-1.3$, standard error 7.3, Table~\ref{tab:extscores}).

\subsection{The Hidden Markov Model in Hierarchical Form}\label{sec:exthmm}

Here the two-state model replaces the two Weibulls of the base variant, on the state-split
table and with the same scoring, year effect, shocks and covariates. Two forms were tried.

In the first, each drive model has its own three rates, each with a normal prior on the log
scale around a population mean. This did not sample: after 1{,}500 warm-up and 1{,}000 retained iterations of four chains on
the data through 2020, $\hat R$ was between about 1.5 and 3.0 for every top-level parameter.
The cause was not established. The likelihood of this model has two basins for a single drive
model (Appendix~\ref{sec:hmm}), and with three free rates for each drive model the
posterior may have many modes.

In the second, each drive model has one random effect, a time scale:
\begin{equation}
\log(\lambda_{1g},q_g,\lambda_{2g})^\top=(\mu_1,\mu_q,\mu_2)^\top+(x_g^\top\gamma+\sigma e_g)\mathbf{1}_3,\qquad e_g\sim\Normal(0,1),
\end{equation}
where $\mathbf{1}_3$ is the vector of three ones: the same scalar shift is added to the
logarithm of each of the three rates. Thus $S_g(t)=S_0(z_gt)$ with $\log z_g=x_g^\top\gamma+\sigma e_g$, and the drive model keeps a
two-state chain of its own. This is the hierarchical counterpart of the constant on time in
Appendix~\ref{sec:hmmper}. It samples without difficulty ($\hat R\le1.011$, one divergent
transition in the nine fold fits).

On the held-out years it is 275 behind the base variant (standard error 116, a ratio of $-2.4$),
and ahead in one year of nine; with the SMART state added to both it is 163 behind (standard
error 96). It over-predicts: 1.31 times the observed failures over the nine years and a mean
log ratio of $+0.30$ (standard error 0.13), against 1.05 and $+0.06$ for the base variant.
Its predictive intervals have similar coverage and are wider (the fleet interval is 2.2 times
its median, against 1.6).

The fitted rates at the covariates' reference values are $\lambda_1=0.010$, $q=0.064$,
$\lambda_2=0.073$ per year. The posterior is centred on the surface $\lambda_2=\lambda_1+q$
(Appendix~\ref{sec:hmm}), on which the hazard is $\lambda_2-q/(1+qt)$ and cannot exceed
$\lambda_2$, but not tightly:
$(\lambda_1+q-\lambda_2)/\lambda_2$ has a 95\% interval of about $-0.3$ to $+0.4$. The spread of the
time scale across drive models is $\sigma=0.68$ on the log scale. All four chains start from
the same point in the wear-out basin, so $\hat R$ would not reveal a second mode.

The year effect is where this model differs most from the base variant
(Table~\ref{tab:exthyper}). Under the base variant it falls almost steadily, to 0.48 of its 2013
level by 2017 and 0.21 by 2026. Under the hidden Markov model it doubles from 2013 to 2015
(2.3), falls to 0.71 by 2020, and is back at 1.08 in 2026. From 2015 to 2020 it falls by a
factor of 3.2, where the base variant falls by 4.8 over the same years. This model also needs a
fleet-wide rate that fell through the second half of the 2010s, but by less. Its ratio of 2026
to 2013 near one depends on the starting year, a partial year of 17{,}000 drive-years that is
the first year of the data (not of the drives' lives). With the SMART state added the same model gives 0.39 for that
ratio. The two distributions divide the record between age and calendar time differently
(Section~\ref{sec:apc}), the held-out years favour the base variant's division by 275 log units,
and neither is tested beyond the first eight years of a drive's life.

\section{Solid-State Drives and Small Hard Drives}\label{sec:classes}

The models so far use hard drives of 1.4~TB and more. The daily files also hold
3{,}936 solid-state drives of 25 models, first seen in 2018, and 4{,}008 hard drives below
1.4~TB or of unreported capacity, of 51 models. Backblaze uses both mainly as boot drives; the
files do not mark the role, so the classes are defined by drive model and capacity. Two
solid-state models whose names do not contain ``SSD'', WDC WDS250G2B0A (191 drives) and
Seagate IronWolf ZA250NM10002 (205), are reclassified here from the shared table's
small-drive category.

\begin{table}[h]
\centering
\caption{The Three Classes of Drive. Drive-years are first-to-last calendar spans, including gaps; failures count drives with a failure flag. The data hard drives are counted in full. The lifetime fits use the 504{,}624 of them with a usable power-on age (Section~\ref{sec:age}), which carry 31{,}881 failures.}\label{tab:classtotal}
\small\begin{tabular}{@{}lrrrrr@{}}
\toprule
Class & Drives & Models & Drive-years & Failures & Failures per drive-year (\%) \\
\midrule
Data hard drives & 504{,}656 & 99 & 2{,}023{,}459 & 31{,}899 & 1.58 \\
Hard drives below 1.4 TB & 4{,}008 & 51 & 18{,}321 & 1{,}432 & 7.82 \\
Solid-state drives & 3{,}936 & 25 & 16{,}784 & 235 & 1.40 \\
\bottomrule
\end{tabular}
\end{table}

\begin{table}[h]
\centering
\caption{Drive-Years and Failure Rates by Year of Power-On Age and Class. Rates are failures per drive-year in percent. Solid-state drives exclude the model HP SSD S700. A dash marks fewer than 100 drive-years.}\label{tab:classage}
\small\begin{tabular}{@{}lrrrrrr@{}}
\toprule
& \multicolumn{2}{c}{Data hard drives} & \multicolumn{2}{c}{Small hard drives} & \multicolumn{2}{c}{Solid-state} \\
Age (yr) & Drive-years & Rate & Drive-years & Rate & Drive-years & Rate \\
\midrule
0--1 & 437{,}139 & 1.09 & 2{,}623 & 4.69 & 3{,}151 & 1.97 \\
1--2 & 417{,}014 & 1.31 & 2{,}757 & 6.06 & 3{,}122 & 0.54 \\
2--3 & 349{,}350 & 1.52 & 2{,}681 & 10.11 & 2{,}872 & 0.77 \\
3--4 & 276{,}443 & 1.68 & 2{,}418 & 11.99 & 2{,}421 & 0.41 \\
4--5 & 219{,}618 & 1.84 & 2{,}351 & 9.53 & 1{,}808 & 0.55 \\
5--6 & 141{,}891 & 2.27 & 1{,}945 & 7.61 & 1{,}068 & 0.28 \\
6--7 & 81{,}332 & 2.70 & 1{,}573 & 7.82 & 289 & 0.00 \\
7--8 & 57{,}598 & 2.53 & 1{,}088 & 4.87 & 28 & --- \\
8--9 & 34{,}679 & 1.82 & 542 & 3.69 & 12 & --- \\
\bottomrule
\end{tabular}
\end{table}

\begin{table}[h]
\centering
\caption{Left-Truncated, Right-Censored Fits on Power-On Age by Class. Fits are pooled and with a free multiplier per drive model. The AIC and the exponents are for the fit with multipliers. Degenerate means that the second Weibull coincides with the first.}\label{tab:classfits}
\footnotesize\begin{tabular}{@{}p{6.4cm}lrrrl@{}}
\toprule
& & \multicolumn{2}{c}{Log-likelihood} & & \\
Class & Distribution & pooled & with drive model & AIC & Exponent(s) \\
\midrule
Hard drives below 1.4 TB (4{,}007 drives, 1{,}432 failures) & Exponential & -5{,}082.0 & -4{,}752.1 & 9{,}578 & --- \\
 & Weibull & -5{,}077.4 & -4{,}729.8 & 9{,}536 & 1.20 \\
 & Two Weibulls & -5{,}074.1 & -4{,}718.3 & 9{,}517 & 0.28, 1.35 \\
\addlinespace
Solid-state drives (3{,}518 drives, 232 failures) & Exponential & -1{,}196.0 & -481.3 & 991 & --- \\
 & Weibull & -993.2 & -478.5 & 987 & 0.85 \\
 & Two Weibulls & -993.2 & -478.5 & 991 & degenerate \\
\addlinespace
Solid-state, without one model (3{,}408 drives, 124 failures) & Exponential & -716.8 & -596.0 & 1{,}218 & --- \\
 & Weibull & -668.4 & -575.2 & 1{,}178 & 0.54 \\
 & Two Weibulls & -668.4 & -575.2 & 1{,}182 & degenerate \\
\bottomrule
\end{tabular}
\end{table}

\emph{Small hard drives} fail at 7.8\% a year, five times the rate of the data drives; the
rate peaks at 12\% in the fourth year of age (Table~\ref{tab:classage}). With a multiplier per
drive model, two competing Weibulls have exponents 0.28 and 1.35, against 0.34 and 1.60 for the
data drives under the same fit, and gain 12 log-likelihood units over a single Weibull for two parameters. Among
drive models with 500 drive-years, annual rates range from 1.4\% to 14\%.

\emph{Solid-state drives.} One model, HP SSD S700 250GB, has 108 failure flags on 110 drives
within 14 drive-years. A rate above 100 failures per 100 drive-years is possible over such
short follow-up: it is an annualized rate, not a one-year failure probability.
The files cannot tell a bad batch from a recording artefact, and the
model is set aside. The remaining 3{,}408 drives with a usable age have 124 failures in
14{,}800 drive-years, 0.84\% a year: 2.0\% in the first year of age and 0 to 0.8\% in each of
the next six. A Weibull with drive-model multipliers has exponent 0.54, a hazard falling with age
throughout the observed range. No failure is recorded after age six (336 drive-years), so the
second Weibull is not identified. The files do not establish the devices' workload. The oldest cohort is eight years old,
and 124 failures support no statement about wear-out in either direction. Nothing here
applies to solid-state drives under a storage workload.

\section{SMART Attributes, Temperature and Load as Predictors}\label{sec:cov}

We score the drive-level quantities of Section~\ref{sec:smartstate} as predictors
and compare the observed SMART states with the hidden states of the two-state model. The main
results are summarized in Section~\ref{sec:covsum}.

\subsection{The SMART States and the Hidden Markov Model}

Set beside the hidden Markov fit of Appendix~\ref{sec:hmm} ($\lambda_1=0.98\%$, $q=5.0\%$, and
$\lambda_2=6.0\%$ a year, the last loosely determined: 4.5\% to 8.4\%), the observed rates of
Section~\ref{sec:smartstate} (Table~\ref{tab:smartyear}) are $0.40\%$ for clear drives,
$3.9\%$ a year for a first warning, and $16.3\%$ for warned drives, the counterparts of $\lambda_1$, $q$
and $\lambda_2$. The hidden states of that fit are therefore not the SMART
states: the second state's failure rate is at most half the observed one ($\lambda_2=16\%$ costs
21 log-likelihood units), and the first state's is more than twice it. A model fitted only to
lifetimes does not establish that its hidden states correspond to SMART warnings.

\subsection{A Regression with Drive-Level Covariates}\label{sec:covreg}

The warning state and three other quantities a drive reports are put into a regression
and scored on held-out years: the counts of the five
warning attributes, temperature (attribute 194), and load cycles (193) and power cycles (12)
per year of power-on time. Every daily file was reduced to one row per drive and month: days
seen, failure flag, mean temperature, and each counter on the drive's last day of the month.
This gives 24{,}474{,}235 drive-months for the data drives
with a usable age, 2{,}003{,}021 drive-years and 31{,}881 failures (31{,}880 of them in cells
of the regression), under the event history of
Section~\ref{sec:population}: the 10{,}151 drive-months after a drive's failure month are not
exposure, the 44 earlier months with a repeated failure flag are not failures, and the failure
month's exposure ends on the failure day.

The model is a Poisson regression for the number of failures in a cell of drive-months, with
exposure as offset, age in twelve bins, manufacturer, and a normal random effect per drive
model (\code{mgcv::bam}). It has no year effect and no shocks.
Covariates enter as factors and are never taken from the month being predicted. Two
information sets are used:
\begin{itemize}[leftmargin=1.4em]
\item \emph{lagged}: the drive's values at the end of the previous month, a forecast one month
  ahead;
\item \emph{frozen}: the drive's values at the end of the previous calendar year, which is what
  is known on 1~January about the year to come. A drive that arrives during the year has the
  level ``new''.
\end{itemize}
Each model is trained through year $Y$ and scored on year $Y+1$ by the Poisson log score of
its cells, for $Y=2017,\dots,2025$. The cells differ between the two information sets, so
scores are comparable within a column of Table~\ref{tab:covscores} and not across.

\begin{table}[h]
\centering
\caption{Held-Out Log Score of Regressions with Drive-Level Covariates. The scores are over nine test years. Gain is over the regression without the covariates, and Years counts the years in which it gains.}\label{tab:covscores}
\footnotesize\setlength{\tabcolsep}{4pt}\begin{tabular}{@{}p{6.4cm}rrrrrr@{}}
\toprule
& \multicolumn{3}{c}{Known at end of previous month} & \multicolumn{3}{c}{Known on 1 January} \\
Covariates & Score & Gain & Years & Score & Gain & Years \\
\midrule
No drive-level covariate & -114{,}526 & --- & --- & -68{,}939 & --- & --- \\
SMART warning & -87{,}173 & $+27{,}353$ & 9 of 9 & -57{,}811 & $+11{,}128$ & 9 of 9 \\
SMART warning, by manufacturer & -86{,}397 & $+28{,}129$ & 9 of 9 & -56{,}864 & $+12{,}075$ & 9 of 9 \\
Counts of the five warning attributes & -78{,}516 & $+36{,}010$ & 9 of 9 & -54{,}556 & $+14{,}383$ & 9 of 9 \\
Temperature & -114{,}102 & $+424$ & 7 of 9 & -68{,}650 & $+289$ & 9 of 9 \\
Temperature relative to the drive model & -114{,}321 & $+205$ & 8 of 9 & -68{,}750 & $+189$ & 7 of 9 \\
Load cycles and power cycles per year & -113{,}674 & $+851$ & 8 of 9 & -68{,}503 & $+436$ & 7 of 9 \\
Counts, relative temperature, load and power cycles & -78{,}182 & $+36{,}344$ & 9 of 9 & -54{,}427 & $+14{,}512$ & 8 of 9 \\
\bottomrule
\end{tabular}
\end{table}

\begin{table}[h]
\centering
\caption{Failure-Rate Ratio of a Warned Drive to a Clear One. Ratios are given age and drive model, with 95\% intervals, from fits to all data.}\label{tab:covflag}
\small\begin{tabular}{@{}lrr@{}}
\toprule
& Warning at end of previous month & Warning on 1 January \\
\midrule
All manufacturers & 27.3 (26.6--28.0) & 9.6 (9.3--9.9) \\
\addlinespace
Seagate & 18.9 (18.3--19.5) & 6.8 (6.6--7.0) \\
HGST & 76.0 (71.4--80.8) & 35.7 (33.6--37.9) \\
Toshiba & 37.1 (34.9--39.4) & 11.6 (10.8--12.5) \\
WDC & 31.8 (23.1--43.8) & 14.9 (9.4--23.6) \\
\bottomrule
\end{tabular}
\end{table}

\emph{The warning attributes.} The warning alone gains 27{,}353 log units one month ahead and
11{,}128 one year ahead, in nine years of nine (Table~\ref{tab:covscores}); the counts of the
five attributes in four levels gain 36{,}010 and 14{,}383. The standard errors of these sums
from the year-to-year spread are 7{,}400 and 3{,}600. For comparison, the drive-model effect
was worth 2{,}569 as a point forecast (Section~\ref{sec:valid}) and the calendar shocks 2{,}050 in predictive score (B3 with drift over B0) on the scale of
Section~\ref{sec:bayesvalid}; the scales differ, since these scores are over finer cells, but
the order of magnitude does not depend on that. A drive warned at the end of last month
fails at 27 times the rate of a clear drive of the same model and age, and a drive warned
on 1~January at 9.6 times the rate over the year (Table~\ref{tab:covflag}). The ratio differs
by manufacturer, from 19 for Seagate to 76 for HGST. Differences in attribute reporting
complicate this comparison: positive readings of attributes 187 and 188 occur mainly on
Seagate drives, with exceptions described in Section~\ref{sec:smartstate}.
Letting the ratio vary by manufacturer gains a further 776 and 947.

\begin{table}[h]
\centering
\caption{Failure-Rate Ratios from the Regression with All Covariates. Ratios have 95\% intervals and come from fits to all data. The reference levels are a count of zero, a temperature more than 4$^\circ$C below the drive model's mean in that month, fewer than 300 load cycles, and less than one power cycle per power-on year.}\label{tab:covratios}
\footnotesize\begin{tabular}{@{}llrr@{}}
\toprule
Covariate & Level & End of previous month & 1 January \\
\midrule
5 reallocated sectors & 1--9 & 4.64 (4.43--4.86) & 3.97 (3.77--4.17) \\
 & 10--99 & 6.53 (6.24--6.82) & 5.59 (5.32--5.87) \\
 & 100+ & 10.35 (9.92--10.81) & 7.40 (7.03--7.78) \\
\addlinespace
187 reported uncorrectable & 1--9 & 3.01 (2.87--3.17) & 2.13 (2.01--2.25) \\
 & 10--99 & 3.71 (3.52--3.91) & 2.62 (2.45--2.80) \\
 & 100+ & 5.14 (4.80--5.51) & 3.13 (2.82--3.48) \\
\addlinespace
188 command timeouts & 1--9 & 1.36 (1.25--1.46) & 1.31 (1.19--1.45) \\
 & 10--99 & 1.87 (1.47--2.36) & 1.45 (1.04--2.03) \\
 & 100+ & 2.01 (1.92--2.12) & 1.52 (1.43--1.63) \\
\addlinespace
197 pending sectors & 1--9 & 18.33 (17.29--19.44) & 8.51 (7.94--9.11) \\
 & 10--99 & 22.31 (20.72--24.03) & 9.12 (8.38--9.94) \\
 & 100+ & 42.41 (38.82--46.34) & 16.64 (14.70--18.84) \\
\addlinespace
198 offline uncorrectable & 1--9 & 0.27 (0.25--0.29) & 0.47 (0.43--0.52) \\
 & 10--99 & 0.32 (0.30--0.35) & 0.56 (0.51--0.62) \\
 & 100+ & 0.37 (0.34--0.41) & 0.63 (0.54--0.72) \\
\addlinespace
Temperature relative to model ($^\circ$C) & $-4$ to $-2$ & 0.94 (0.90--0.98) & 0.95 (0.91--0.99) \\
 & $-2$ to 2 & 0.95 (0.92--0.98) & 0.98 (0.95--1.01) \\
 & 2 to 4 & 0.98 (0.94--1.02) & 1.05 (1.00--1.09) \\
 & above 4 & 1.06 (1.03--1.10) & 1.18 (1.14--1.22) \\
\addlinespace
Load cycles per power-on year & 300--1k & 1.17 (1.10--1.24) & 1.21 (1.14--1.28) \\
 & 1k--3k & 1.21 (1.13--1.30) & 1.20 (1.13--1.29) \\
 & 3k--10k & 1.16 (1.08--1.24) & 1.16 (1.08--1.25) \\
 & 10k--30k & 1.16 (1.08--1.25) & 1.06 (0.99--1.15) \\
 & 30k+ & 1.41 (1.30--1.54) & 1.08 (0.98--1.18) \\
\addlinespace
Power cycles per power-on year & 1--3 & 1.30 (1.22--1.38) & 1.27 (1.20--1.35) \\
 & 3--10 & 1.65 (1.56--1.75) & 1.57 (1.48--1.66) \\
 & 10--30 & 1.80 (1.67--1.93) & 1.63 (1.51--1.75) \\
 & 30+ & 2.30 (2.10--2.52) & 1.95 (1.77--2.14) \\
\bottomrule
\end{tabular}
\end{table}

In the joint fit (Table~\ref{tab:covratios}) pending sectors (197) carry the largest ratio,
18 to 42 one month ahead, then reallocated sectors (5), 4.6 to 10. Offline uncorrectable
sectors (198) have a ratio below one given the others; attributes 197 and 198 usually rise
together, and the coefficient says only that a drive with both is not as much worse as the
product of the two would make it. The regression with counts over-predicts the test years by
7\% to 11\% in total, where the regression without covariates under-predicts by 3\%.

\begin{table}[h]
\centering
\caption{Unadjusted Failure Rates by Covariate Level. Each covariate is taken at the end of the previous month.}\label{tab:covraw}
\footnotesize\begin{tabular}{@{}llrrr@{}}
\toprule
Covariate & Level & Drive-years & Failures & Rate (\%/yr) \\
\midrule
Mean temperature ($^\circ$C) & $<$22 & 141{,}674 & 3{,}172 & 2.24 \\
 & 22--26 & 294{,}768 & 5{,}288 & 1.79 \\
 & 26--30 & 452{,}431 & 6{,}318 & 1.40 \\
 & 30--34 & 472{,}212 & 6{,}345 & 1.34 \\
 & 34--38 & 371{,}929 & 5{,}481 & 1.47 \\
 & 38+ & 246{,}775 & 4{,}491 & 1.82 \\
\addlinespace
Load cycles per power-on year & $<$300 & 605{,}898 & 5{,}527 & 0.91 \\
 & 300--1k & 680{,}146 & 11{,}180 & 1.64 \\
 & 1k--3k & 248{,}351 & 4{,}025 & 1.62 \\
 & 3k--10k & 224{,}697 & 4{,}101 & 1.83 \\
 & 10k--30k & 159{,}560 & 3{,}179 & 1.99 \\
 & 30k+ & 42{,}638 & 1{,}864 & 4.37 \\
\addlinespace
Power cycles per power-on year & $<$1 & 151{,}111 & 1{,}290 & 0.85 \\
 & 1--3 & 786{,}675 & 11{,}352 & 1.44 \\
 & 3--10 & 712{,}019 & 12{,}177 & 1.71 \\
 & 10--30 & 221{,}932 & 4{,}092 & 1.84 \\
 & 30+ & 92{,}570 & 1{,}323 & 1.43 \\
\bottomrule
\end{tabular}
\end{table}

\emph{Temperature.} Unadjusted, the failure rate is highest for the coolest and the warmest
drives (Table~\ref{tab:covraw}); that mixes drive models, which run at different
temperatures. Given age and drive model the rate rises steadily with temperature: relative to
drives below 22$^\circ$C the ratio is 1.05 at 22--26$^\circ$C, 1.16 at 30--34$^\circ$C, and 1.76
(1.67 to 1.85) at 38$^\circ$C and above. Measured against the drive model's own mean
temperature in the same month, a drive more than 4$^\circ$C warmer fails at 1.32 times the rate
of one more than 4$^\circ$C cooler. As a predictor temperature is weak: it gains 424 log units
one month ahead and 289 one year ahead, one to two hundredths of what the warning counts gain, and
once those are in the model the ratio for the warmest group is 1.06 one month ahead and 1.18
one year ahead. The files do not give the position of a drive in its chassis or the airflow,
so temperature here may stand for location.

\emph{Load cycles and power cycles.} Drives with more power cycles per year of power-on time
fail more often: alone, the ratio rises from 1.55 at one to three a year to 2.8 at ten or more,
relative to fewer than one; with the warning counts in the model it is 1.30 to 2.30. Load
cycles show a step from the lowest group to all others (about 1.5 alone, 1.2 jointly); one
month ahead there is a second step above 30{,}000 a year, which the year-ahead fit does not
show. Together the two gain 851 log units one month ahead and
436 one year ahead. A power cycle is as likely to be a consequence of trouble in a pod as a
cause of trouble in a drive, and nothing here separates the two.

\section{Tracking the Best Forecaster}\label{sec:track}

Forecasters are compared by their totals over the folds in Section~\ref{sec:valid}, but their
ranking can change over time. We therefore evaluate mixtures of their count forecasts, in the
vocabulary of prediction with expert advice: an \emph{expert} is one of the candidate
forecasters, already fitted; a \emph{round} is one test period (a year or a quarter); the
\emph{game} is the sequence of rounds, in which a combined forecaster announces a distribution
before each round and is scored on it; and an \emph{oracle} is a comparator chosen with
knowledge of the outcomes, such as the single expert with the best total in hindsight. Herbster and
Warmuth's fixed-share algorithm~\cite{herbster1998tracking}, a descendant of the weighted
majority algorithm~\cite{littlestone1994weighted}, keeps a weight on each of $N$ experts and
forecasts with the weighted mixture of their distributions. After each round it multiplies each
weight by the probability that expert gave the outcome, then redistributes a fraction $\omega$
of the total weight evenly:
\begin{equation}
v_i = \frac{w_i\,P_i(d)}{\sum_j w_j\,P_j(d)},\qquad w_i' = (1-\omega)\,v_i + \frac{\omega}{N}.
\end{equation}
(Herbster and Warmuth share to the other $N-1$ experts; sharing to all $N$, as here, is the
same update with their sharing fraction set to $\omega(N-1)/N$.) With $\omega=0$ this is Bayesian model
averaging with a uniform prior: its log loss is within $\log N$ of the best expert's, and it can
concentrate on one expert. With $\omega>0$ it can follow a change of leader, and its cumulative
log loss is within a bound of that of the best sequence of experts with a given number of
switches. With $\omega=1$ the weights stay uniform (the experts themselves are still fitted), so
comparing it with smaller $\omega$ measures what the weight updates add to a uniform mixture.
A proper score does not guarantee that mixing improves a forecast.

The game is played on each drive model's failure count in a test period. A point forecaster
enters as a Poisson distribution with its summed cell means; a predictive forecaster enters as
the negative binomial of its gamma posterior for the drive model's total, which has the same
mean as its point twin and a larger dispersion. The quarterly game has 34 rounds and 21
experts: the three simple forecasts (constant rate, bathtub, per-year Weibull); M1, M2 and the
hidden Markov baseline each pooled, with plain frailty, with covariates and with covariates and
year calibration, as point forecasts; and the plain and covariate frailty versions of the same
three as predictive forecasts. The yearly game has 9 rounds and 31 experts: the same with M0
in place of the hidden Markov baseline, the five additive models A1 to A5, and the Bayesian
variants B0, B1 with drift, and B3 without year effect, with random walk and with drift.

The Bayesian variants enter only as point forecasts at their posterior means. What was saved of
their predictive distributions per drive model (quantiles, the distribution function at the
outcome, and the PIT) would have let them into the log-score game as Monte Carlo
probabilities, but not into the mixture's distribution function needed for
Table~\ref{tab:trackcalib}, so for consistency they were kept out of both. As a Poisson at its
mean, a Bayesian variant loses its extra dispersion and scores lower, as every point forecaster
does: the 25 point experts are 1{,}300 to 15{,}900 log units behind the best predictive expert,
with the bathtub last and B3 with drift the best of them.

Two games are played. The fleet game has one weight vector, and an expert's loss in a round is
its loss summed over drive models. The per-model game has a weight vector for each drive model,
started uniform when the model first appears. Drive models for which some expert has no
forecast in a round are left out of that round.

\begin{table}[h]
\centering\small
\caption{Tracking Forecasters Against the Best Single Expert. Each entry is the log score of the tracking forecaster minus that of the best single expert in hindsight (M2 + covariates, predictive), on each drive model's total failures per test period, summed over rounds. Positive is better. The oracle rows are chosen with knowledge of the outcomes and are restricted as stated. A mixture can exceed every oracle held to one expert at a time, but not the one that picks the best expert in every round.}\label{tab:track}
\footnotesize\setlength{\tabcolsep}{2.4pt}
\begin{tabular}{@{}lrrrr@{}}
\toprule
& \multicolumn{2}{c}{Quarterly, 34 rounds} & \multicolumn{2}{c}{Yearly, 9 rounds} \\
Forecaster & fleet game & per-model game & fleet game & per-model game \\
\midrule
Best single expert in hindsight (M2 + covariates, predictive in both) & 0 & -- & 0 & -- \\
Static mixture ($\omega=0$) & -3 & +400 & -3 & +783 \\
Fixed share, $\omega=1/T$ & +394 & +1{,}786 & +518 & +1{,}841 \\
Fixed share, $\omega=0.05$ & +399 & +1{,}815 & +515 & +1{,}806 \\
Fixed share, $\omega=0.1$ & +406 & +1{,}852 & +518 & +1{,}836 \\
Fixed share, $\omega=0.25$ & +415 & +1{,}886 & +522 & +1{,}879 \\
Fixed share, $\omega=0.5$ & +419 & +1{,}877 & +525 & +1{,}906 \\
Fixed share, $\omega=0.75$ & +417 & +1{,}829 & +526 & +1{,}914 \\
Uniform mixture every round ($\omega=1$) & +402 & +1{,}725 & +527 & +1{,}905 \\
Follow the leader & -58 & -- & -408 & -- \\
Oracle: best sequence with 1 switch & +275 & -- & +415 & -- \\
Oracle: best sequence with 2 switches & +333 & -- & +481 & -- \\
Oracle: best sequence with 3 switches & +355 & -- & +523 & -- \\
Oracle: best expert in each round & +491 & -- & +556 & -- \\
Oracle: best expert per drive model & -- & +519 & -- & +923 \\
\bottomrule
\end{tabular}
\end{table}

\begin{table}[h]
\centering\small
\caption{Calibration in the Per-Model Game. The table gives the coverage (\%) of the central 50, 80 and 95\% intervals of each forecaster's distribution for a drive model's failures in the test period, and the shares of randomized PIT values below 0.025 and above 0.975. It covers all model-rounds, 1{,}455 quarterly and 425 yearly.}\label{tab:trackcalib}
\footnotesize\setlength{\tabcolsep}{4pt}
\begin{tabular}{@{}lrrrrrrrrrr@{}}
\toprule
& \multicolumn{5}{c}{Quarterly} & \multicolumn{5}{c}{Yearly} \\
\cmidrule(lr){2-6}\cmidrule(lr){7-11}
Forecaster & 50 & 80 & 95 & $<$2.5 & $>$97.5 & 50 & 80 & 95 & $<$2.5 & $>$97.5 \\
\midrule
M2 + covariates, predictive & 62 & 73 & 82 & 10 & 11 & 49 & 63 & 73 & 14 & 16 \\
Static mixture ($\omega=0$) & 62 & 74 & 84 & 8 & 11 & 50 & 68 & 77 & 12 & 15 \\
Fixed share, $\omega=0.25$ & 66 & 81 & 91 & 6 & 6 & 53 & 74 & 88 & 7 & 7 \\
Uniform mixture ($\omega=1$) & 67 & 81 & 91 & 8 & 5 & 58 & 79 & 90 & 7 & 5 \\
\bottomrule
\end{tabular}
\end{table}

\begin{figure}[h]
\centering
\includegraphics[width=\textwidth]{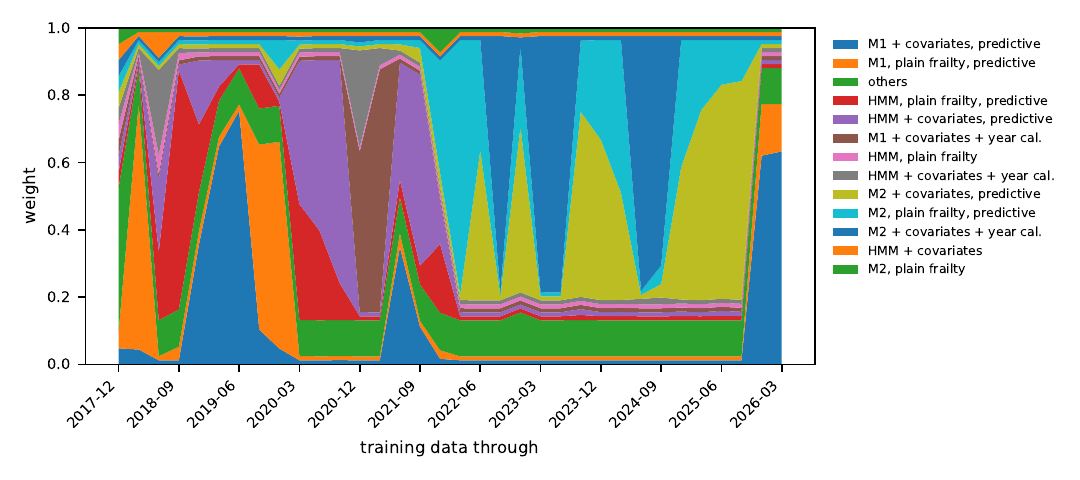}
\caption{Weights of the fixed-share forecaster ($\omega=0.25$) in the quarterly fleet game,
labelled by the end of the training data; the test quarter is the one that follows. Experts
that never reach 5\% after the uniform start are pooled as ``others''.}\label{fig:track}
\end{figure}

In the fleet game (Table~\ref{tab:track}) the static mixture ends about $\log N$ behind the best
expert (3 log units in each game; $\log N$ is 3.0 and 3.4), as Bayesian model averaging does
when one expert's cumulative lead is large. Every fixed-share setting with $\omega\ge1/T$ beats
the best single expert, by 394 to 420 log units in the quarterly game and 518 to 527 in the
yearly one. In the quarterly game this exceeds the best sequence of experts with three switches
chosen in hindsight (355) and falls 70 to 100 short of the oracle that picks the best expert in
every round (491). In the yearly game it matches the three-switch oracle (523) only for
$\omega\ge0.5$, and the per-round oracle (556) is about 30 ahead. In the per-model game the
gain is 1,725 to 1,886 log units in the quarterly game, three and a half times that of the best
expert for each drive model chosen in hindsight (519), and 1,841 to 1,914 in the yearly game,
twice that oracle's 924. Following the leader, which switches to the expert with the best
record so far, loses 58 and 408 log units.

The uniform mixture ($\omega=1$) gains 402 of the 420 in the quarterly fleet game and is
the best setting in the yearly fleet game (527). In the per-model games it is 161 log units behind
the best $\omega$ in the quarterly game and 9 behind in the yearly one (totals over the 79
drive models). Nearly all of the gain therefore comes from mixing the experts, and little from
updating the weights. The ranking does change (nine different experts are best in at least one
of the 34 quarters, the M2~+~covariates predictive forecaster in eight), but in this experiment
choosing the leader from past scores lowers the cumulative score.

Table~\ref{tab:trackcalib} compares the mixed distributions with the best expert's on the same
model-rounds. The best single expert's intervals are too narrow: its 95\% interval contains the
outcome 82\% of the time in the quarterly game and 73\% in the yearly one, and 10 to 16\% of
outcomes fall in each 2.5\% tail. The 95\% intervals of the fixed-share ($\omega=0.25$) and uniform mixtures
cover 88 to 91\%, with 5 to 8\% in each tail, and their 50\% intervals cover 53 to 67\%:
despite the better log scores, the outer intervals still under-cover and the inner ones
over-cover. The static mixture's 95\% intervals cover 84\% (quarterly) and 77\% (yearly). These figures pool all drive
models, including small ones whose intervals are easy to cover, and use the closed integer
interval of a discrete distribution. They are therefore not comparable with the coverages of
Section~\ref{sec:calib}, which are for drive-model-years with at least five expected failures,
with intervals read from simulated quantiles.

Figure~\ref{fig:track} shows the weights after each training period. The hidden Markov and
single-Weibull predictive forecasters held most of the weight through 2021, except that the
year-calibrated HMM point forecast had the largest single weight after the quarter ending
2018-06 (0.25), and the year-calibrated M1 point forecast after 2020-12 and 2021-03 (0.48 and
0.72). The two-Weibull predictive forecasters held most of it from 2022 to 2025, with the
year-calibrated M2 point forecast on top after five quarters in that span (0.68 to 0.76), and
the single-Weibull predictive forecasters in the last two quarters. The year-calibrated experts
are best in 7 of the 34 quarters.

A count from a mixture can be sampled by selecting an expert according to its weight and
drawing from that expert's count distribution; this does not specify a joint lifetime process
for drives or replacements. Within the candidates and periods tested, the uniform and
fixed-share mixtures improve the log score over every single expert, and tail coverage over the
best single expert, the only one whose coverage was measured. Whether the mixtures approach the
calibration of the Bayesian shock models is not settled: the shock models' saved distribution
functions at the outcome and PIT values would allow a tail comparison on a common footing, but
it was not made.

\section{Statistical Background}\label{sec:primer}

These data need ideas from survival analysis and Bayesian statistics, set out here with the
reasons for them; the formal definitions are in the sections that use them. Standard texts are
Lawless~\cite{lawless2003} and Meeker, Escobar and Pascual~\cite{meeker2022smrd} for lifetime
data, Aalen, Borgan and Gjessing~\cite{aalen2008} for frailty and event-history models, and
Gelman et al.~\cite{gelman2013bda} for Bayesian inference.

\subsection{Hazard, Survival, and What the Record Can Show}\label{sec:primerhaz}

Let $T$ be the age at which a drive fails. Its \emph{survival function} $S(t)=\Pr(T>t)$ is the
probability of still running at age $t$. Its \emph{hazard} $h(t)$ is the failure rate among
drives that have survived to $t$: for a short time $\mathrm{d}t$,
$\Pr(t<T\le t+\mathrm{d}t\mid T>t)\approx h(t)\,\mathrm{d}t$. The \emph{cumulative hazard} is
$H(t)=\int_0^t h(u)\,\mathrm{d}u$, and the three are tied by $S(t)=e^{-H(t)}$. The annualized
failure rate that Backblaze publishes, failures divided by drive-years, estimates the hazard
averaged over the drives and ages in the count. A hazard of 0.02 per year means about 20
failures a year among 1{,}000 drives of that age, to the extent that the hazard is flat over the
year, which it is not in the first weeks of life.

A hazard that falls with age describes early failures (``infant mortality''); a constant
hazard, a memoryless process with the same chance of failing in the next month at every age;
and a rising hazard, wear-out. The Weibull distribution,
$H(t)=(t/\eta)^\beta$, covers all three with one exponent: $\beta<1$, $\beta=1$
(the exponential distribution), and $\beta>1$. The scale $\eta$ is the age by which $1-e^{-1}$, or 63\%,
of drives have failed. For a gamma multiplier with mean one we call $\alpha$ (or $\kappa$) the
\emph{precision}: its variance is $1/\alpha$, so a larger precision means less variation.

Two features of the record determine the form of every likelihood in this paper. A drive still running when
it is last seen, because the data end or because it was removed in a migration, is
\emph{right-censored}: we know only that $T$ exceeds its last age. Dropping such drives, or
their running time, would make drives look worse.
A drive that first appears already in service, at power-on age $a_0$, is
\emph{left-truncated}: it is in the data only because it survived to $a_0$, and drives of the
same purchase that failed earlier were never seen. Its record can say nothing about the hazard
before $a_0$, so its contribution is conditioned on survival to $a_0$. Treating its age at
first sighting as zero, as a fit on ``age since first sighting'' does, mixes old drives in
with new ones and changes the fitted exponent, by as much as 0.12 in the early years of the record
(Section~\ref{sec:families}).

\subsection{Competing Risks and Mixtures}\label{sec:primercr}

Two different ways of combining distributions recur.

In a \emph{competing-risk} distribution every drive is exposed to several failure mechanisms at once,
and it fails at the first. If mechanism $j$ alone would give lifetime $T_j$ and the $T_j$ are
independent, then $T=\min_j T_j$, survival functions multiply, and hazards add:
$H(t)=\sum_j H_j(t)$. The independence is an assumption of the distribution, not an observation: the
record shows one failure per drive and never the mechanism. Two
competing Weibulls, one with $\beta_1<1$ and one with $\beta_2>1$, give a hazard that first falls
and then rises. A drive is sampled by drawing each $T_j$ and taking the minimum, which is how
the bathtub of Section~\ref{sec:elerathlit} is built.

In a \emph{mixture} each drive belongs to one of several subpopulations, and only that
subpopulation's distribution applies to it. Survival functions average:
$S(t)=p\,S_1(t)+(1-p)\,S_2(t)$. A mixture of components with constant but different hazards has a
falling hazard, because the weaker subpopulation fails first; more generally, a mixture of
components whose hazards do not increase has a hazard that does not increase~\cite{proschan1963}. A pooled
fleet of drive models with different but constant rates therefore shows apparent infant
mortality that no individual drive has, which is one reason the models separate drive models
before interpreting the age-specific hazard.

The limited-failure-population model of Appendix~\ref{sec:laws3} lies between the two: a fraction
$\pi$ of drives carries a defect that can kill it early, and all drives wear out. Its early
cumulative hazard levels off at $-\log(1-\pi)$, where that of a competing Weibull keeps
growing.

\subsection{Likelihood and Held-Out Scores}\label{sec:primerlik}

The \emph{likelihood} is the probability the model assigns to the observed data, as a
function of the parameters. A failed drive contributes the probability of failing on the day it
did, given that it was running when first seen; a censored drive, the probability of surviving
to its last day. The log-likelihood $\ell$ is the sum over drives, and the maximum-likelihood estimate is the
parameter value that maximizes it. A difference of one unit of
$\ell$ means the data are $e\approx2.7$ times as probable under one parameter value as under
another, and 100 units means $e^{100}$ times. A model always fits at least as well as one it
contains as a special case, so in-sample comparisons are penalized for parameter count (the
Akaike information criterion, $2k-2\ell$).

The comparisons that decide which models are carried forward are made on data the model has
not seen. A model fitted to the data through December of one year forecasts every cell (drive
model, month, age bin) of the next year, and the forecast is scored by the logarithm of the
probability it gave to the counts that occurred: the \emph{log score}. It is a proper scoring
rule~\cite{gneiting2007}: a forecaster maximizes its expected score only by reporting the
probabilities it believes, so neither spreading them out nor false confidence gains. (A mixture
of several forecasters, as in Appendix~\ref{sec:track}, is a different forecast, not a
misreport, and a proper score can reward it.)
A \emph{point forecast} gives only an expected count for each cell and is scored as a Poisson
distribution with that mean. A \emph{predictive distribution} also says how uncertain the count
is; it scores worse than a point forecast if it is too wide, and much better when the count
falls far from the expectation, where a Poisson with the wrong mean gives very small
probabilities. Scores of two models are compared year by year; the year-to-year spread of the
differences gives a standard error, so a total difference is judged against how much it varies
between years.

\subsection{Heterogeneity Between Drive Models: Random Effects}\label{sec:primerre}

The fleet contains about a hundred drive models whose failure rates differ by more than an order
of magnitude. There are three ways to treat the difference.
\begin{itemize}[leftmargin=1.4em]
\item \emph{Complete pooling}: one distribution for all drives. It ignores the difference, and its
  parameters describe whatever mixture the fleet happened to hold (Section~\ref{sec:families}).
\item \emph{No pooling}: a separate rate for each drive model. This works with much data, but
  a drive model with 2 failures in its first 100 drive-years gets a very uncertain 2\% per year,
  and one not yet seen gets no estimate.
\item \emph{Partial pooling}: each drive model's rate is a multiplier $Z_g$ on a common
  age curve, and the multipliers are themselves treated as draws from a population of drive
  models. This is a \emph{random effect}, called a \emph{frailty} in survival
  analysis~\cite{vaupel1979,hougaard2000}.
\end{itemize}
The population of multipliers is taken to be a gamma distribution with mean $\mu_g$ and precision
$\alpha$. The mean depends on what is known before a drive model fails at all (manufacturer,
helium status, capacity, year of introduction); the precision says how much drive models with the same attributes
still differ (the coefficient of variation is $1/\sqrt\alpha$). The gamma distribution is used
because it is the conjugate prior of a Poisson rate: if a drive model has had $D$ failures with
an exposure of $M$ (measured in expected failures at $Z=1$), the distribution of its multiplier
given those data is again gamma, with shape $\alpha+D$ and rate $\alpha/\mu_g+M$, and mean
\[
\E[Z_g\mid D,M]=\frac{\alpha+D}{\alpha/\mu_g+M}.
\]
This is a weighted average of the prior mean $\mu_g$ and the drive model's own ratio $D/M$,
with weights $\alpha/\mu_g$ and $M$: the more exposure, the more its own record counts. With
the fitted precision, $\alpha\approx1.32$, and $\mu_g=1$, a drive model with $D=2$ failures where
$M=1$ was expected gets $\E[Z_g]=(1.32+2)/(1.32+1)=1.43$, not the raw 2.0; one with $D=200$ where
$M=100$ was expected gets 1.99. This pull toward the mean is \emph{shrinkage}. It makes the estimates for small drive models
less noisy and on held-out years forecasts far better than complete pooling
(Section~\ref{sec:valid}); a drive model never seen gets the population distribution itself.

\subsection{Empirical Bayes}\label{sec:primereb}

The population parameters ($\alpha$ and the coefficients in $\mu_g$), together with the age
curve, must be estimated. In the \emph{empirical-Bayes} approach~\cite{morris1983,gaver1987}
they are estimated from the data. We estimate them by maximum likelihood after the multipliers
have been integrated out: the
likelihood of the parameters is the probability of all drive models' data averaged over the
possible values of each $Z_g$. For the gamma population this integral is closed-form
(Section~\ref{sec:hier}). The fitted population then serves as the prior for each drive model,
and the formula above gives its posterior.

A forecast for a drive model's count in a future cell is a Poisson count whose mean is
uncertain because $Z_g$ is: a \emph{negative binomial} count, with variance $m+m^2/a$ for a gamma
with precision $a$ and mean $m$. This empirical-Bayes predictive distribution carries the
uncertainty about each drive model's hazard multiplier. It omits the uncertainty in the population
parameters, which are plugged in as if known, and anything that changes from year to year;
the second omission makes its intervals far too narrow (Section~\ref{sec:calib}).

\subsection{Bayesian Inference}\label{sec:primerbayes}

A Bayesian model treats every unknown, including the age-curve parameters and the population
parameters, as a random quantity with a \emph{prior} distribution that states what is believed
before the data are seen. Bayes's theorem combines the prior with the likelihood:
\[
p(\theta\mid\text{data})=\frac{p(\text{data}\mid\theta)\,p(\theta)}{p(\text{data})},
\]
where the denominator, the \emph{marginal likelihood}, is the likelihood averaged over the
prior. The \emph{posterior} is a joint distribution over all the unknowns. Unlike a single best
estimate, it gives how uncertain the wear-out exponent is, how that uncertainty is correlated with
the year effect's, and the range of failure counts to expect next year with all of these
allowed for.

Most priors in this paper are weak: wide distributions that make absurd values (a wear-out
exponent of 50, say) improbable and leave the rest to the data.
The exceptions are the parameters of the year effect, which are learned from only a handful of
yearly steps, so their priors matter (Section~\ref{sec:bayesmodel}).
The model is \emph{hierarchical}: the drive-model multipliers have the gamma population as
their prior, and that population's parameters have priors of their own. The empirical-Bayes
model of Section~\ref{sec:hier} is close to the simplest Bayesian variant (B0, with no year
effect and no shocks) with the top level fixed at its maximum-likelihood estimate, though the
two are fitted to differently tabulated data (Section~\ref{sec:bayesmodel}); the Bayesian
variants that add year effects and shocks contain more than either.

\textbf{Computing the posterior.} With hundreds of unknowns the posterior has no closed form
and cannot be integrated on a grid. It is represented by a sample of a few thousand parameter
vectors; any quantity of interest (a survival probability, a forecast count, a ratio of two
drive models' rates) is computed for each draw, and the spread of those values is its posterior
uncertainty.

The draws come from Markov chain Monte Carlo, a sequence of parameter vectors, each drawn given
the one before, whose long-run share of time in any region equals that region's posterior
probability. We use Hamiltonian Monte Carlo~\cite{neal2011hmc} with the No-U-Turn
sampler~\cite{hoffman2014nuts}, which follows trajectories steered by the gradient of the log
posterior (computed by automatic differentiation), so that each step can move far in a
high-dimensional space while staying in probable regions; the No-U-Turn rule extends each
trajectory until it starts to double back. An initial \emph{warm-up} phase tunes the step size
and the scale of each parameter, and its draws are discarded.

Three diagnostics check that the chains have explored the whole
posterior~\cite{gelman2013bda,vehtari2021rhat,betancourt2017}:
\begin{itemize}[leftmargin=1.4em]
\item $\hat R$ compares the spread between several chains, usually started at different
  points, with the spread within them. It is 1 if they have converged to the same distribution.
  Vehtari et al.~\cite{vehtari2021rhat} recommend using a sample only if $\hat R<1.01$.
\item The \emph{effective sample size} (ESS) allows for the correlation of successive draws:
  1{,}000 correlated draws may carry the information of only 100 independent ones. Vehtari et al.\
  recommend an effective sample size above 400, in the tails as well as the bulk.
\item A \emph{divergence} is a step on which the numerical trajectory broke down, usually
  where the posterior curves too sharply to follow. It warns that part of the posterior may not
  have been visited and estimates may be biased~\cite{betancourt2017}. Divergence counts are
  reported with the results, and fits with many are treated as unreliable.
\end{itemize}
Table~\ref{tab:bayesdiag} and Appendix~\ref{sec:ext} report all three.

\textbf{Predictions and simulation.} The \emph{posterior predictive} distribution of next
year's failures takes a posterior draw, draws next year's random quantities (the year effect,
the shocks) given it, and then draws the counts; repeated over draws, it carries every source
of uncertainty in the model. A simulator does the same: each simulated system takes one
posterior draw and draws drive lifetimes from it, which carries parameter uncertainty into
the simulation.

\textbf{Comparing Bayesian forecasts.} The held-out log score of a Bayesian model is the
logarithm of the posterior predictive probability of the test year. That probability is the
ratio of two marginal likelihoods, that of the training and test data together and that of the
training data alone. The average of the predictive probability over posterior draws is
unbiased for the probability, but when that probability varies by many orders of magnitude
between draws, as it does here, the average rests on one or two draws and its logarithm, the
score, is biased downward. The ratio is therefore estimated from the two posteriors, by bridge
sampling~\cite{meng1996bridge} or by importance sampling where bridge sampling fails (Section~\ref{sec:bayesvalid}).

\subsection{Shared Shocks and Calendar Time}\label{sec:primershock}

Failures of different drives are not independent: in some years a drive model does much worse
than its age and history predict, and all its drives share that. A \emph{shock} is a random
multiplier common to a group, here all drives of one drive model in one calendar year. Given the
shock the drives fail independently; averaged over it, their failure counts vary more than a
Poisson count and are correlated with each other. With a gamma shock of mean 1 and precision
$\kappa$, a group's count is again negative binomial with variance $m+m^2/\kappa$. A simulator
reproduces the correlation by drawing one shock per group and year and applying it to every
drive in the group.

The fleet-wide level also changes with calendar year. It is modelled as a \emph{random walk
with drift} on the log scale: each year's level is the previous year's plus an average trend
plus a random step. The sizes of the trend and of the steps are learned from the years in the
training data, as few as five in the earliest fold. A forecast continues the trend, and its
uncertainty grows with the horizon as unseen steps accumulate.

\subsection{Hidden Markov Distributions}\label{sec:primerhmm}

A hidden Markov (Coxian phase-type) lifetime distribution describes a drive as a small
continuous-time Markov chain: it starts in a first state, fails from it at rate $\lambda_1$ or moves to a second state
at rate $q$, and in the second state fails at rate $\lambda_2$. The state is \emph{hidden}: the
record shows only whether the drive has failed. Each state's waiting time is exponential, so a
lifetime is sampled by simulating the chain with two or three exponential variates, and the
survival function is given by the matrix exponential of the chain's rate
matrix~\cite{neuts1981}: a sum of two exponentials when $\lambda_2\ne\lambda_1+q$, and
$e^{-\lambda_2t}(1+qt)$ on the surface $\lambda_2=\lambda_1+q$, where the fits in this paper
land. The hazard
moves monotonically from $\lambda_1$ toward $\min(\lambda_2,\lambda_1+q)$, so it is bounded,
unlike a Weibull wear-out term, which grows without bound. The hidden states have a possible
physical interpretation (healthy, then degraded) that a Weibull lacks. This does not make
the transition and second-state rates identifiable from lifetimes alone: distinct positive
rate triples can give exactly the same distribution (Appendix~\ref{sec:hmm}). That appendix
also examines whether the data prefer the distribution.

When the hazard of a hidden Markov model is multiplied by a factor, whether a constant or one that
changes with calendar year, the result is not the same chain with scaled rates (which would scale time, not the hazard),
and the chain's jumps can no longer be simulated directly. Two exact methods remain:
\emph{inversion}, which solves $H(T)=E$ for an exponential variate $E$ numerically, and
\emph{thinning}~\cite{lewis1979thinning}, which proposes candidate failure times at a rate
that bounds the true hazard and accepts each with probability equal to the ratio of the true
hazard to the bound. The cost of both is in Section~\ref{sec:sim}.

\subsection{Calibration}\label{sec:primercal}

\begingroup\widowpenalty=10000
A predictive distribution is \emph{calibrated} if its intervals contain the outcome as often
as they claim: about 95\% of observed drive-model-year counts inside their 95\% intervals. If
only half are, the intervals are too narrow, and a simulator built on the model would
understate how much a fleet's failures can vary. A finer
check is the \emph{probability integral transform}: the predictive probability of a count at or
below the one observed. For a correct model these values are uniform on $(0,1)$; too many near
0 and 1 means the predictive distributions are too narrow. For counts the transform is
randomized within the observed count's probability so that uniformity still
holds; Czado et al.~\cite{czado2009} describe this and a nonrandomized alternative.\par
\endgroup

\section{Boosted Forecasters: Background, the Lexis-Cell Hazard, Diagnostics and Cost}\label{sec:boostdiag}

In support of Section~\ref{sec:boost} we give the origins of boosting, the full account of the
boosted hazard on the Lexis cells, the diagnostic refits of the landmark ensemble that
forecasts without the SMART record, and the cost of a query.

\subsection{Background}\label{sec:boostbg}

AdaBoost and the Hedge algorithm come from the same paper~\cite{freund1997decision}. Hedge
generalizes the weighted majority algorithm, and the fixed-share algorithm of
Appendix~\ref{sec:track}~\cite{herbster1998tracking} adapts Vovk's related aggregating
algorithm. Hedge keeps multiplicative weights on a fixed set of experts,
multiplies each weight after every outcome by how well that expert did, and forecasts with the
weighted mixture. AdaBoost keeps multiplicative weights on the training examples: after each
round it raises the weight of the examples the latest weak learner got wrong, fits a new weak
learner to the reweighted examples, and adds it to the ensemble with a weight set by its error
rate. In this paper the fixed-share game of Appendix~\ref{sec:track} combines forecasters that
are already fitted, while a boosted ensemble is a single forecaster that could enter that game
as one more expert. Statistically, AdaBoost fits an additive model $F(x)=\sum_m f_m(x)$ one
small tree at a time so as to reduce the exponential loss $\mathrm{E}\,e^{-yF(x)}$ for labels
$y=\pm1$, whose population minimizer is half the log odds, so the ensemble's output is half a
logit~\cite{friedman2000additive}. Gradient boosting (Section~\ref{sec:boosttheory}) keeps the
stagewise additive fit and lets the loss be any differentiable function.

\subsection{A Boosted Hazard on the Lexis Cells}\label{sec:boostlexis}

The ensembles summarized in Section~\ref{sec:boostlexissum} use the Poisson loss: for a count $D$ with exposure $E$ and $F=\log\mu-\log E$, the loss is the
negative log-likelihood $Ee^{F}-DF$ and its negative gradient is $D-Ee^{F}$, so each tree is
fitted to the difference between observed and expected failures, cell by cell, which is the
quantity the Lexis likelihood of Section~\ref{sec:lexis} scores. Here $\log E$ is the
\emph{offset}, a term added to the ensemble's output with a fixed coefficient of one, so that
the trees model the rate and the exposure enters as it does in the Poisson regressions of
Section~\ref{sec:valid}.

Four ensembles are fitted to the HDD cells of the Lexis table, each with Poisson loss and log
exposure as the offset, on the rolling-origin folds of Section~\ref{sec:valid}: train through
December of year $Y$, test on year $Y+1$, for $Y$ from 2017 to 2025. Two are fixed-round
ensembles kept as references (learning rate 0.05, 400 rounds, 15 leaves, at least 30
cells per leaf, $L_2$ penalty 1; \code{HistGradientBoostingRegressor}) on log age, log
capacity, vintage and manufacturer, with and without the drive-model identity. Two are
LightGBM ensembles with the same tree size and penalty, feature subsampling of 0.9 and
LightGBM's smoothing of categorical splits, the drive-model identity, and the number of rounds
chosen by early stopping on the last training year (fitted to the years before it), then
refitted through it, which chose 219 to 570 rounds without the year feature and 226 to 743 with it, against the fixed 400
\unskip: one with the same
features, one with the calendar year added. (The fixed ensemble with the identity keeps the
250 largest drive models by exposure as categories, a cap that never binds on 99 harmonized models.) Training cells are the Lexis cells summed over
months within a year; the test cells are the monthly cells, as for every other candidate.

The two fixed
ensembles score $-31{,}930$ with the drive-model identity and $-31{,}984$ without it, 438 and
493 behind the empirical-Bayes point forecast. The hierarchy wins four of nine years against
the better of them. Early stopping and LightGBM's handling of the drive-model identity move
the ensemble without a year feature to $-31{,}696$. Adding the calendar year moves it to
$-31{,}143$: 348 ahead of the empirical-Bayes point forecast (five of nine years; 1.0 standard
errors of the sum), 292 ahead of the year-calibrated point forecast (1.0 standard errors),
374 behind additive A2 (which wins seven of nine years, 1.8 standard errors), and 1{,}697
behind the empirical-Bayes predictive.

Splitting the cells by whether the drive model was in the training data shows where these
differences lie. On the cells of previously seen drive models, which hold 24{,}849 of the
24{,}975 failures, the tuned ensemble with the year is 47 ahead of A2 and 330 ahead of the
empirical-Bayes point forecast. On new-model cells it is 421 behind A2, accounting for its
entire overall deficit of 374. There are 126 failures observed (55 in 2020); predicted: the tuned ensemble with the year 795 (465 in 2020); without the year 790 (422 in 2020); the empirical-Bayes point forecast 769 (611 in 2020); additive A2 251 (143 in 2020); the fixed covariates-only ensemble 247 (62 in 2020)
\unskip.
These totals show substantial over-prediction for new models by both the tuned ensembles
and the hierarchy's prior mean. Relative to the best score on those cells,
the nine predictive rows are 0 to 10 below the best candidate; the 6 point forecasts that give a new drive model a pooled rate (the constant rate, the per-year Weibull, the pooled distributions and the covariates-only ensemble) are 48 to 120 below; the 18 point forecasts with a per-model term are 60 (Additive A2) to 1{,}322 (M0, point) below, the tuned ensembles 480 and 476 beside the empirical-Bayes point forecast's 498 and the fixed ensemble with the identity 146; the bathtub, which carries nothing, is 278 below
\unskip. Carrying the uncertainty of the
multiplier helps the predictive rows, but the 126 failures in nine years provide limited
evidence about the ordering of the point forecasts for new models.

The year feature gains 553 log units on the cells over nine years, winning seven of nine
with a ratio to the standard error of 2.9. Its predicted-over-observed ratio still swings
with the fleet-wide rate: 1.39 in 2018, 0.78 in 2019, 2.62 in 2020, 0.75 in 2022 and 1.31
in 2025. The ensemble without the year feature swings similarly
(1.50 in 2018, 0.90 in 2019, 2.56 in 2020, 0.72 in 2022, 1.33 in 2025
\unskip). The Poisson score of the nine yearly
totals also improves with the year feature, from $-1{,}486$ to $-1{,}416$. This is a
different score from the sum over monthly Lexis cells. The summed predicted-over-observed
ratio changes from 1.17 without the year to 1.13 with it, against 0.99 to 1.00 for the
uncalibrated and year-calibrated M2 hierarchy and 1.13 for A2.

The scores of the year-calibrated hierarchy and the tuned trees are not clearly separated
against the variation across these nine years. The same holds for A2's 722-unit advantage
over the uncalibrated hierarchy (1.7 standard errors), of which 439 comes from new-model
cells. This does not establish equivalence, and no feature ablation separates the value of
the trees' interactions from the effect of their treatment of new drive models. The
1{,}697-unit gap between the tuned ensemble and the empirical-Bayes predictive is of the
size of the hierarchy's own 2{,}046-unit point-to-predictive gap. The explanation of
Section~\ref{sec:valid} applies: the predictive row is a distribution with dispersion,
whereas the ensemble supplies a plug-in Poisson forecast with no distribution over its
leaf values.

\subsection{Diagnostics of the Ensemble Without the SMART Record}

The boosted ensemble given only the lifetime models' inputs ranks drives worse than the
gamma-frailty hierarchy at one year in the 2021 to 2023 cohorts
(Section~\ref{sec:boostlandmark}). These are the refits and tables made to look for the cause. They were run on three
cohorts chosen after the main result was seen, and the test years are the development years of
the paper.

\begin{table}[h]
\centering\small
\caption{Tuning of the Boosted Landmark Forecasters by Yearly Landmark, 365-Day Horizon. The AUC is on complete follow-up, for the empirical-Bayes predictive forecast and for the base and SMART ensembles stopped on log loss and on AUC. Rounds are those chosen by early stopping. The last column is the first validation origin of the inner temporal split.}\label{tab:boostrounds}
\footnotesize\setlength{\tabcolsep}{3.5pt}
\begin{tabular}{@{}rrrrrrrrrrl@{}}
\toprule
& \multicolumn{3}{c}{AUC complete, log-loss stopped} & \multicolumn{2}{c}{AUC-stopped} & \multicolumn{4}{c}{Rounds} & \\
\cmidrule(lr){2-4}\cmidrule(lr){5-6}\cmidrule(lr){7-10}
Year & EB & base & SMART & base & SMART & base & SMART & base & SMART & Validation from \\
\midrule
2018 & 0.689 & 0.684 & 0.738 & 0.683 & 0.733 & 117 & 65 & 2 & 1 & 2016-04-01 \\
2019 & 0.597 & 0.607 & 0.666 & 0.553 & 0.620 & 109 & 31 & 4 & 1 & 2017-04-01 \\
2020 & 0.604 & 0.589 & 0.694 & 0.580 & 0.689 & 31 & 152 & 487 & 236 & 2018-04-01 \\
2021 & 0.669 & 0.556 & 0.724 & 0.541 & 0.725 & 46 & 51 & 13 & 44 & 2019-04-01 \\
2022 & 0.690 & 0.617 & 0.750 & 0.607 & 0.747 & 1 & 19 & 3 & 16 & 2020-04-01 \\
2023 & 0.679 & 0.623 & 0.758 & 0.591 & 0.753 & 20 & 53 & 1 & 20 & 2021-04-01 \\
2024 & 0.648 & 0.608 & 0.759 & 0.614 & 0.759 & 9 & 44 & 1 & 44 & 2022-04-01 \\
2025 & 0.666 & 0.638 & 0.789 & 0.648 & 0.786 & 6 & 35 & 45 & 310 & 2023-04-01 \\
\bottomrule
\end{tabular}
\end{table}

\begin{table}[h]
\centering\small
\caption{Diagnostic Refits of the Base Ensemble on Three Cohorts. Entries are the AUC on complete follow-up at one year for age alone, the hierarchy, the base ensemble as fitted, and its refits. The refits are stopped on AUC, run for 300 forced rounds, drop the calendar year, drop year and vintage, replace the drive model's identity by its smoothed label rate in the training rows or by its failure rate per drive-year from the Lexis cells before the landmark, or fit on the rows with complete follow-up only. The last two columns are the AUC on all drives of the hierarchy and of the base ensemble as fitted.}\label{tab:boostdiag}
\scriptsize\setlength{\tabcolsep}{3pt}
\begin{tabular}{@{}rrrrrrrrrrrrr@{}}
\toprule
& & & & & & \multicolumn{2}{c}{Without} & \multicolumn{2}{c}{Model as rate} & & \multicolumn{2}{c}{AUC all} \\
\cmidrule(lr){7-8}\cmidrule(lr){9-10}\cmidrule(lr){12-13}
Year & Age & Hier. & Base & AUC-stop & 300 r. & year & year, vint. & label & Lexis & Complete & Hier. & Base \\
\midrule
2018 & 0.615 & 0.689 & 0.684 & 0.683 & -- & -- & -- & -- & -- & -- & 0.638 & 0.650 \\
2019 & 0.429 & 0.597 & 0.607 & 0.553 & -- & -- & -- & -- & -- & -- & 0.585 & 0.606 \\
2020 & 0.489 & 0.604 & 0.589 & 0.580 & -- & -- & -- & -- & -- & -- & 0.580 & 0.534 \\
2021 & 0.568 & 0.669 & 0.556 & 0.541 & 0.558 & 0.553 & 0.550 & 0.585 & 0.552 & 0.604 & 0.614 & 0.507 \\
2022 & 0.614 & 0.690 & 0.617 & 0.607 & 0.601 & 0.591 & 0.590 & 0.502 & 0.623 & 0.613 & 0.689 & 0.617 \\
2023 & 0.658 & 0.679 & 0.623 & 0.591 & 0.620 & 0.583 & 0.562 & 0.571 & 0.595 & 0.611 & 0.666 & 0.619 \\
2024 & 0.646 & 0.648 & 0.608 & 0.614 & -- & -- & -- & -- & -- & -- & 0.630 & 0.620 \\
2025 & 0.649 & 0.666 & 0.638 & 0.648 & -- & -- & -- & -- & -- & -- & 0.670 & 0.643 \\
\bottomrule
\end{tabular}
\end{table}

Table~\ref{tab:boostdiag} reports the refits on the 2021 to 2023 cohorts, where the base
ensemble's AUC is 0.556, 0.617 and 0.623 against the hierarchy's 0.669, 0.690 and 0.679. Early
stopping on the log loss had halted these ensembles at 46, 1 and 20 rounds; 300 forced rounds
give 0.558, 0.601 and 0.620, and stopping on the validation AUC (13, 3 and 1 rounds) moves
the AUC by $-0.015$, $-0.010$ and $-0.032$. Dropping the calendar year moves it by $-0.003$,
$-0.026$ and $-0.040$, and dropping year and vintage by $-0.006$, $-0.027$ and $-0.061$.
Replacing the categorical identity by the model's smoothed failure rate in the training rows
moves it by $+0.029$, $-0.115$ and $-0.052$. Replacing it by the model's failure rate per
drive-year from the Lexis cells through the eve of the landmark (the hierarchy's own
information about the model, which in this diagnostic reaches training rows that predate it)
moves it by $-0.004$, $+0.006$ and $-0.028$. Fitting on the rows with complete follow-up only,
so that removal cannot pass as survival, gives 0.604, 0.613 and 0.611. That closes two fifths of
the 2021 gap and none of the 2022 or 2023 gaps; it is the largest single
change among the refits in 2021 and below the ensemble
as fitted in 2022 and 2023. The best refit of each year leaves the ensemble 0.06 to 0.07 short of the
hierarchy.

\begin{table}[p]
\centering\small
\caption{Age Profile of the 2021 and 2023 Cohorts and of the Base Ensemble's Training Rows. Left: the cohort by power-on age, with the drives, the share removed unfailed within the year, the observed failure rate on complete follow-up and on all drives (removed drives counting as non-failures, the ensembles' target), and the mean forecasts of the hierarchy and of the base and SMART ensembles. Right: the training rows in the same bins (quarterly origins from 2014 to one year before the landmark), with the rows, the share removed within the window, the label rate (failed within 365 days before any removal), and the hazard in failures per drive-year of exposure. All rates are in percent.}\label{tab:boostage}
\footnotesize\setlength{\tabcolsep}{3.5pt}
\begin{tabular}{@{}lrrrrrrrrrrr@{}}
\toprule
& \multicolumn{7}{c}{2021 cohort} & \multicolumn{4}{c}{Training rows} \\
\cmidrule(lr){2-8}\cmidrule(lr){9-12}
Age (years) & Drives & Rem. & Obs. & Obs. all & Hier. & Base & SMART & Rows & Rem. & Label & Hazard \\
\midrule
0.0--0.5 & 32{,}707 & 0 & 0.62 & 0.62 & 0.72 & 1.24 & 1.17 & 279{,}148 & 1 & 1.47 & 1.48 \\
0.5--1.0 & 20{,}555 & 0 & 0.77 & 0.76 & 0.84 & 0.93 & 1.11 & 277{,}495 & 2 & 1.71 & 1.74 \\
1.0--2.0 & 24{,}926 & 11 & 1.25 & 1.11 & 0.92 & 0.86 & 1.02 & 491{,}637 & 5 & 2.22 & 2.29 \\
2.0--3.0 & 17{,}853 & 77 & 2.50 & 0.57 & 1.83 & 1.04 & 1.25 & 376{,}691 & 7 & 1.76 & 1.84 \\
3.0--4.0 & 29{,}877 & 17 & 1.16 & 0.97 & 1.25 & 0.77 & 1.21 & 254{,}462 & 19 & 1.67 & 1.84 \\
4.0--5.0 & 20{,}510 & 0 & 1.08 & 1.08 & 1.42 & 0.63 & 1.03 & 118{,}309 & 52 & 1.46 & 2.10 \\
5.0--6.0 & 14{,}038 & 0 & 1.81 & 1.81 & 2.75 & 0.86 & 1.54 & 24{,}456 & 79 & 1.21 & 2.61 \\
6.0--7.0 & 1{,}990 & 0 & 0.40 & 0.40 & 0.75 & 0.41 & 0.43 & 308 & 91 & 0.97 & 4.08 \\
7.0--8.0 & 2 & 0 & 0.00 & 0.00 & 0.91 & 0.37 & 0.30 & 165 & 61 & 0.00 & 0.00 \\
\midrule
& \multicolumn{7}{c}{2023 cohort} & \multicolumn{4}{c}{Training rows} \\
\midrule
0.0--0.5 & 15{,}445 & 0 & 0.57 & 0.57 & 0.50 & 1.49 & 1.57 & 502{,}612 & 1 & 1.13 & 1.14 \\
0.5--1.0 & 13{,}427 & 0 & 0.61 & 0.61 & 0.80 & 1.59 & 1.79 & 480{,}145 & 2 & 1.35 & 1.37 \\
1.0--2.0 & 62{,}571 & 0 & 0.81 & 0.81 & 0.92 & 1.40 & 1.28 & 750{,}512 & 8 & 1.84 & 1.92 \\
2.0--3.0 & 54{,}015 & 0 & 1.67 & 1.67 & 1.19 & 1.74 & 1.47 & 539{,}070 & 17 & 1.52 & 1.68 \\
3.0--4.0 & 21{,}441 & 0 & 3.63 & 3.62 & 1.16 & 1.49 & 1.58 & 395{,}243 & 15 & 1.47 & 1.60 \\
4.0--5.0 & 3{,}854 & 1 & 3.81 & 3.79 & 0.90 & 1.36 & 1.53 & 301{,}274 & 20 & 1.40 & 1.60 \\
5.0--6.0 & 24{,}179 & 9 & 3.14 & 2.86 & 1.40 & 1.49 & 1.95 & 140{,}532 & 14 & 1.66 & 1.84 \\
6.0--7.0 & 20{,}096 & 4 & 2.16 & 2.08 & 1.64 & 1.44 & 1.74 & 36{,}037 & 1 & 2.13 & 2.17 \\
7.0--8.0 & 13{,}336 & 44 & 4.96 & 2.76 & 3.06 & 1.98 & 2.48 & 5{,}652 & 2 & 0.58 & 0.59 \\
8.0+ & 2{,}833 & 60 & 0.89 & 0.35 & 1.04 & 1.01 & 0.66 & 90 & 72 & 0.00 & 0.00 \\
\bottomrule
\end{tabular}
\end{table}

Table~\ref{tab:boostage} compares the two forecasters' mean rates by age with the observed
rate over all drives, the ensemble's target, in the two cohorts, with the ensemble's training
rows in the same bins. Where fewer than a tenth of a bin's drives were removed within the year,
the two forecasters differ most in two places. In the youngest bins the ensemble's rate is
above the observed rate and above the hierarchy's in both cohorts (above its training rows'
label rate for that age in 2023 and below it in 2021). In the five-to-six-year bin of 2021 the
ensemble is below the observed rate and the hierarchy above it, so the ensemble ranks those old
drives below the young ones and the hierarchy ranks them above. In the middle ages of 2023 both
forecasters are at about half the observed rates or less. No decomposition of the AUC gap by age
was made, and bin means cannot supply one: the AUC counts pairs of a failure and a survivor,
which may lie in the same bin or in different ones.

The five-to-six-year rows of the 2021 training set were few and mostly truncated: 24{,}500
among 1.8 million, 79\% retired inside their window, with a label rate (1.2\%) under half the
hazard (2.6\%). The refit on complete follow-up removes the truncation but also most of those
rows, and raises the 2021 AUC from 0.556 to 0.604, two fifths of the way to the hierarchy's 0.669;
the label thus accounts for part of the gap in that cohort, and not the rest. The Lexis-rate
refit hands the ensemble one number per model through the eve of the landmark and no new rows.
That information is deliberately retrospective, since the Lexis cells of the last year before
the landmark include outcomes the training rows do not; it recovers 0.006 of the 0.073 gap in
2022 and nothing in 2021 or 2023 (whole-cohort figures, not localized by age). The ensemble's
training origins end a year before the landmark, so the drive models deployed in that year
reach it as missing categories and the drives deployed in it not at all, while the hierarchy's
multipliers use every drive-day to the eve of the landmark. These diagnostics do not distinguish
restrictions of the model's form (one joint profile of age, model and year learned from the rows
the past provided, against a shared age distribution and a per-model multiplier) from differences in the
training information available to the two forecasters. Each refit changed one factor on three
cohorts, so interactions among the tuning, the information and the outcome definition are not
excluded; the rows are repeated observations of the same drives at overlapping origins within
shared fleet years, so their number is not a measure of evidence; and the three cohorts were
chosen after the main result was seen. That the two forecasters agree at 90 days is consistent
with a gap that opens where there is a year to extrapolate over, but no 90-day age table was made to test it.

\subsection{Cost of a Query}\label{sec:boostspeed}

A boosted forecaster is evaluated by walking every tree from root to leaf and summing the
leaf values, so its cost is the number of trees times their depth, with no closed form and no
sampling. Table~\ref{tab:boosttiming} in Appendix~\ref{sec:timingapp} gives the measured cost
through the LightGBM library's Python binding on one thread, for the ensembles of the 2025
cohort at one year. It was measured as the samplers were (Appendix~\ref{sec:speed}): every
round a fresh process pinned to one Cortex-X925 core of a GB10, the single-row calls on
2{,}000 drives drawn afresh in each round, and pilot-bench sessions with the same fixed
requirements, one session per ensemble and call on each of \frag{three} machines. In a
batch a tree costs \frag{12 to 18}~ns across the six ensembles, so a drive
costs \frag{0.11}~$\mu$s for the base ensemble, which early stopping
left with \frag{6} trees, \frag{0.53}~$\mu$s for
the SMART ensemble of \frag{35} trees, and
\frag{5.2}~$\mu$s for the AUC-stopped SMART ensemble of
\frag{310} trees. One drive per call costs \frag{16 to 28}~$\mu$s,
at least \frag{15.7}~$\mu$s more than the same drive in a batch: the per-call
overhead of the library and the binding. The number of trees is set by the stopping rule and
varies by two orders of magnitude between cohorts (Table~\ref{tab:boostrounds}). The costs
include the Python binding, so no cost ratio with the compiled samplers of
Appendix~\ref{sec:speed} is claimed. A forecast is also not a
lifetime: it is a probability of failing within $\tau$ for a drive in a known state, so a
simulator that used an ensemble would evaluate it once per drive per horizon step, with the
drive's SMART state simulated or held.

\section{Cost of Sampling and Prediction}\label{sec:timingapp}

We timed every lifetime sampler on two processors (Table~\ref{tab:benchall}), the benchmark
summarized in Section~\ref{sec:speedsum}, and the extensions, and we timed the prediction of the
boosted landmark forecasters (Table~\ref{tab:boosttiming}, discussed in
Appendix~\ref{sec:boostspeed}).

\subsection{The Benchmark of the Lifetime Samplers}\label{sec:speed}

Every sampler is implemented in a Rust program that uses WyRand as its generator, the
inverse-CDF convention $U\in(0,1]$, and the numerically stable form
$a\,[\exp\{\log(1+E/H(a))/\beta\}-1]$ for the remaining life of an aged Weibull drive. Each
draw picks a drive model at random, so table lookups are included. One round of a measurement
is a fresh process, pinned to one core and given its own seed, that makes $2\times10^6$ draws
($2\times10^5$ for the two slowest samplers) and reports the mean time per draw. Rounds are run
and analysed by pilot-bench under requirements fixed before any reading: a 95\% confidence
interval of the mean whose full width is at most 1\% of the mean, estimated from at least 200
batches of consecutive rounds, each batch large enough that the lag-1 autocorrelation of the
batch means lies within $\pm0.1$ (pilot-bench's strict preset), and a session limit of
1{,}800~s. Pilot-bench also checks the series for a change point and leaves out rounds it
takes for outliers. The cost of a session here is instead the mean of all its rounds: on the i5
up to \frag{0.3}\% of a session's rounds take twice its median time or
more, and pilot-bench's mean is up to \frag{2.55}\% lower than the mean of
all rounds (up to \frag{0.12}\% on the GB10).

Within one session the interval is narrow: the median 95\% half-width over all sessions is
\frag{0.20}\% of the mean and the largest \frag{0.50}\%. It
does not show how much a measurement varies between sessions, machines, cores or random seeds,
so each sampler was measured in several sessions on each machine, alternating between two cores,
each session with its own seeds and with the samplers in a different random order. The number
of sessions was set by a rule fixed before the first: on each platform, the 95\% half-width from
the spread of the session means ($t$ with $n-1$ degrees of freedom) must be at most 0.5\% of the
mean in every cell, with at least three and at most ten sessions per machine. On the
performance cores of \frag{three} NVIDIA GB10 machines (DGX Spark, Arm Cortex-X925
at 3.9~GHz, performance governor) this took \frag{3} sessions per
machine, \frag{nine} in all; on an Intel Core i5-8259U (4 cores, up to 3.8~GHz,
turbo enabled, one session at a time so that the clock does not depend on the other cores'
load), the second architecture, it took \frag{ten}. Table~\ref{tab:benchall} gives
the mean of the session means of each platform and the half-width of the 95\% interval from
their spread. Sessions took \frag{200 to 5{,}024} rounds (median \frag{200}).
\frag{No session reached the limit.} On the GB10 the between-session half-width is at most
\frag{0.15}\% of the mean (median \frag{0.05}\%),
and the largest difference between the means of two machines is
\frag{0.3}\%; on the i5 it is at most \frag{0.58}\%, with one cell above the target at the limit of ten sessions.
The machines were checked to be idle before their sessions; the i5 also serves a small
Mattermost instance. During one GB10 session another job ran on other cores (one-minute load
average up to \frag{1.70} at the start of a measurement);
the \frag{3} measurements that started with a load above 1.5 differ from the
other sessions of their sampler by at most \frag{0.07}\%, and are kept.

The i5 takes \frag{1.3 to 2.6} times as long as the GB10, least for the samplers
that invert a cumulative hazard numerically and most for those dominated by power functions.
Samplers within about \frag{57}\% of each other on one machine can change places
on the other; the large differences below hold on both (bisection, for example, is
\frag{23} times slower than the closed form on the GB10 and
\frag{24} times on the i5).

For the fixed distributions and the empirical-Bayes and Bayesian models, the measurements give four
results.
\begin{itemize}[leftmargin=1.4em]
\item A multiplier per drive model adds one table lookup to the distribution it is built on. A draw from
  the empirical-Bayes model or from B3 at posterior means takes \frag{30 to 31}~ns for a
  new drive and \frag{93 to 97}~ns for an aged one, against \frag{27}
  and \frag{93}~ns for the same two competing Weibulls at multiplier one:
  about \frag{10} million aged-drive lifetimes per second on one core.
\item Choosing a posterior draw and drawing one year effect and one drive-model-year shock
  brings an aged-drive lifetime to \frag{147}~ns.
\item Redrawing shocks in every year of a drive's life costs about \frag{154}~ns per year simulated: a
  normal and a gamma variate and two conditional Weibull draws for each year until the drive
  fails or reaches a cap of 60 years. With the drift continued for three years (the default) a
  new drive is simulated for \frag{24.5} years on average and a lifetime costs
  \frag{3.8}~$\mu$s. The cost
  therefore depends on the cap and on the drift assumption; with the drift continued
  indefinitely most drives reach the cap (Section~\ref{sec:apc}). A simulator with shared
  shocks would draw them once per drive model and year for the whole system and pay only the
  two conditional Weibull steps per drive and year. The \frag{154}~ns figure includes a normal and a
  gamma variate per drive-year that such a simulator would not draw, so it is an upper estimate
  for the shared scheme, which was not timed separately.
\item For an aged drive under the bathtub, bisection on the cumulative hazard costs
  \frag{2.4}~$\mu$s. The bathtub is a competing-risk model, so the remaining life
  is the minimum of three conditional Weibull draws, each in closed form:
  \frag{103}~ns, \frag{23} times faster, for the same
  distribution (mean remaining lives of \frag{7.9396} and
  \frag{7.9391} years over $2\times10^6$ draws).
\end{itemize}
A model with no closed form, such as the additive models of Section~\ref{sec:valid} or the
boosted hazard of Appendix~\ref{sec:boostlexis}, can be served from a table of its cumulative hazard
at the cost of one table per drive model: a binary search and an interpolation take about
\frag{29 to 34}~ns in the Rust benchmark and 30 to 56\unskip~ns in the C++ library, comparable with the closed
forms. Three of the lifetime distributions of Section~\ref{sec:families} and Appendix~\ref{sec:laws}, the gamma,
lognormal and generalized gamma, have no closed-form conditional draw. Inverting their
cumulative hazards by bisection at every draw costs 0.8 to 5.3\unskip\
$\mu$s, against 12 to 128\unskip\ ns for the distributions with a closed form
(the exponentiated Weibull, the Weibull mixture, GLFP and Weibull--Gompertz have closed-form
conditional draws, by inversion, component selection or a minimum). The C++ library ships the
three distributions with tabulated inverses, at 30 to 56\unskip\ ns per draw
(Appendix~\ref{sec:raid}).

\subsection{The Extensions}\label{sec:extspeed}

A persistent vault multiplier, one gamma variate per 1{,}200 drives, adds at most
\frag{1.0}~ns to the model at its posterior means and at most
\frag{0.6}~ns to a posterior draw with a year shock and a model-year
shock (the four-factor structure of Appendix~\ref{sec:extvault}, \frag{74}~ns
new and \frag{148}~ns aged).
The SMART two-state drive costs \frag{80}~ns for a new drive and
\frag{136}~ns for an aged one, about \frag{2.6} and
\frag{1.5} times the model at posterior means. It draws a time to the
warning and, when the warning comes first (\frag{43}\% of new
drives and \frag{34}\% of aged ones with the benchmark's rates,
counted over $2\times10^6$ draws; a further \frag{7.4}\% of aged
drives start flagged), a second conditional lifetime. It uses one constant warning rate, one multiplier and, for an
aged drive, a starting state drawn with the fleet's warned share. It illustrates what a
transition model would cost: it combines the pooled warning rate of Section~\ref{sec:smartstate}
with a frozen-state hazard multiplier from Appendix~\ref{sec:extsmart}, rather than jointly
fitted transition hazards. Its marginal lifetime
does not reproduce the lifetime distribution (Section~\ref{sec:covsum}).

Taken alone the hidden Markov model is of phase type and is the cheapest lifetime in
Table~\ref{tab:benchall} after the exponential, \frag{12}~ns new and \frag{25}~ns aged: two or three
exponentials and no power function. The fitted models multiply its hazard by the year effect
and the shocks, and the chain cannot then be simulated with scaled rates: scaling the rates by
$c$ gives $S(ct)$, not $S(t)^c$, and although $S(t)^c$ is of phase type for integer $c$ (the
minimum of $c$ independent copies), the representation grows with $c$ and does not exist for
real-valued, time-varying multipliers. Two exact methods remain. One inverts the cumulative
hazard numerically (bracketing, then safeguarded Newton steps to a relative tolerance of
$10^{-12}$); it costs \frag{410 to 550}~ns with a constant multiplier and
\frag{550 to 660}~ns with the shocks drawn. The other uses the fact that the hazard of this distribution lies between $\lambda_1$
and $\min(\lambda_2,\lambda_1+q)$, whichever is smaller, and never exceeds the larger,
$h_{\max}$: candidate failure times are drawn from a Poisson process at the multiplier times
$h_{\max}$ and each is accepted with probability $h(t)/h_{\max}$ (thinning). That costs
\frag{48 to 52}~ns with a constant multiplier and \frag{106 to 111}~ns with
the shocks drawn, against \frag{74 to 147}~ns for a posterior draw of two competing
Weibulls with its shocks. Sampled by thinning with its shocks, the hidden Markov model costs
\frag{less than the Weibulls for an aged drive and more for a new one}.
At a multiplier of one both exact methods agree with the phase-type draw: over
\frag{20} runs of four million draws each, the mean lifetimes are
\frag{26.45, 26.45, 26.45} years for new drives and \frag{24.39, 24.39, 24.40} for aged ones
(phase-type draw, inversion, thinning), and no difference from the phase-type draw exceeds
\frag{1.6} standard errors.

\begin{table}[p]
\centering
\caption{Time to Draw One Remaining Lifetime. Times are in nanoseconds for a new drive (age 0) and for an aged drive (age uniform on 0.1 to 8 years). They are means over \frag{nine} pilot-bench sessions on \frag{three} GB10 machines and over \frag{ten} sessions on an Intel i5-8259U, each with the half-width of a 95\% confidence interval from the spread between sessions. The last column is the mean lifetime drawn for new drives in years, a check on the code and not a prediction.}\label{tab:benchall}
\scriptsize\setlength{\tabcolsep}{3pt}
\begin{tabular}{@{}p{5.6cm}rrrrrrrrr@{}}
\toprule
& \multicolumn{4}{c}{GB10 (Cortex-X925), \frag{9} sessions} & \multicolumn{4}{c}{Intel i5-8259U, \frag{10} sessions} & \\
\cmidrule(lr){2-5}\cmidrule(lr){6-9}
& \multicolumn{2}{c}{New} & \multicolumn{2}{c}{Aged} & \multicolumn{2}{c}{New} & \multicolumn{2}{c}{Aged} & Mean \\
Sampler & ns & $\pm$ & ns & $\pm$ & ns & $\pm$ & ns & $\pm$ & life \\
\midrule
\multicolumn{9}{@{}l}{\emph{Fixed distributions}} \\
One uniform variate (cost of the generator) & 2.2 & 0.0034 & 2.6 & 0.0034 & 3.7 & 0.0065 & 4.8 & 0.012 & 0.5 \\
Exponential, mean 35 years & 6.0 & 0.0049 & 6.8 & 0.0023 & 12.2 & 0.032 & 13.3 & 0.060 & 35.0 \\
Single Weibull (the 2025 fleet-year fit) & 14.8 & 0.011 & 45.3 & 0.034 & 37.0 & 0.049 & 86.7 & 0.16 & 34.8 \\
Three-Weibull bathtub: minimum of three closed-form draws & 23.8 & 0.011 & 102.9 & 0.036 & 56.9 & 0.11 & 200.4 & 0.20 & 10.9 \\
The same distribution, aged drives by bisection on the cumulative hazard & 23.0 & 0.031 & 2{,}401.7 & 0.90 & 57.4 & 0.10 & 4{,}872.4 & 2.7 & 10.9 \\
Two competing Weibulls: the empirical-Bayes baseline at multiplier one & 27.4 & 0.011 & 92.8 & 0.036 & 71.4 & 0.097 & 172.7 & 0.14 & 12.1 \\
\multicolumn{9}{@{}l}{\emph{Empirical-Bayes model}} \\
Multiplier of the drive model at its posterior mean & 30.4 & 0.016 & 96.7 & 0.034 & 79.9 & 0.12 & 177.8 & 0.15 & 20.6 \\
Multiplier drawn from its posterior & 56.6 & 0.026 & 129.3 & 0.056 & 131.9 & 0.15 & 231.7 & 0.31 & 23.9 \\
Tabulated cumulative hazard: binary search and interpolation & 28.5 & 0.029 & 33.5 & 0.016 & 56.3 & 0.030 & 57.1 & 0.058 & 19.3 \\
\multicolumn{9}{@{}l}{\emph{Bayesian model}} \\
Parameters and multipliers at posterior means & 30.6 & 0.015 & 93.2 & 0.038 & 80.1 & 0.080 & 177.5 & 0.17 & 17.3 \\
One posterior draw, one year shock, one model-year shock, held for the whole life & 74.3 & 0.061 & 147.3 & 0.086 & 169.0 & 0.19 & 270.5 & 0.25 & 27.1 \\
The fitted model: year effect and model-year shock advanced every year of life & 3{,}773.1 & 1.9 & 3{,}402.3 & 1.2 & 7{,}296.1 & 11 & 6{,}554.9 & 6.9 & 24.0 \\
Posterior means, times a persistent multiplier per vault & 31.5 & 0.024 & 94.2 & 0.037 & 81.5 & 0.058 & 180.2 & 0.23 & 20.0 \\
One posterior draw with both shocks, times a vault multiplier & 74.3 & 0.037 & 147.9 & 0.095 & 169.7 & 0.23 & 270.1 & 0.25 & 30.9 \\
Two-state illustration: time to flag, then a multiplied hazard (not a fitted model) & 80.5 & 0.036 & 135.9 & 0.051 & 168.5 & 0.22 & 254.0 & 0.26 & 12.7 \\
\multicolumn{9}{@{}l}{\emph{Two-state hidden Markov model}} \\
The distribution alone, rates per drive model (phase-type draw) & 11.8 & 0.0067 & 25.2 & 0.013 & 25.0 & 0.072 & 43.7 & 0.063 & 26.5 \\
Hazard times its fitted 2026 level, by thinning & 52.0 & 0.033 & 47.8 & 0.046 & 85.1 & 0.099 & 78.6 & 0.085 & 25.0 \\
Hazard times year and model-year shocks, by thinning & 110.8 & 0.14 & 106.0 & 0.040 & 185.9 & 1.1 & 177.1 & 0.23 & 39.8 \\
Hazard times its fitted 2026 level, by inversion & 408.8 & 0.20 & 551.8 & 0.18 & 545.5 & 0.53 & 711.3 & 0.59 & 25.0 \\
Hazard times year and model-year shocks, by inversion & 549.1 & 0.23 & 658.4 & 0.24 & 717.0 & 0.44 & 866.9 & 0.40 & 39.6 \\
\bottomrule
\end{tabular}
\end{table}

\begin{table}[h]
\centering\small
\caption{Prediction Cost of the Boosted Landmark Forecasters. The cohort is that of 2025 at one year (300{,}789 drives). Costs are measured through the LightGBM library's Python binding on one thread, pinned to one Cortex-X925 core of a GB10, for the whole cohort in one call and for one drive per call (2{,}000 drives drawn afresh in every round). Each cost is the mean over \frag{three} machines of one pilot-bench session each, under the requirements of Appendix~\ref{sec:speed}, with the half-width of a 95\% interval from the spread between the machines. The single-row cost includes the binding's per-call overhead.}\label{tab:boosttiming}
\footnotesize
\begin{tabular}{@{}llrrlrr@{}}
\toprule
Set & Stopped on & Trees & Features & Call & $\mu$s/row & $\pm$ \\
\midrule
base & log loss & 6 & 9 & batch & 0.11 & 0.00034 \\
base & log loss & 6 & 9 & single row & 15.86 & 0.041 \\
base & AUC & 45 & 9 & batch & 0.56 & 0.00055 \\
base & AUC & 45 & 9 & single row & 16.62 & 0.051 \\
flag & log loss & 26 & 11 & batch & 0.37 & 0.00022 \\
flag & log loss & 26 & 11 & single row & 16.33 & 0.039 \\
flag & AUC & 301 & 11 & batch & 4.26 & 0.021 \\
flag & AUC & 301 & 11 & single row & 24.88 & 0.15 \\
smart & log loss & 35 & 36 & batch & 0.53 & 0.00054 \\
smart & log loss & 35 & 36 & single row & 17.12 & 0.061 \\
smart & AUC & 310 & 36 & batch & 5.21 & 0.012 \\
smart & AUC & 310 & 36 & single row & 28.32 & 0.16 \\
\bottomrule
\end{tabular}
\end{table}

\section{How Modeling Assumptions Change Simulated Data-Loss Risk}\label{sec:raid}
\newcommand{\rot}[1]{\rotatebox{90}{#1}}

We compare the loss probabilities produced by the lifetime samplers in an illustrative RAID
simulation. The comparison is of the consequences of the specified lifetime distributions and
simulation assumptions; it is not an estimate of the data-loss probability of Backblaze's
operating system. The simulator is supplied with the accompanying C++ library.

\subsection{The Simulation}

An array of $D$ data drives and $P$ parity drives loses data when more than $P$ of its drives
are failed at the same time. Three layouts are run: RAID~5 with $7+1$, RAID~6 with $8+2$, and the
$17+3$ code of Backblaze's own storage vaults. Each trial starts the array with $D+P$ drives of
the same age and runs it for a ten-year mission. A failure starts a repair whose duration
follows a Weibull distribution with exponent 1.5 and mean 168 hours; means of 24 and 720 hours
are sensitivity cases. At repair completion an age-zero replacement begins service. The
replacement cannot fail during reconstruction, and repairs proceed in parallel without
resource limits. A trial ends at the first loss or at ten years. Latent sector errors, placement and decoder behavior are omitted.

The quantity reported is the probability of data loss within the mission, estimated as the
share of trials with a loss and given with a Wilson 95\% interval; a cell with no loss in any
trial is reported as an upper bound. The program also reports
$T/\hat p-T+\bar t_{\mathrm{loss}}$, where $T$ is the mission length and $\bar t_{\mathrm{loss}}$
the mean loss time among trials with a loss: a renewal estimate under independent resets to
the initial array state after a loss or a completed mission. It is not the mean loss time of
an array that continues aging beyond $T$, and it is undefined when no loss is observed. Every
cell uses
100,000,000\unskip\ trials for the RAID~6 and $17+3$ layouts, ten million for RAID~5, and
20,000,000\unskip\ for the rebuild-time sensitivity; the whole matrix is
764\unskip\ cells and ran on four 20-core Arm machines in under three hours.
Every cell was run by one build of the library (revision \code{49ea2f6\unskip}),
with each cell's seed derived from its specification. Each result row carries digests of the
sources, of the simulator binary and of the run settings, and the figure and table builder
refuses rows that are malformed or that come from more than one build.
Failures per drive-year are computed over the nominal exposure of the whole mission; a lost
array stops early, which understates them by under 2\% in the fleet table and by up to half for
the drive models of Table~\ref{tab:raidallmodels} that lose most arrays.

Each lifetime model enters the simulation through one call, the remaining life of a drive that
has survived to a given age (Section~\ref{sec:sim}). The statistical models with a drive-model
multiplier are run for six drive models that span the fitted range, for a drive model absent
from the fit (a 20~TB Seagate introduced in 2027, drawn from the prior), and for the fleet
population of the timing benchmark, in which every drive of the array draws its drive model at
random from the 99 fitted ones. The Bayesian model with shocks (B3) is run in the three
per-drive forms of Section~\ref{sec:sim} and in the shared-context form. For an array of one
named drive model, the shared context shares a posterior draw, calendar path and model-year
shocks across the array. In the uniform-mixture experiment, positions assigned the same
drive model share its shock, but the shock size is computed for one drive rather than the
number of positions assigned that model. The baselines that the fitted models replace are the
fixed three-component benchmark of the simulators, its common-scale refit, the April~2026 per-year Weibull and
the landmark comparator Weibull for 2025.

For a drive model absent from the fit, the fully shared context draws one multiplier and one
model-year shock path and reuses them across positions and replacements.

\subsection{Failure Probabilities for a New Drive}

\begin{figure}[tbp]
\centering
\includegraphics[width=\textwidth]{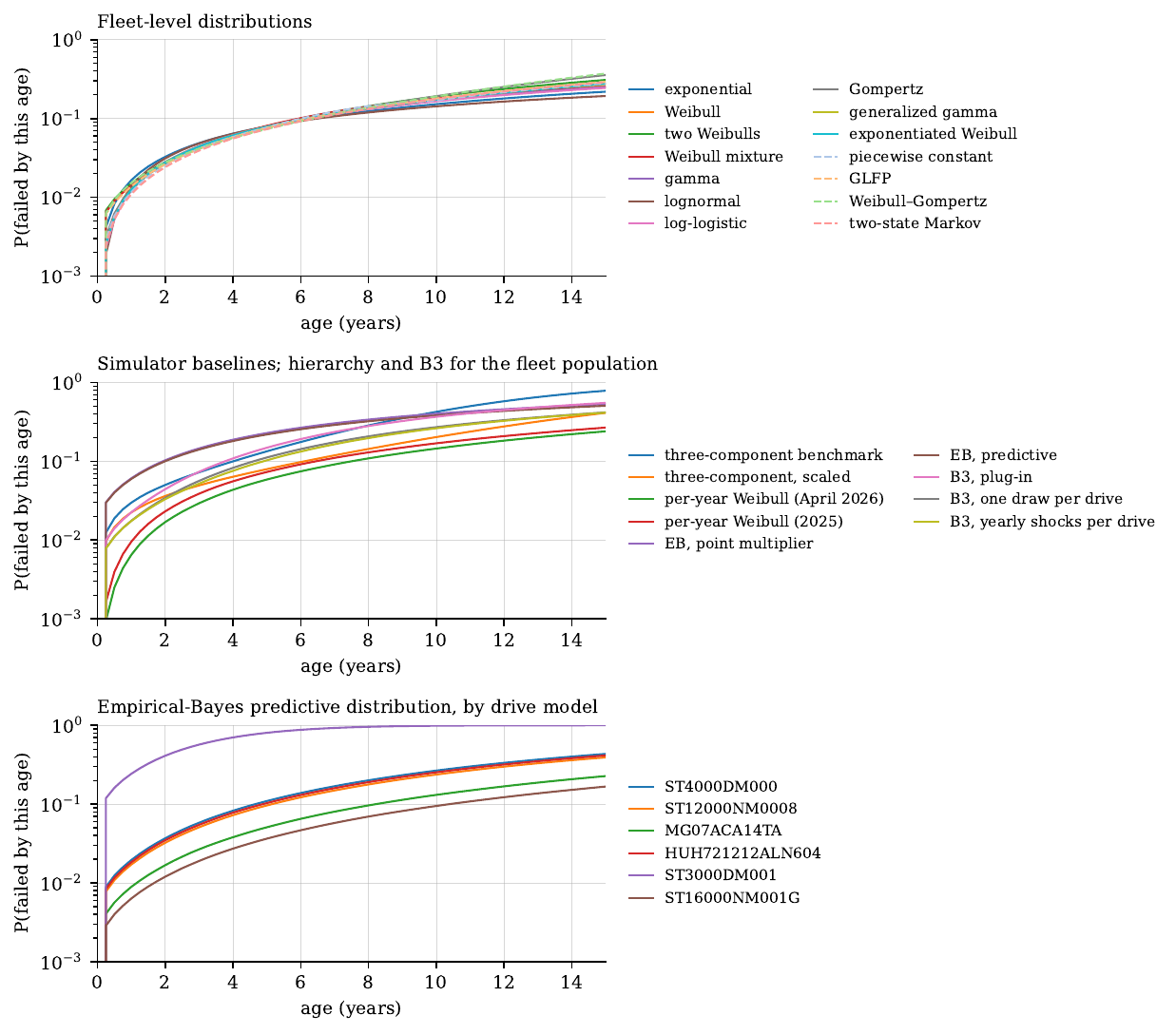}
\caption{Probability that a new drive has failed by a given age (log scale), from two million
draws of each sampler. Top: the fleet-level distributions fitted to all data drives on power-on age.
Middle: the simulators' baselines (the fixed three-component benchmark, its common-scale refit, the April~2026
per-year Weibull, the landmark comparator Weibull for 2025) and the hierarchy and the Bayesian
model for the fleet population of the timing benchmark (a drive model drawn uniformly from the
99 fitted ones for each drive). Bottom: the empirical-Bayes predictive distribution for six drive
models.}\label{fig:raidsurv}
\end{figure}

Figure~\ref{fig:raidsurv} shows the failure curves the simulation draws from. The fourteen
fleet-level distributions differ by a factor of 1.5 in the probability of failure within the first year
(the exponential highest, the two-state model lowest), by 1.1 to 1.15 at four to six years, where
the data are densest, and by 1.35 at ten years. Beyond ten they separate further, at ages
where the wear-out exponent is weakly determined (Section~\ref{sec:apc}). The fixed
three-component benchmark lies above every fitted distribution from the first day: in the simulation it gives
4.75\unskip\% failures per drive-year against
2.04\unskip\% for the two-Weibull distribution fitted to this fleet, a ratio
of 2.33\unskip. The spread between drive models (bottom
panel) exceeds the spread between fleet-level distributions: the ST3000DM001's multiplier on the
hierarchy's baseline is 6.9\unskip\ against
0.16 to 0.49\unskip\ for the other five (Table~\ref{tab:models}), and it
has lost 24\unskip\% of its drives at one year when the others
have lost between 0.6\% and 2.0\%\unskip. In the fleet population of the
middle panel, each drive's model is drawn uniformly from the 99 fitted ones, so the small
drive models with large multipliers weigh as much as the large ones with small multipliers.
That population fails at 4.8\unskip\% per drive-year against the
2.04\unskip\% of the exposure-weighted fleet. The hierarchy's rows for
that population in the tables below are therefore not a fleet average. They describe an array
whose drives are of unknown, uniformly chosen models; a simulation of real arrays should name
its drive model.

\subsection{Loss Probabilities}

\begin{table}[tbp]
\centering\small
\caption{Probability of Data Loss Within a Ten-Year Mission for Arrays of New Drives. Results are by lifetime model and layout, with the model's failures per drive-year in the RAID~5 runs. A value preceded by $<$ is the upper 95\% limit of a cell with no loss. The hierarchy and the Bayesian model are run for the fleet population of the timing benchmark, in which each drive's model is drawn uniformly. In the shared context each position keeps its drive model across replacements, positions of the same model share its shock, and the calendar path is shared.}\label{tab:raidfleet}
\begin{tabular}{@{}lrrrr@{}}
\toprule
Model & Failures / drive-yr (\%) & RAID~5, $7+1$ & RAID~6, $8+2$ & $17+3$ \\
\midrule
exponential & 1.64 & $2.9\times10^{-3}$ & $5.9\times10^{-6}$ & $6.0\times10^{-8}$ \\
Weibull & 1.82 & $3.6\times10^{-3}$ & $8.3\times10^{-6}$ & $1.6\times10^{-7}$ \\
two Weibulls & 2.04 & $4.8\times10^{-3}$ & $1.7\times10^{-5}$ & $5.6\times10^{-7}$ \\
Weibull mixture & 1.93 & $4.3\times10^{-3}$ & $1.2\times10^{-5}$ & $4.3\times10^{-7}$ \\
gamma & 1.80 & $3.5\times10^{-3}$ & $8.3\times10^{-6}$ & $2.0\times10^{-7}$ \\
lognormal & 1.54 & $2.6\times10^{-3}$ & $5.2\times10^{-6}$ & $1.1\times10^{-7}$ \\
log-logistic & 1.78 & $3.5\times10^{-3}$ & $7.8\times10^{-6}$ & $2.1\times10^{-7}$ \\
Gompertz & 2.05 & $4.8\times10^{-3}$ & $1.5\times10^{-5}$ & $3.8\times10^{-7}$ \\
generalized gamma & 1.86 & $3.8\times10^{-3}$ & $9.6\times10^{-6}$ & $2.1\times10^{-7}$ \\
exponentiated Weibull & 1.86 & $3.8\times10^{-3}$ & $9.4\times10^{-6}$ & $2.1\times10^{-7}$ \\
piecewise constant & 1.82 & $3.8\times10^{-3}$ & $9.6\times10^{-6}$ & $1.4\times10^{-7}$ \\
GLFP & 1.93 & $4.3\times10^{-3}$ & $1.3\times10^{-5}$ & $3.4\times10^{-7}$ \\
Weibull–Gompertz & 2.04 & $5.0\times10^{-3}$ & $1.7\times10^{-5}$ & $7.2\times10^{-7}$ \\
two-state Markov & 1.88 & $4.0\times10^{-3}$ & $1.0\times10^{-5}$ & $2.2\times10^{-7}$ \\
three-component benchmark & 4.75 & $3.0\times10^{-2}$ & $2.5\times10^{-4}$ & $1.8\times10^{-5}$ \\
three-component, scaled & 2.19 & $5.8\times10^{-3}$ & $2.1\times10^{-5}$ & $7.7\times10^{-7}$ \\
per-year Weibull (April 2026) & 1.52 & $2.7\times10^{-3}$ & $5.5\times10^{-6}$ & $1.1\times10^{-7}$ \\
per-year Weibull (2025) & 1.80 & $3.6\times10^{-3}$ & $9.1\times10^{-6}$ & $2.3\times10^{-7}$ \\
EB, point multiplier & 5.08 & $3.2\times10^{-2}$ & $4.1\times10^{-4}$ & $9.9\times10^{-5}$ \\
EB, predictive & 4.79 & $2.8\times10^{-2}$ & $3.8\times10^{-4}$ & $9.8\times10^{-5}$ \\
EB with SMART state & 5.36 & $3.3\times10^{-2}$ & $2.8\times10^{-4}$ & $2.8\times10^{-5}$ \\
SMART observed-state & 2.16 & $5.6\times10^{-3}$ & $1.9\times10^{-5}$ & $4.3\times10^{-7}$ \\
B0, posterior draw & 4.92 & $3.1\times10^{-2}$ & $4.6\times10^{-4}$ & $1.3\times10^{-4}$ \\
B3, plug-in & 4.18 & $2.1\times10^{-2}$ & $1.3\times10^{-4}$ & $7.0\times10^{-6}$ \\
B3, one draw per drive & 3.00 & $1.0\times10^{-2}$ & $4.5\times10^{-5}$ & $1.7\times10^{-6}$ \\
B3, yearly shocks per drive & 2.88 & $9.8\times10^{-3}$ & $4.3\times10^{-5}$ & $1.5\times10^{-6}$ \\
B3, shared context & 3.28 & $1.8\times10^{-2}$ & $3.2\times10^{-4}$ & $1.2\times10^{-4}$ \\
transition model (new drives) & 3.27 & $1.2\times10^{-2}$ & $5.4\times10^{-5}$ & $2.3\times10^{-6}$ \\
\bottomrule
\end{tabular}
\end{table}

\begin{figure}[tbp]
\centering
\includegraphics[width=\textwidth]{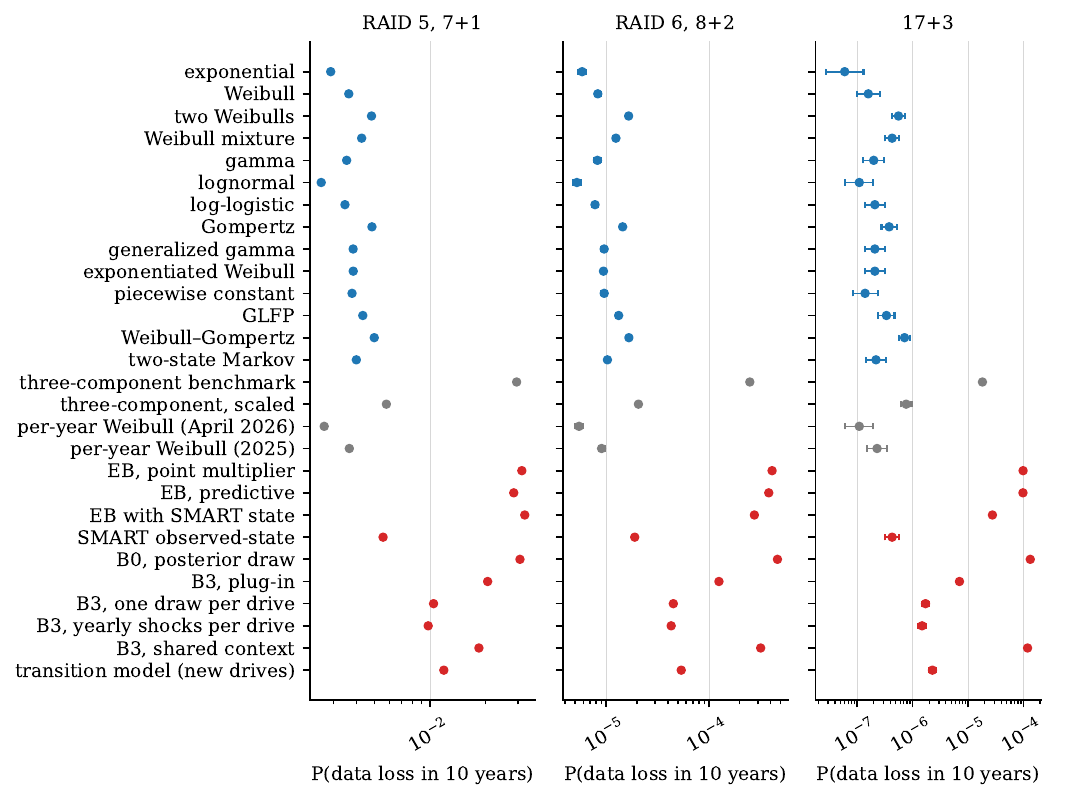}
\caption{The probability of data loss in ten years (log scale) under every model of
Table~\ref{tab:raidfleet}, for the three layouts; 95\% intervals, an arrowhead for cells with
no loss (upper limit). Blue: the fleet-level distributions; grey: the simulators' baselines; red: the
hierarchy and the Bayesian model for the fleet population.}\label{fig:raidploss}
\end{figure}

Table~\ref{tab:raidfleet} and Figure~\ref{fig:raidploss} give the loss probabilities for the
fleet-level distributions, the baselines, and the hierarchical and Bayesian models for the fleet
population. Across the fourteen fleet-level distributions the RAID~5 loss probability spans
$2.6\times10^{-3}$ to $5.0\times10^{-3}$\unskip, the RAID~6 probability
$5.2\times10^{-6}$ to $1.7\times10^{-5}$\unskip, and the $17+3$ probability
$6.0\times10^{-8}$ to $7.2\times10^{-7}$\unskip: factors of
1.9, 3.2 and 12.0\unskip\ (the last resting on
6\unskip\ losses in its smallest cell, so it is known only to
within a factor of two). The order of the distributions follows their failures per drive-year: the
three distributions with the highest rates (two Weibulls, Weibull--Gompertz, Gompertz, at 2.0\% to
2.1\%) take the top three places in RAID~5 and RAID~6 and three of the top four in $17+3$,
and the lognormal and exponential, at 1.5\% to 1.6\%, are the last two in every column. The
factor grows with the number of parities because a loss needs $P+1$ failures to coincide, so
a ratio of hazards enters roughly to the power $P+1$: the distributions' failure rates span
1.5\% to 2.1\%\unskip, a ratio of 1.33, whose second, third and fourth
powers are 1.8, 2.4 and 3.1\unskip. The observed spreads exceed these
powers; the approximation does not separate the contributions of age-dependent hazards,
replacement histories and Monte Carlo variation. The fixed three-component benchmark, with
2.33\unskip\ times the fitted failure rate, gives
6.1\unskip, 15.1\unskip\
and 32.6\unskip\ times the two-Weibull distribution's loss probability
in the three layouts, against 5.4, 12.7 and 29.6\unskip\ for the powers of
the rate ratio. Its common-scale refit (a single factor on the three scales, fitted to this
fleet in Section~\ref{sec:families}) still exceeds the highest fitted distribution by
1.07 to 1.24\unskip. That refit keeps the three Weibull exponent
parameters, and the comparison does not separate their effect from that of the resulting
failure rate. The
two per-year Weibulls, the simulators' April~2026 fit and the landmark comparator, give
1.3 to 5.1\unskip\ times lower loss probabilities than the two-Weibull
distribution, the factor growing from RAID~5 to $17+3$.

\begin{table}[tbp]
\centering\footnotesize\setlength{\tabcolsep}{4pt}
\caption{RAID~6 $8+2$ Loss Probability in Ten Years by Drive Model, New Drives. Rows are the models with a drive-model effect; the last is the fitted transition model under continued service.}\label{tab:raidmodels}
\begin{tabular}{@{}l*{6}{r}@{}}
\toprule
Model & \rot{ST4000DM000} & \rot{ST12000NM0008} & \rot{MG07ACA14TA} & \rot{HUH721212ALN604} & \rot{ST3000DM001} & \rot{ST16000NM001G} \\
\midrule
EB cr2, point & $4.5\times10^{-5}$ & $3.1\times10^{-5}$ & $4.8\times10^{-6}$ & $3.9\times10^{-5}$ & $4.7\times10^{-2}$ & $2.0\times10^{-6}$ \\
EB cr2, predictive & $4.6\times10^{-5}$ & $3.1\times10^{-5}$ & $4.4\times10^{-6}$ & $3.8\times10^{-5}$ & $4.7\times10^{-2}$ & $1.8\times10^{-6}$ \\
EB Weibull & $2.9\times10^{-5}$ & $1.7\times10^{-5}$ & $2.3\times10^{-6}$ & $2.3\times10^{-5}$ & $2.2\times10^{-2}$ & $7.8\times10^{-7}$ \\
EB exponential & $2.3\times10^{-5}$ & $1.3\times10^{-5}$ & $1.7\times10^{-6}$ & $2.0\times10^{-5}$ & $2.0\times10^{-2}$ & $3.7\times10^{-7}$ \\
EB GLFP & $4.3\times10^{-5}$ & $2.9\times10^{-5}$ & $4.4\times10^{-6}$ & $3.8\times10^{-5}$ & $4.5\times10^{-2}$ & $1.5\times10^{-6}$ \\
EB two-state & $3.6\times10^{-5}$ & $2.3\times10^{-5}$ & $3.7\times10^{-6}$ & $2.9\times10^{-5}$ & $2.7\times10^{-2}$ & $1.3\times10^{-6}$ \\
per-model wear-out & $3.1\times10^{-5}$ & $3.1\times10^{-5}$ & $1.1\times10^{-5}$ & $1.9\times10^{-4}$ & $8.0\times10^{-2}$ & $4.7\times10^{-6}$ \\
EB with SMART & $1.0\times10^{-4}$ & $7.4\times10^{-5}$ & $2.0\times10^{-5}$ & $4.8\times10^{-5}$ & $2.1\times10^{-2}$ & $5.7\times10^{-6}$ \\
B0 draw & $4.8\times10^{-5}$ & $3.0\times10^{-5}$ & $4.8\times10^{-6}$ & $3.8\times10^{-5}$ & $6.1\times10^{-2}$ & $1.8\times10^{-6}$ \\
B3 plug-in & $1.8\times10^{-5}$ & $5.0\times10^{-5}$ & $1.0\times10^{-5}$ & $3.0\times10^{-5}$ & $1.6\times10^{-3}$ & $7.0\times10^{-6}$ \\
B3 draw & $8.1\times10^{-6}$ & $2.2\times10^{-5}$ & $4.6\times10^{-6}$ & $1.3\times10^{-5}$ & $5.7\times10^{-4}$ & $3.2\times10^{-6}$ \\
B3 yearly & $7.5\times10^{-6}$ & $1.9\times10^{-5}$ & $4.5\times10^{-6}$ & $1.2\times10^{-5}$ & $5.2\times10^{-4}$ & $2.7\times10^{-6}$ \\
B3 context & $5.3\times10^{-4}$ & $1.0\times10^{-3}$ & $3.9\times10^{-4}$ & $7.4\times10^{-4}$ & $9.8\times10^{-3}$ & $2.9\times10^{-4}$ \\
transition model & $1.0\times10^{-5}$ & $2.2\times10^{-5}$ & $3.4\times10^{-6}$ & $1.7\times10^{-5}$ & $1.3\times10^{-2}$ & $1.8\times10^{-6}$ \\
\bottomrule
\end{tabular}
\end{table}

\begin{table}[tbp]
\centering\footnotesize\setlength{\tabcolsep}{4pt}
\caption{RAID~6 $8+2$ Loss Probability in Ten Years by Drive Model, Drives Aged Four Years. Rows are as in Table~\ref{tab:raidmodels}, for arrays whose drives are four years old at the start of the mission.}\label{tab:raidmodelsaged}
\begin{tabular}{@{}l*{6}{r}@{}}
\toprule
Model & \rot{ST4000DM000} & \rot{ST12000NM0008} & \rot{MG07ACA14TA} & \rot{HUH721212ALN604} & \rot{ST3000DM001} & \rot{ST16000NM001G} \\
\midrule
EB cr2, point & $8.3\times10^{-5}$ & $5.5\times10^{-5}$ & $8.4\times10^{-6}$ & $6.7\times10^{-5}$ & $4.8\times10^{-2}$ & $3.1\times10^{-6}$ \\
EB cr2, predictive & $8.2\times10^{-5}$ & $5.6\times10^{-5}$ & $9.4\times10^{-6}$ & $6.8\times10^{-5}$ & $4.8\times10^{-2}$ & $3.0\times10^{-6}$ \\
EB Weibull & $4.0\times10^{-5}$ & $2.4\times10^{-5}$ & $3.2\times10^{-6}$ & $3.3\times10^{-5}$ & $2.6\times10^{-2}$ & $8.4\times10^{-7}$ \\
EB exponential & $2.3\times10^{-5}$ & $1.3\times10^{-5}$ & $1.6\times10^{-6}$ & $1.9\times10^{-5}$ & $2.0\times10^{-2}$ & $5.0\times10^{-7}$ \\
EB GLFP & $7.6\times10^{-5}$ & $5.2\times10^{-5}$ & $7.8\times10^{-6}$ & $6.5\times10^{-5}$ & $4.7\times10^{-2}$ & $2.9\times10^{-6}$ \\
EB two-state & $6.6\times10^{-5}$ & $4.2\times10^{-5}$ & $6.7\times10^{-6}$ & $5.6\times10^{-5}$ & $3.5\times10^{-2}$ & $2.2\times10^{-6}$ \\
per-model wear-out & $2.9\times10^{-5}$ & $6.6\times10^{-5}$ & $1.9\times10^{-5}$ & $6.5\times10^{-4}$ & $5.8\times10^{-1}$ & $2.2\times10^{-6}$ \\
EB with SMART & $2.2\times10^{-4}$ & $1.7\times10^{-4}$ & $4.5\times10^{-5}$ & $9.1\times10^{-5}$ & $2.7\times10^{-2}$ & $1.7\times10^{-5}$ \\
B0 draw & $8.4\times10^{-5}$ & $5.6\times10^{-5}$ & $9.2\times10^{-6}$ & $6.8\times10^{-5}$ & $6.1\times10^{-2}$ & $3.3\times10^{-6}$ \\
B3 plug-in & $5.0\times10^{-5}$ & $1.3\times10^{-4}$ & $3.0\times10^{-5}$ & $8.1\times10^{-5}$ & $2.9\times10^{-3}$ & $1.9\times10^{-5}$ \\
B3 draw & $2.2\times10^{-5}$ & $5.5\times10^{-5}$ & $1.4\times10^{-5}$ & $3.5\times10^{-5}$ & $1.1\times10^{-3}$ & $9.4\times10^{-6}$ \\
B3 yearly & $1.9\times10^{-5}$ & $4.9\times10^{-5}$ & $1.2\times10^{-5}$ & $3.1\times10^{-5}$ & $10.0\times10^{-4}$ & $7.8\times10^{-6}$ \\
B3 context & $7.0\times10^{-4}$ & $1.3\times10^{-3}$ & $5.2\times10^{-4}$ & $9.6\times10^{-4}$ & $1.1\times10^{-2}$ & $3.9\times10^{-4}$ \\
\bottomrule
\end{tabular}
\end{table}

\begin{figure}[tbp]
\centering
\includegraphics[width=\textwidth]{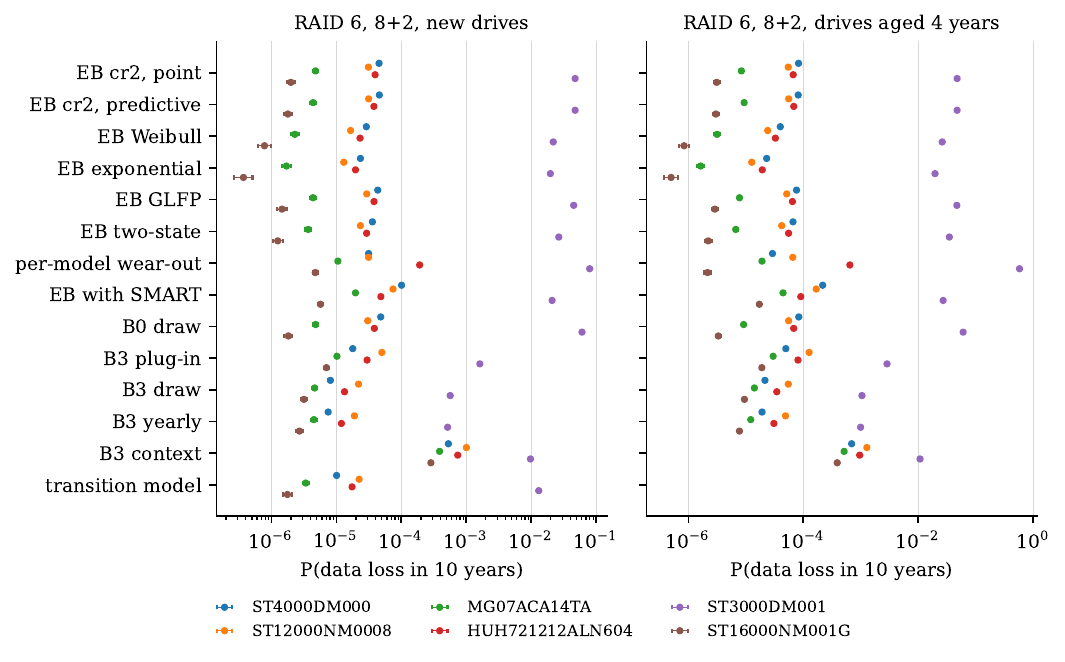}
\caption{RAID~6 $8+2$ loss probability in ten years by drive model (colour) under the
thirteen models with a drive-model multiplier, for arrays of new drives (left) and of drives
aged four years (right).}\label{fig:raidmodels}
\end{figure}

Tables~\ref{tab:raidmodels} and~\ref{tab:raidmodelsaged} and Figure~\ref{fig:raidmodels} give
the RAID~6 results by drive model. The drive model is the largest single factor in the
example: under the empirical-Bayes model the six models span $2.0\times10^{-6}$ (ST16000NM001G)
to $4.7\times10^{-2}$ (ST3000DM001), four orders of magnitude, and the other five models
span a factor of 23. Within a drive model the five hierarchical distributions (two Weibulls,
Weibull, exponential, GLFP, two-state) agree to within a factor of
1.9 to 4.8\unskip\ for new arrays and
2.4 to 6.0\unskip\ for aged ones, the exponential baseline lowest
in every case because its hazard neither falls in early life nor rises with age. The point and
predictive forms of the empirical-Bayes model differ by at most
11\unskip\% for these heavily observed models, whose
multipliers are known to a few per cent (Table~\ref{tab:models}). The per-model wear-out model
(Appendix~\ref{sec:perwear}) departs most from the shared age distribution: for the HGST
HUH721212ALN604, ST16000NM001G, Toshiba MG07ACA14TA and ST3000DM001 it gives
5.1, 2.6, 2.4 and 1.7\unskip\ times the empirical-Bayes loss probability for new
arrays, 0.7\unskip\ for the ST4000DM000 and
1.00\unskip\ for the ST12000NM0008; for arrays aged four years it
puts the ST3000DM001 at
0.58\unskip\ where the shared distribution gives
0.048\unskip. The per-model wear-out model failed as an extrapolator in the held-out comparison
(Section~\ref{sec:valid}), and its simulated loss probabilities differ substantially from the shared
distribution's.

\begin{figure}[h]
\centering
\includegraphics[width=\linewidth]{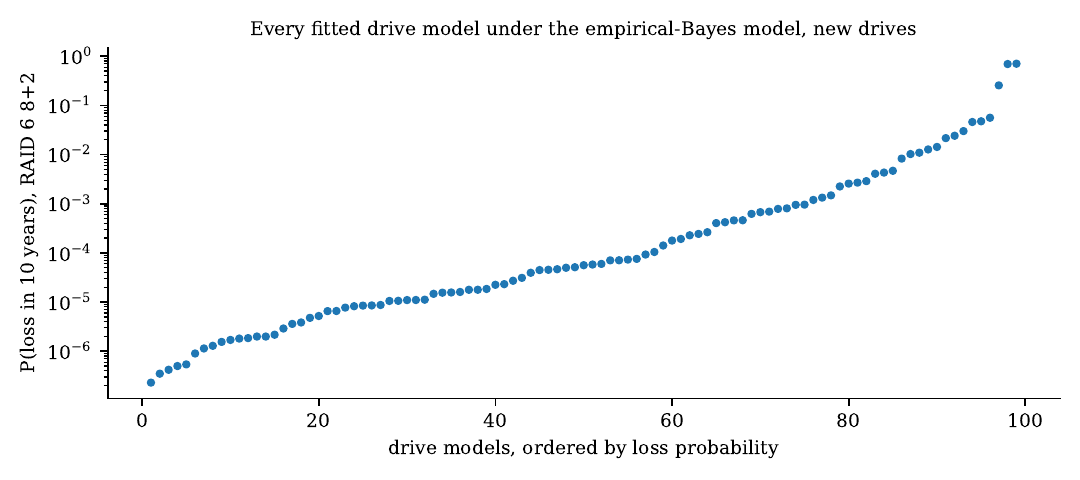}
\caption{Every fitted drive model in the same RAID~6 $8+2$ array of new drives under the empirical-Bayes model at its posterior-mean multiplier, ordered by loss probability in ten years (log scale). The simulation's own 95\% intervals are narrower than the points.}\label{fig:raidallmodels}
\end{figure}

The six drive models above are an illustration. The matrix also runs every fitted drive model
through the same RAID~6 $8+2$ array of new drives under the empirical-Bayes model at its
posterior-mean multiplier (Figure~\ref{fig:raidallmodels}, Table~\ref{tab:raidallmodels}).
Every one of the 99\unskip\ drive models loses data in at least one of
its $10^8$ trials. The loss probability runs from $2.3\times10^{-7}$\unskip\
(HGST HMS5C4040BLE640\unskip) to $7.0\times10^{-1}$\unskip\
(ST1500DL003\unskip), 6.5\unskip\ orders
of magnitude, with a median of $5.6\times10^{-5}$\unskip; the middle half of the
drive models lie between $8.5\times10^{-6}$ to $9.6\times10^{-4}$\unskip. The five highest are Seagate
drive models of 1.5 to 3~TB (Table~\ref{tab:raidallmodels}). Some models near the ends of the
spread have few drives, and the cells use each model's posterior-mean multiplier: the
second highest, the ST2000DL003, has 17 drives, 3.5 drive-years and 8 failures, and the 95\%
interval of its multiplier is 10.9 to 41; the second lowest, the HGST HUH721010ALE600, has 20
drives and no failure, with an interval of 0.005 to 0.32 (the full table behind
Table~\ref{tab:models}). For such models the predictive form would spread the loss
probability over that interval, which the point and predictive rows of
Table~\ref{tab:raidmodels}, for six heavily observed models, do not show. The spread is
therefore that of the fitted multipliers, shrinkage included, not of well-measured failure
rates. Within it, the drive model a simulation assumes decides its loss probability before
any other choice in this appendix does. A reported loss probability that does not name the
drive model, or place it in this spread, has left out the quantity that determines it.

\begin{table}[h]
\centering\footnotesize\setlength{\tabcolsep}{5pt}
\caption{RAID~6 $8+2$ Loss Probability in Ten Years for Every Fitted Drive Model, New Drives, Empirical-Bayes Distribution. The five lowest and five highest of the fitted drive models, with losses over $10^8$ trials and the failures per drive-year in the run.}\label{tab:raidallmodels}
\begin{tabular}{@{}lrrr@{}}
\toprule
Drive model & $P(\text{loss})$ & Losses & Failures per drive-year (\%) \\
\midrule
HGST HMS5C4040BLE640 & $2.3\times10^{-7}$ & 23 & 0.46 \\
HGST HUH721010ALE600 & $3.5\times10^{-7}$ & 35 & 0.61 \\
HGST HMS5C4040ALE640 & $4.2\times10^{-7}$ & 42 & 0.60 \\
ST8000NM000A & $5.0\times10^{-7}$ & 50 & 0.73 \\
WDC WUH721414ALE6L4 & $5.4\times10^{-7}$ & 54 & 0.71 \\
ST3000DM001 & $4.7\times10^{-2}$ & 4{,}749{,}143 & 30.49 \\
ST320005XXXX & $5.6\times10^{-2}$ & 5{,}606{,}769 & 32.16 \\
ST2000DL001 & $2.6\times10^{-1}$ & 25{,}530{,}484 & 49.78 \\
ST2000DL003 & $6.9\times10^{-1}$ & 69{,}382{,}527 & 51.72 \\
ST1500DL003 & $7.0\times10^{-1}$ & 70{,}488{,}996 & 51.36 \\
\bottomrule
\end{tabular}
\end{table}

The Bayesian model's yearly sampler gives lower loss probabilities than the empirical-Bayes model for four of the six
drive models, by a factor of 6\unskip\ for the ST4000DM000 and
of 92\unskip\ for the ST3000DM001; it ties for the Toshiba
MG07ACA14TA and is 1.5\unskip\ times higher for the
ST16000NM001G (the one-draw form is within 20\% of the yearly form throughout). Two fitted
effects could contribute to the lower values; these runs do not measure their contributions
separately. First, the calendar level $Q_{2026}$ is 0.18\unskip, below the
level of most years in which the three older models (ST4000DM000, ST3000DM001, HGST
HUH721212ALN604) were observed, so the fitted model assigns a lower calendar multiplier to a
drive installed in 2027. The drift projected for the first three years lowers it further in
the yearly and context forms, and by one year's drift in the one-draw form. The Toshiba MG07ACA14TA, ST12000NM0008 and ST16000NM001G
were observed mainly in years whose level was already low, and for them the yearly sampler
is 1.0, 0.6 and 1.5 times the empirical-Bayes model. Second, the calendar levels and
model-year shocks attribute part of the ST3000DM001's multiplier to the years in which it ran
(Section~\ref{sec:bayesfit}; its 2014 collapse is described in Section~\ref{sec:families}),
where the empirical-Bayes multiplier carries it as a permanent property of the model. Whether
a 2027 array of that model would see the second effect is the extrapolation question of
Section~\ref{sec:apc}; these runs show the size of the difference, not its source. Aging the array
four years raises the empirical-Bayes loss probability by
1.7 to 2.1\unskip\ across the other five models, leaves the exponential
baseline unchanged, and moves the per-model wear-out model's values by three and seven for
the HGST and the ST3000DM001 and by 0.46\unskip\ for the
ST16000NM001G.

The SMART-state illustration (EB with SMART) gives 0.4 to 4.5\unskip\
times the empirical-Bayes model's loss probability across the six drive models. It is not a
fitted transition model: its warning rate and state multiplier are the manufacturer's from the
observed-state analysis and the Bayesian SMART variant, its clear-state hazard is the
empirical-Bayes hazard divided by the manufacturer's exposure-weighted state multiplier, and
a population that starts clear accumulates warnings over the mission, so its hazard is below
the marginal at first and above it later. The row illustrates the effect of a state-dependent
hazard on an array; a simulator that tracks SMART should draw from the fitted transition model of
Section~\ref{sec:mstate}. Its row in Table~\ref{tab:raidmodels}, for new drives under continued
service, gives 0.2 to 0.9\unskip\ times the empirical-Bayes model's loss
probability for the 6\unskip\ drive models with losses under both.
We do not run it for aged drives: a drive installed at four years reaches the oldest age bin
within the mission, where the fitted failure hazards are close to zero.

\begin{figure}[tbp]
\centering
\includegraphics[width=\textwidth]{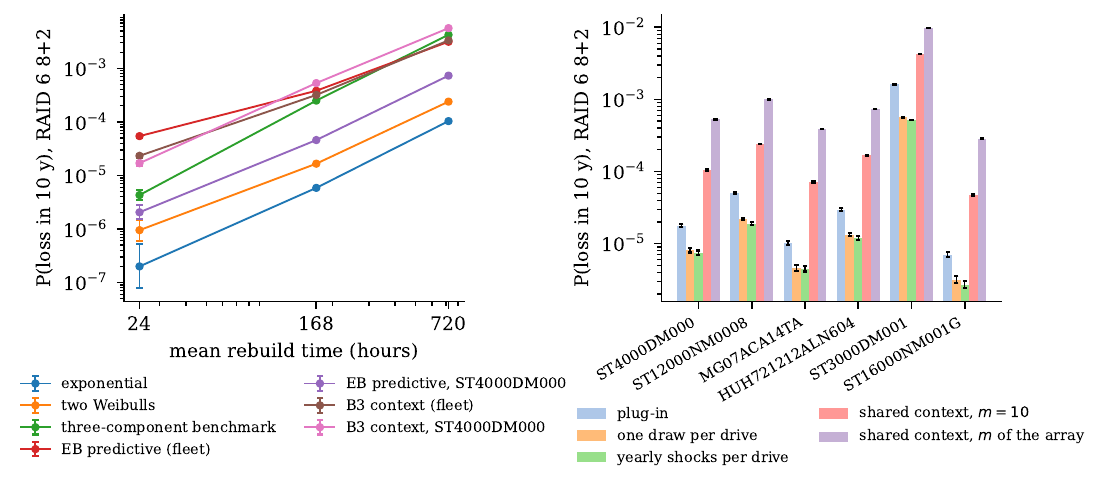}
\caption{Left: RAID~6 loss probability against the mean rebuild time for seven models. Right: the five ways of
drawing from the Bayesian model with shocks (B3), by drive model, RAID~6 of new drives: the
plug-in distribution, one posterior draw held for each drive's life, shocks redrawn every year for each
drive independently, and the shared context in which the array's drives see the same calendar
path and the same shock, with the shock size $m$ set to the interface's default of 10 or to
the array's own expected failures in its first year.}\label{fig:raidb3}
\end{figure}

Figure~\ref{fig:raidb3} compares five B3 sampling implementations and three rebuild times.
For the six named drive models, the plug-in distribution and the two independent per-drive samplers
give loss probabilities within a factor of 3.1\unskip\ of
one another. With $m=10$, the shared context gives
8 to 18\unskip\ times the independent yearly sampler's
RAID~6 loss probability. With $m$ computed from the array's first-year failure probability,
the ratio is 19 to 106\unskip. The two settings of $m$ differ
by 2.3 to 6.0\unskip.

Those comparisons mix several differences. The independent yearly sampler restarts its
calendar path at each replacement and draws a posterior row for each drive; the context keeps
the array's calendar clock and one posterior row; and the two use different shock sizes.
Table~\ref{tab:raidshare} holds all of these fixed except the dependence. Its four arms of
the context sampler share the calendar clock, the lifetime parameters, the posterior
convention, the cap, the replacement policy and the marginal distribution of the shock. They differ
only in what the array's drives share: nothing (each drive its own posterior row, calendar
path and shocks), the posterior row, the row and the calendar path, or all three with one
shock per drive model and year. Within one shock-size specification, the four arms give a single drive the same marginal
distribution, since everything one drive sees is unchanged. Across the two specifications it is not
the same distribution. For a conditional cumulative hazard $H$ and a shock $U\sim\Gam(\kappa,\kappa)$,
the survival is $E[e^{-UH}]=(1+H/\kappa)^{-\kappa}$, and $\kappa$ depends on $m$. The failures
per drive-year of a drive model's four arms agree within
0.3\unskip\% at the array's $m$ and
0.1\unskip\% at $m=10$, and those of all eight within
2.3\unskip\%; agreement in this rate does not make the
two specifications' distributions equal.
For the three named drive models, relative to the arm that shares nothing, sharing the
posterior row changes RAID~6 loss probability by a factor of
1.3\unskip, sharing the calendar path as well by
5.2 to 7.1\unskip, and sharing the model-year shock as well by
51.0 to 93.1\unskip\ at the array-derived shock size and
11.0 to 15.3\unskip\ at $m=10$. Sharing the posterior row changes the
loss probability little. Sharing the calendar path, a fleet-wide level that the array's drives
then see together year by year, multiplies it by 4.2 to 5.5 over the row alone, and sharing
the model-year shock by a further 2.1 to 2.2 at $m=10$ and 9.7 to 13.0 at the array's $m$.
For the fleet population, positions that happen to draw the same drive model share its shock
and the rest have their own, so there the shared path is nearly the whole of the dependence.

\begin{table}[h]
\centering\footnotesize\setlength{\tabcolsep}{3pt}
\caption{The Matched Dependence Experiment. Entries are RAID~6 $8+2$ loss probabilities in ten years under the Bayesian model with shocks, with everything held equal except what the array's drives share. The left group takes the shock size $m$ from the array's own first-year expected failures, and the right group sets $m=10$. Within each group the arms share the one-drive marginal distribution; the two groups' distributions differ. The last column is the failures per drive-year of the nothing-shared arm at the array's $m$. One build and one set of settings produced every entry (Appendix~\ref{sec:repro}).}\label{tab:raidshare}
\begin{tabular}{@{}l*{8}{r}r@{}}
\toprule
& \multicolumn{4}{c}{array $m$} & \multicolumn{4}{c}{$m=10$} & \\
Drive model & none & row & +path & +shock & none & row & +path & +shock & Fail./dr-yr (\%) \\
\midrule
ST4000DM000 & $7.2\times10^{-6}$ & $9.4\times10^{-6}$ & $4.6\times10^{-5}$ & $5.3\times10^{-4}$ & $7.7\times10^{-6}$ & $9.9\times10^{-6}$ & $5.1\times10^{-5}$ & $1.1\times10^{-4}$ & 1.49 \\
ST12000NM0008 & $2.0\times10^{-5}$ & $2.4\times10^{-5}$ & $1.0\times10^{-4}$ & $1.0\times10^{-3}$ & $2.2\times10^{-5}$ & $2.6\times10^{-5}$ & $1.1\times10^{-4}$ & $2.4\times10^{-4}$ & 2.11 \\
MG07ACA14TA & $4.1\times10^{-6}$ & $5.6\times10^{-6}$ & $2.9\times10^{-5}$ & $3.8\times10^{-4}$ & $4.7\times10^{-6}$ & $6.0\times10^{-6}$ & $3.3\times10^{-5}$ & $7.1\times10^{-5}$ & 1.26 \\
fleet population & $6.8\times10^{-5}$ & $8.1\times10^{-5}$ & $2.8\times10^{-4}$ & $3.2\times10^{-4}$ & $7.7\times10^{-5}$ & $9.2\times10^{-5}$ & $3.0\times10^{-4}$ & $3.1\times10^{-4}$ & 3.35 \\
\bottomrule
\end{tabular}
\end{table}

What the experiment cannot settle is the size of the shared shock. The fitted rule
$\kappa=\kappa_0m^{\xi}$ relates dispersion to expected counts across the fleet's model-year
cells; applying it to a ten-drive array is an additional assumption, the rule is not
consistent under subdivision (Section~\ref{sec:limits}), and the array-derived $m$ used here
is a declared scale (the expected first-year failures of the installed drives at unit shock,
held for the mission), not the fitted model-year mean. The two scales run here bracket a
factor of 2.3 to 6.0\unskip\ in the loss probability. The intervals in the
tables are simulation error given the implementation and the fitted inputs; they do not
include uncertainty about the lifetime distribution, the shock scale or the array mechanisms the
simulation omits.

The uniform-mixture comparison changes further assumptions. The context keeps each drive
position's model across replacements and gives it a shock sized for one drive. The
independent sampler redraws the drive model at every replacement. Their ratio of
7.4\unskip\ therefore combines changes in composition,
shock scale and dependence.

Lengthening the mean rebuild time from 24 to 168 hours multiplies RAID~6 loss probability by
7 to 58\unskip\ across the seven models with a loss at 24 hours.
The two-Weibull distribution's 24-hour cell has only
19\unskip\ losses. The ordering also changes: the
uniform-mixture population has the largest loss probability at 24 hours and the fourth
largest at 720 hours.

\subsection{Sampling Cost}

\begin{table}[tbp]
\centering\footnotesize
\caption{Nanoseconds per Draw of the C++ Samplers. Times are means over the pilot-bench sessions on \frag{two} GB10 machines described in the text, one session per sampler, age and machine, every round a fresh process pinned to one Cortex-X925 core; each session's 95\% interval is at most 1\% of its mean. The parentheses say how the distribution is drawn. Table is the tabulated inverse shipped with the library, and exact bisection is the same distribution inverted on its cumulative hazard at every draw.}\label{tab:cpptiming}
\begin{tabular}{@{}lrr@{}}
\toprule
Sampler & New drive & Aged 4 years \\
\midrule
exponential & 11.8 & 12.4 \\
Weibull & 21.2 & 51.3 \\
two Weibulls & 29.5 & 89.8 \\
three Weibulls & 40.4 & 128 \\
log-logistic & 31.0 & 45.6 \\
Gompertz & 23.2 & 23.6 \\
piecewise constant & 19.4 & 16.2 \\
two-state Markov (phase-type) & 14.5 & 38.3 \\
three-component benchmark & 38.8 & 128 \\
three-component, common-scale refit & 38.8 & 125 \\
three-component, exponents fixed, scales refitted & 38.7 & 124 \\
per-year Weibull (April 2026) & 19.7 & 49.2 \\
gamma (table) & 45.5 & 47.4 \\
lognormal (table) & 29.9 & 31.1 \\
Weibull mixture (component given survival) & 23.8 & 77.2 \\
generalized gamma (table) & 51.4 & 53.7 \\
exponentiated Weibull (inverse distribution function) & 52.5 & 95.8 \\
GLFP (susceptibility indicator, minimum) & 19.1 & 55.6 \\
Weibull--Gompertz (minimum of component draws) & 32.9 & 67.1 \\
gamma (exact bisection) & 2,315 & 2,856 \\
lognormal (exact bisection) & 870 & 843 \\
Weibull mixture (exact bisection) & 788 & 836 \\
generalized gamma (exact bisection) & 3,961 & 5,341 \\
exponentiated Weibull (exact bisection) & 3,592 & 4,120 \\
GLFP (exact bisection) & 2,937 & 3,696 \\
Weibull--Gompertz (exact bisection) & 1,373 & 1,857 \\
EB two Weibulls, point multiplier & 31.7 & 97.1 \\
EB two Weibulls, predictive & 49.9 & 122 \\
EB Weibull & 22.1 & 52.3 \\
EB exponential & 11.9 & 12.2 \\
EB GLFP (table) & 54.9 & 55.7 \\
EB two-state Markov (Newton inversion) & 183 & 172 \\
per-model wear-out & 57.5 & 134 \\
EB with SMART state & 91.7 & 177 \\
SMART observed-state & 11.8 & 32.7 \\
B0 plug-in & 31.1 & 95.5 \\
B0 posterior draw & 48.6 & 123 \\
B3 plug-in & 30.4 & 91.5 \\
B3 one draw per drive & 91.7 & 163 \\
B3 yearly shocks per drive & 3,805 & 3,329 \\
transition model of the SMART state (exact, piecewise constant) & 111 & 70.1 \\
\bottomrule
\end{tabular}
\end{table}

The timings in Table~\ref{tab:cpptiming} are means over \frag{two} GB10 machines
(\frag{hathor, sekhmet}) of one pilot-bench session per sampler, age and machine
(\frag{164} sessions), run on the code supplied with the paper. The GB10 has two kinds of
core, Cortex-X925 and the slower Cortex-A725, and an unpinned process is timed on whichever the
scheduler chooses; every round here is a fresh process pinned to one Cortex-X925 core
(\frag{3.9}~GHz, performance governor) with its own seed, and makes $2\times10^6$ draws
($2\times10^5$ for the yearly B3 sampler and the two-state model, $2\times10^4$ for exact
bisection). The requirements, fixed before any reading, are those of
Appendix~\ref{sec:speed}: a 95\% interval whose full width is at most 1\% of the mean, from at
least 200 batches whose means have lag-1 autocorrelation within $\pm0.1$, and a session limit of
1{,}800~s. Sessions took \frag{200 to 210,000} rounds (median \frag{200}), and the
widest interval is \frag{1.00}\% of the mean. \frag{One session reached the limit with its interval at 0.03\% of the mean, and is kept. The session had 28 complete batches, below the required 200.}

One sampler needs a different treatment. In about one process in \frag{253}
(\frag{0.36 to 0.41}\% of the rounds) the transition sampler (\code{ms}) runs
\frag{14 to 21} times slower on the Cortex-X925. The cost depends on where
the kernel places the process's stack, not on the seed or the draws: a slow round's seed rerun
is fast and gives the same draws, no run is slow with address-space randomization off, and
shifting the stack in 16-byte steps shows one slow placement in every 4~KiB. Pilot-bench
takes these rounds for outliers in some sessions and not in others, so its mean is not the
cost of a draw. The \code{ms} cells are therefore timed over every round, with nothing left out,
until the same interval requirement is met (\frag{84,000 to 210,000} rounds); the
mean includes the slow placements in the proportion the kernel produces them. As for the Rust benchmark, the cost of every other cell is the mean of all its
session's rounds, which differs from pilot-bench's own mean by at most
\frag{0.07}\%. The two machines' means for a sampler differ by at most
\frag{0.6}\%. The distributions with a closed-form
conditional draw (exponential, Weibull and its competing-risk sums, log-logistic, Gompertz, the
piecewise hazard, the phase-type distribution) cost 12 to 128\unskip~ns per draw,
from the exponential to the aged three-component benchmark. Four more distributions are drawn in closed form
without inverting a cumulative hazard, at 19 to 96\unskip\ ns: the
exponentiated Weibull by its inverse distribution function; the Weibull mixture by choosing
the component given survival and drawing its conditional Weibull; GLFP by drawing the
susceptibility indicator given survival and taking the minimum of the conditional component
draws; and Weibull--Gompertz as the minimum of its two conditional component draws.

Three distributions (gamma, lognormal, generalized gamma), and the hierarchy's GLFP under a drive-model
multiplier, have no closed-form conditional draw and are drawn from tables. Bisection on the
cumulative hazard to a relative tolerance of $10^{-13}$ costs
0.8 to 5.3\unskip\ $\mu$s per draw. The shipped implementation instead
inverts a table of cumulative hazards on 8,192 geometric age points from $10^{-4}$ years to
the 60-year cap, plus the origin, by binary search and linear interpolation, at
30 to 56\unskip\ ns per draw. The library's checks compare tabulated
and exact quantiles for all four tables over seven ages, five hazard multipliers spanning the
fitted drive-model range and six tail probabilities down to $10^{-6}$, and require relative
error below 0.1\% for lives over four days and absolute error below $10^{-4}$ years for
shorter ones; the array comparison below is the rare-event check for the three distributions. The
timings compare the supplied algorithms, not the minimum achievable cost for each family:
gamma and lognormal lifetimes could also be drawn directly and conditioned by rejection, which
the library does not implement.

The yearly B3 sampler costs 3.3 to 3.8\unskip~$\mu$s per draw and
advances through calendar years. This per-drive timing does not measure the cost per draw
when several drives reuse one context.

In the array the table and the exact inversion agree. Table~\ref{tab:raidexact} runs the three
tabulated distributions through the RAID~6 simulation with the tabulated inverse and with exact
bisection at every draw, from the same seed. Each thread draws all its trials from one stream,
so the two implementations consume the same random numbers until a table error moves some
draw across an event boundary (the mission's end, another event, or the cap); after that, the
thread's later trials diverge. The comparison is therefore a common-random-numbers test up to
the first such divergence, not a draw-by-draw identity. Over the 3\unskip\
distributions the loss counts differ by at most 1\unskip\ trial in
$10^8$, where two runs of the exact implementation from different seeds under the same
revision differ by up to 0.7\unskip\ standard errors of
the loss count. The tabulated cells here are the fleet table's cells (Table~\ref{tab:raidfleet})
themselves, not reruns: \code{:tab} names the default draw and the run script gives it the
fleet cell's seed, so their loss counts are identical.

\begin{table}[h]
\centering\small
\caption{Tabulated Inverse Against Exact Bisection in the RAID~6 $8+2$ Simulation. Arrays are of new drives over ten years. The table gives the loss probability, losses over trials, and failures per drive-year under each implementation of the same distribution.}\label{tab:raidexact}
\begin{tabular}{@{}lrrrrrr@{}}
\toprule
& \multicolumn{2}{c}{Table} & \multicolumn{2}{c}{Exact bisection} & \multicolumn{2}{c}{Failures / drive-yr (\%)} \\
Distribution & $P$(loss) & losses / trials & $P$(loss) & losses / trials & table & exact \\
\midrule
gamma & $8.3\times10^{-6}$ & 825 / 100M & $8.3\times10^{-6}$ & 825 / 100M & 1.7992 & 1.7992 \\
lognormal & $5.2\times10^{-6}$ & 521 / 100M & $5.2\times10^{-6}$ & 521 / 100M & 1.5428 & 1.5428 \\
generalized gamma & $9.6\times10^{-6}$ & 959 / 100M & $9.6\times10^{-6}$ & 958 / 100M & 1.8672 & 1.8672 \\
\bottomrule
\end{tabular}
\end{table}

\subsection{Consequences for Simulation}

In these experiments, changing the fitted fleet lifetime distribution changes loss probability by
about a factor of two for RAID~5 and twelve for the $17+3$ code. The fixed three-component benchmark gives six
to thirty-three times the two-Weibull distribution's loss probability; the simulators' per-year Weibull
gives 1.8 to 5.1\unskip\ times lower probabilities. These are comparisons
among extrapolated models; there are no observed array losses to compare them with.
The six named drive models span four orders of magnitude under the empirical-Bayes model.
Simulations should therefore specify the hardware population as well as the lifetime distribution.
With everything else held equal, letting an array's drives share the fitted model-year shock
multiplies its RAID~6 loss probability by 11.0 to 15.3\unskip\ at the
default shock size and 51.0 to 93.1\unskip\ at the array-derived one;
sharing the posterior row alone changes it little, and sharing the row and the calendar path
by 5.2 to 7.1\unskip. The size of the model-year shock for a small
array remains an assumption the fleet data do not fix.

\section{Reproducibility Notes}\label{sec:repro}

The corrected definitions are in the main text. How the results were produced and checked, and
the corrections made to earlier rounds of the analysis, are recorded here.

\subsection{Provenance of the Tables}

Tables and text fragments are generated from declared result files. The Lexis comparisons
use observed-day exposure; the interval fits and the analyses of removals and drive classes
use calendar spans. The hidden Markov baseline's Lexis scores come from its observed-day
rerun. Numbers typed into the prose are not generated.
Both pipelines stop at the first failing command, and the table builder refuses to fall back
to packed exposure unless told to. A manifest kept with the results (\code{results/manifest.json})
records the code revision the results were collected under, the raw-data cutoff, the
exposure mode, and the digest and size of every table the fits read, of every summary file
on the data host that the paper's tables are generated from, and of every file under the
repository's results directory. It was regenerated after the last result of this revision
was collected.
Large per-drive intermediate files on the data host (cohort files, day masks,
posterior samples) are not digested. The dependency of each generated
table on its inputs is documented in the table builders but not recorded mechanically.

\subsection{Corrections to the Prospective Evaluation}

An earlier round of the prospective evaluation computed the SMART-change features against
the wrong month, fitted the per-year Weibull baseline with a malformed likelihood, and
admitted to a cohort drives absent from the landmark's daily file (now excluded,
Section~\ref{sec:landmark}). Only the corrected run is reported in the article. The
comparison below is between that earlier run and the current one, so it also carries the
later rounds' changes (the event-history rule of Section~\ref{sec:population} and the as-of
age rule of Section~\ref{sec:landmark}). Against the earlier run,
the corrected cohort rule removed 9{,}795 of 1{,}497{,}339 drive-landmark observations from the eight yearly one-year cohorts; the corrected features changed the one-month differences of 380 and the three-month differences of 243 of the 7{,}824{,}383 feature rows shared by the two rounds, and 383 rows now carry a positive gap where before every gap was zero; on the drives common to both rounds of each yearly cohort, the AUC on complete follow-up moved by -0.141 to +0.021 for the per-year Weibull (the likelihood fix), -0.000 to +0.000 for the empirical-Bayes forecast, -0.018 to +0.023, -0.018 to +0.013 and -0.013 to +0.005 for the base, flag and SMART ensembles, and the summed log losses over the eight cohorts by -81, -3, -111, +352 and +613 units
\unskip.

\IfFileExists{paper/repro_A.tex}{\subsection{Exact and Packed Exposure in the Lexis Table}\label{sec:repropacked}

An earlier version of the Lexis table placed each month's observed days consecutively from the
month's first observed day (``packed'' exposure) instead of dividing each run of observed days
among the bins it falls in (Section~\ref{sec:lexis}). Compared with the exact table, the packed
construction puts 417{,}169 drive-days (0.056\% of the exposure) in a different cell, changes
5.3\% of the cells, leaves three failures on days with no exposure, and overstates the
exposure in single-day bins 2 to 17 by up to 1.9\%, all in one direction, so that the early
hazard there is understated by that much; from the first quarter on no bin changes by more than
0.5\%. Refitted on the packed table, the empirical-Bayes model's nine-year held-out totals move
by 1 to 4 log units out of 29{,}000 to 45{,}000 and B3's early-life exponent $\beta_1$ by 0.11
posterior standard deviations, with no other hyperparameter moving by more than 0.05. All
Lexis comparisons in this paper use observed-day placement rather than packed exposure.

\subsection{Gamma Parametrization in the Code}

NumPy's gamma sampler takes a scale, the reciprocal of the rate, while R and NumPyro take a
rate. The sampling paths were checked for this against the convention of
Section~\ref{sec:notation}.
}{}
\IfFileExists{paper/repro_D.tex}{\subsection{Metadata and Failure-Flag Audits of the Landmark Cohorts}\label{sec:repro_landmark}

Three fields of the shared drive table used by the landmark forecasters
(Section~\ref{sec:landmark}) are taken from the drive's full history rather than reconstructed
at the landmark $L$: the drive-model name, which is the last observed; the power-on age at first
sighting, read from the earliest usable SMART~9 reading; and the drive model's capacity, the
modal reported capacity over the model's history, a per-model constant. The audits below
measure how far these choices reach.

\textbf{January cohorts.} Over the eight yearly cohorts at one year, 0 of 1{,}487{,}544 drive-landmark observations have a drive-model name that changed later, and none has its first usable power-on reading in a month after the landmark: every drive's age source predates or falls on its landmark, so the age used was computable on $L$; excluding the observations with either exception leaves 1{,}487{,}544 of 1{,}487{,}544 and moves the mean per-cohort AUC on all drives by less than 0.0001, and the summed log losses of the empirical-Bayes, SMART-informed and SMART-ensemble forecasts by +0, +0 and +0 units, the excluded drives' own contribution. A further 2 drives present on these eight landmarks are outside every cohort because no usable power-on reading exists for them anywhere in the record (0 to 1 per landmark); they are absent from the training data of every forecaster as well
\unskip.

\textbf{Quarterly origins.} Across the additional quarterly evaluation origins,
255
\unskip\ drive-origin rows had an age-source month
later than the origin month. These origins also supply training rows for later forecasts, so
those drives are excluded from the origin's cohort and feature rows, and every forecaster was
refitted and rescored without them (the scores moved by less than their fold-to-fold
variation; Section~\ref{sec:landmark}). Two further audits cover the remaining fields at
every origin, yearly and quarterly.

\textbf{Capacity.} On each landmark day the capacity each present drive reported was compared
with its drive model's modal capacity and with the class that capacity implies. Over the
50\unskip\ origins and 7,824,383\unskip\
feature rows, 438\unskip\ rows report no valid capacity on
the day and cannot be compared. Among the 7,823,945\unskip\ that
can, a capacity differing from the modal one by more than 1\% occurs in
1\unskip\ row (a class change in
1\unskip), and a drive-model name on the day that differs
from the full-history name in 48\unskip\ rows.

\textbf{Age source.} 49\unskip\ drives in the corpus take
their age from the month-minimum fallback rather than a first-day reading,
27\unskip\ of them HDD data drives with a usable age.
All 27\unskip\ of those failed, after a median of
7\unskip\ days in the corpus
(25\unskip\ at most);
0\unskip\ were present on an origin day, so
0\unskip\ feature rows carry a fallback age. A longer-lived
fallback drive would carry its fallback age into every later cohort; the result is a fact about
these drives, not a property of the design.

\textbf{Failure flags.} A failure is dated by the last flagged row of the drive
(Section~\ref{sec:data}). Of the 33,566\unskip\ drives with
a failure flag, 57\unskip\ carry more than one flagged
row. Of these, 57\unskip\ have flags more than a day
apart, 48\unskip\ more than thirty days and
35\unskip\ more than a year apart, and
1,158\unskip\ reported rows after their
last flag. Over the eight January cohorts at 365 days (Table~\ref{tab:flagcohort}),
17\unskip\ drives would have been excluded under a
first-flag rule because their first flag precedes the landmark;
25\unskip\ of the
20,552\unskip\ failures within the horizon have a first flag
on a different day from the last, 8\unskip\ of them
before the landmark; and 23\unskip\ drives first
flagged inside the horizon have their last flag beyond it. A first-flag rule would therefore
remove 17\unskip\ of the
1,487,544\unskip\ cohort drives (the
8\unskip\ early-flagged failures among them) and add
23\unskip\ failures to the
20,552\unskip\ within the horizons. Table~\ref{tab:firstflag}
reports the scores recomputed under that rule, using the existing forecasts. The
training-side consequence (refitting the learners with the exception rows removed) has not
been run.

\begin{table}[h]
\centering\small
\caption{Failure-flag episodes and the January cohorts at 365 days: drives in the cohort with
any failure flag in their record; those a first-flag rule would exclude (first flag before the
landmark); failures within the horizon; those whose first and last flags fall on different
days, those whose first flag precedes the landmark, and drives first flagged within the horizon
whose last flag falls beyond it.}\label{tab:flagcohort}
\begin{tabular}{@{}lrrrrrr@{}}
\toprule
Cohort & Flagged drives & Excluded (first flag) & Failures & First $\neq$ last & First before $L$ & Last beyond horizon \\
\midrule
2018 & 7,830 & 1 & 984 & 0 & 0 & 1 \\
2019 & 8,213 & 1 & 1,892 & 0 & 0 & 0 \\
2020 & 10,163 & 1 & 1,122 & 0 & 0 & 0 \\
2021 & 12,977 & 2 & 1,510 & 0 & 0 & 7 \\
2022 & 14,839 & 4 & 2,858 & 2 & 2 & 1 \\
2023 & 12,920 & 3 & 3,989 & 3 & 3 & 6 \\
2024 & 10,003 & 1 & 4,199 & 1 & 0 & 7 \\
2025 & 6,241 & 4 & 3,998 & 19 & 3 & 1 \\
\bottomrule
\end{tabular}
\end{table}
}{}
\IfFileExists{paper/repro_E.tex}{\subsection{The Fleet Example Script}\label{sec:reproE}

The accompanying \code{bb\_bayes\_fleet\_example.R} simulates a ten-year, two-model fleet
with replacements at the start of the year after failure, against a shared Bayesian context
(Section~\ref{sec:query}). The repaired scheduler queues failures from every installation
cohort. All seven deterministic checks in \code{bb\_fleet\_example\_test.R} pass, including
delayed failures and overlapping replacement cohorts. These checks establish the scheduling
behavior; they do not validate the fitted hazards or a real fleet's replacement policy.
}{}

\section{Technical Qualifications of the Analysis}\label{sec:qualifications}

Section~\ref{sec:practical} discusses the practical meaning and principal limitations of the
results. This appendix retains the detailed qualifications of the data conventions, model
variants, evaluation, and timing measurements, including the numerical sensitivity checks
and comparisons not yet made.

\subsection{Data Conventions and Provenance}
\begin{itemize}[leftmargin=1.4em]
\item \emph{Removal reasons are inferred.} ``Bulk'' is a threshold on same-day, same-model
  removals. With a threshold of 100 there are 28{,}451 single removals, of which 23\% are
  warned, and the single-removal row of Table~\ref{tab:retiresens} would be larger.
\item \emph{Exposure conventions differ} between the interval likelihood (calendar span) and the
  Lexis table (observed rows). After 2014 the two agree within 2.5\%.
\item \emph{The rows after a failure flag are read as not exposure.} One event history
  (Section~\ref{sec:population}) is applied in every analysis: the editor's seventh-round
  audit found the earlier scripts inconsistent on this point, and all of them were corrected
  and rerun (Appendix~\ref{sec:repro}). We do not know why the drives reappear, and
  Table~\ref{tab:postfail} shows that the reading does not move the fleet-level estimates.
\item \emph{Provenance.} We generate tables and text fragments from one declared set of
  result files (Appendix~\ref{sec:repro}). We checked the numbers typed into the prose against
  them, but those numbers are not generated. The table builders document the dependency of
  each generated table on its inputs, but we do not record it mechanically.
\item \emph{Classes are assigned from drive-model names and capacity.} The record does not
  give a drive's role.
\end{itemize}

\subsection{Model Assumptions and Unresolved Fits}
\begin{itemize}[leftmargin=1.4em]
\item \emph{Extrapolation in age.} Section~\ref{sec:apc}. We check forecast levels to three
  years ahead and no further.
\item \emph{The first days and old age.} By age band (Table~\ref{tab:age}) observed over expected
  failures is between 0.93 and 1.14 from day 4 to year 8. In the first three days the model
  expects too few failures (76 against 107 observed, on 367 drive-years), and from 8 to 10
  years too many (1{,}066 against 770).
\item \emph{The rule for the shock precision $\kappa_{gy}$ is not consistent under subdivision} of a group. The vault-style
  rule of Appendix~\ref{sec:extvault} is consistent only under a joint construction with a
  shared common factor. Neither the fitted marginal gamma nor the samplers implement that
  construction.
\item \emph{Shocks are annual and independent across years.} In this model a drive model's
  shock in one year says nothing about the next, so an episode that lasts two years is two
  surprises. Calendar years are an arbitrary grouping, and we model nothing within a year.
\item \emph{New drive models.} The prior mean over-predicts. For drive models in their first
  year, 126 failures occurred where the empirical-Bayes model forecast 769
  (Section~\ref{sec:newmodels}) and the Bayesian variants 676 to 874 (Section~\ref{sec:calib}).
  Most of the excess is two drive models in one year, and the cause is not established.
\item \emph{Per-model wear-out} has higher predictive scores, reproduced over three runs,
  but some training fits and B5 frozen-cutoff target fits fail the mixing screen (Table~\ref{tab:pmsdiag}). B4 and B5
  (Table~\ref{tab:pms}) over-predict: their median fleet forecasts are 1.50 to 1.73 times the
  observed failures over the nine held-out years, against 0.96 for B3 with drift
  (Section~\ref{sec:pmsmedians}). The excess is
  located (large drive models, the recent test years; Section~\ref{sec:pms}) but not
  explained. The empirical-Bayes version (Appendix~\ref{sec:perwear}) did not
  converge. In B4 the predictive fleet-count mean is infinite in all nine held-out folds;
  the finite numerical point forecasts are not estimates of that mean. B5's count mean has
  not been established to be finite (Section~\ref{sec:newshapemean}). The reported B5 scores
  use the corrected frozen-cutoff targets; this resolves the earlier target mismatch, not
  the remaining posterior-sampling limitations.
\item \emph{The models of Appendix~\ref{sec:laws3} share a baseline hazard function across drive models.} We did
  not fit the hierarchical limited-failure-population model of Mittman et al., with
  parameters per drive model, so nothing here compares their model with ours.
\item \emph{The per-drive-model constants of Appendix~\ref{sec:hmmper} are unshrunk} and
  limited to drive models with twenty or more failures. A hierarchical version is in
  Appendix~\ref{sec:exthmm}.
\item \emph{The covariate regression is not the Bayesian model.} It has no year effect and
  no shocks, treats cells as independent Poisson counts, and bins every covariate. Its
  coefficients are associations. The first month of a drive and any month after a gap carry
  the level ``unknown''. Rates per power-on year inherit the wrap of attribute 9 on some
  Seagate models (Section~\ref{sec:age}). The regression does not model the transition into the warned
  state. The transition model of Section~\ref{sec:mstate} fits it by age, calendar year and
  drive model, but without the regression's other covariates.
\item \emph{The extensions of Appendix~\ref{sec:ext}} are scored by a sum of marginal scores
  whose Monte Carlo error is not small for a few drive-model-years. We check them one year
  ahead only. They take the SMART state as known on 1~January and do not model its evolution
  within and beyond the year, so a forecast two years ahead cannot use the state as it
  stands. The vault multiplier is not in any likelihood. The benchmark's SMART and vault
  samplers use single constants where the fits have values by manufacturer.
\item \emph{Empirical Bayes.} Hyperparameters are point estimates. Their uncertainty is not
  propagated into the posteriors of Table~\ref{tab:models}.
\item \emph{The sensitivity fit is a fixed-effects version} of the empirical-Bayes model, without
  covariates, shocks, or a year effect.
\item \emph{Binned ages.} The likelihood treats the hazard as constant within each of 81 age
  bins at its mean over the bin.
\end{itemize}

\subsection{Evaluation Design and Numerical Precision}
\begin{itemize}[leftmargin=1.4em]
\item \emph{The prospective evaluation} (Section~\ref{sec:landmark}) forecasts the drives in
  service on a date. It says nothing about deployments after that date, which would need a
  schedule the data do not contain. Its hazard forecasters have no retirement model, so we
  score them on the recorded endpoint and charge that omission to them. The boosted
  classifiers, whose label is failure before removal, and the fitted transition model of
  Section~\ref{sec:mstate}, which has removal hazards, account for removal. The
  empirical-Bayes forecaster with SMART is not a transition model. It uses a warning ratio
  estimated from the 1~January cohorts before the landmark and one fleet-wide rescaling that
  holds the cohort's posterior-mean hazard fixed (its expected failures fall by about 2\%).
  Two fields of the drive table are taken from the full history: the last drive-model name
  and the capacity. The
  capacity is a per-model constant, which we checked against each origin's reported capacity
  (Section~\ref{sec:landmark}). A drive whose power-on age could not have been known on an
  origin, because its first usable power-on reading is in a later month, is not in that
  origin's cohort or feature rows, for scoring or for training
  (255
\unskip\ drive-origin rows, all at quarterly
  origins). These audits are at the resolution of a month: we have not audited the source
  dates within the origin's month.
\item \emph{The boosted forecasters} (Section~\ref{sec:boost}) read one monthly SMART record
  per drive and its changes, not the daily trajectory. Their label is the operator's outcome,
  failed before removal, which is not the hazard. Their training origins end a horizon before
  the landmark, so drives and drive models deployed in that interval are absent from their
  rows. The inner split that chooses their number of rounds holds the fitting rows' windows
  closed before the first validation origin. Their rows are, however, repeated observations
  of the same drives at overlapping origins, so the validation loss is not an
  independent-sample estimate. We ran the diagnostics of Appendix~\ref{sec:boostdiag} on three
  cohorts after we had seen the main result.
\item \emph{Model development used the same nine test years} that we report. We held no
  period back untouched.
\item \emph{We chose the variant after seeing the nine test years.} We compared ten
  variants on the same folds and preferred B3 with drift for its forecast level. The gain
  from shocks does not depend on that choice. The preference for drift over no drift does. It
  rests on the bias of nine yearly forecasts and on the longer horizons, which reuse the same
  years. Nothing here shows B3 with drift to forecast the level better than B0.
\item \emph{Quarterly folds overlap in training data}, so their scores are correlated. The
  standard errors treat them as independent.
\item \emph{We check calibration one year ahead only}, on nine years, and for five of the
  Bayesian variants. Table~\ref{tab:calib} forms its group ``five or more expected'' from each
  variant's own expectation. Table~\ref{tab:calibcommon} repeats the comparison on one
  common population defined from training data with interval widths and both tails. Its standard errors come from
  a bootstrap over the nine test years, which is the only replication the design offers.
\item \emph{Monte Carlo and quadrature error in the score.} For B1 and B3, the bridge estimates
  carry a standard error of 0.15 to 0.3 on the nine-year sum. We computed it from four
  chains, so it is itself rough. The importance-sampling estimate for B0 carries 0.06. The
  paired standard errors in Table~\ref{tab:bayespairs} do not include these errors, which
  matter only for the smallest differences in that table. The score has three
  discretizations: 24 Gauss--Hermite nodes for the year shock, a 241-point grid for each drive
  model's log multiplier, and 200 prior quantiles for new drive models. We checked them by
  doubling all three (48, 481, 400) and rescoring every fold of four shock variants (B1 with
  drift; B3 with no year effect, with a random walk, and with drift) from the saved
  posteriors (Table~\ref{tab:finescores}). The nine-year sums move by at most
  0.04\unskip\ log units and no fold by more than
  0.03\unskip. That is about a seventh of the bridge standard
  error of the nine-year sum (in the worst fold, 1.6 times that fold's own standard error), and two orders of
  magnitude below the 15- and 26-unit differences interpreted in Section~\ref{sec:bayesvalid}.
  Posterior Monte Carlo error and the nine test years limit those comparisons, not
  quadrature. These checks do not cover the per-model wear-out variants: their revised
  integration is checked on selected draws at two origins, and their bridges can have weak
  posterior overlap. Aggregate Monte Carlo errors omit cross-fold covariance and do not
  measure quadrature error or shared sampling bias (Section~\ref{sec:pms}).
\item \emph{Divergent transitions.} The fits of B3 with drift had 12 among 40{,}000 retained
  draws, and those of B1 with drift had 25. Neither had a divergence in the fit to all data
  (Table~\ref{tab:bayesdiag}).
\end{itemize}

\subsection{Scope of the Timing Measurements}
\begin{itemize}[leftmargin=1.4em]
\item \emph{The benchmark} measures one program, on two processors (\frag{three} GB10 machines and
  one Intel i5), with one compiler. It measures the cost of the arithmetic, not of a
  simulator's event loop. Other processors and compilers may order samplers that are within
  about 20\% of each other differently. We timed the boosted forecasters through a Python
  binding on a shared machine, so no cost ratio between them and the compiled samplers is
  established.
\item \emph{The context interface} shares one posterior draw, one calendar path and one shock
  per drive model and year within a system. The Rust benchmark program does not implement it.
  We therefore did not time the shared scheme's cost per drive and year separately. That cost
  omits the normal and gamma variates the \frag{154}~ns figure includes.
\end{itemize}

\subsection{Comparisons Still to Be Made}

The following comparisons remain:
\begin{itemize}[leftmargin=1.4em]
\item a regularized discrete-time model of failure and retirement, fitted from observations
  before each origin, which would forecast the recorded endpoint with a retirement model and
  let partly followed drives contribute;
\item the additive model A2 with a predictive distribution and shared calendar variation, so
  that its comparison with the hierarchy is like for like;
\item a hierarchical limited-failure-population model with a defect fraction and wear-out
  parameters per drive model, the model of Mittman et al. We fitted the GLFP model with a
  multiplicative drive-model effect, and per-model wear-out only in the two-Weibull form (B4
  and B5);
\item a plateau or damped calendar trend as an alternative future calendar trend, with the
  frailty's distribution (gamma against lognormal) as a sensitivity.
\end{itemize}

\end{document}